\documentclass[%
reprint,
nofootinbib,
 amsmath,amssymb,
 aps,
prd,
]{revtex4-1}
\usepackage{multirow}
\usepackage{graphicx}
\usepackage{dcolumn}
\usepackage{bm}
\usepackage{amsmath}
\usepackage{amssymb}
\usepackage{mathrsfs}
\usepackage{amsfonts}
\usepackage{color, soul}
\usepackage{makecell}
\numberwithin{equation}{subsection}
\usepackage{etoolbox}

\def\oversim#1#2{\lower0.5ex\vbox{\baselineskip0pt\lineskip0pt
                 \lineskiplimit0pt\everycr{}\tabskip0pt
                 \halign{$\mathsurround0pt #1\hfil##\hfil$\crcr #2\crcr\sim\crcr}}}

\begin{document}

\preprint{arXiv:1807.03762}

\title{A Bayesian analysis of sneutrino DM in the NMSSM with Type-I seesaw mechanism}

\author{Junjie Cao$^{1,2}$, Jie Li$^1$, Yusi Pan$^1$, Liangliang Shang$^1$,Yuanfang Yue$^1$,Di Zhang$^1$}

\affiliation{ $^1$ College of Physics and Materials Science,Henan Normal University, Xinxiang 453007, China \\
   $^2$  Center for High Energy Physics, Peking University, Beijing 100871, China
  }


\begin{abstract}
  In the Next-to-Minimal Supersymmetric Standard Model (NMSSM) with extra heavy neutrino superfields, neutrino may acquire
  its mass via a seesaw mechanism and sneutrino may act as a viable dark matter (DM) candidate. Given the strong tension between
  the naturalness for $Z$ boson mass and the DM direct detection experiments for customary neutralino DM candidate, we augment the NMSSM with Type-I seesaw
  mechanism, which is the simplest extension of the theory to predict neutrino mass, and study the scenarios of sneutrino DM. We construct likelihood function
  with LHC Higgs data, B-physics measurements, DM relic density and its direct and indirect search limits, and perform a comprehensive scan over the parameter
  space of the theory by Nested Sampling method. We adopt both Bayesian and frequentist statistical quantities to illustrate the favored parameter space of the
  scenarios, the DM annihilation mechanism as well as the features of DM-nucleon scattering. We find that the scenarios are viable over broad parameter regions, especially
  the Higgsino mass $\mu$ can be below about $250 {\rm GeV}$ for a significant part of the region, which predicts $Z$ boson mass in a natural way. We also find that
  the DM usually co-annihilated with the Higgsinos to get the measured relic density, and consequently the DM-nucleon scattering rate is naturally suppressed to
  coincide with the recent XENON-1T results even for light Higgsinos. Other issues, such as the LHC search for the Higgsinos, are also addressed.
\end{abstract}


\maketitle

\section{Introduction}
\label{intro}
A large number of cosmological and astrophysical observations have firmly established the existence of non-baryonic
DM \cite{Ade:2015xua,Bertone:2016nfn,Iocco:2015xga,Springel:2006vs}. Among the possible candidates, the weakly interactive
massive particles (WIMPs) are most attractive since they naturally lead to right DM abundance\cite{Bertone:2004pz,Arcadi:2017kky}.
So far this type of DM candidates are still compatible with the more and more stringent constraints from DM direct and indirect search experiments,
which was recently emphasized in \cite{Leane:2018kjk}.As the most popular ultraviolet-complete Beyond Standard Model (BSM), the Minimal Supersymmetric
Standard Model (MSSM) predicts two WIMP-like DM candidates in the form of sneutrino\cite{Hagelin:1984wv} or neutralino\cite{Jungman:1995df} when R-parity
is imposed. For the left-handed sneutrino as the lightest supersymmetric particle (LSP), its interaction with the Z boson predicts a very small relic
abundance relative to its measured value as well as an unacceptably large DM-nucleon scattering rate, which was firstly noticed by T. Falk {\it et. al.}~\cite{Falk:1994es}
after considering the Heidelberg--Moscow DM direct detection (DD) experiment~\cite{Beck:1994ex} and later emphasized in~\cite{Arina:2007tm}. Consequently, the neutralino
DM as the solely viable WIMP candidate has been intensively studied over the past decades. However, with the rapid progress in DM DD
experiments (such as PandaX-II\cite{Tan:2016zwf,Fu:2016ega}, LUX\cite{Akerib:2016vxi} and XENON-1T\cite{Aprile:2017iyp,Aprile:2018dbl}) in recent years,
it was found that the candidate became disfavored by the experiments \cite{Baer:2016ucr,Huang:2017kdh,Badziak:2017the} assuming that it is fully
responsible for the measured DM relic density and that the Higgsino mass $\mu$ is of ${\cal{O}}(10^2 {\rm GeV})$, which is favored by $Z$ boson mass.
This situation motivates us to consider DM physics in extended MSSM.

Besides the DM puzzle, the non-vanishing neutrino mass is another firm evidence on the existence of new physics\cite{Mohapatra:2005wg}. Seesaw mechanism is
the most popular way to generate the mass, and depending on the introduction of heavy neutrino fields, several variants of this mechanism, such as
Type-I\cite{Minkowski:1977sc}, -II\cite{Schechter:1980gr}, -III\cite{Foot:1988aq} and Inverse\cite{Mohapatra:1986aw,Mohapatra:1986bd} seesaw, have been
proposed. Among these variants, the Type-I mechanism is the most economical one where only right-handed neutrino field is introduced. In the simplest
supersymmetric realization of this mechanism, namely the MSSM with Type-I mechanism, a pure right-handed sneutrino LSP\cite{Gopalakrishna:2006kr,McDonald:2006if,Lee:2007mt}
or a mixed left- and right-handed sneutrino LSP\cite{ArkaniHamed:2000bq,Arina:2007tm,Hooper:2004dc} may act as a viable DM candidate. For the former case,
the coupling of the candidate to ordinary matter is extremely suppressed either by neutrino Yukawa couplings or by the mass scale of the right-handed neutrino.
As a result, its self- and co-annihilation cross sections are so tiny that it has to be non-thermal to avoid an overclosed universe\cite{Gopalakrishna:2006kr,McDonald:2006if}.
For the latter case, a significant chiral mixture of the sneutrinos requires an unconventional supersymmetry breaking mechanism\cite{ArkaniHamed:2000bq}.
Furthermore, since the couplings of the DM candidate with the SM particles are determined by the mixing, it is difficult to predict simultaneously the right DM
abundance and a suppressed DM-nucleon scattering rate required by the DM DD experiments\cite{Thomas:2007bu,Belanger:2010cd,Dumont:2012ee,Kakizaki:2016dza}\footnote{Numerically
speaking, the mixing angle should satisfy $\sin \theta_{\tilde{\nu}} \sim 0.02$ for $m_{\tilde{\nu}} = 100 {\rm GeV}$ to predict the right DM relic density by the $Z$ boson mediated
annihilation, which corresponds to the scattering rate at the order of $10^{-45} {\rm cm^{-2}}$\cite{Dumont:2012ee}. Such a rate has been excluded by the latest
XENON-1T experiment, which, on the other side, limits $\sin \theta_{\tilde{\nu}} < 0.01$ by the recent calculation in~\cite{Kakizaki:2016dza}. Moreover, we find that the correlation between the relic density and the scattering rate is underestimated in FIG.1 of \cite{Arina:2013zca}.}. These facts reveal that the DM physics in Type-I MSSM is also unsatisfactory. This situation is also applied
to the inverse seesaw extension of the MSSM \cite{BhupalDev:2012ru}.

The situation may be changed greatly if one embeds the seesaw mechanism into the NMSSM, which, as one of most economical extensions of the MSSM, is characterized by predicting one gauge singlet Higgs superfield $\hat{S}$\cite{Ellwanger:2009dp}. It has long been known that the field $\hat{S}$ plays an extraordinary role in the model: solving the $\mu$ problem of the MSSM\cite{Ellwanger:2009dp}, enhancing the theoretical prediction about the mass of the SM-like Higgs boson\cite{Hall:2011aa,Ellwanger:2011aa,Cao:2012fz} as well as enriching the phenomenology of the NMSSM (see for example \cite{Cao:2013gba,Ellwanger:2014hia,Cao:2014efa,Cao:2015loa,Cao:2016nix,Ellwanger:2018zxt}). In this context we stress that, in the seesaw extension of the NMSSM, it is also responsible for heavy neutrino mass and the annihilation of sneutrino DM\cite{Chang:2017qgi,Cao:2017cjf,Chang:2018agk}, and consequently makes the sneutrino DM compatible with various measurements. The underlying reason for the capability is that the newly introduced heavy neutrino fields in the extension are singlets under the gauge group of the SM model, so they can couple directly with $\hat{S}$. We also stress that the seesaw extension is essential not only to generate neutrino mass, but also to enrich greatly the phenomenology of the NMSSM given the very strong constraint of the recent DM DD experiments on neutralino DM in the NMSSM\cite{Cao:2016cnv}.

In our previous work\cite{Cao:2017cjf}, we augmented the NMSSM with inverse seesaw mechanism by introducing two types of gauge singlet chiral superfields $
\hat{\nu}_R$ and $\hat{X}$, which have lepton number $-1$ and 1 respectively, and sketched the features of sneutrino DM. We found that,
due to the assignment of the superfield's charge under the SM gauge group,  the scalar component fields of $\hat{\nu}_R$, $\hat{X}$ and $\hat{S}$
compose a secluded DM sector, which can account for the measured DM relic abundance, and also be testable by future
DM indirect detect experiments and LHC experiments.
Since this sector communicates with the SM sector mainly through the small singlet-doublet Higgs mixing, the DM-nucleon scattering
rate is naturally suppressed, which is consistent with current DM direct search results.  We note that these features should be applied to the Type-I seesaw
extension of the NMSSM due to the similarities of the two theoretical frameworks \footnote{Generally speaking, the property of the sneutrino
	DM in the Type-I extension differs from that in the inverse seesaw extension only in two aspects. One is that in the Type-I extension, the sneutrino DM is a roughly pure right-handed sneutrino state, while in the inverse seesaw extension, it is a mixture of $\tilde{\nu}_L$, $\tilde{\nu}_R$ and $\tilde{X}$ with the last two components being dominated. Consequently, the DM physics in the inverse seesaw extension is more complex\cite{Chang:2017qgi,Cao:2017cjf,Chang:2018agk}. The other is that in both the extensions, CP-even and CP-odd sneutrino states split in mass due to the presence of lepton number violating interactions. In the Type-I extension the splitting may be quite large, while in the other model, it tends to be smaller than about $1~{\rm GeV}$. }. We also note that, in comparison with the inverse seesaw extension,
the sneutrino sector in the Type-I seesaw extension involves less parameters, and thus is easier to be fully explored in practice. So as a preliminary work in our series of studies on sneutrino DM in different seesaw extensions of the NMSSM, we focus on the Type-I extension, and perform a rather sophisticated scan
over the vast parameter space of the model by Nested Sampling method\cite{Feroz:2008xx}. The relevant likelihood function is constructed from LHC Higgs data,  B-physics measurements, DM relic density and its direct and indirect search limits, and statistic quantities are used to analyse the scan results. As far as we know,
such an analysis of the Type-I seesaw extended NMSSM has not been done before, and new insights about the model are obtained. For example,
we find that for most samples in the $1 \sigma$ regions of the posterior probability distribution function (PDF) on the plane of the sneutrino DM mass $m_{\tilde{\nu}_1}$ versus  the Higgsino mass $\mu$, the two masses are nearly degenerate. This is because the singlet Higgs field can mediate the transition between the DM pair and the Higgsino pair, which implies that the DM and the Higgsinos can be in thermal equilibrium in early Universe before their freeze-out. If their mass splitting is less than about $10\%$, the number density of the Higgsinos can track that of the DM during freeze-out, and consequently the Higgsinos played an important role in determining DM relic density\cite{Coannihilation} (in literature such a phenomenon was called coannihilation \cite{Griest:1990kh}). As a result, even for very weak  couplings of the DM with SM particles, the DM may still reach the correct relic density by coannihilating with the Higgsino-dominated particles.  Such a possibility is not discussed in literatures. We also find that in many cases, the model does not need fine tuning to be consistent with the DM DD limits on the DM-nucleon scattering rate. This is a great advantage of the theory in light of the tightness of the limits, but it is not emphasized in literatures.

\begin{table*}[t]
	\begin{center}
		\begin{tabular}{|c|c|c|c|c|c|}
			\hline \hline
			SF & Spin 0 & Spin \(\frac{1}{2}\) & Generations & \((U(1)\otimes\, \text{SU}(2)\otimes\, \text{SU}(3))\) \\
			\hline
			\(\hat{q}\) & \(\tilde{q}\) & \(q\) & 3 & \((\frac{1}{6},{\bf 2},{\bf 3}) \) \\
			\(\hat{l}\) & \(\tilde{l}\) & \(l\) & 3 & \((-\frac{1}{2},{\bf 2},{\bf 1}) \) \\
			\(\hat{H}_d\) & \(H_d\) & \(\tilde{H}_d\) & 1 & \((-\frac{1}{2},{\bf 2},{\bf 1}) \) \\
			\(\hat{H}_u\) & \(H_u\) & \(\tilde{H}_u\) & 1 & \((\frac{1}{2},{\bf 2},{\bf 1}) \) \\
			\(\hat{d}\) & \(\tilde{d}_R^*\) & \(d_R^*\) & 3 & \((\frac{1}{3},{\bf 1},{\bf \overline{3}}) \) \\
			\(\hat{u}\) & \(\tilde{u}_R^*\) & \(u_R^*\) & 3 & \((-\frac{2}{3},{\bf 1},{\bf \overline{3}}) \) \\
			\(\hat{e}\) & \(\tilde{e}_R^*\) & \(e_R^*\) & 3 & \((1,{\bf 1},{\bf 1}) \) \\
			\(\hat{s}\) & \(S\) & \(\tilde{S}\) & 1 & \((0,{\bf 1},{\bf 1}) \) \\
			\(\hat{\nu}\) & \(\tilde{\nu}_R^*\) & \(\nu_R^*\) & 3 & \((0,{\bf 1},{\bf 1}) \) \\
			\hline \hline
		\end{tabular}
	\end{center}


\caption{Field content of the NMSSM with Type-I seesaw mechanism.}
\label{table1}
\end{table*}

The Type-I extension of the NMSSM was firstly proposed in \cite{Kitano:1999qb}, and its DM physics was sketched in \cite{Cerdeno:2008ep,Cerdeno:2009dv}. Since then a lot of works appeared to study the phenomenology of the model\cite{Cerdeno:2011qv,Wang:2013jya,Chatterjee:2014bva,Cerdeno:2014cda,Huitu:2014lfa,Tang:2014hna,Cerdeno:2015ega,Gherghetta:2015ysa,Cerdeno:2017sks,Chatterjee:2017nyx}. For example, the spectral features of the $\gamma$-ray from DM annihilations were investigated in \cite{Cerdeno:2011qv,Chatterjee:2014bva,Cerdeno:2014cda,Cerdeno:2015ega,Gherghetta:2015ysa},  and
the Higgs physics was discussed in \cite{Wang:2013jya,Huitu:2014lfa}. We note that some of these studies focused on the parameter regions which predict a relatively large DM-nucleon scattering rate\cite{Cerdeno:2008ep,Cerdeno:2009dv}. These regions have now been excluded by the DM DD experiments, and thus the corresponding results are out of date. We also note that some other works were based on multi-component DM assumption, so they used the upper limit of the DM relic abundance as the criterion for parameter selection\cite{Cerdeno:2011qv,Chatterjee:2014bva,Cerdeno:2014cda,Cerdeno:2015ega}. Obviously, the conclusions obtained in this way are less definite than those with single DM candidate assumption. Given the incompleteness of the research in this field and also the great improvements of experimental limits on DM property in recent years, we are encouraged to carry out a comprehensive study on the key features of the model in this work.

This paper is organized as follows. In section \ref{Section-Model}, we introduce briefly the basics of the NMSSM with Type-I seesaw extension and the property of the sneutrino DM.  In Section \ref{CP-even},  we perform a comprehensive scan over the vast parameter space of the model by considering various experimental measurements, and adopt statistic quantities to show the favored parameter space, DM annihilation mechanisms as well as the features of DM-nucleon scattering for the cases that the sneutrino DM is CP-even and CP-odd respectively. Collider constraints on sneutrino DM scenarios are discussed in Section \ref{LHC-search}, and conclusions are presented in Section \ref{Section-Conclusion}.

\section{\label{Section-Model}NMSSM with Type-I Mechanism}
In this section we first recapitulate the basics of the NMSSM with Type-I mechanism,  including its Lagrangian and Higgs sector, then we concentrate on sneutrino sector by analyzing sneutrino mass matrix, the annihilation channels of sneutrino DM and its scattering with nucleon.

\subsection{The Lagrangian of the Model}
As the simplest seesaw extension of the NMSSM, the NMSSM with Type-I seesaw mechanism introduces three generations of right-handed neutrino fields to generate neutrino mass. Consequently, the extension differs from the NMSSM only in neutrino/sneutrino sector, and the sneutrino DM as the lightest supersymmetric state in this sector couples mainly to Higgs bosons. With the field content presented in Table \ref{table1}, the relevant superpotential and soft breaking terms are\cite{Cerdeno:2008ep,Cerdeno:2009dv}
\begin{eqnarray}
W &=& W_F+\lambda \hat{s} \hat{H}_u \cdot  \hat{H}_d
+\frac{1}{3} \kappa \hat{s}^3+\bar{\lambda}_{\nu} \hat{s} \hat{\nu} \hat{\nu}
+ Y_\nu \,\hat{l} \cdot \hat{H}_u \,\hat{\nu},   \nonumber \\
L_{soft} &=&  m_{H_d}^2 |H_d|^2 +m_{H_u}^2 |H_u|^2 +m_S^2 |S|^2 + \bar{m}_{\tilde\nu}^{2} \tilde{\nu}_{R}\tilde{\nu}^*_{R}
\nonumber \\
&& + ( \lambda A_\lambda S H_u\cdot H_d  + \frac{1}{3} \kappa A_\kappa S^3 + \bar{\lambda}_{\nu} \bar{A}_{\lambda_{\nu}}S \tilde{\nu}^*_{R} \tilde{\nu}^*_{R}
\nonumber \\
&& + Y_\nu A_{\nu} \tilde{\nu}^*_{R} \tilde{l} H_u + \mbox{h.c.})  + \cdots
\label{superpotential}
\end{eqnarray}
where $W_F$ denotes the superpotential of the MSSM without $\mu$ term, and a $Z_3$ symmetry is introduced to forbid the appearance of any dimensional parameters in $W$. In above formulae, the coefficients $\lambda$ and $\kappa$ parameterize the interactions among the Higgs fields, $Y_\nu$ and $\bar{\lambda}_\nu$ are neutrino Yukawa couplings
with flavor index omitted, $m_i$ ($i=H_u, H_d, \cdots$) denote soft breaking masses, and $A_i$ ($i=\lambda, \kappa, \cdots$) are soft breaking coefficients for trilinear terms.

\subsection{Higgs sector}

Since the soft breaking squared masses $m_{H_u}^2$, $m_{H_d}^2$ and $m_S^2$ are related with the vacuum expectation values of the fields $H_{u}$, $H_d$ and $S$,
$ \langle H_u \rangle = v_u/\sqrt{2}$, $ \langle H_d \rangle = v_d/\sqrt{2}$ and $\langle S \rangle = v_s/\sqrt{2}$,
by the minimization conditions of the Higgs potential after the electroweak symmetry breaking\cite{Ellwanger:2009dp}, it is conventional to
take $\lambda$, $\kappa$, $\tan \beta \equiv v_u/v_d$, $A_\lambda$, $A_\kappa$ and $\mu \equiv \lambda v_s/\sqrt{2} $ as theoretical input
parameters in Higgs sector. The elements of the squared mass matrix for CP-even Higgs fields in the basis ($S_1 \equiv \cos \beta {\rm Re}[H_u^0] - \sin \beta {\rm Re}[H_d^0]$,
$S_2 \equiv \sin \beta {\rm Re}[H_u^0] + \cos \beta {\rm Re}[H_d^0]$, $S_3 \equiv {\rm Re}[S]$)
are then given by \cite{Cao:2012fz}
\begin{eqnarray}
{\cal M}^2_{11}&=& \frac{2 \mu (\lambda A_\lambda + \kappa \mu)}{\lambda \sin 2 \beta} + \frac{1}{2} (2 m_Z^2- \lambda^2v^2)\sin^22\beta, \nonumber \\
{\cal M}^2_{12}&=&-\frac{1}{4}(2 m_Z^2-\lambda^2v^2)\sin4\beta, \nonumber \\
{\cal M}^2_{13}&=&-\sqrt{2} ( \lambda A_\lambda + 2 \kappa \mu) v \cos 2 \beta, \nonumber \\
{\cal M}^2_{22}&=&m_Z^2\cos^22\beta+ \frac{1}{2} \lambda^2v^2\sin^22\beta,\nonumber  \\
{\cal M}^2_{23}&=& \frac{v}{\sqrt{2}} \left[2 \lambda \mu - (\lambda A_\lambda + 2 \kappa \mu) \sin2\beta \right], \nonumber \\
{\cal M}^2_{33}&=& \frac{\lambda A_\lambda \sin 2 \beta}{4 \mu} \lambda v^2   + \frac{\mu}{\lambda} (\kappa A_\kappa +  \frac{4 \kappa^2 \mu}{\lambda} ), \label{Mass-CP-even-Higgs}
\end{eqnarray}
where $S_2$ represents the SM Higgs field and ${\cal M}^2_{22}$ is its squared mass at tree level without considering the mixing
among $S_i$. Similarly, the elements for CP-odd Higgs fields in the basis ($A \equiv \cos \beta {\rm Im}[H_u^0] - \sin \beta {\rm Im}[H_d^0]$, ${\rm Im}[S]$)
are
\begin{eqnarray}
{\cal M}^2_{P,11}&=& \frac{2 \mu (\lambda A_\lambda + \kappa \mu)}{\lambda \sin 2 \beta}, \nonumber  \\
{\cal M}^2_{P,22}&=& \frac{(\lambda A_\lambda + 3 \kappa \mu) \sin 2 \beta }{4 \mu} \lambda v^2  - \frac{3 \mu}{\lambda} \kappa A_\kappa, \nonumber  \\
{\cal M}^2_{P,12}&=& \frac{v}{\sqrt{2}} ( \lambda A_\lambda - 2 \kappa \mu). \label{Mass-CP-odd-Higgs}
\end{eqnarray}
As a result, the model predicts three CP-even Higgs mass eigenstates $h_i$ with $i=1,2,3$ and two CP-odd mass eigenstates $A_1$ and $A_2$, which are the mixtures of the real and imaginary parts of the fields $H_u^0$, $H_d^0$ and $S$, respectively. Throughout this paper, we label these eigenstates in an ascending mass order, i.e.
$m_{h_1} < m_{h_2} < m_{h_3}$ and $m_{A_1} < m_{A_2}$.

So far the measurement on the property of the $125 {\rm GeV}$ Higgs boson at the LHC indicates that it is quite SM-like, which restricts the mixing of the $S_2$ field with the other fields to be small. This implies from the definition of the $S_1$ and $S_2$ fields that the SM-like Higgs boson is $ {\rm Re}[H_u^0]$ dominated if $\tan \beta \gg 1$, and the heavy neutral doublet dominated mass eigenstates are mainly composed by $H_d^0$ field. In Section III and Appendixes, we will discuss in detail the implication of the Higgs physics on the input parameters. We remind that it is also popular to adopt the basis (${\rm Re}[H_d^0]$, ${\rm Re}[H_u^0]$, ${\rm Re}[S]$) in studying the property of the CP-even Higgs bosons since the couplings of the basis with SM fermions are simple.

The model also predicts a pair of charged Higgs $H^\pm$, $H^\pm = \cos \beta H_u^\pm + \sin \beta H_d^\pm$, and they are approximately degenerate in mass with the
heavy neutral states. The LHC search for extra Higgs bosons together with the indirect
constraints from $B$-physics have required $m_{H^\pm} \gtrsim 0.8 {\rm TeV}$, which is quite similar to MSSM case\cite{Bagnaschi:2017tru}. These doublet-like states couple
with the sneutrino DM via the interaction $\lambda \bar{\lambda}_\nu H_u \cdot H_d \tilde{\nu}^\ast_R \tilde{\nu}^\ast_R + h.c.$, which is
induced by the F-term of the Lagrangian. In the limit $m_{H^\pm} \to \infty$, only the SM-like Higgs boson plays a role in DM physics, but
since its interaction with sneutrino pair is proportional to $\lambda \bar{\lambda}_\nu v_d$, its effect is usually unimportant.

As for the singlet-dominated Higgs bosons, collider constraints on them are rather weak and consequently they may be light. These states couple with
sneutrino pair with three or four scalar interaction induced by the $\bar{\lambda}_{\nu}\,\hat{s}\,\hat{\nu}\,\hat{\nu}$ term in the superpotential
and its soft breaking term. Consequently, they can act as the annihilation product of the sneutrino DM or mediate the annihilation, and thus play an important role in DM annihilation.

\subsection{Sneutrino Sector}

In the NMSSM with Type-I seesaw extension, the active neutrino mass matrix is given by  $ m_{\nu} = \frac{1}{2}Y_{\nu} v_u M^{-1} Y_{\nu}^T v_u$ with $M = \sqrt{2} \bar{\lambda}_\nu v_S $ denoting the heavy neutrino mass matrix\cite{Kitano:1999qb}. Since the mass scale of the active neutrinos is $\sim 0.1 {\rm eV}$, the magnitude of $Y_\nu$ should be about $10^{-6}$ for the scale of $M$ around $100 {\rm GeV}$.
In order to reproduce neutrino oscillation data, $m_\nu$ must be flavor non-diagonal, which can be realized by assuming that the Yukawa coupling $Y_\nu$ is non-diagonal, while $\bar{\lambda}_\nu$ is diagonal. If one further assumes that the soft breaking parameters in sneutrino sector, such as $m_{\tilde{l}}$, $\bar{m}_{\tilde\nu}$ and $\bar{A}_{\lambda_{\nu}}$, are flavor diagonal, the flavor mixings of sneutrinos are then extremely suppressed by the off-diagonal elements of $Y_\nu$. In this case, it is enough to only consider one generation case in studying the properties of the sneutrino DM, which is what we will do.  In the following, we use the symbols $\lambda_\nu$, $A_{\lambda_\nu}$ and $m_{\tilde{\nu}}$ to denote the 33 element of $\bar{\lambda}_\nu$, $\bar{A}_{\lambda_\nu}$ and $\bar{m}_{\tilde{\nu}}$ respectively.

After decomposing sneutrino fields into CP-even and CP-odd parts:
\begin{equation}
\tilde{\nu}_L \equiv \frac{1}{\sqrt{2}}(\tilde{\nu}_{L1} + i
\tilde{\nu}_{L2}) ,
\quad\quad
\tilde{\nu}_R \equiv \frac{1}{\sqrt{2}}(\tilde{\nu}_{R1} + i \tilde{\nu}_{R2}) ,
\end{equation}
one can write down the sneutrino mass matrix in the basis ($\tilde{\nu}_{L1}$, $\tilde{\nu}_{R1}$, $\tilde{\nu}_{L2}$, $\tilde{\nu}_{R2}$)
as follows \eqref{sneutrino_matrix}

\begin{widetext}
\begin{equation}
{\cal M}^2_{\tilde{\nu}} =\begin{pmatrix}
m_{L\bar{L}}^2
& \frac{m_{LR}^2+m_{L\bar{R}}^2 + {\rm c.c} }{2}
&  0
& i \frac{m_{LR}^2-m_{L\bar{R}}^2 - {\rm c.c} }{2}  \\
\frac{m_{LR}^2+m_{L\bar{R}}^2 + {\rm c.c} }{2}
& m_{R\bar{R}}^2 + m_{RR}^2+m_{RR}^{2*}
& i \frac{m_{LR}^2-m_{L\bar{R}}^2 - {\rm c.c} }{2}
& i (m_{RR}^2 - m_{RR}^{2*})  \\
0
& i \frac{m_{LR}^2-m_{L\bar{R}}^2 - {\rm c.c} }{2}
& m_{L\bar{L}}^2
& \frac{-m_{LR}^2+m_{L\bar{R}}^2 + {\rm c.c} }{2} \\
i \frac{m_{LR}^2-m_{L\bar{R}}^2 - {\rm c.c} }{2}
& i (m_{RR}^2 - m_{RR}^{2*})
& \frac{-m_{LR}^2+m_{L\bar{R}}^2 + {\rm c.c} }{2}
& m_{R\bar{R}}^2 - m_{RR}^2 - m_{RR}^{2*}
\end{pmatrix}
\label{sneutrino_matrix}
\end{equation}
\end{widetext}
where
\begin{eqnarray}
m_{L\bar{L}}^2
&\equiv& m_{\tilde{l}}^2 + |Y_\nu v_u|^2 + \frac{1}{8} (g_1^2 + g_2^2) (v_d^2 - v_u^2), \nonumber \\
m_{LR}^2
&\equiv& Y_\nu\left(-\lambda v_s v_d \right)^{\ast}
+ Y_\nu A_{Y_\nu} v_u , \nonumber\\
m_{L\bar{R}}^2
&\equiv& Y_\nu v_u \left(-\lambda v_s \right)^{\ast} , \nonumber\\
m_{R\bar{R}}^2
&\equiv& m_{\tilde{\nu}}^2 +|2\lambda_{\nu} v_s |^2
+ |Y_\nu v_u|^2 , \nonumber\\
m_{RR}^2
&\equiv& \lambda_\nu \left( A_{\lambda_\nu} v_s+(
\kappa v_s^2-\lambda v_d v_u )^{\ast} \right) .
\label{sn:mrr}
\end{eqnarray}
If all the parameters in the matrix are real, namely there is no CP violation, the real and imaginary parts of the sneutrino fields will not mix and the mass term can be split into two parts
\begin{eqnarray}
&&\frac{1}{2}(\tilde{\nu}_{Li}, \tilde{\nu}_{Ri})
\left(
\begin{array}{cc}
m_{L\bar{L}}^2           &  \pm m_{LR}^2+m_{L\bar{R}}^2 \\
\pm m_{LR}^2+m_{L\bar{R}}^2  &  m_{R\bar{R}}^2 \pm 2m_{RR}^2 \\
\end{array}
\right)
\left(
\begin{array}{c}
\tilde{\nu}_{Li}  \\
\tilde{\nu}_{Ri} \\
\end{array}
\right), \nonumber
\end{eqnarray}
where $i=1$ ($i=2$) for CP-even (CP-odd) states, and the minus signs in the matrix are for the CP-odd states.
From this formula, one can learn that the chiral mixings of the sneutrinos are proportional to $Y_{\nu}$, and hence can be ignored safely. So
sneutrino mass eigenstate coincides with chiral state. In our study, the sneutrino DM corresponds to the lightest right-handed sneutrino.
One can also learn that the mass splitting between the CP-even and CP-odd right-handed states  is given by $\Delta m^2 \equiv m_{even}^2 - m_{odd}^2 = 4 m^2_{RR}$, which
implies that  the CP-even state $\tilde{\nu}_{R1}$ is lighter than the CP-odd state $\tilde{\nu}_{R2}$ if $m^2_{RR} < 0$ and vice versa. This implies that the sneutrino
DM may be either CP-even or CP-odd. In this work, we consider both the possibilities.

Once the form of the sneutrino DM $\tilde{\nu}_1$ is given, one can determine its coupling strength with Higgs bosons. For example, for a CP-even $\tilde{\nu}_1$ we have \begin{small}
\begin{eqnarray}
C_{\tilde{\nu}_1\tilde{\nu}_1 h_i}&=&
\frac{\lambda\lambda_{\nu}M_W}{g}(sin\beta Z_{i1} + cos\beta Z_{i2}) - \nonumber\\
&&\left[
\frac{\sqrt{2}}{\lambda} \left(2\lambda_{\nu}^2 + \kappa\lambda_{\nu}\right) \mu  - \frac{\lambda_{\nu} A_{\lambda_\nu}}{\sqrt{2}}
\right]Z_{i3}, \label{Csnn} \\
C_{\tilde{\nu}_1 \tilde{\nu}_1 A_m A_n} &=& -  \frac{1}{2} \lambda \lambda_\nu \cos \beta \sin \beta Z^\prime_{m1} Z^\prime_{n1} \nonumber \\
&& - (\lambda_\nu^2 - \frac{1}{2} \lambda_\nu \kappa )  Z^\prime_{m2} Z^\prime_{n2}, \label{CsnnAA}
\end{eqnarray}
\end{small}
where $Z_{ij}$ with $i,j=1,2,3$ ($Z^\prime_{mn}$ with $m,n=1,2$) are the elements of the matrix to diagonalize the CP-even Higgs mass matrix in the basis (${\rm Re}[H_d^0]$, ${\rm Re}[H_u^0]$, ${\rm Re}[S]$) (the CP-odd Higgs mass matrix in Eq.(\ref{Mass-CP-odd-Higgs})), and for a CP-odd $\tilde{\nu}_1$, its coupling strengthes can be obtained from the expressions by the substitution $\lambda_\nu \to -\lambda_\nu$.   As is expected, the first coupling is suppressed by a factor $\lambda \lambda_\nu \cos \beta$ if $h_i$ as the SM-like Higgs boson is $Re(H_u^0)$ dominant, and all the couplings may be moderately large if the Higgs bosons are singlet dominant.

\subsection{DM Relic density}
It is well known that the WIMP's relic abundance is related to its thermal averaged annihilation cross section at the time of freeze-out~\cite{Jungman:1995df}.  In order to obtain the WIMP's abundance one should solve following Boltzmann equation
\begin{equation}
\frac{dY}{dT}=\sqrt{\frac{\pi g_{\ast}(T)}{45}}M_{p}\left\langle \sigma v\right\rangle(Y^{2}-Y^{2}_{eq})\,,
\label{relicabundance}
\end{equation}
where $g_{\ast}$ is the effective number of degrees of freedom at thermal equilibrium, $M_{p}$ is the Plank mass, $Y$ and $Y_{eq}$ are the relic abundance and the thermal equilibrium abundance respectively, and $\left\langle \sigma v\right\rangle$ is WIMP's relativistic thermal averaged annihilation cross section with $v$ denoting the relative velocity between the annihilating particles. $\left\langle \sigma v\right\rangle$ is related to the particle physics model by\cite{Belanger:2013oya}
\begin{equation}
\left\langle \sigma v\right\rangle=\frac{\displaystyle\sum_{i,j}g_{i}g_{j}\displaystyle\int_{ (m_{i}+m_{j})^{2} }ds\sqrt{s}K_{1}(\frac{\sqrt{s}}{T})p_{ij}^{2}\displaystyle\sum_{k,l}\sigma_{ij;kl}(s)}{2T(\displaystyle\sum_{i}g_{i}m^{2}_{i}K_{2}(m_{i}/T) )^{2}}\,,
\label{evolutionabundance}
\end{equation}
where $g_{i}$ is the number of degrees of freedom, $\sigma_{ij;kl}$ is the cross section for annihilation of a pair of particles with masses $m_{i}$ and $m_{j}$ into SM particles $k$ and $l$, $p_{ij}$ is the momentum of incoming particles in their center of mass frame with squared total energy $s$, and $K_i$ ($i=1,2$) are modified Bessel functions.

The present day abundance is obtained by integrating Eq.(\ref{relicabundance}) from $T=\infty$ to $T=T_{0}$, where $T_{0}$ is the temperature of the Universe today\footnote{Generally speaking, it is accurate enough to integrate the Boltzmann equation from $T=m_{\rm DM}$ to $T=T_0$ in getting the DM abundance~\cite{Gondolo:1990dk}.
In the code microMEGAs, it actually starts the integration from the temperature $T_1$ defined by $Y(T_1) = 1.1 Y_{eq}(T_1)$~\cite{Belanger:2004yn}.
This temperature is moderately higher than the freeze-out temperature $T_f \simeq m_{\rm DM}/25$, which is defined by the equation $Y(T_f) = 2.5 Y_{eq}(T_f)$~\cite{Gondolo:1990dk}.}. And WIMP relic density can be written as~\cite{Belanger:2013oya}
\begin{equation}
\Omega h^{2}=2.742\times 10^{8}\frac{M_{WIMP}}{\mbox{GeV}} Y(T_{0})\,.
\label{solution}
\end{equation}

In the NMSSM with Type-I seesaw mechanism, possible annihilation channels of the sneutrino DM include\cite{Cerdeno:2008ep,Cerdeno:2009dv}
\begin{itemize}
	\item[(1)]  $\tilde{\nu}_1 \tilde{\nu}_1 \rightarrow V V^\ast$, $VS$, $f \bar{f} $ with $V$, $S$ and $f$ denoting a vector boson ($W$ or $Z$), a Higgs boson and
	a SM fermion, respectively.  This kind of annihilations proceed via s-channel exchange of a CP-even Higgs boson.
	
	\item[(2)] $\tilde{\nu}_1 \tilde{\nu}_1 \rightarrow S S^\ast $
	via $s$-channel Higgs exchange, $t/u$-channel sneutrino exchange, and relevant scalar quartic couplings.
	
	\item[(3)]  $\tilde{\nu}_1 \tilde{\nu}_1 \rightarrow \nu_R \bar{\nu}_R$
	via $s$-channel Higgs exchange and $t/u$-channel neutralino exchange.
	\item[(4)]
	$\tilde{\nu}_1 \tilde{\nu}_1^\prime \rightarrow A_i^{(\ast)} \rightarrow X Y$ and $\tilde{\nu}_1^\prime \tilde{\nu}_1^\prime \rightarrow X^\prime Y^\prime$ with
    $\tilde{\nu}_1^\prime$ denoting a right-handed sneutrino with an opposite CP number to that of $\tilde{\nu}_1$, and
    $X^{(\prime)}$ and $Y^{(\prime)}$ denoting any possible light state.
	These annihilation channels are important in determining the relic density only when the CP-even and CP-odd states are nearly
	degenerate in mass.
	\item[(5)]
	$\tilde{\nu}_1 \tilde{H} \rightarrow X Y$ and $\tilde{H} \tilde{H}^\prime \rightarrow X^\prime Y^\prime$ with $\tilde{H}$ and $\tilde{H}^\prime$ being
    any Higgsino dominated neutralino or Higgsino dominated chargino.
	These annihilation channels are called coannihilation\cite{Coannihilation,Griest:1990kh}, and they become important if
	the mass splitting between Higgsino and $\tilde{\nu}_1$ is less than about $10\%$.
\end{itemize}
The expressions of $\sigma v$ for some of the channels are presented in \cite{Cerdeno:2009dv}. One can learn from them that the parameters in sneutrino sector, such as $\lambda_\nu$, $A_\nu$ and
$m_{\nu}^2$, as well as the parameters in Higgs sector are involved in the annihilations\footnote{We note that the works \cite{Cerdeno:2008ep,Cerdeno:2009dv} failed
	to consider the annihilation channels (4) and (5). Especially, the channel (5) is testified to be most important by our following study. We also note that in calculating $\sigma v$ of the channel $\tilde{\nu}_1 \tilde{\nu}_1 \to A_i A_j$, the authors neglected the contribution from the $t/u$-channel sneutrino exchange. }.

\subsection{DM Direct detection}

Since $\tilde{\nu}_1$ in this work is a right-handed scalar with definite CP and lepton numbers, its scattering with nucleon $N$ ($N=p,n$) proceeds only by exchanging CP-even Higgs bosons. In the non-relativistic limit, this process is described by an effective operator ${\cal{L}}_{\tilde{\nu}_1 N} = f_N \tilde{\nu}_1 \tilde{\nu}_1 \bar{\psi}_N \psi_N$ with the coefficient $f_N$ given by  \cite{Han:1997wn}
\begin{eqnarray}
	f_N &=&  m_N \sum_{i=1}^3 \frac{C_{\tilde{\nu}_1 \tilde{\nu}_1 h_i} C_{N N h_i}}{m_{h_i}^2}
	\\ \nonumber
	 &=&
    m_N \sum_{i=1}^3 \frac{C_{\tilde{\nu}_1 \tilde{\nu}_1 h_i}}{m_{h_i}^2} \frac{(-g)}{2 m_W} \left ( \frac{Z_{i2}}{\sin\beta} F^{(N)}_u +  \frac{Z_{i1}}{\cos \beta} F^{(N)}_d \right ), \\
\end{eqnarray}
where $C_{N N h_i}$ denotes the Yukawa coupling of the Higgs boson $h_i$ with nucleon $N$, and $F^{(N)}_u=f^{(N)}_u+\frac{4}{27}f^{(N)}_G$ and $F^{(N)}_d=f^{(N)}_d+f^{(N)}_s+\frac{2}{27}f^{(N)}_G$ are nucleon form factors
with $f^{(N)}_q=m_N^{-1}\left<N|m_qq\bar{q}|N\right>$
(for $q=u,d,s$) and $f^{(N)}_G=1-\sum_{q=u,d,s}f^{(N)}_q$. Consequently, the spin-dependent cross section for the scattering vanishes, and
the spin independent cross section is given by\cite{Han:1997wn}
\begin{small}
\begin{eqnarray}
\sigma^{\rm SI}_{\tilde{\nu}_1-N} &=& \frac{\mu^2_{\rm red}}{4 \pi m_{\tilde{\nu}_1}^2} f_N^2 = \frac{4 F^{(N)2}_u \mu^2_{\rm red} m_N^2}{\pi} (\sum_i \xi_i)^2,  \label{SI-expression}
\end{eqnarray}
\end{small}
where $\mu_{\rm red}= m_N/( 1+ m_N^2/m_{\tilde{\nu}_1}^2) $ is the reduced mass of the nucleon with $m_{\tilde{\nu}_1}$, and $\xi_i$ with $i=1,2,3$ are defined by
\begin{eqnarray}
\xi_i = -\frac{g}{8 m_W} \frac{C_{\tilde{\nu}_1 \tilde{\nu}_1 h_i}}{m_{h_i}^2 m_{\tilde{\nu}_1}} \left ( \frac{Z_{i2}}{\sin\beta} + \frac{Z_{i1}}{\cos \beta} \frac{F^{(N)}_d}{F^{(N)}_u} \right )
\end{eqnarray}
to facilitate our analysis. Obviously, $\xi_i$ represents the $h_i$ contribution to the cross section.

In our numerical calculation of the DM-proton scattering rate $\sigma^{\rm SI}_{\tilde{\nu}_1-p}$, we use the default setting of the package micrOMEGAs \cite{Belanger:2013oya,micrOMEGAs-1,micrOMEGAs-3}
for the nucleon form factors, $\sigma_{\pi N} = 34 {\rm MeV}$ and $\sigma_0 = 42 {\rm MeV}$, and
obtain $F_u^{(p)} \simeq 0.15$ and  $F_d^{(p)} \simeq 0.14$ \footnote{Note that different choices of the pion-nucleon sigma term $\sigma_{\pi N}$ and $\sigma_0$ can induce an
	uncertainty of ${\cal{O}} (10\%)$ on $F_u^{(p)}$ and $F_d^{(p)}$.  For example, if one takes  $\sigma_{\pi N} = 59 {\rm MeV}$ and $\sigma_0 = 57 {\rm MeV}$,
	which are determined from \cite{Alarcon:2011zs,Ren:2014vea,Ling:2017jyz} and \cite{Alarcon:2012nr} respectively, $F_u^{(p)} \simeq 0.16$ and  $F_d^{(p)} \simeq 0.13$.}. In this case,
Eq.(\ref{SI-expression}) can be approximated by
\begin{small}
\begin{eqnarray}
\sigma^{\rm SI}_{\tilde{\nu}_1-p} & \simeq &  \frac{4 F^{(p)2}_u \mu^2_{\rm red} m_p^2  }{\pi}
 \label{SI-simplify1} \\
 && \left \{ \frac{g}{8 m_W} \sum_i \left [ \frac{C_{\tilde{\nu}_1 \tilde{\nu}_1 h_i}}{m_{h_i}^2 m_{\tilde{\nu}_1}} (\frac{Z_{i2}}{\sin\beta} + \frac{Z_{i1}}{\cos \beta} ) \right ] \right \}^2.  \nonumber
\end{eqnarray}
\end{small}
From this approximation and also the expression of $C_{\tilde{\nu}_1 \tilde{\nu}_1 h_i}$ in Eq.(\ref{Csnn}), one can get following important features about
the scatting of the sneutrino DM with nucleon:
\begin{itemize}
	\item For each of the $h_i$ contributions, it depends not only on the parameters in Higgs sector, but also on the parameters in sneutrino sector such as $\lambda_\nu$ and $A_{\lambda_\nu}$. This feature lets the theory have a great degree of freedom to adjust the contribution size so that the severe cancellation among the contributions
	can be easily achieved. By contrast, in the NMSSM with neutralino as a DM candidate, the contribution depends on, beside the DM mass, only the parameters in Higgs sector,
	and consequently it is not easy to reach the blind spot for DM-nucleon scattering due to the tight constraints on the Higgs parameters from LHC experiments\cite{Crivellin:2015bva,Badziak:2015exr,Han:2016qtc,Badziak:2017uto}.
	\item Each of the $h_i$ contributions can be suppressed. To be more specific, the $Re(H_d^0)$ dominated Higgs is usually at TeV scale, so its contribution
	is suppressed by the squared mass; the $Re(H_u^0)$ dominated scalar corresponds to the SM-like Higgs boson, and its coupling with $\tilde{\nu}_1$ can be suppressed by $\cos\beta$ and/or $\lambda_\nu$, or by the accidental cancellation among different terms in $C_{\tilde{\nu}_1 \tilde{\nu}_1 h_i}$. In most cases, the contribution from the singlet-dominated scalar is most important, but such a contribution is obviously suppressed by the doublet-singlet mixing in the scalar.
\end{itemize}

We emphasize that these features make the theory compatible with the strong constraints from the DM DD experiments in broad parameter spaces. In order to parameterize the degree of the cancellation among the contributions, we define the fine tuning quantity $\Delta_{FT}$ as
\begin{eqnarray}
\Delta_{FT}=\mathop{\max}_i \left \{\frac{\xi_i^2}{(\sum_i \xi_i )^2}\right \}.  \label{Fine-tuning}
\end{eqnarray}
Obviously, $\Delta_{FT} \sim 1$  is the most ideal case  for any DM theory in predicting an
experimentally allowed $\sigma^{\rm SI}_{\tilde{\nu}_1-p}$, and the larger $\Delta_{FT}$ becomes, the more unnatural the theory is.

\section{\label{CP-even} Numerical Results}

In our study of the sneutrino DM scenarios, we utilize the package \textsf{SARAH-4.11.0} \cite{sarah-1,sarah-2,sarah-3} to build the model, the codes  \textsf{SPheno-4.0.3} \cite{spheno} and \textsf{FlavorKit}\cite{Porod:2014xia} to generate the particle spectrum and compute low energy
flavor observables, respectively, and the package \textsf{MicrOMEGAs 4.3.4}\cite{Belanger:2013oya,micrOMEGAs-1,micrOMEGAs-3}
to calculate DM observables by assuming that the lightest sneutrino is the sole DM candidate in the universe.
We also consider the bounds from the direct searches for Higgs bosons at LEP, Tevatron and LHC by the packages
\textsf{HiggsBounds-5.0.0}\cite{HiggsBounds} and \textsf{HiggsSignal-2.0.0} \cite{HiggsSignal}. Note that in calculating the
radiative correction to the mass spectrum of the Higgs bosons, the code \textsf{SPheno-4.0.3}
only includes full one- and two-loop effects using a diagrammatic approach with vanishing external
momenta~\cite{spheno}. This leaves an uncertainty less than about $3~{\rm GeV}$ for the SM-like Higgs boson mass.

\subsection{Scan strategy}

\begin{table*}
	\begin{center}
		\resizebox{0.6\textwidth}{!}{
		\begin{tabular}{|c|c|c|c|c|c|}
			\hline
			Parameter& Value & Parameter & Value& Parameter& Value \\ \hline
			$m_{\tilde{q}}$ & $2{\rm TeV} \times \bf{1}$ & $A_{u,c,d,s}$ & 2{\rm TeV} & $A_\lambda$ & 2{\rm TeV} \\
			$m_{\tilde{l}}$  & $0.4{\rm TeV} \times \bf{1}$ & $A_{e,\mu,\tau}$ & 0.4{\rm TeV} & $y_\nu$ &  $10^{-6} \times \bf{1}$ \\
			$M_1$ & 0.4{\rm TeV} & $M_2$ & 0.8{\rm TeV} & $M_3$ & 2.4{\rm TeV}\\
            $[\bar{\lambda}_\nu]_{11,22}$ & 0.3 & $[\bar{A}_{\lambda_\nu}]_{11,22}$ & 0 & $[\bar{m}_{\tilde{\nu}}]_{11,22}$ &  0.4{\rm TeV} \\
            \hline
		\end{tabular}
        }
		\end {center}


		\caption{Fixed parameters in the seesaw extension of the NMSSM, where $m_{\tilde{q}}$ for $\tilde{q} = \tilde{Q}, \tilde{U}, \tilde{D}$ and $m_{\tilde{l}}$ for $\tilde{l} = \tilde{L}, \tilde{E}$ denote soft masses of squarks and sleptons, respectively, with $\bf{1}$ being unit matrix in flavor space,  and $M_i$ with $i=1,2,3$ are soft masses for gauginos. $A_i$ with $i=u,c,d,s,e,\mu,\tau$ are coefficients of soft trilinear terms for a specific flavor, and the other flavor changing trilinear coefficients are assumed to be zero. }
		\label{benchmark}
	\end{table*}
	
	The previous discussion indicates that only the parameters in Higgs and sneutrino sectors are involved in the DM physics. We perform a sophisticated scan over these parameters in following ranges:
\begin{eqnarray}\label{parameter-scan}
&&  0 < \lambda \leq 0.7, |\kappa| \leq 0.7, 0 < \lambda_{\nu} \leq 0.7, 1 \leq tan\beta \leq 60, \nonumber
\\ \nonumber
&& |A_{\lambda_\nu}|,|A_\kappa| \leq 1 {\rm TeV}, 100 {\rm GeV} \leq \mu \leq 300{\rm GeV}, \nonumber
\\
&& 0 \leq m_{\tilde{\nu}} \leq 300{\rm GeV},|A_t| \leq 5 {\rm TeV},
 \label{scan-ranges}
\end{eqnarray}
	by setting $A_b = A_t$ and the other less important parameters in Table \ref{benchmark}. All the parameters are defined at the scale $Q=1 {\rm TeV}$. Our interest in the parameter space is due to following considerations:
	\begin{itemize}
		\item Since the masses of the heavy doublet-dominated Higgs bosons are usually at TeV scale, which is favored by direct searches for extra Higgs boson at the LHC, their effects on the DM physics are decoupled and thus become less important. We fix the parameter $A_\lambda$, which is closely related with $m_{H^\pm}$\cite{Ellwanger:2009dp}, at $2 {\rm TeV}$ as an example to simplify the calculation given that the scan is very time-consuming for our computer clusters. Consequently, we find that the Higgs bosons are heavier than about $1 {\rm TeV}$ for almost all samples. We will discuss the impact of the other choices of $A_\lambda$ on the results in Appendices of this work.
		\item It is well known that the radiative correction from top/stop and bottom/sbottom loops to the Higgs mass spectrum plays an important role for SUSY to coincide with relevant experimental measurements at the LHC. In our calculation, we include such an effect by fixing $m_{\tilde{Q}_3} = m_{\tilde{U}_3} = m_{\tilde{D}_3} = 2 {\rm TeV}$, setting $A_t = A_b$ and varying $A_t$ over a broad region, $|A_t| \leq 5 {\rm TeV}$. We remind that within the region, the color and charge symmetries of the theory remain unbroken\cite{Park:2010wf,Chowdhury:2013dka}, and the LHC search for third generation squarks does not impose any constraints.
		\item The upper bounds of the parameters $\lambda$, $\kappa$, $\lambda_\nu$ and $\tan \beta$ coincide with the perturbativity of the theory up to Planck scale\cite{Ellwanger:2009dp}.
		\item Because the Higgsino mass $\mu$ is directly related to $Z$ boson mass, naturalness prefers $\mu \sim {\cal{O}}(10^2 {\rm GeV})$\cite{Ellwanger:2009dp}. So we require $100 {\rm GeV} \leq \mu \leq 300 {\rm GeV}$ in this work, where the lower bound comes from the LEP search for chargino and neutralinos, and the upper bound is imposed by hand. We will discuss the effect of a wider range of $\mu$ on the results in Appendix B.
		\item Since the sneutrino DM must be lighter than the Higgsino, its soft breaking mass $m_{\tilde{\nu}}$ is therefore upper bounded by about $300 {\rm GeV}$, and $|A_{\lambda_\nu}| \lesssim 1 {\rm TeV}$ is favored by naturalness.
		\item In order to get correct EWSB and meanwhile predict the masses of the singlet dominated scalars around $100 {\rm GeV}$,   $|A_\kappa|$
		can not be excessively large from naturalness argument (see Appendix A).
		\item We require the other dimensional parameters, such as $M_i$ with $i=1,2,3$ and $m_{\tilde{l}}$, sufficiently large so that their prediction on sparticle spectrum is consistent with the results of the direct search for sparticles at the LHC.
	\end{itemize}
	
	In order to make the conclusions obtained in this work as complete as possible, we adopt the MultiNest algorithm introduced in \cite{Feroz:2008xx} in the scan, which is implemented in our code \textsf{EasyScan\_HEP}\cite{es}, with totally more than $2 \times 10^8$ physical samples computed \footnote{To be more specific, we performed five independent scans for each case mentioned in the text. In each scan, we set the {\it nlive} parameter in MultiNest (which denotes the number of active or live points used to determine the iso-likelihood contour in each iteration~\cite{Feroz:2008xx,Feroz:2013hea}) at 2000. }.  The output of the scan includes the Bayesian evidence defined by
	\begin{eqnarray}
	Z(D|M) \equiv \int{P(D|O(M,\Theta)) P(\Theta|M) \prod d \Theta_i}, \nonumber
	\end{eqnarray}
	where $P(\Theta|M)$ is called prior
	probability density function (PDF) for the parameters $\Theta = (\Theta_1,\Theta_2,\cdots)$ in the model $M$, and $P(D|O(M, \Theta))\equiv \mathcal{L}(\Theta)$ is the likelihood function for the theoretical predictions on the observables $O$ confronted with their experimentally measured values $D$. Computationally, the evidence is an average likelihood, and it depends on  the priors of the model's parameters. For different scenarios in one theory, the larger $Z$ is, the more readily the corresponding scenario agrees with the data. The output also includes weighted and unweighted parameter samples which are subject to the posterior PDF $P(\Theta|M,D)$. This PDF is given by
	\begin{equation}\label{eq:bayes}
	\begin{aligned}
	P(\Theta|M,D)=\frac{P(D|O(M,\Theta))P(\Theta|M)}{Z},
	\end{aligned}
	\end{equation}
	and it reflects the state of our knowledge about the parameters $\Theta$ given the experimental data $D$, or alternatively speaking, the updated prior PDF after considering the impact of the experimental data. This quantity may be sensitive to the shape of the prior, but the sensitivity can be counterbalanced by sufficient data and thus lost in certain conditions~\cite{Bayes}. Obviously, one can infer from the distribution of the samples the underlying physics of the model.
	
	In the scan, we take flat distribution for all the parameters in Eq.(\ref{scan-ranges})\footnote{Dimensional parameters usually span wide ranges, and if one sets them flat distributed, the efficiency of the scan is not high in general. Due to this reason, special treatments of such kind of parameters were frequently adopted in literatures~\cite{Kowalska:2012gs,Athron:2017fxj}. In Appendix B of this work, we will show that the prior setting of simple log distribution for $A_\kappa$ can overemphasize the low $|A_\kappa|$ region in the scan, and consequently deform the shape of posterior distributions.}.  We will discuss the influence of different prior PDFs on our results in Appendix B and verify that the flat distribution can predict the right posterior distributions. The likelihood function we construct contains
	\begin{eqnarray}
	\mathcal{L}(\Theta)&=&\mathcal{L}_{Higgs} \times  \mathcal{L}_{Br(B_s\rightarrow \mu^+\mu^-)} \times \mathcal{L}_{Br(B_s \rightarrow X_s\gamma)} \nonumber \\
&&	\times \mathcal{L}_{\Omega_{\widetilde{\nu}_1}} \times \mathcal{L}_{DD} \times \mathcal{L}_{ID},  \label{Likelihood}
	\end{eqnarray}
	where each contribution on the right side of the equation is given as follows:
	\begin{itemize}
		\item For the likelihood function of the Higgs searches at colliders, we set
		\begin{equation}
		\mathcal{L}_{Higgs}=e^{-\frac{1}{2} (\chi^2 + A^2)},
		\end{equation}
		where $\chi^2$ comes from the fit of the theoretical prediction on the property of the SM-like Higgs boson to relevant LHC data with its value calculated by the code HiggsSignal \cite{HiggsSignal}, and $A^2$ reflects whether the parameter point is allowed or excluded by the direct searches for extra Higgs at colliders. In practice,
we take a total (theoretical and experimental) uncertainty of $3 {\rm GeV}$ for the SM-like Higgs boson mass in calculating the $\chi^2$ with the package HiggsSignal, and set
by the output of the code HiggsBounds~\cite{HiggsBounds} either $A^2=0$ for the case of experimentally allowed at $95\%$ confidence level (C. L.) or $A^2=100$ for the other cases.
		
		Since the Bayesian evidence for the scenario where the lightest Higgs boson $h_1$ acts as the SM-like Higgs boson is much larger than that with the next lightest Higgs boson $h_2$ as the SM-like Higgs boson~\cite{Ellwanger:2011aa,Cao:2012fz} (see the discussion at the end of this work), we focus on the scenario where $h_1$ corresponds to the Higgs boson discovered at the LHC, and present the results according to the CP property of the sneutrino DM $\tilde{\nu}_1$.
		
		\item For the second, third and fourth contributions, i.e. the likelihood functions about the measurements of $Br(B_s \to \mu^+ \mu^-)$, $Br(B_s \to X_s \gamma)$ and DM relic density $\Omega_{\widetilde{\nu}_1}$, they are Gaussian distributed, i.e.
		\begin{equation}
		\mathcal{L}=e^{-\frac{[\mathcal{O}_{th}-\mathcal{O}_{exp}]^2}{2\sigma^2}},
		\end{equation}
		where $\mathcal{O}_{th}$ denotes the theoretical prediction of the observable $\mathcal{O}$, $\mathcal{O}_{exp}$ represents
		its experimental central value and $\sigma$ is the total (including both theoretical and experimental) uncertainty. We take the experimental information about the B physics measurements from latest particle data book~\cite{PDG} and the relic density $\Omega_{\widetilde{\nu}_1}$ from~\cite{Aghanim:2018eyx}.
		
		\item For the likelihood function of DM DD experiments $\mathcal{L}_{DD}$, we take a Gaussian form with a mean value of zero~\cite{Matsumoto:2016hbs}:

        \vspace{-1cm}

		\begin{equation}
		\mathcal{L}=e^{-\frac{\sigma^2}{2\delta_{\sigma}^2}},
		\end{equation}
		where $\sigma$ stands for DM-nucleon scattering rate, and $\delta_{\sigma}$ is evaluated by $\delta_{\sigma}^2 = UL_\sigma^2/1.64^2 + (0.2\sigma)^2$ with
		$UL_\sigma$ denoting the upper limit of the latest XENON1T results on $\sigma$  at $90\%$ C. L.~\cite{Aprile:2018dbl}, and $0.2 \sigma$ parameterizing
		theoretical uncertainties.
		
		\item For the likelihood function of DM indirect search results from drawf galaxies $\mathcal{L}_{ID}$, we use the data of Fermi-LAT collaboration~\cite{Fermi-Web}, and adopt the likelihood function proposed in~\cite{Likelihood-dSph,Zhou}.
	\end{itemize}

About the likelihood function in Eq.(\ref{Likelihood}), we have three more explanations. The first is that we do not consider the constraint on line signal of $\gamma$-ray from Fermi-LAT data since it is rather weak~\cite{Li:2018xzs}. The second is that we do not include the likelihood function of the LHC search for Higgsinos since the involved Monte Carlo simulation is time consuming, and meanwhile the most favored parameter region satisfy the constraint automatically. We will address this issue in Section IV. And the last point is
that during the scan, one usually encounters the nonphysical situation where the squared mass of any scalar particle is negative or $\tilde{\nu}_1$ is not the LSP. In this case,
we set the likelihood function to be sufficiently small, e.g. $e^{-100}$.

	\begin{figure*}[htbp]
		\centering
		\resizebox{0.7\textwidth}{!}{
		\includegraphics{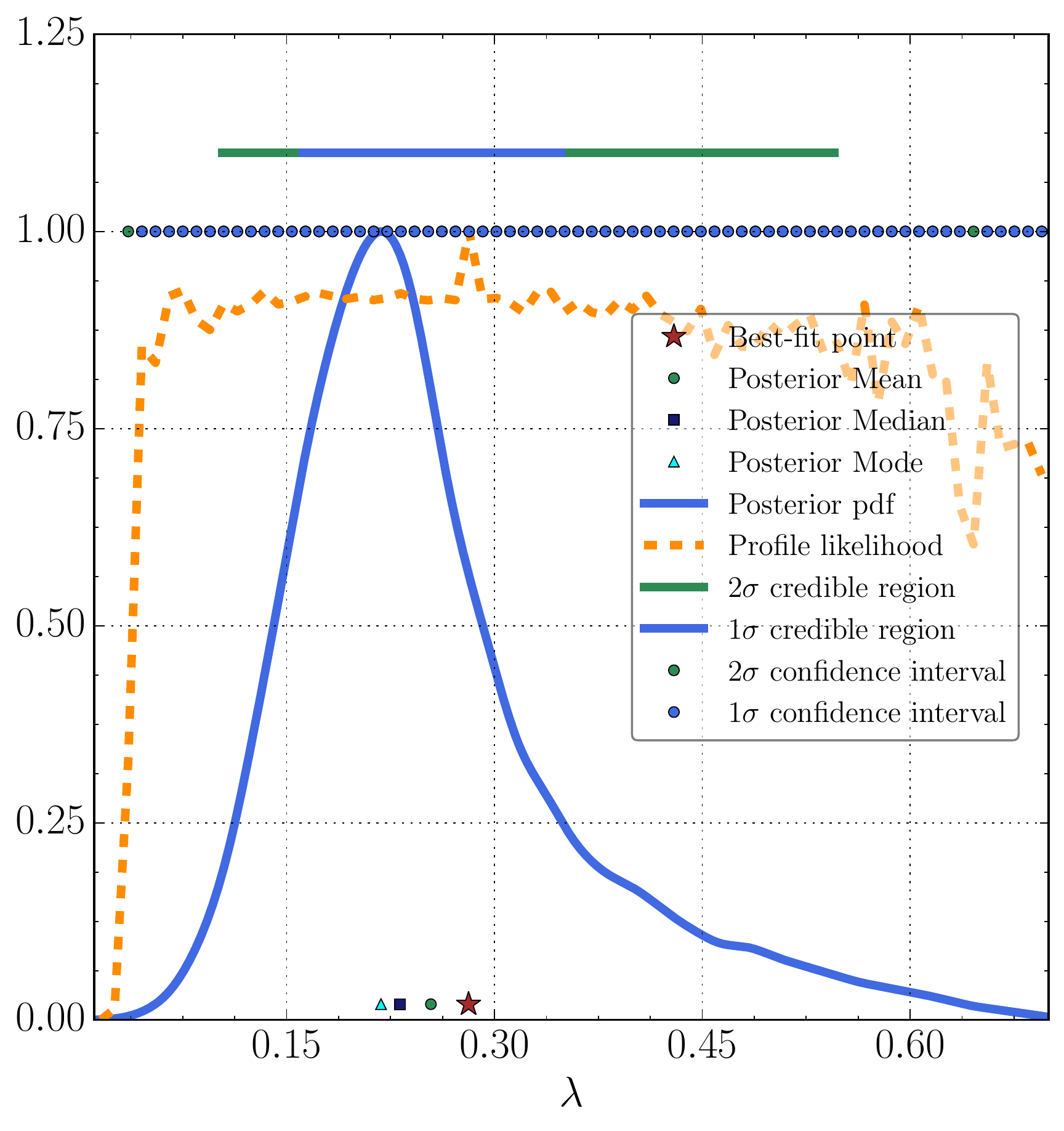}
		\includegraphics{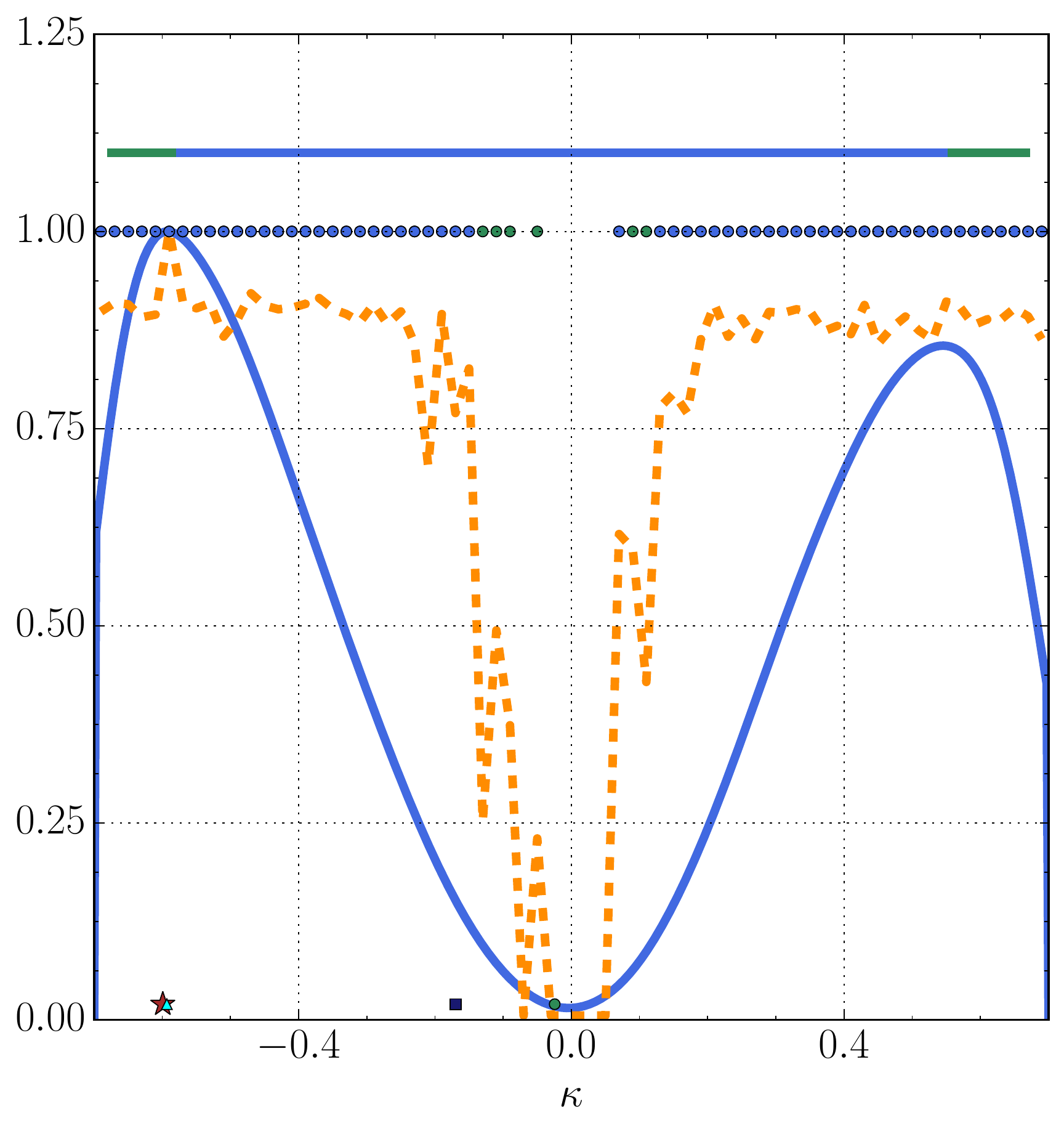}
		\includegraphics{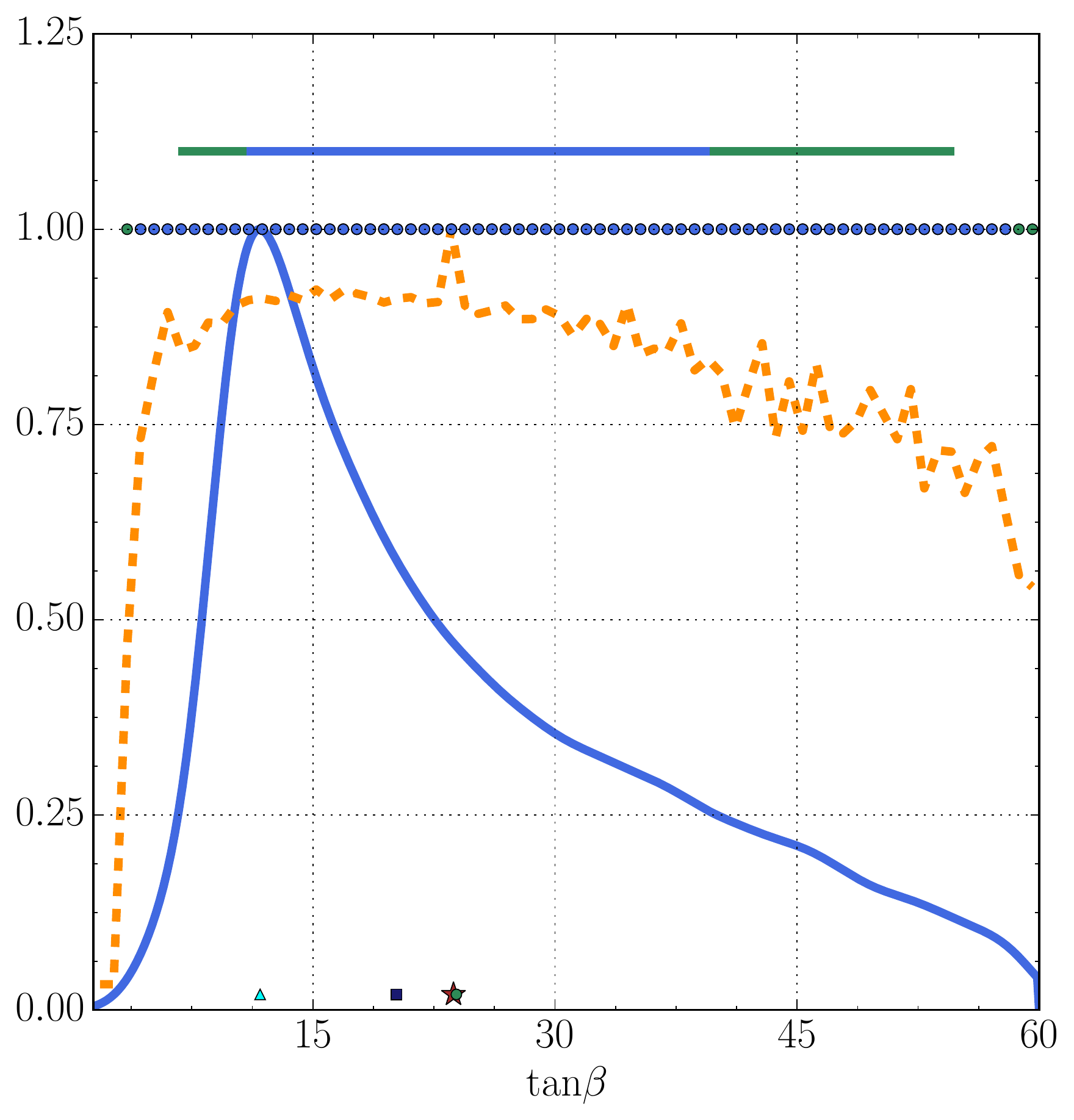}
		}
		\resizebox{0.7\textwidth}{!}{
		\includegraphics{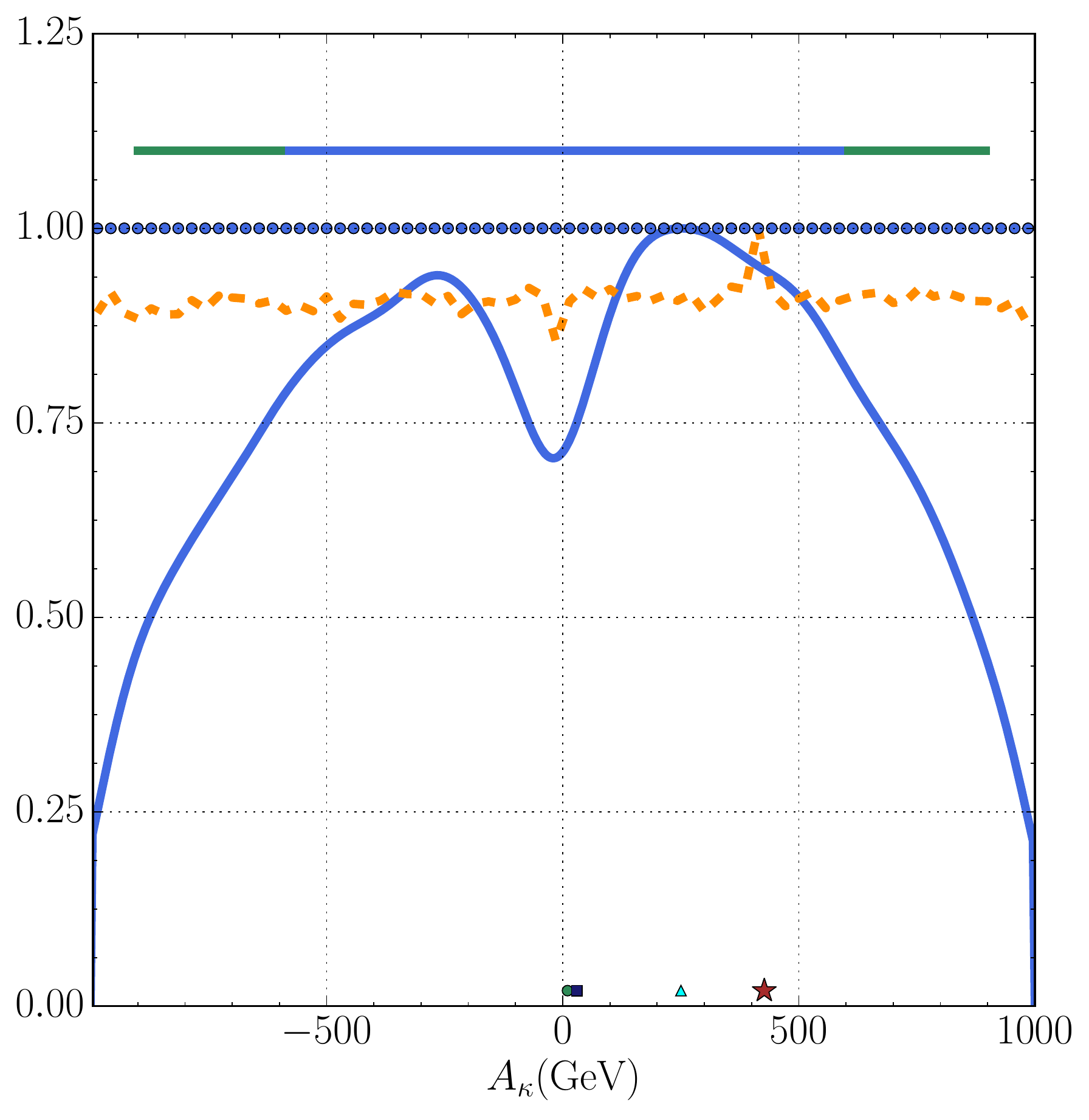}
		\includegraphics{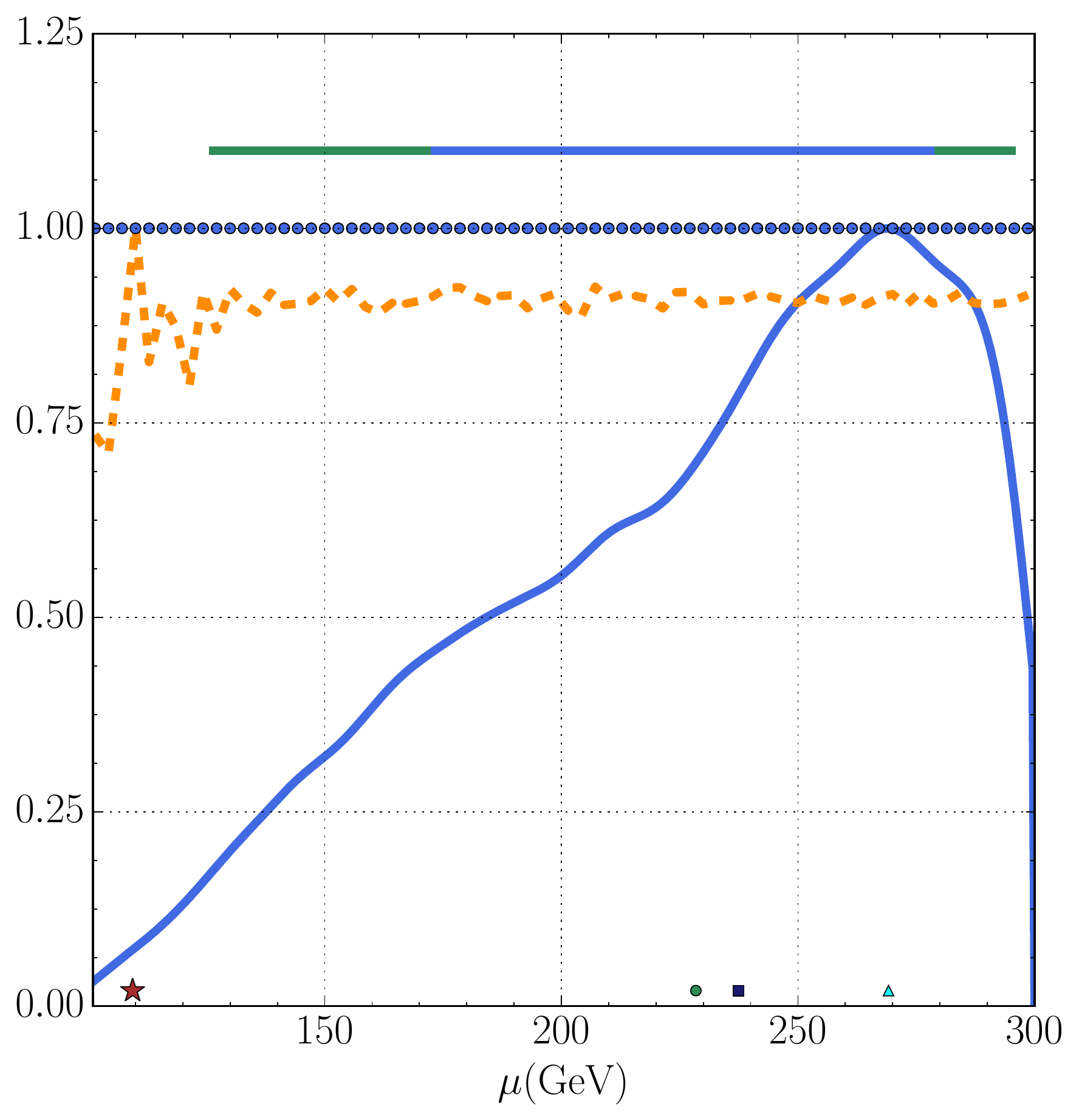}
		\includegraphics{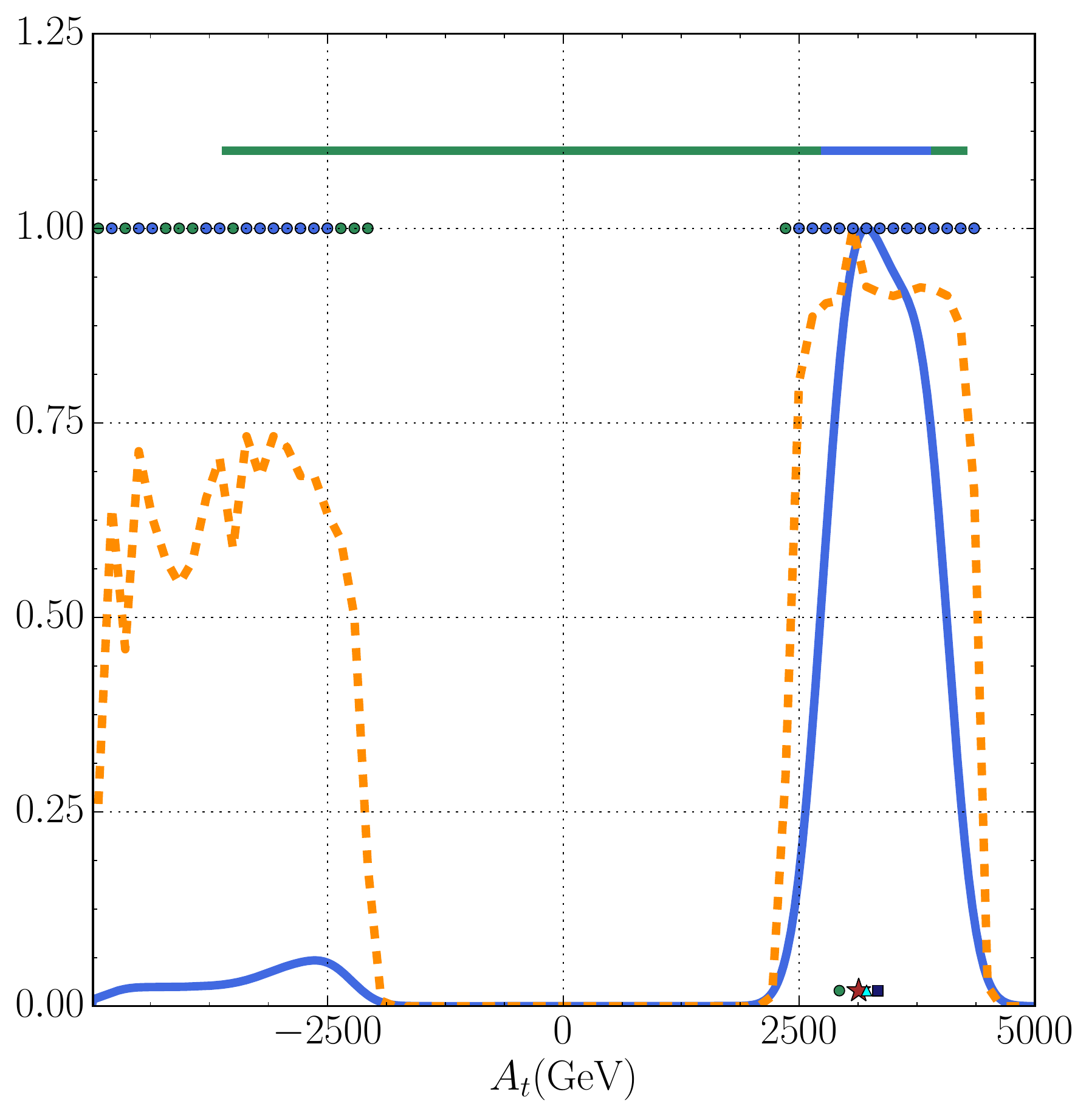}
		}
		\resizebox{0.7\textwidth}{!}{
		\includegraphics{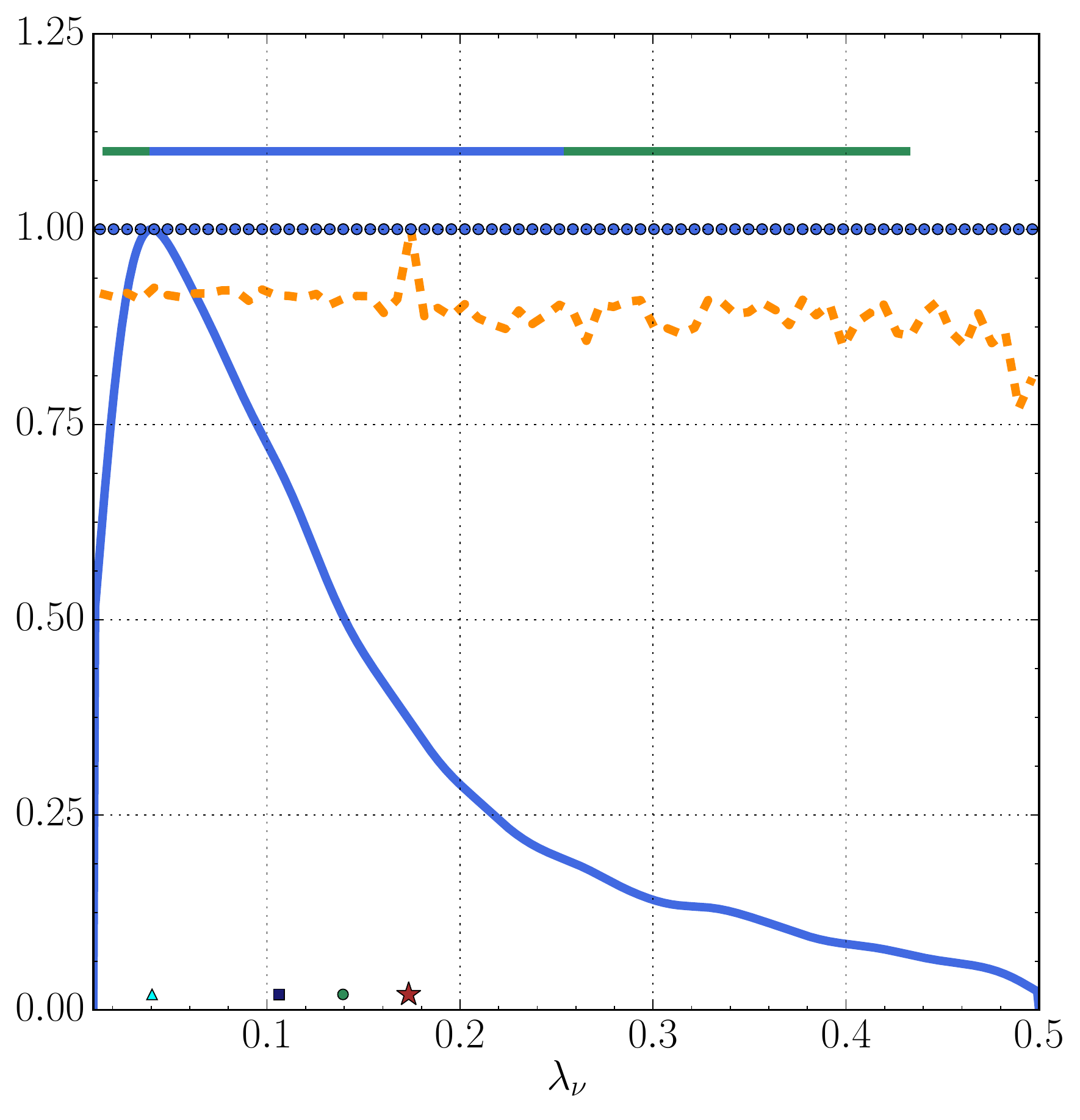}
		\includegraphics{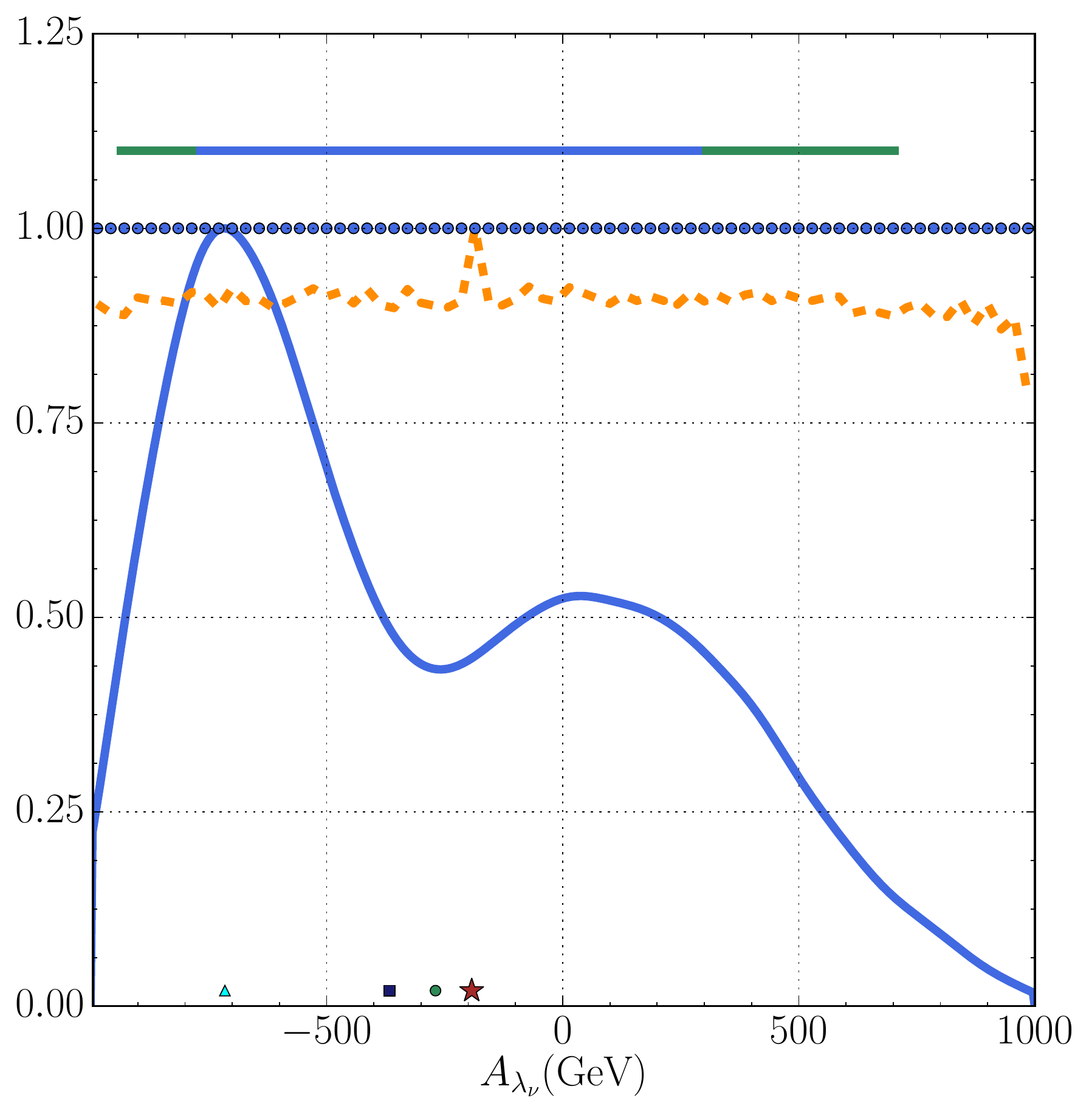}
		\includegraphics{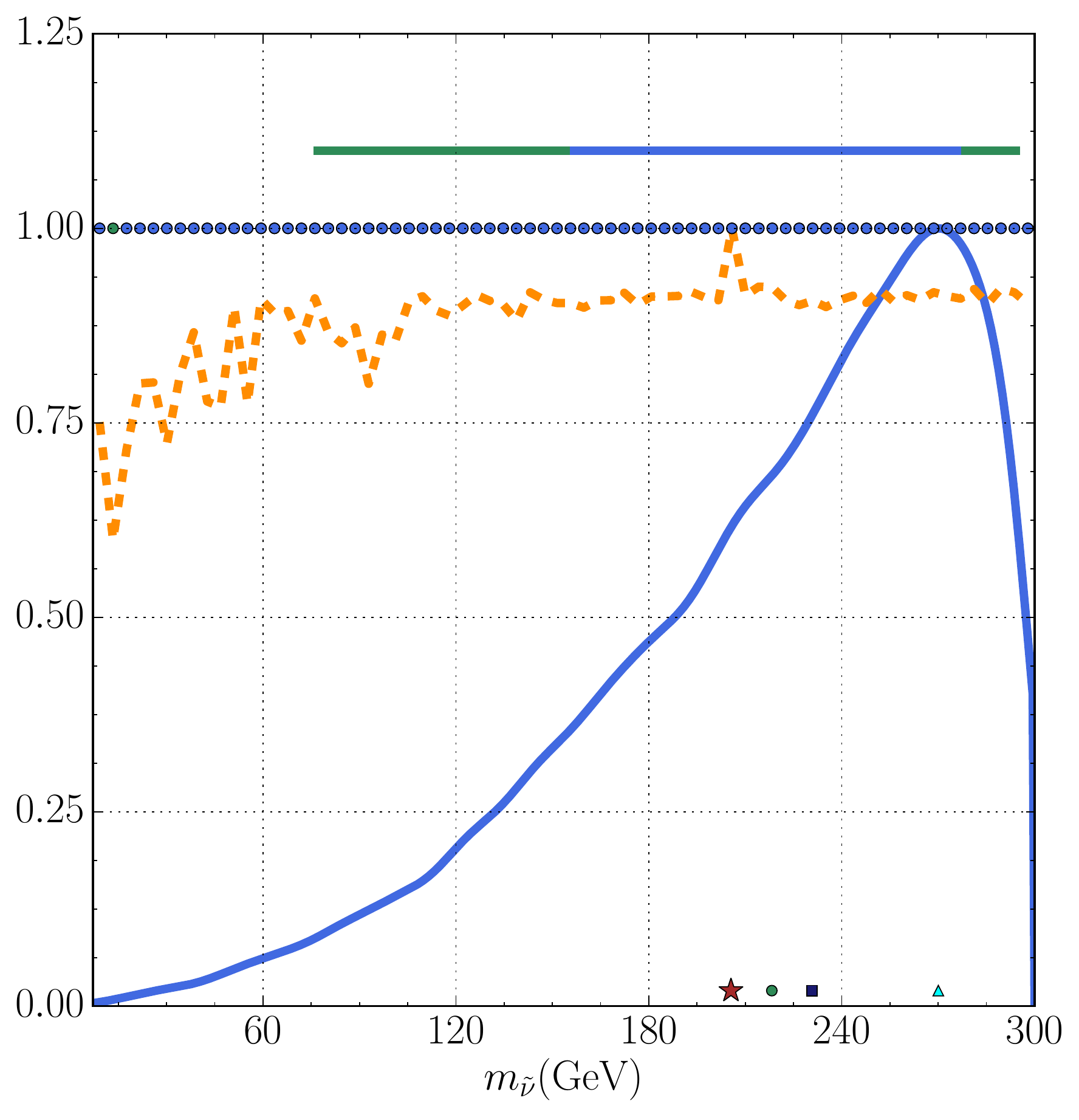}
		}

       \vspace{-0.4cm}

		\caption{One-dimensional marginal posterior PDFs and PLs for the parameters $\lambda$, $\kappa$, $\tan \beta$, $A_\kappa$, $\mu$, $\lambda_\nu$, $A_{\lambda_{\nu}}$
           and $m_{\tilde{\nu}}$ respectively. Credible regions, confidence intervals and other statistic quantities are also presented. \label{fig1} }
	\end{figure*}	

	\begin{figure*}[htbp]
		\centering
		\resizebox{0.8\textwidth}{!}{
		\includegraphics{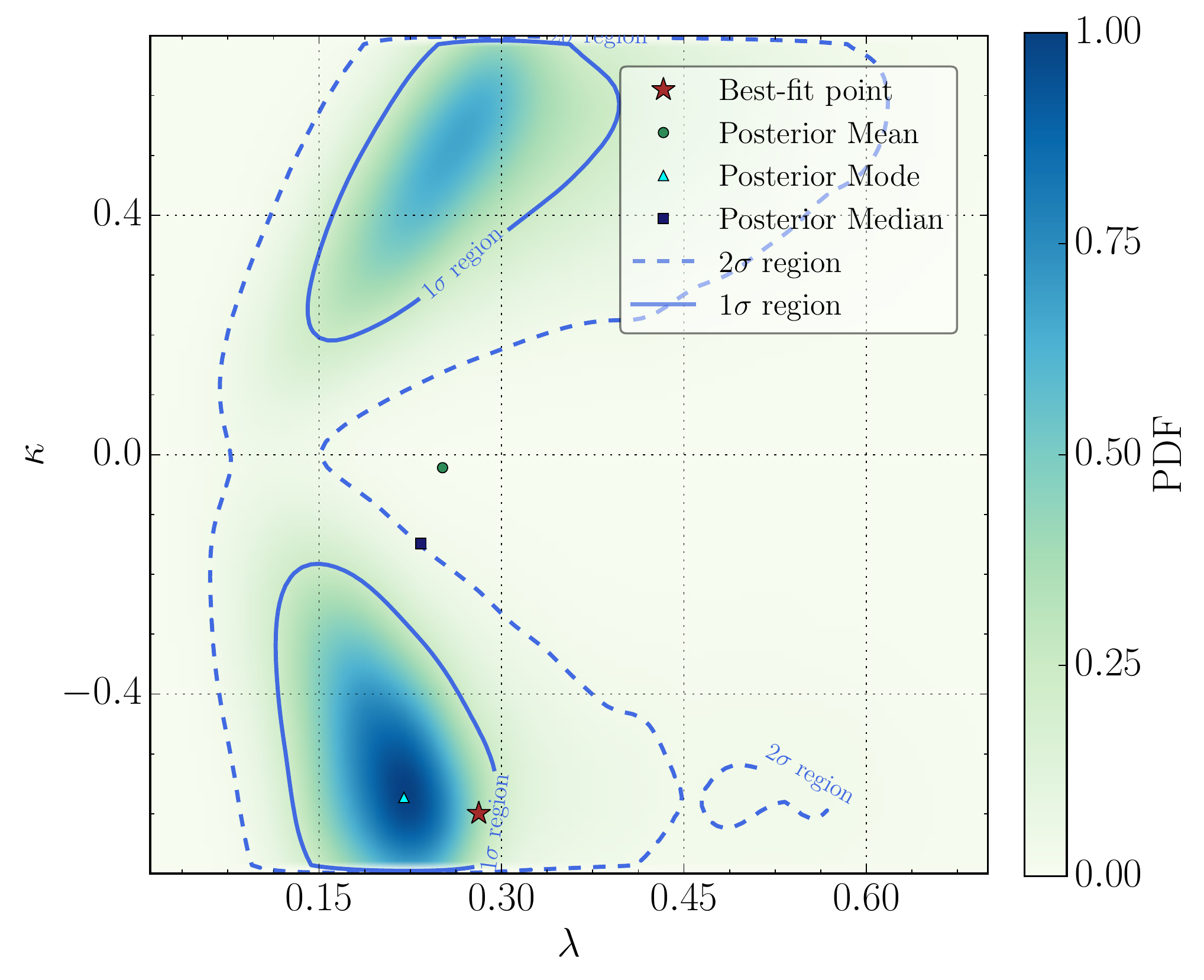}
		\includegraphics{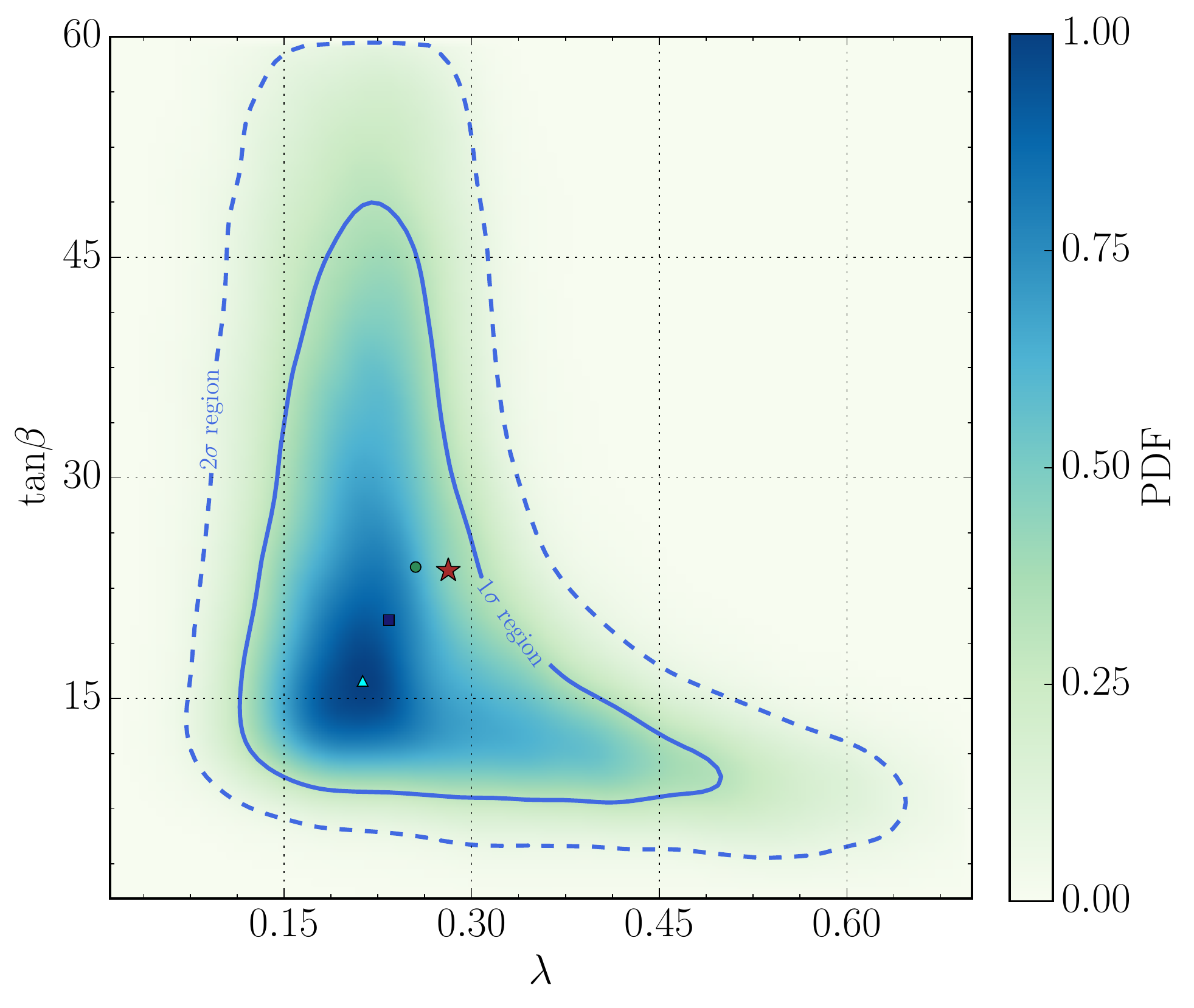}
		\includegraphics{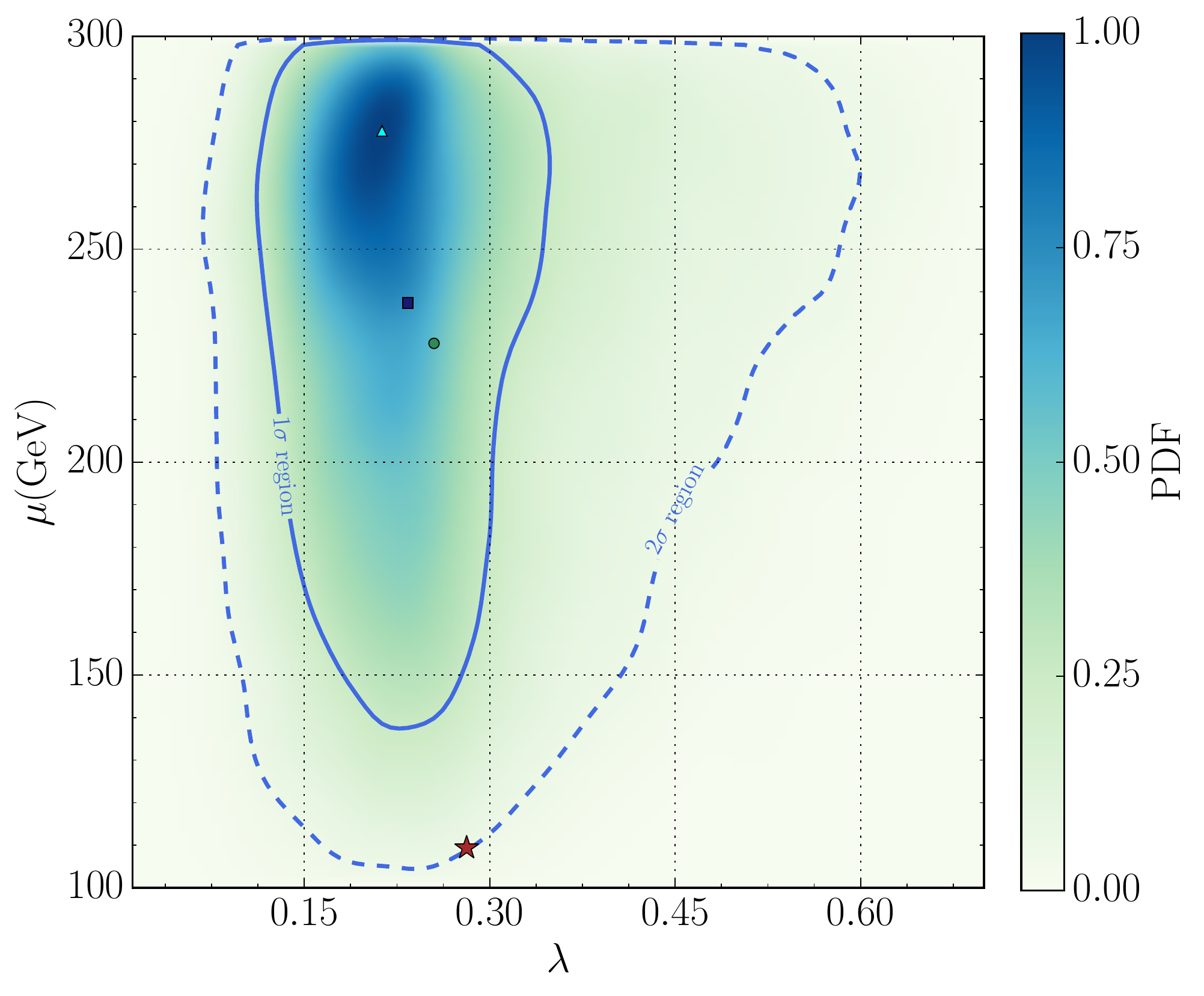}
		}
		\resizebox{0.8\textwidth}{!}{
		\includegraphics{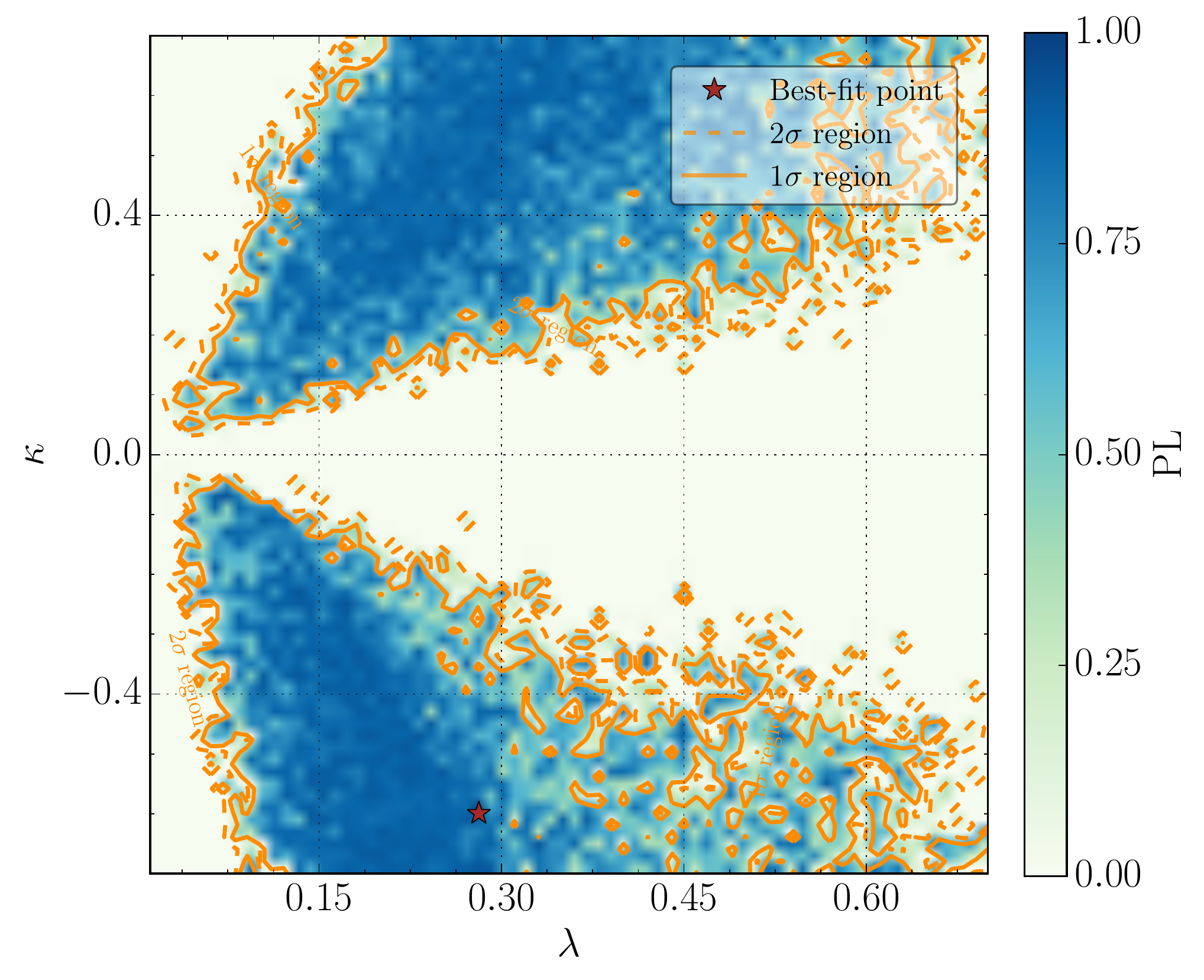}
		\includegraphics{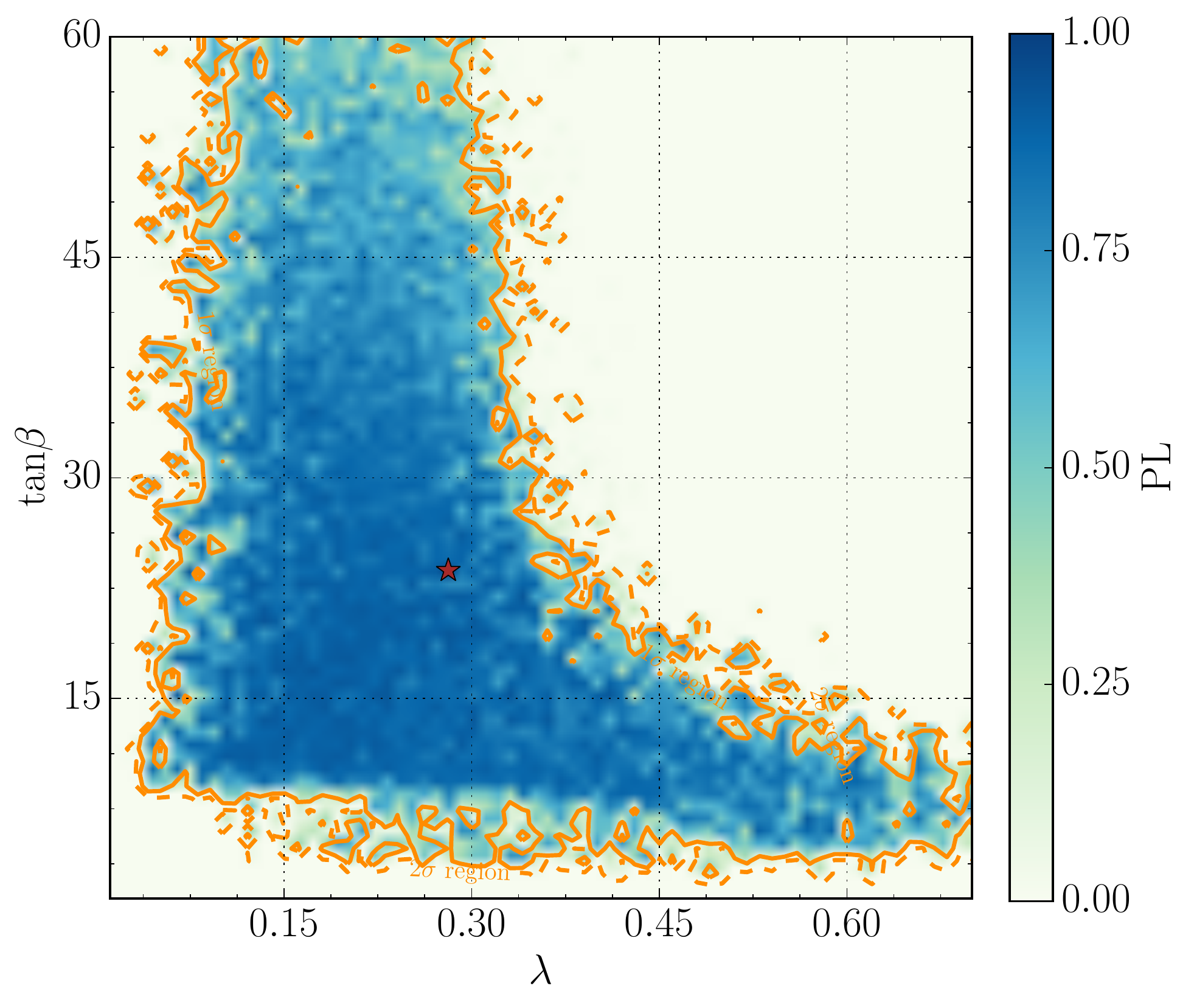}
		\includegraphics{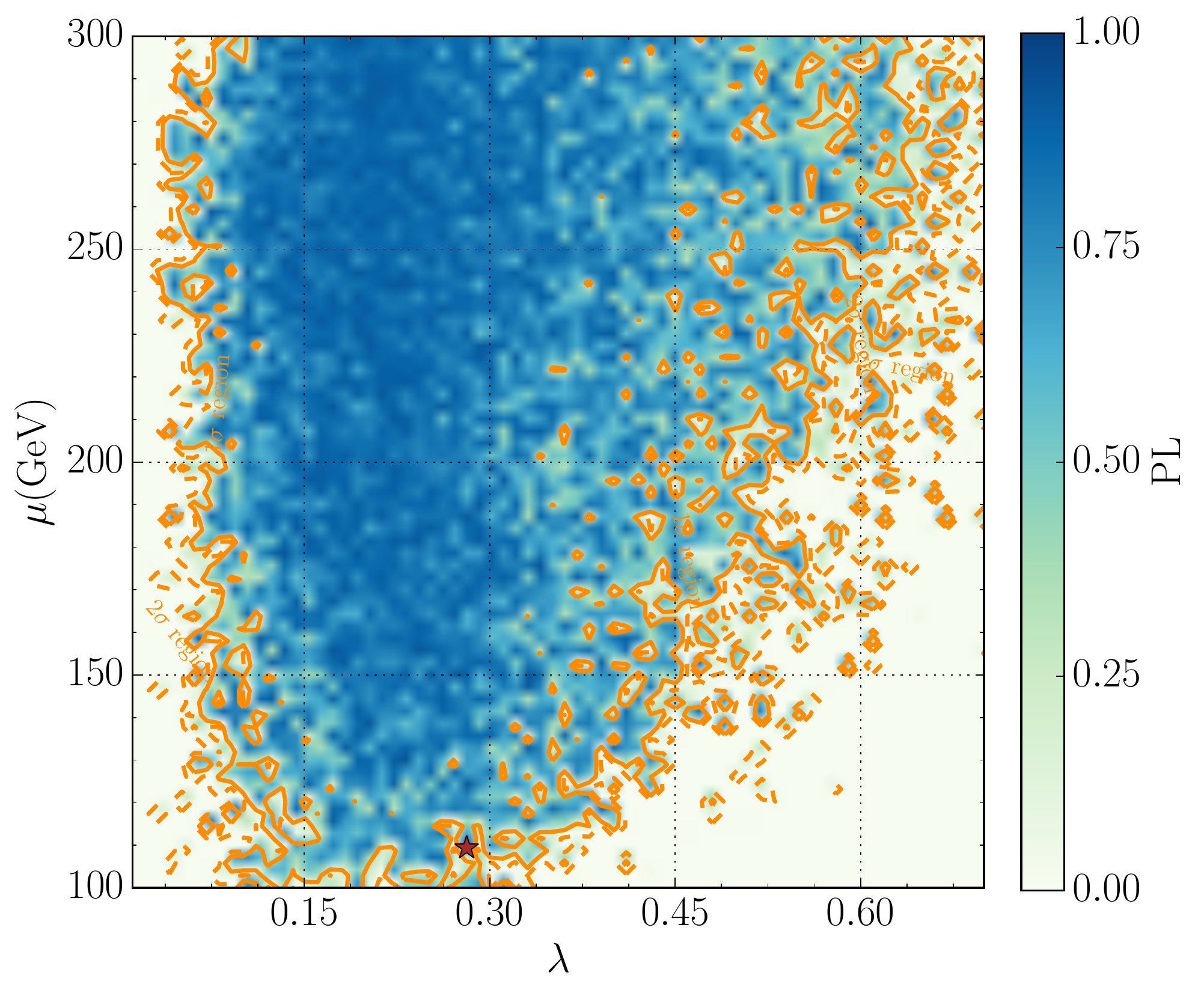}
		}

       \vspace{-0.4cm}

		\caption{ {\bf{Top panels:}} Two-dimensional marginal posterior PDFs, which are projected on $\kappa-\lambda$, $\tan \beta -\lambda$,
			and $\mu-\lambda$ planes respectively.  $1 \sigma$ and $2 \sigma$ credible regions together with the posterior mean,
			median and mode are also shown. {\bf{Bottom panels:}} Similar to the top panels, but for two dimensional profile likelihoods. The best fit
			point as well as $1 \sigma$ and $2 \sigma$ confidence intervals are also plotted. \label{fig2} }
	\end{figure*}

Given the posterior PDF and the likelihood function, one can obtain statistic quantities such as marginal posterior PDF and profile likelihood function (PL).
The marginal posterior PDF for a given set of parameters $(\Theta_A,\Theta_B,\cdots)$ is defined by integrating the posterior PDF $P(\Theta|M,D)$ in
Eq.(\ref{eq:bayes}) over the rest model parameters. For example, the one-dimensional (1D) and two-dimensional (2D) marginal posterior PDFs are given by

    \begin{widetext}	
	\begin{eqnarray}
	P(\Theta_A|D)&=&\int{P(\Theta |M,D)d\Theta_1 d\Theta_2 \cdots d\Theta_{A-1} d\Theta_{A+1} \cdots  \cdots },  \label{stat:Bayes} \\
	P(\Theta_A,\Theta_B|D)&=&\int{P(\Theta |M,D)d\Theta_1 d\Theta_2 \cdots d\Theta_{A-1} d\Theta_{A+1} \cdots d\Theta_{B-1} d\Theta_{B+1} \cdots }. \nonumber
	\end{eqnarray}
	\end{widetext}
	In practice, these PDFs are calculated by the sum of weighted samples in a chain with user-defined bins, and their densities as a function of $\Theta_A$ and $(\Theta_A,\Theta_B)$ respectively reflect the preference of the samples obtained in the scan. On the other hand, the frequentist PL is defined as the largest likelihood value in a certain parameter space. Take the 1D and 2D PLs as an example, we get them by the procedure
	\begin{small}
	\begin{eqnarray}
&&	\mathcal{L}(\Theta_A)=\mathop{\max}_{\Theta_1,\cdots,\Theta_{A-1},\Theta_{A+1},\cdots}\mathcal{L}(\Theta), \label{stat:Frequen} \\
&&	\mathcal{L}(\Theta_A,\Theta_B)=\mathop{\max}_{\Theta_1,\cdots,\Theta_{A-1},\Theta_{A+1},\cdots, \Theta_{B-1}, \Theta_{B+1},\cdots}\mathcal{L}(\Theta). \nonumber
	\end{eqnarray}
	\end{small}
Obviously, PL reflects the preference of a theory on the parameter space, and for a given point on $\Theta_A-\Theta_B$ plane, the value of $\mathcal{L}(\Theta_A,\Theta_B)$ represents the capability of the point in the theory to account for experimental data by varying the other parameters. In the following, we display our results by these quantities. We also use other statistic quantities, such as $1 \sigma$ and $2 \sigma$ credible regions (CRs) for the marginal PDF, $1 \sigma$ and $2 \sigma$ confidence intervals (CIs) for the PL, and posterior mean, median and mode, to illustrate the features of the sneutrino DM scenario. The definition of these quantities can be found in the appendix  of~\cite{Fowlie:2016hew},
	and we use the package Superplot~\cite{Fowlie:2016hew} with kernel density estimation to get them. Note that all the quantities depend on the parameter space in Eq.(\ref{scan-ranges}), which is inspired by the physics of the theory.

	\subsection{Favored parameter regions}

In this subsection, we discuss the favored parameter space of the scan for the case that the sneutrino DM is a CP-even scalar. In FIG.\ref{fig1}, we show the 1D marginal PDFs and
PLs for the parameters $\lambda$, $\kappa$, $\tan \beta$, $A_\kappa$, $\mu$, $A_t$, $\lambda_\nu$, $A_{\lambda_\nu}$ and
$m_{\tilde{\nu}}$ respectively. We also present 2D results in FIG.\ref{fig2} on $\kappa-\lambda$, $\tan \beta -\lambda$, and $\mu-\lambda$ planes with color bar
representing marginal posterior PDF for the top panels and PL for the bottom panels.  The $1\sigma$ and $2\sigma$ CRs, the $1\sigma$ and $2\sigma$ CIs
	together with the best fit point, posterior means, medians and modes are also shown in these panels. From FIG.\ref{fig1}, one can learn following features:

	\begin{itemize}
        \item The marginal PDF of $\lambda$ is peaked around $0.2$ and its $1\sigma$ CR corresponds to the region $0.15 \lesssim \lambda \lesssim 0.37$. The underlying reason for this conclusion is that the parameter $\lambda$ affects both the mass and the couplings of the SM-like Higgs boson with the other SM particles~\cite{Cao:2012fz}. A large $\lambda$ tends to enhance the singlet component in the SM-like Higgs boson (see the expression of ${\cal M}^2_{23}$ in Eq.(\ref{Mass-CP-even-Higgs}) of Section II and also the discussion in Appendix A), and is thus disfavored by the LHC Higgs data\footnote{In fact, as one can learn from the results in \cite{Cao:2012fz}, the parameter space of the NMSSM is rather limited for the large $\lambda$ case to predict a SM-like Higgs boson with mass around $125 {\rm GeV}$.}. On the other hand, a small $\lambda$ is also tightly limited since it can enhance the $\tilde{\nu}_1 \tilde{\nu}_1 h_i$ coupling (see Eq.(\ref{Csnn}) in Section II), and consequently the rate of the DM-nucleon scattering.

            In the second part of Appendix A, we show the impact of different experimental data on various marginal PDFs and PLs with the results presented from FIG.\ref{fig14} and FIG.\ref{fig15} and summarized in Table \ref{table4}. The marginal PDF of $\lambda$ in FIG.\ref{fig14} reveals that it is mainly determined by the Higgs and DM observables, which verifies our conjectures.

        \item The posterior PDFs of $\kappa$ and $A_\kappa$ are maximized around $-0.6$ and $200 {\rm GeV}$ respectively, and the PDFs show an approximate reflection symmetry. Since the underlying reason for the behavior is somewhat complicated, we  will discuss it in an elaborated way in Appendix A and B.

		\item A relatively large $\tan \beta$, i.e. $ 10 \lesssim \tan \beta \lesssim 38$ for $1 \sigma$ CR, is preferred by the marginal PDF. This is due to at least two facts.
         One is that, since $\lambda$ is usually less than about $0.37$ so that its contribution to the squared mass of the SM-like Higgs boson is sub-dominant, a large $\tan \beta$ is helpful to enhance the tree level MSSM prediction of the mass~\cite{Ellwanger:2009dp}. The other is that the heavy doublet dominated Higgs bosons in our model contribute to the process $B_s \to \mu^+ \mu^-$, and the effect can be enhanced by a factor of $\tan^6 \beta$~\cite{Bobeth:2001sq}. So a too large $\tan \beta$ is not favored by B physics.  We note that these two facts are actually reflected in FIG.\ref{fig14}, where we investigate the effect of various observables on the shape of the PDF and PL for $\tan \beta$.

        \item A moderately large $\mu$ ranging from about $170 {\rm GeV}$ to $280 {\rm GeV}$ for $1 \sigma$ CR is favored by the posterior PDF. The reason is twofold. One is that due to the specific choice of $A_\lambda$ and the third generation squark masses, Higgs physics and B physics observables are insensitive to this parameter, and consequently $\mu$ is evenly distributed in the range from about $100 {\rm GeV}$ to $300{\rm GeV}$ (We will discuss this issue in detail in Appendix A, B and C). The other reason is that since the sneutrino DM tends to coannihilate with the Higgsinos to get its right relic density, which requires their mass splitting to be less than about $10\%$ (see FIG.\ref{fig3} below), a large $\mu$ provides a broader parameter space for the annihilation.

            In Appendix C of this work, we will show that any modification of the scan range in Eq.(\ref{scan-ranges}) or the setting of $A_\lambda$ may alter the marginal PDF of $\mu$, and in some cases moderately low values of $\mu$ are preferred.

         \item $|A_t| \gtrsim 2 {\rm TeV}$ is preferred to enhance the SM-like Higgs boson mass by large radiative correction from stop loops~\cite{Cao:2012fz}. Meanwhile a positive $A_t$ is more favored by B physics since stop-chargino loops can mediate the transition of bottom quark to strange quark. These reasons are illustrated in the third row of FIG.\ref{fig15}.

		\item For each input parameter, the range covered by the $1\sigma$ CR is significantly less than that of the $1 \sigma$ CI. This difference is caused by the fact that the
         marginal PDF defined in Eq.(\ref{stat:Bayes}) of this section is determined not only by the likelihood function, but also by the parameter space. By contrast, the PL is affected only by the likelihood function.

        In practice, one usually encounters the phenomenon that with the increase of the setting {\it nlive} in the MultiNest algorithm, the predicted PLs increase so that they become more or less constants over the whole scanned parameter range (see for example the $\kappa \sim 0$ region of the PL in the third row of FIG.\ref{fig16} in Appendix B). The reason is that, with a low setting of {\it nlive}, the scan is not elaborate enough so that it misses the fine tuning cases where the theoretical prediction of a ceratin physical observable approaches to its measured value by strong cancellation of different contributions or by other subtle mechanisms. This situation can be improved by increasing the sampling in the scan\footnote{Note that for the same reason, the best point in a scan also depends on the setting {\it nlive}. In general, the larger {\it nlive} one takes, the more elaborated the scan becomes. Consequently, parameter points with larger likelihood values may be found. }, but it can not change the fact that the case is rare, and hence the corresponding posterior PDF is suppressed. In a word, the volume effect usually makes the marginal PDFs and corresponding PLs quite different.
	\end{itemize}

	\begin{figure*}[htbp]
		\centering
		\resizebox{0.8\textwidth}{!}{
		\includegraphics{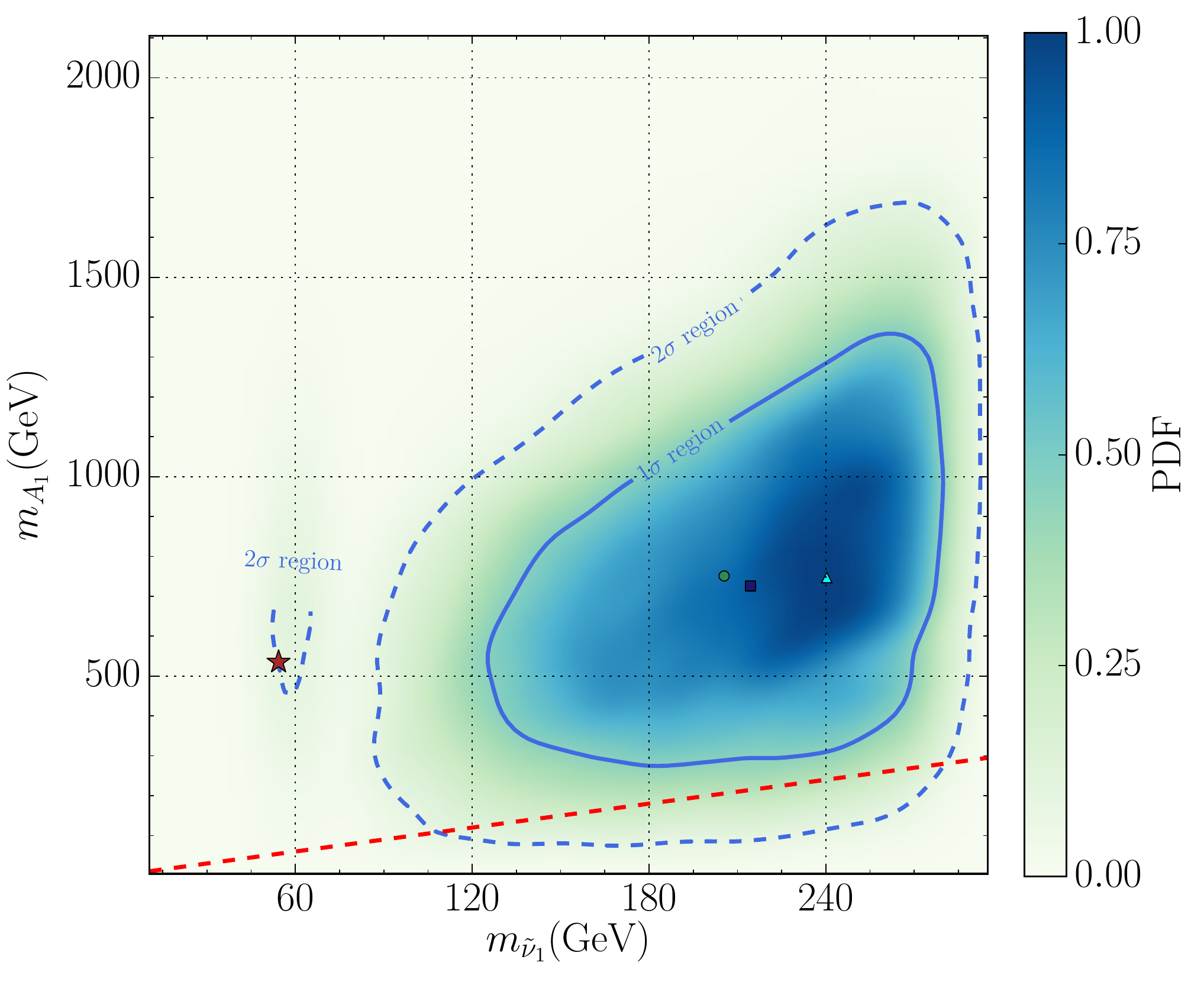}
		\includegraphics{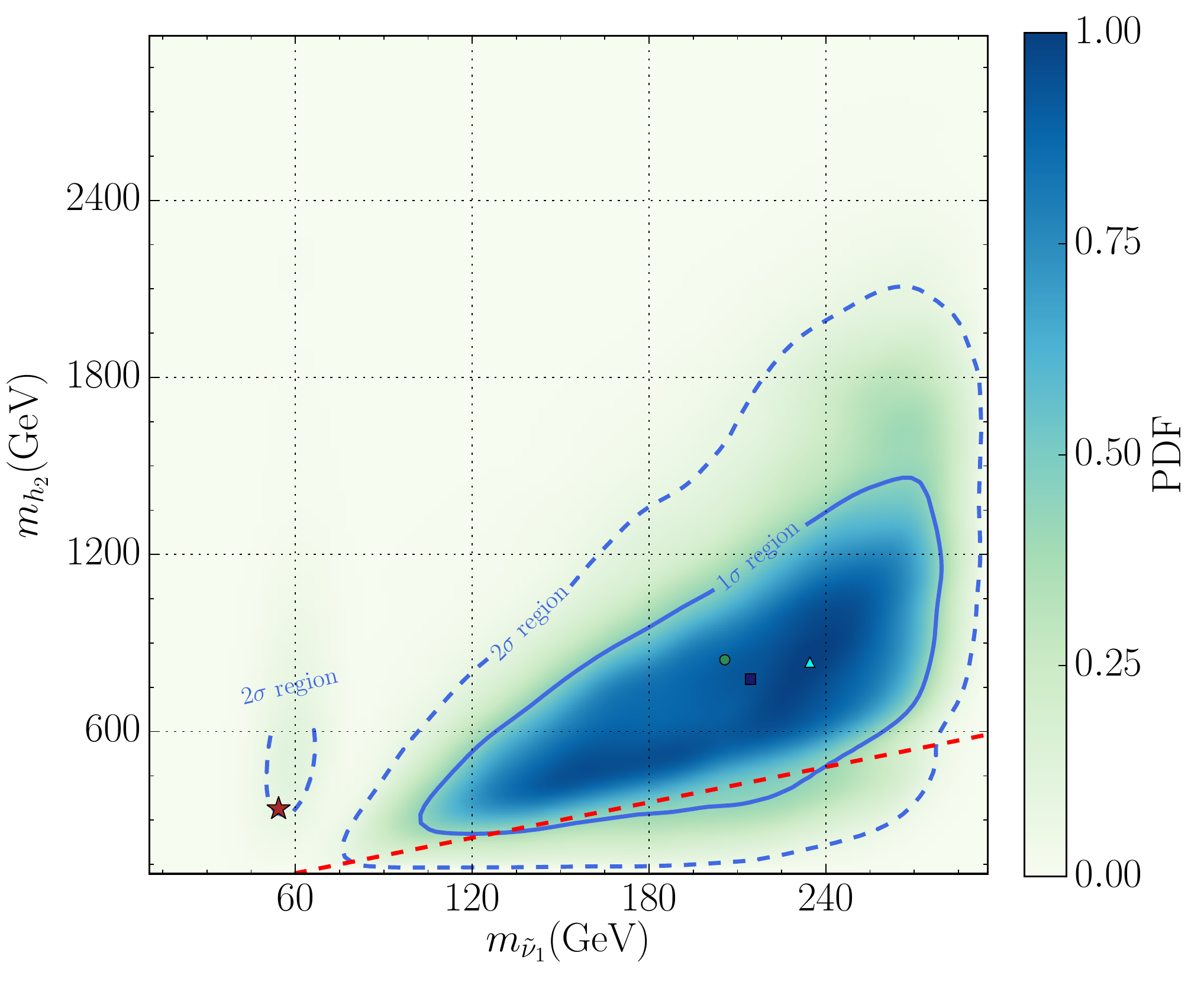}
		\includegraphics{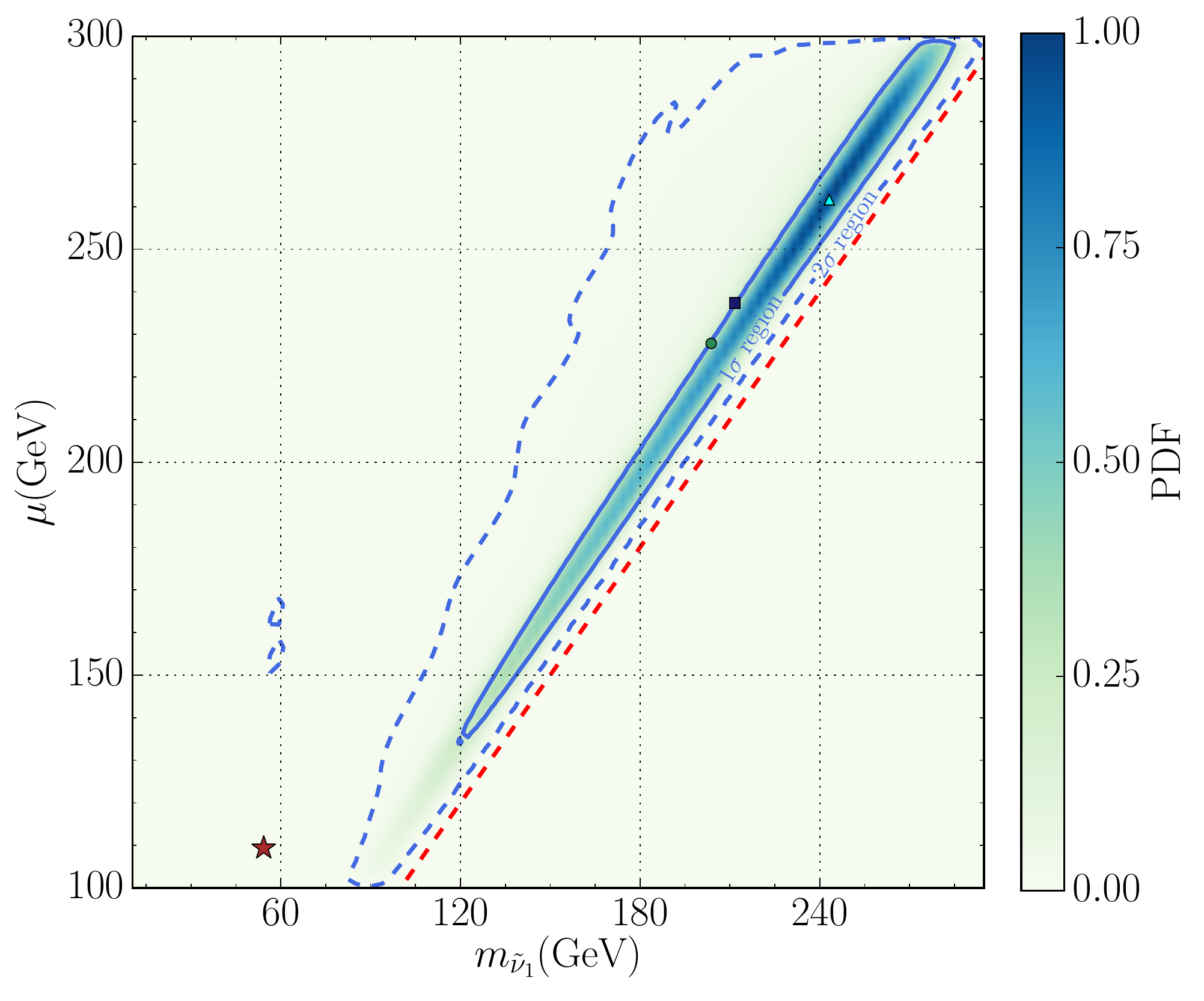}
		}
		\resizebox{0.8\textwidth}{!}{
		\includegraphics{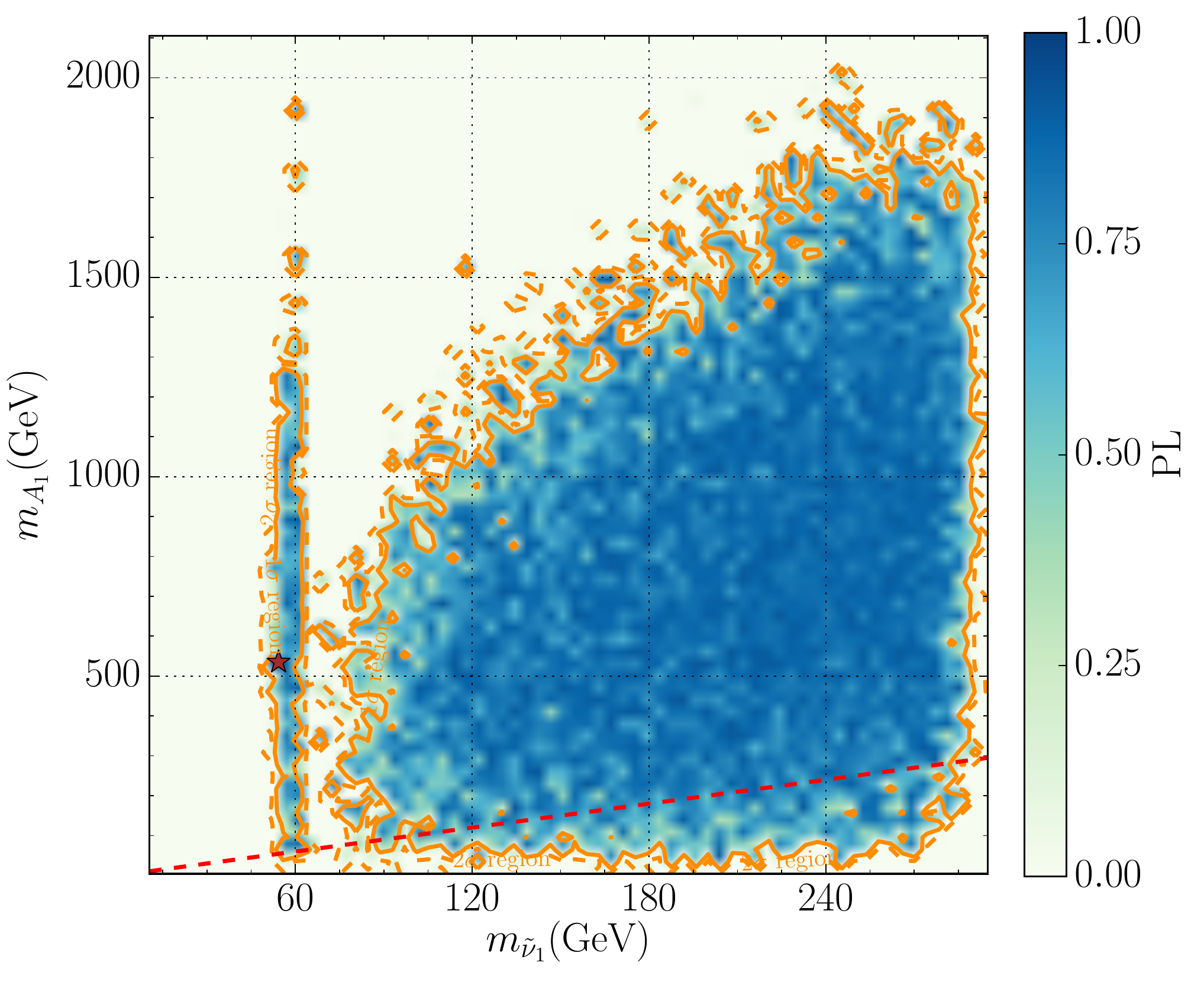}
		\includegraphics{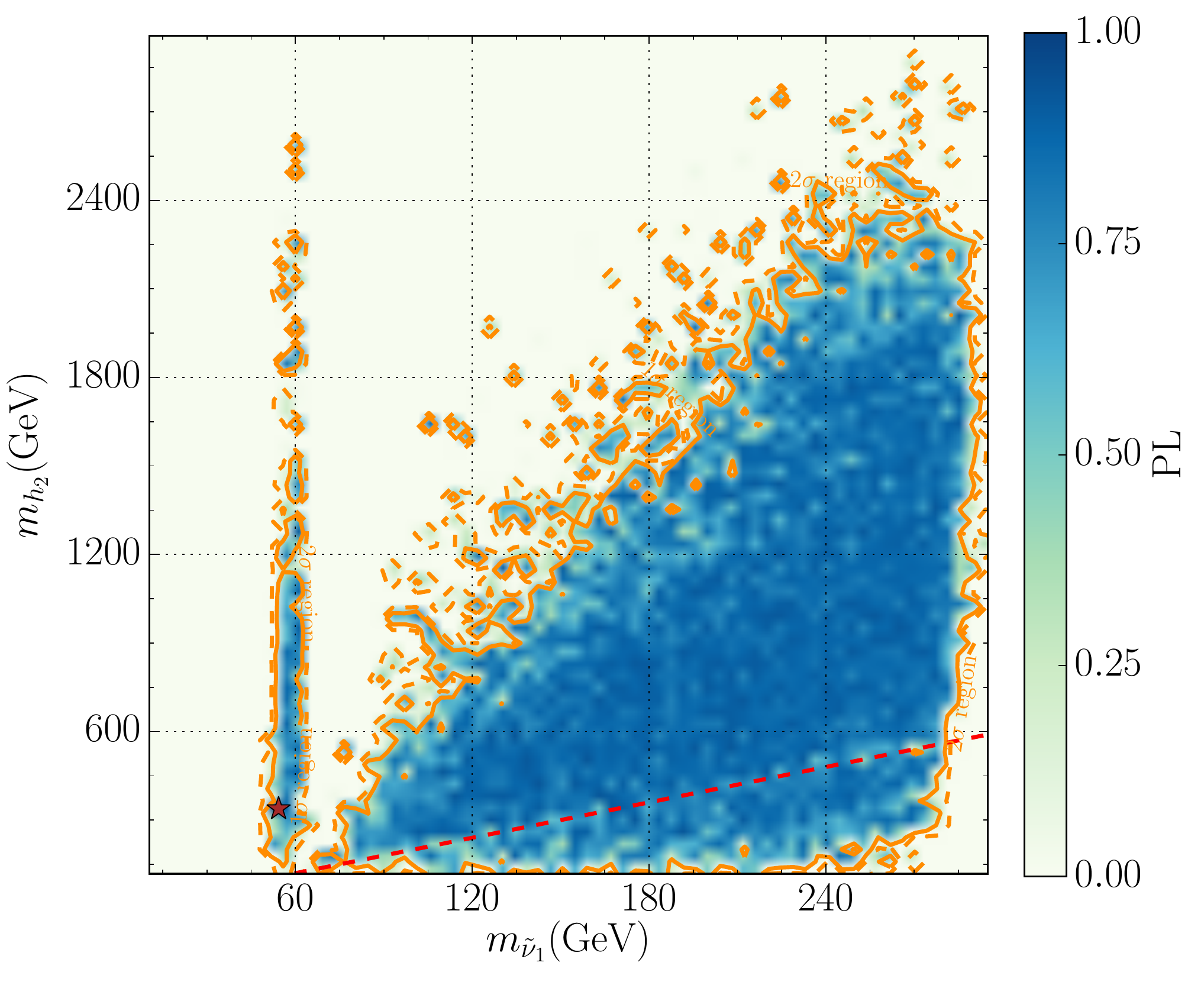}
		\includegraphics{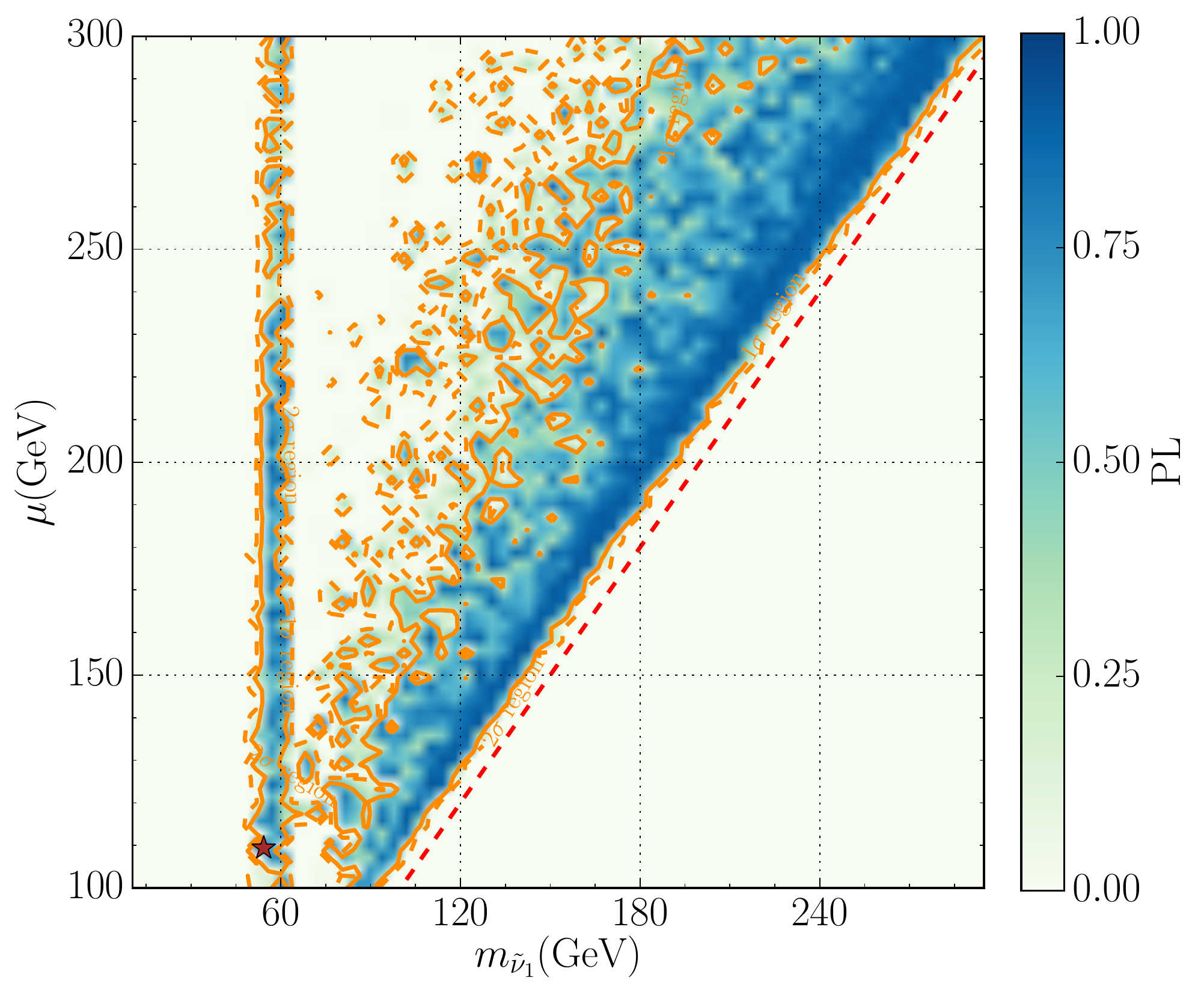}
		}	

       \vspace{-0.4cm}

		\caption{Similar to FIG.\ref{fig2}, but projected on $m_{A_1}-m_{\tilde{\nu}_1}$,  $m_{h_2}-m_{\tilde{\nu}_1}$ and  $\mu-m_{\tilde{\nu}_1}$ planes
			with the red dashed lines corresponding to the cases of $m_{A_1} = m_{\tilde{\nu}_1}$,  $m_{h_2} = 2 m_{\tilde{\nu}_1}$
            and  $\mu = m_{\tilde{\nu}_1}$, respectively. \label{fig3}}
	\end{figure*}

	\begin{figure*}[tbp]
		\centering
		\resizebox{0.58\textwidth}{!}{
		\includegraphics{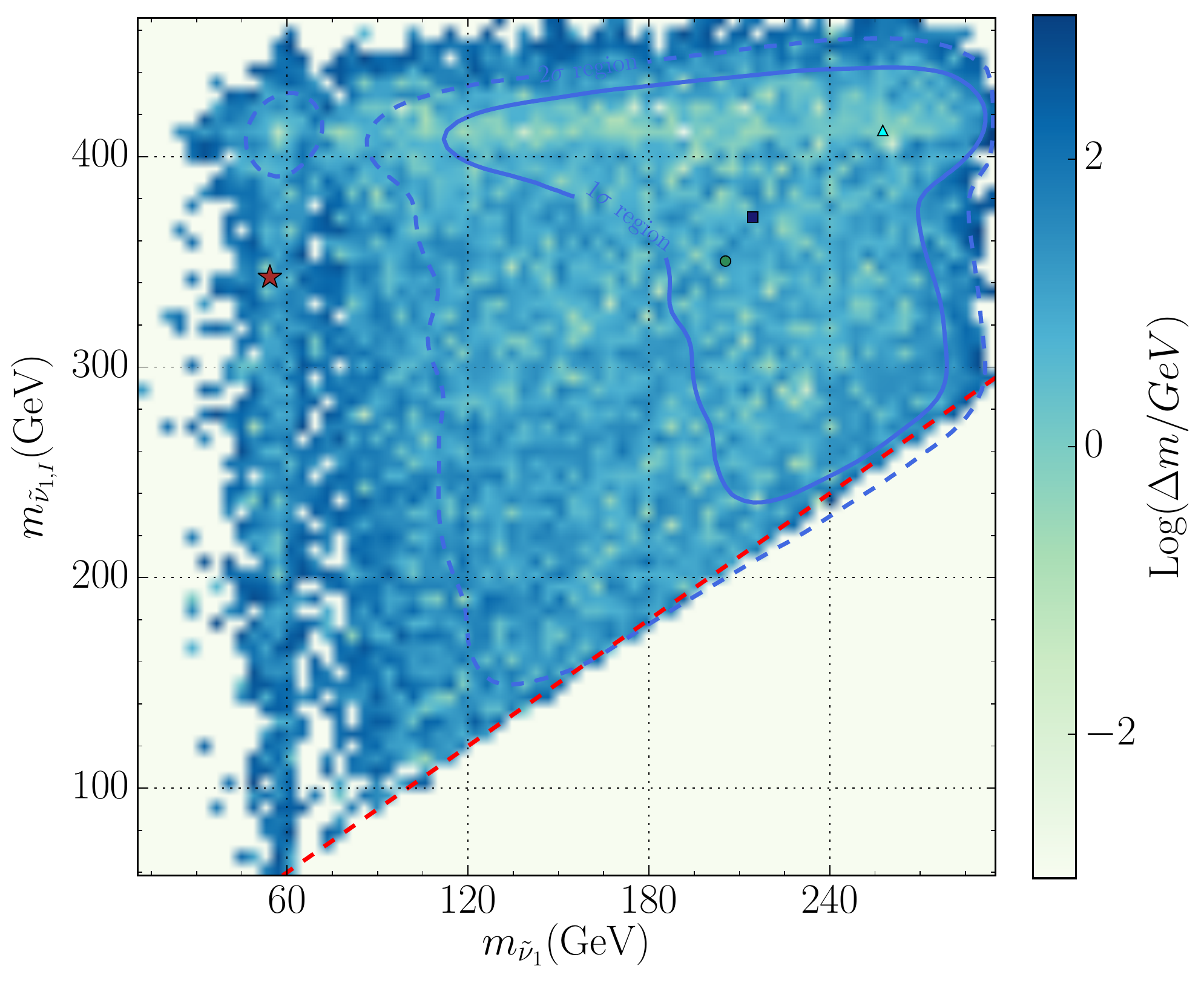}
		\includegraphics{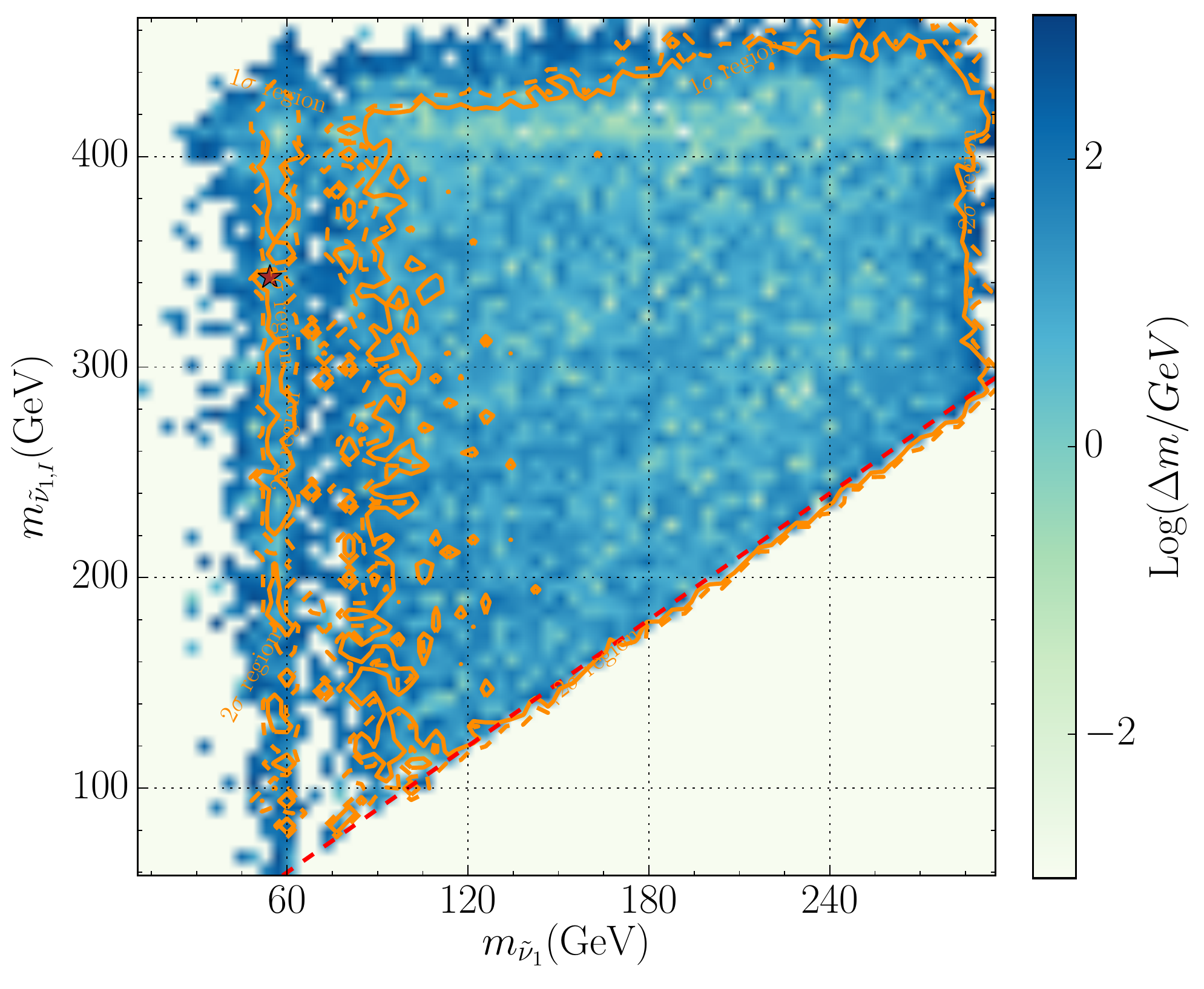}
		}

       \vspace{-0.4cm}

		\caption{The distribution of the mass splitting $\Delta m$ projected on $ m_{\tilde{\nu}_{1, I}} - m_{\tilde{\nu}_1}$ plane, which is split into $70 \times 70 $
          equal boxes, and $\Delta m$ in each box is defined by $\Delta m \equiv {\bf{min}} ( | m_{A_1} - m_{\tilde{\nu}_1} - m_{\tilde{\nu}_{1, I}} |)$ for the
          samples allocated in the box. Different credible and confidence boundaries are also plotted for the left and right panel, respectively, with the red dashed lines corresponding to the case of $ m_{\tilde{\nu}_{1, I}} = m_{\tilde{\nu}_1}$.   \label{fig4}}
	\end{figure*}

FIG.\ref{fig2} shows the correlation of $\lambda$ with the other parameters. From this figure, one can see that the $1\sigma$ and $2\sigma$ CIs on $\tan \beta -\lambda$ plane roughly overlap. One reason for this is that the observables considered in this work, especially the SM-like Higgs boson mass, are sensitive to the two parameters, and so is the likelihood function in Eq.(\ref{Likelihood}) of this section. Consequently, any shift of the parameters can alter the likelihood value significantly. One can also see that the $2 \sigma$ CIs in the figure are usually isolated. This is because the marginal PDFs in these regions are relatively small so that fewer samples in the scan concentrate on the regions. We checked that, with the increase of the {\it nlive}, the regions will be connected and both the $1\sigma$ and $2 \sigma$ CIs usually expand since the improved scan can survey more fine tuning cases. In the second part of Appendix C in this work, we show the impact of different {\it nlive}s on the CIs.

Before we end this subsection, we want to emphasize two points. One is that the posterior PDFs and the PLs are determined by the total likelihood function, not by any individual contribution. For a good theory, the optimal parameter space for one component of the likelihood function should be compatible with those for the other components of the function so that the Bayesian evidence of the theory is not suppressed. In Appendix A, we show the impacts of different experimental measurements on the distribution of the parameters in Higgs sector. By comparing the figures presented in this Appendix, one can learn that the Higgs observables play an important role in determining most of the results, and the other observables are concordant in selecting favorable parameter space of the Type-I seesaw extension. This implies that once the mass spectrum in the Higgs sector is determined, one can adjust the parameters in the sneutrino sector to be consistent with the DM measurements. This is a simplified way to get good parameter points in the model. The other is that the posterior marginal PDFs are affected by the choice of the prior distribution for the input parameters, and they may be quite different if one compares the results of two distinct prior PDFs with insufficient sampling in the scan. In Appendix B, we investigate the difference induced by the flat distribution adopted in this work and the log distribution
usually used in the scan by Markov chain method. Our study shows that although the predictions of the log distribution are stable with the increase of the setting {\it nlive}, the prior distribution overemphasizes the low $|A_\kappa|$ and low $\mu$ regions, and thus deforms the shape of the posterior PDFs from what the underlying physics predicts.
By contrast, the flat distribution is able to predict the right results, but at the cost of a long time calculation by clusters when one sets a large {\it nlive}.

	\begin{figure*}[htbp]
		\centering
		\resizebox{0.58\textwidth}{!}{
		\includegraphics{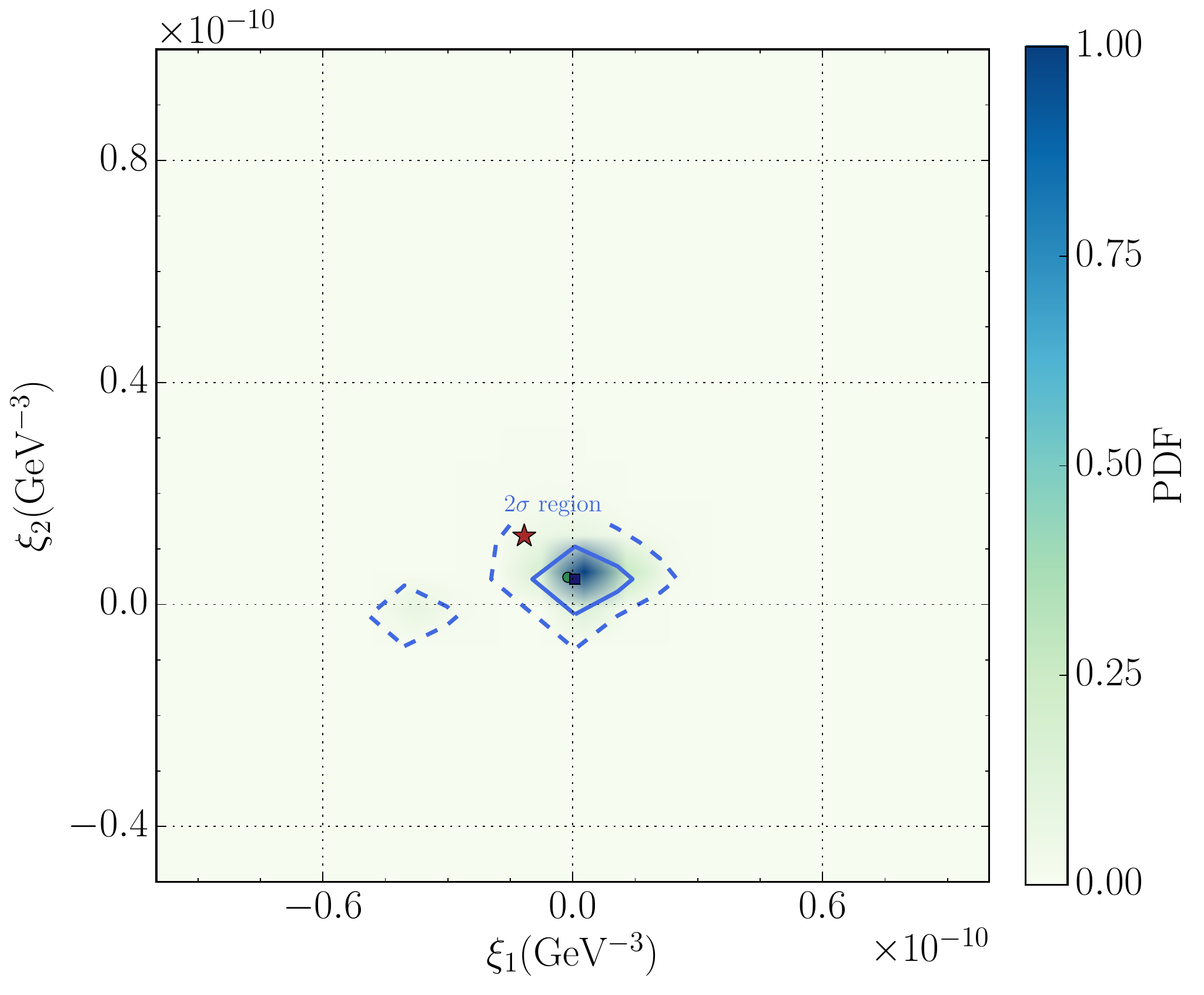}
		\includegraphics{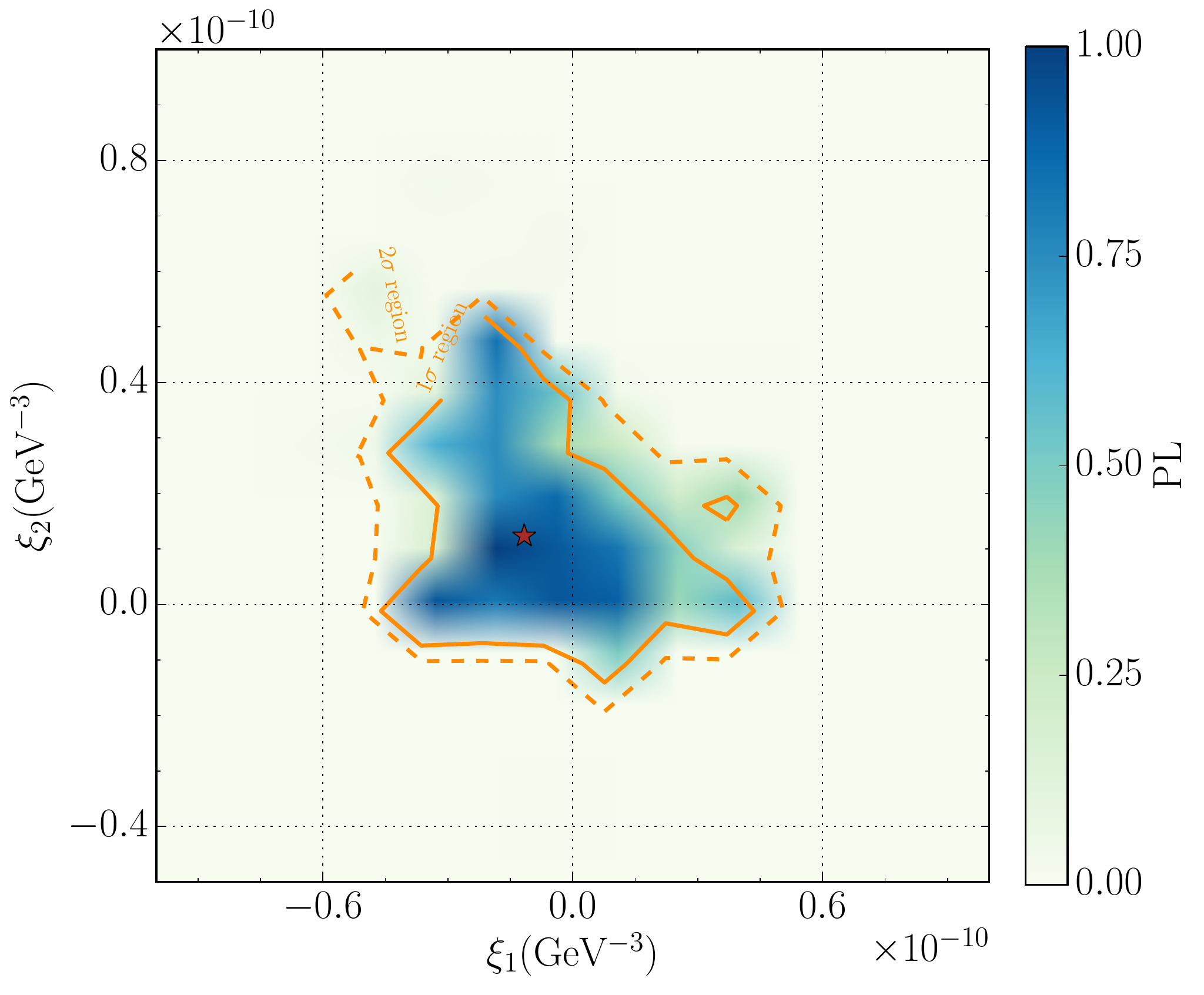}
		}
		\resizebox{0.58\textwidth}{!}{
		\includegraphics{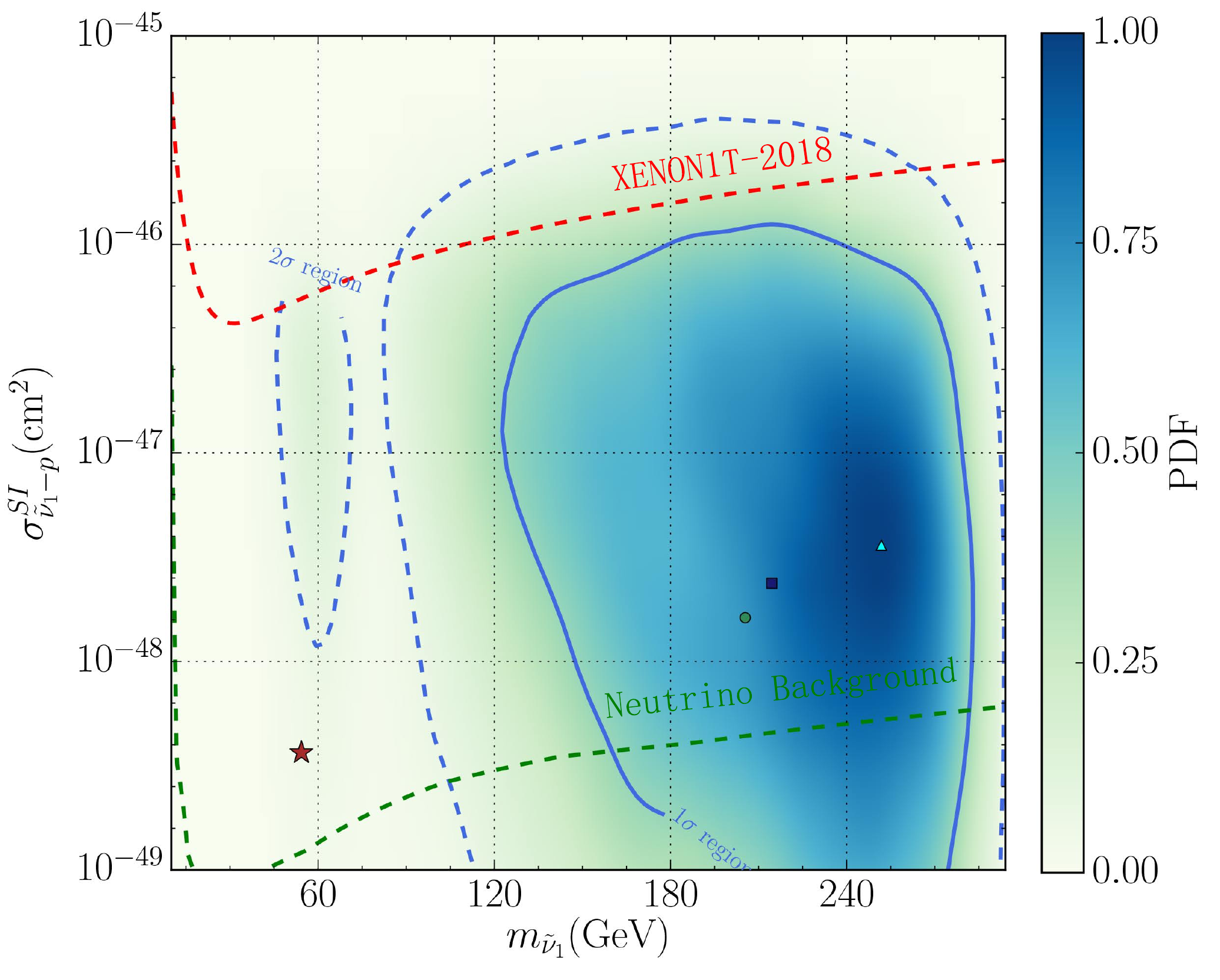}
		\includegraphics{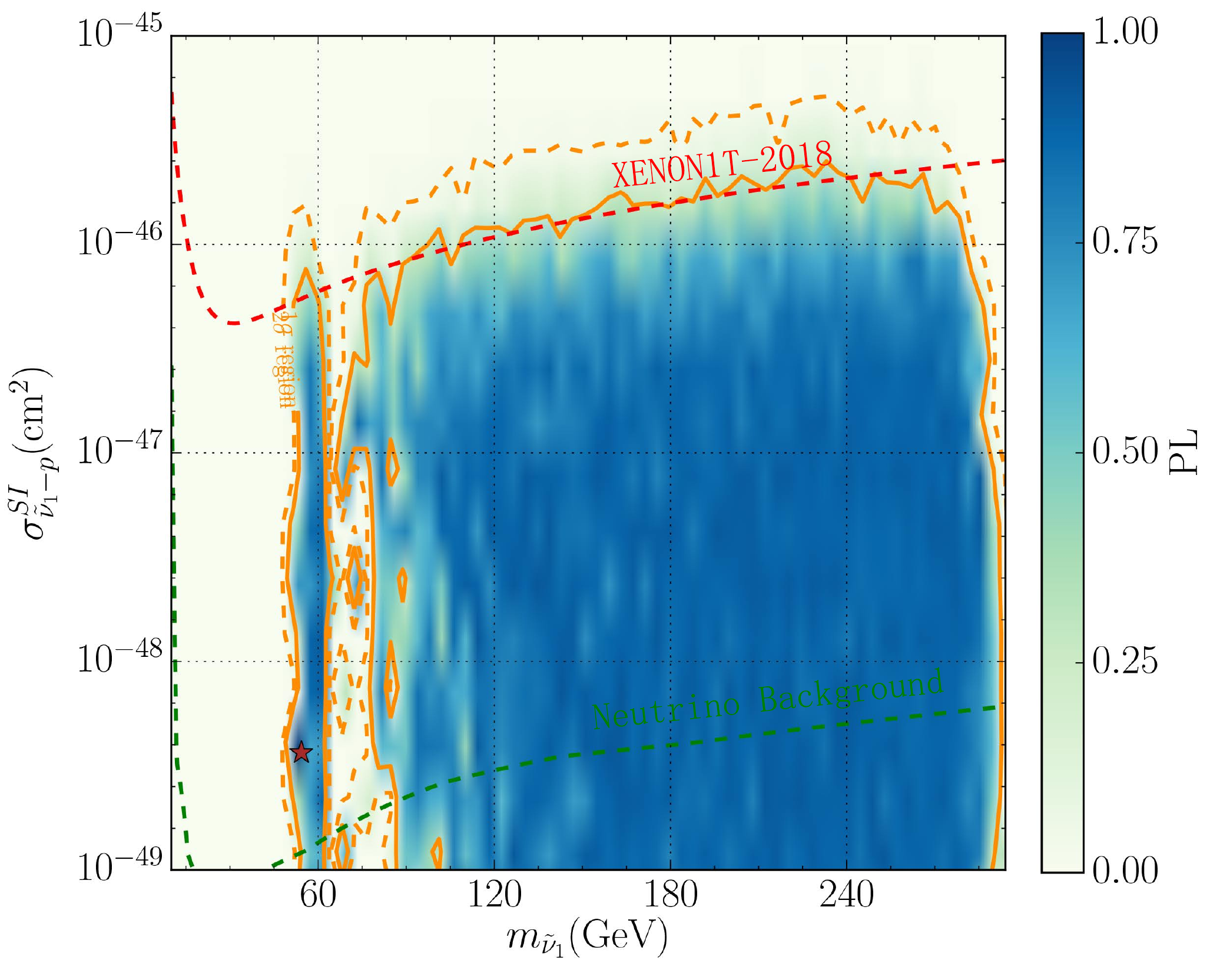}
		}

       \vspace{-0.4cm}

		\caption{Similar to FIG.\ref{fig2}, but projected on $\xi_2 - \xi_1$ plane (top panel) and on $\sigma_{{\tilde{\nu}_1}-p}^{SI}-m_{\tilde{\nu}_1}$ plane (bottom panel). The
         red line in the bottom panels denotes the bound from the latest XENON-1T experiment, and the green line represents the neutrino floor.
			\label{fig5}}
	\end{figure*}
	
		\begin{figure*}[htbp]
		\centering
		\resizebox{0.58\textwidth}{!}{
		\includegraphics{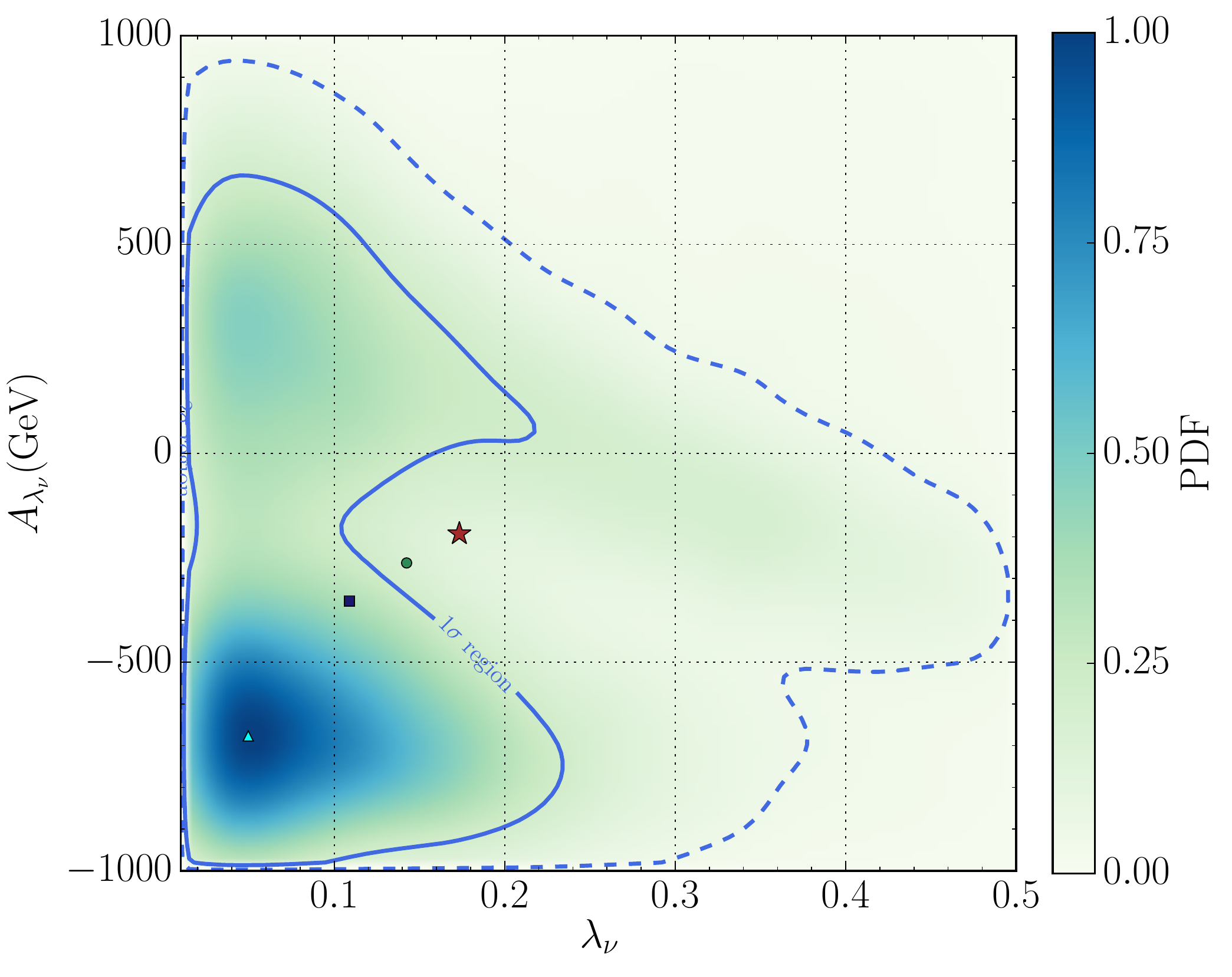}
		\includegraphics{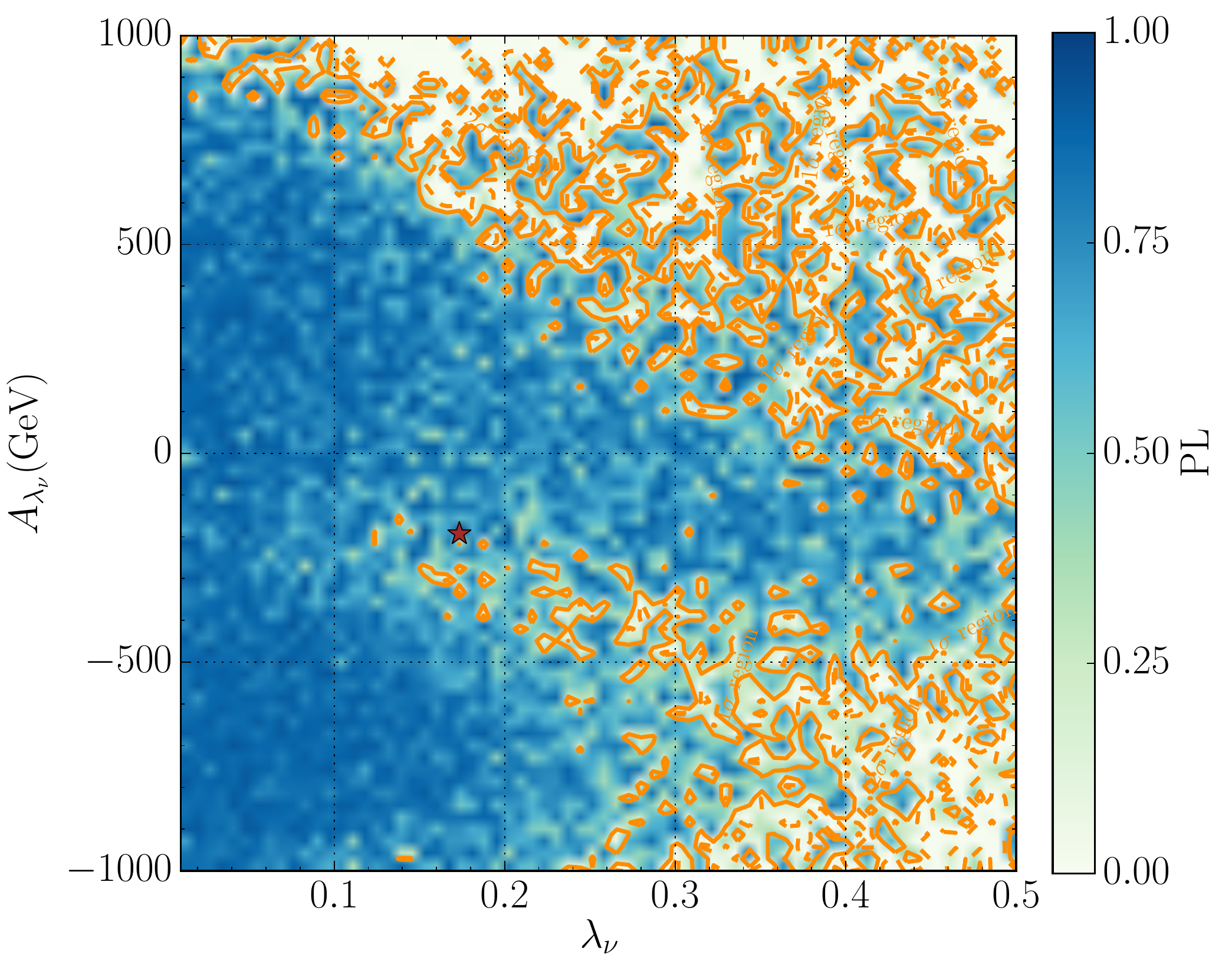}
		}
		\resizebox{0.58\textwidth}{!}{
		\includegraphics{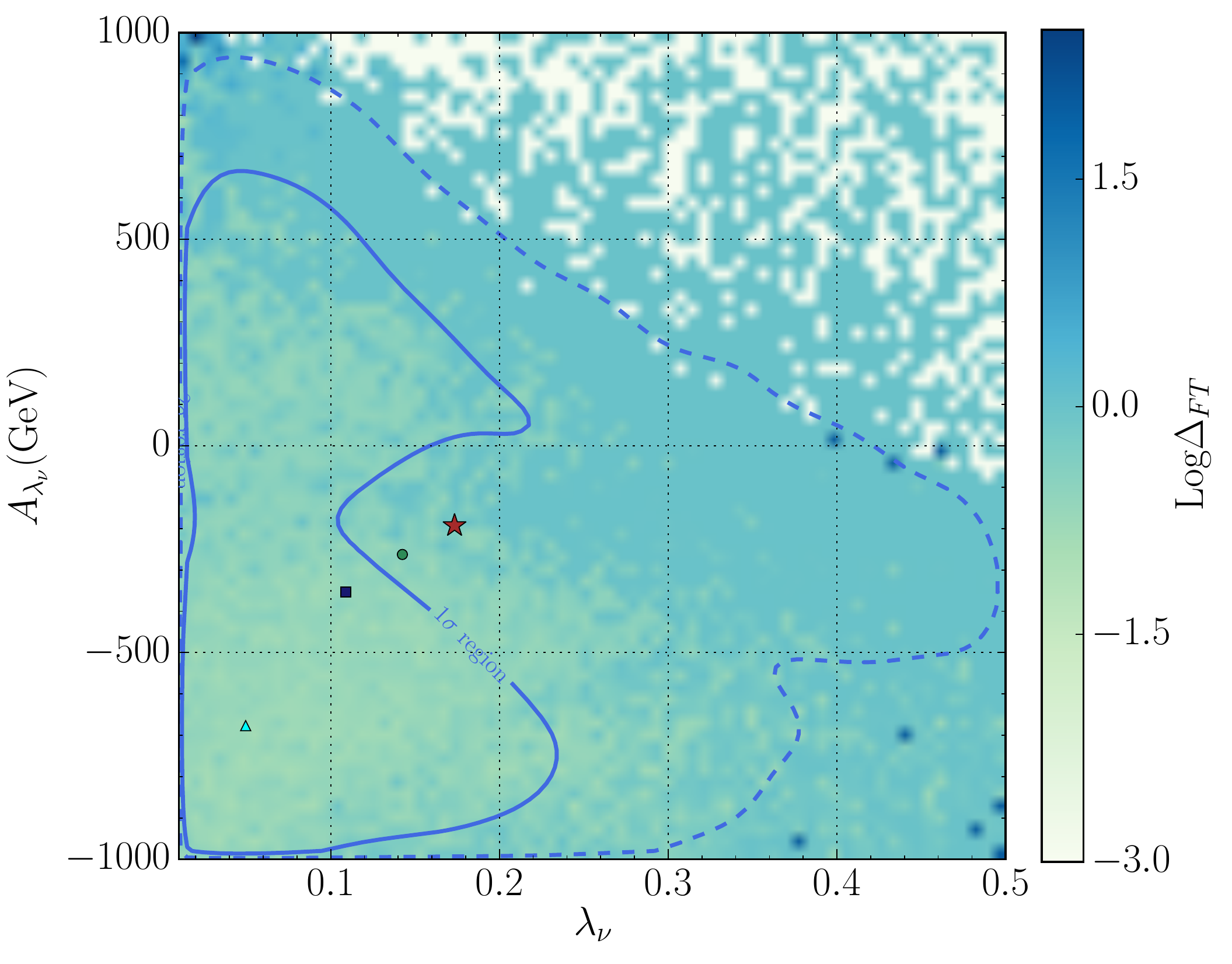}
		\includegraphics{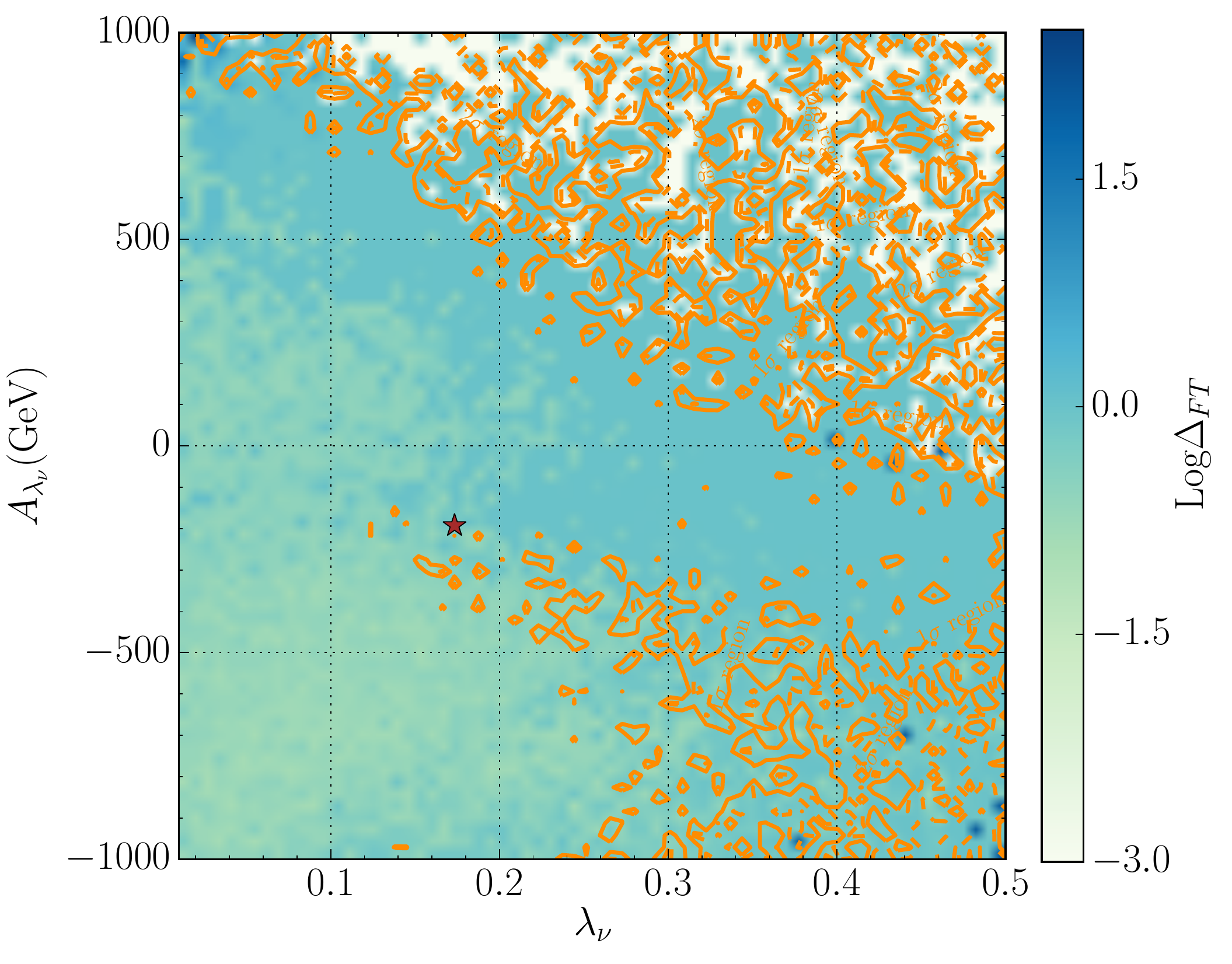}
		}

       \vspace{-0.4cm}

		\caption{{\bf{Upper row:}} Similar to FIG.\ref{fig2}, but projected on $A_{\lambda_\nu} - \lambda_\nu$ plane.  {\bf{Lower row:}} The minimum value of the fine tuning $\Delta_{FT}$ required to satisfy the $90\%$ upper bound of the XENON1T-2018 experiment on DM-nucleon scattering rate, which are projected on $A_{\lambda_\nu} - \lambda_\nu$ plane. The minimum value of $\Delta_{FT}$ is obtained in following way: we first split the plane into $70 \times 70$ equal boxes, then we only consider the samples in each box that satisfy the constraints of the XENON1T-2018 experiment, and pick out the minimum prediction of $\Delta_{FT}$. \label{fig6}}
	\end{figure*}

	\begin{figure*}[htbp]
		\centering
		\resizebox{0.7\textwidth}{!}{
		\includegraphics{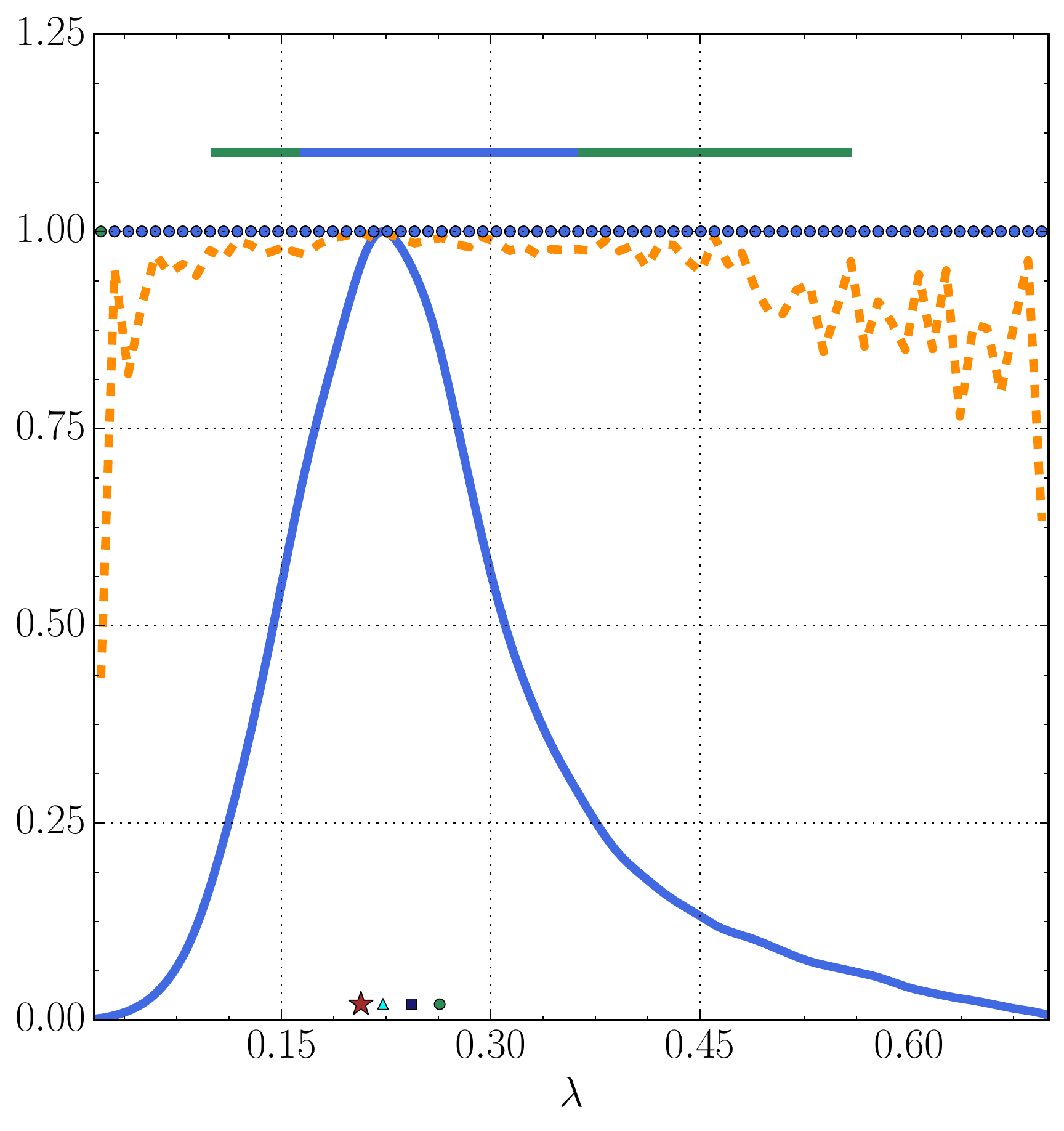}
		\includegraphics{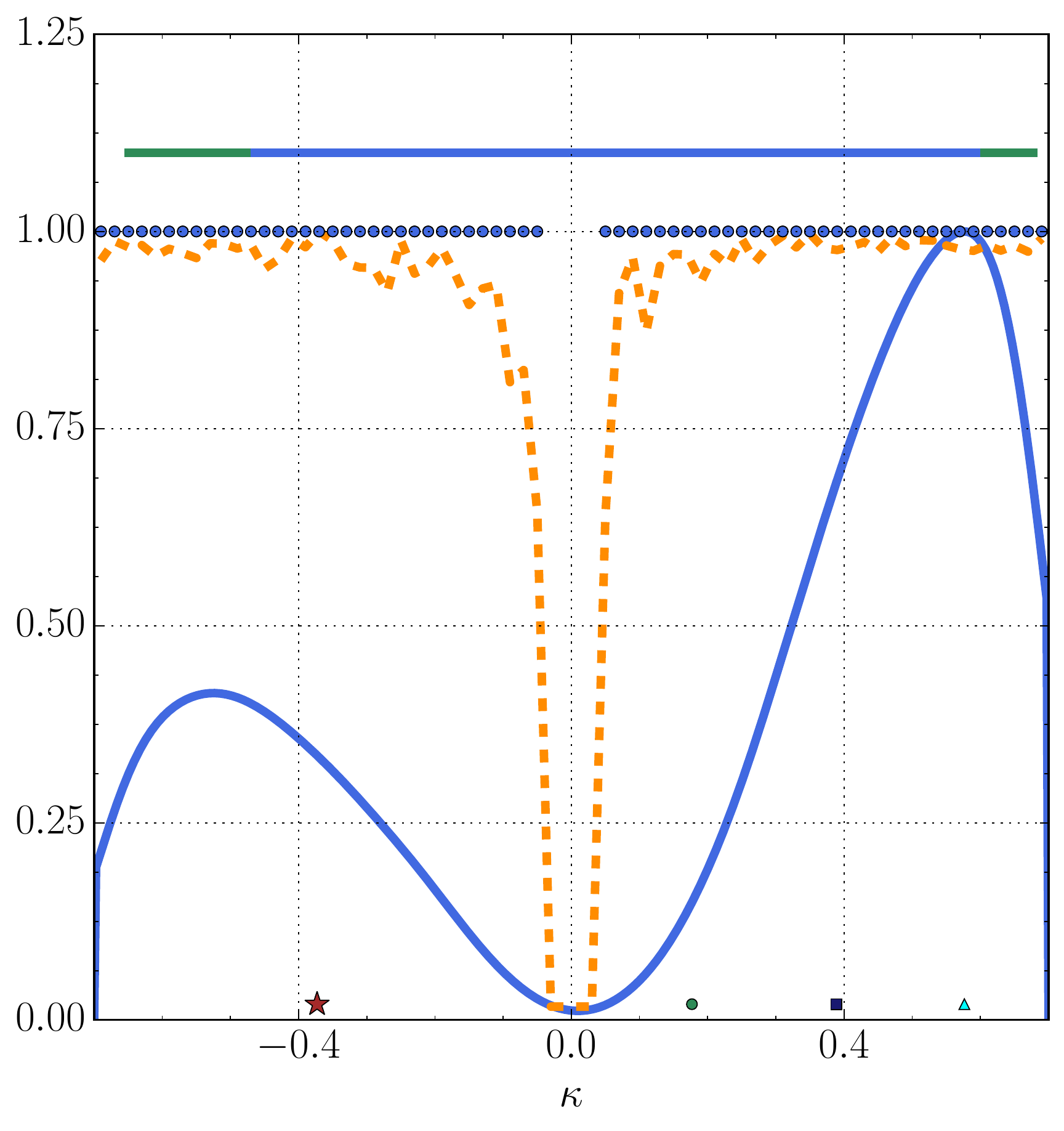}
		\includegraphics{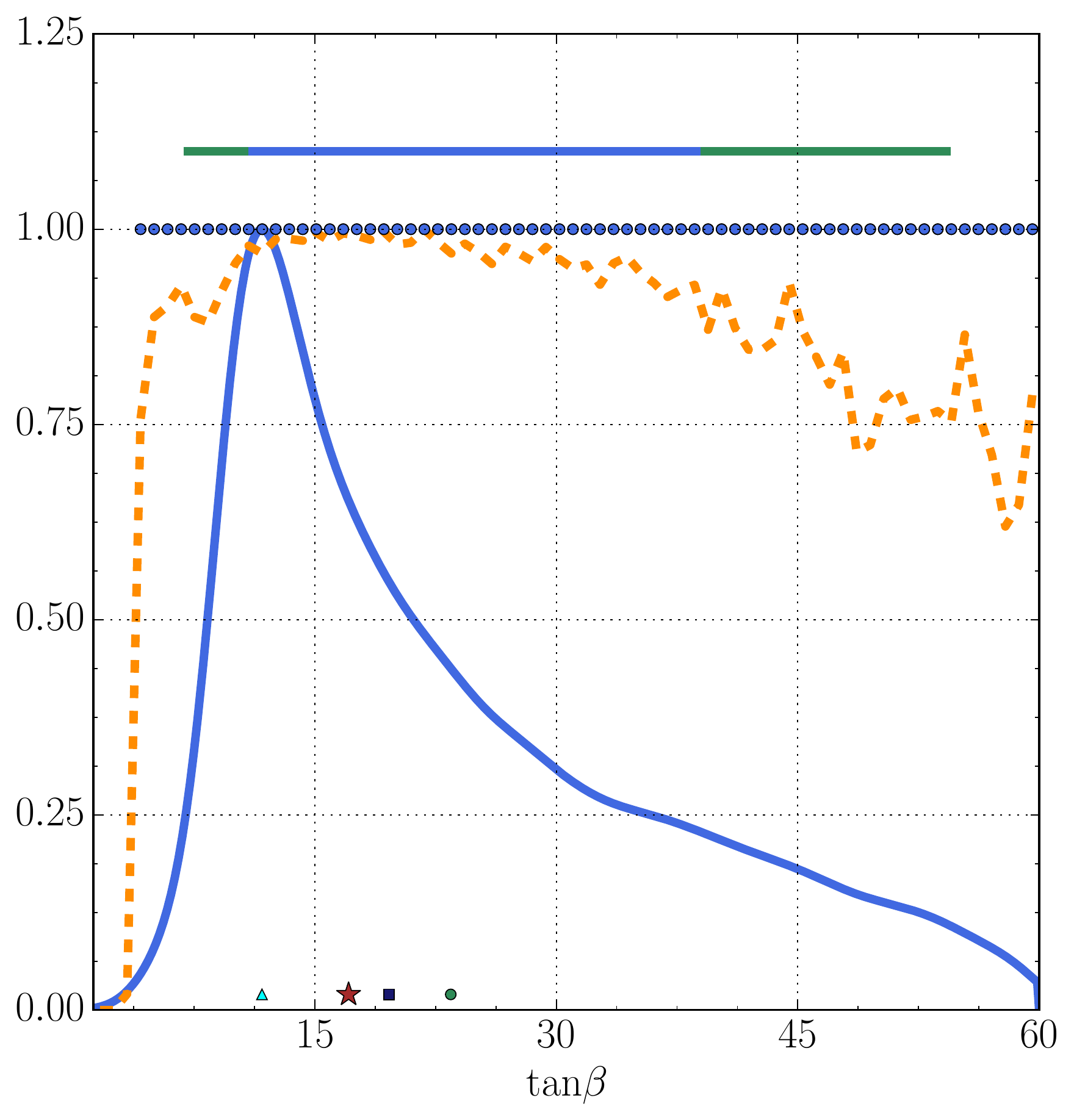}
		}
		\resizebox{0.7\textwidth}{!}{
		\includegraphics{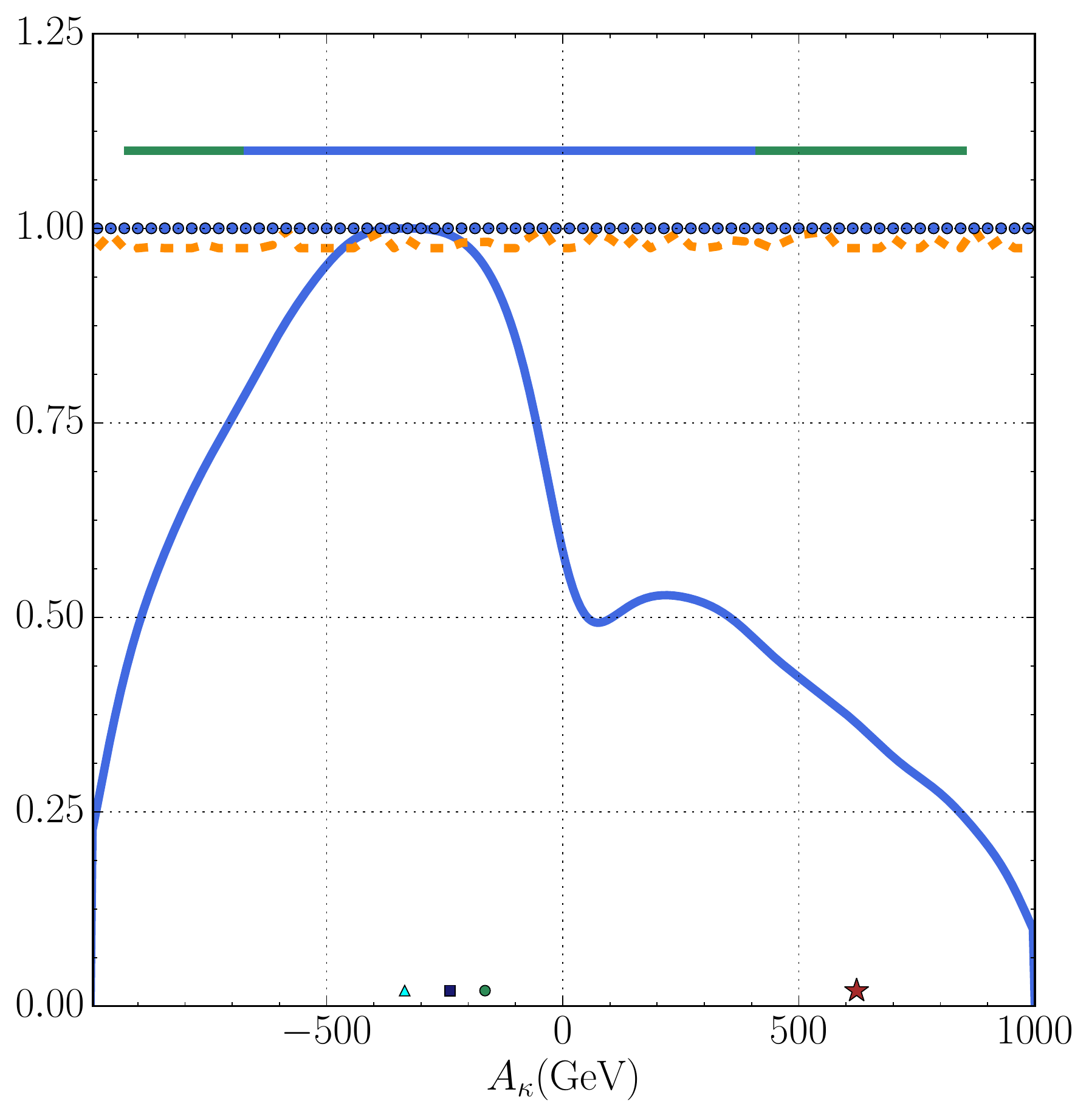}
		\includegraphics{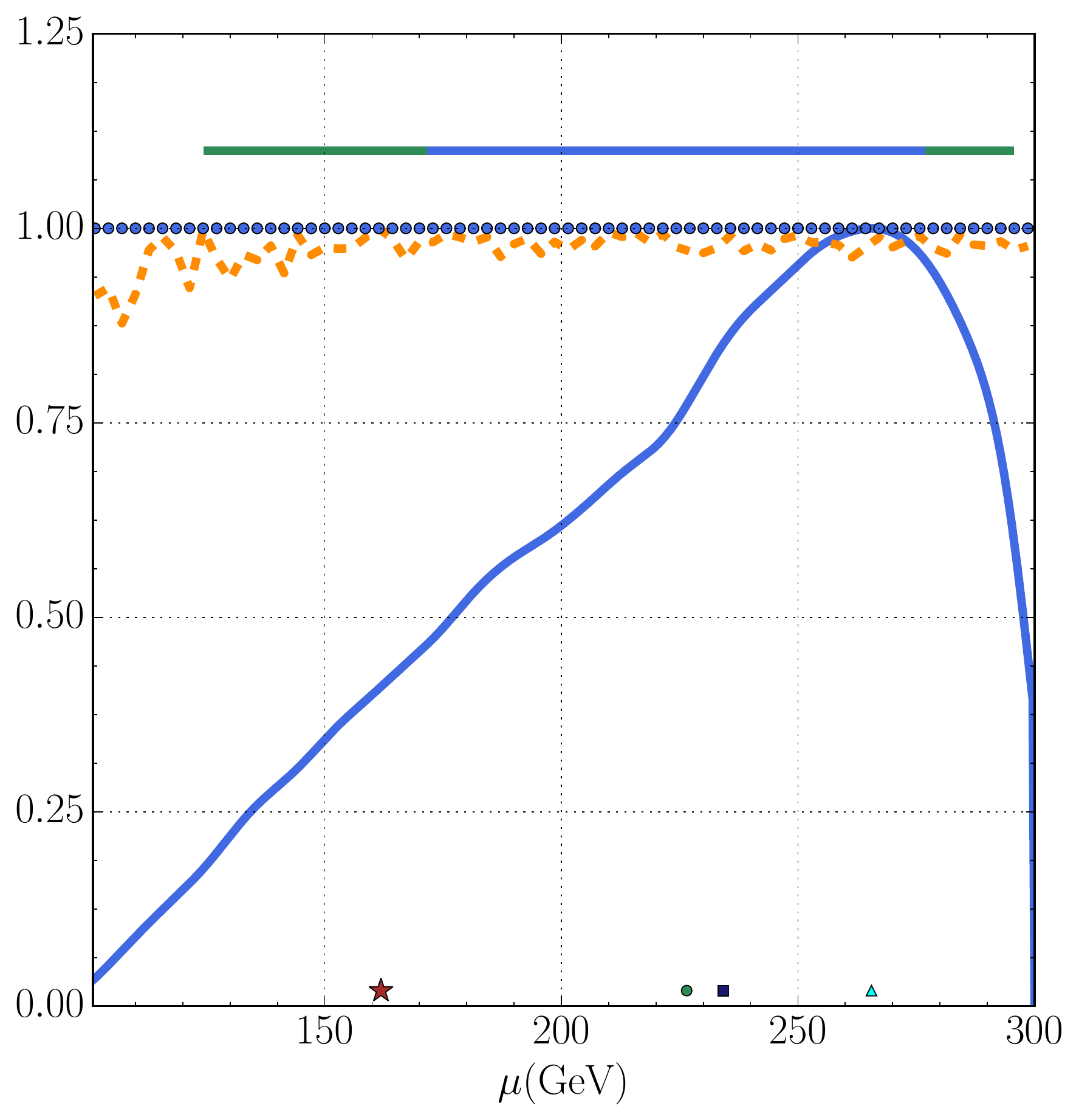}
		\includegraphics{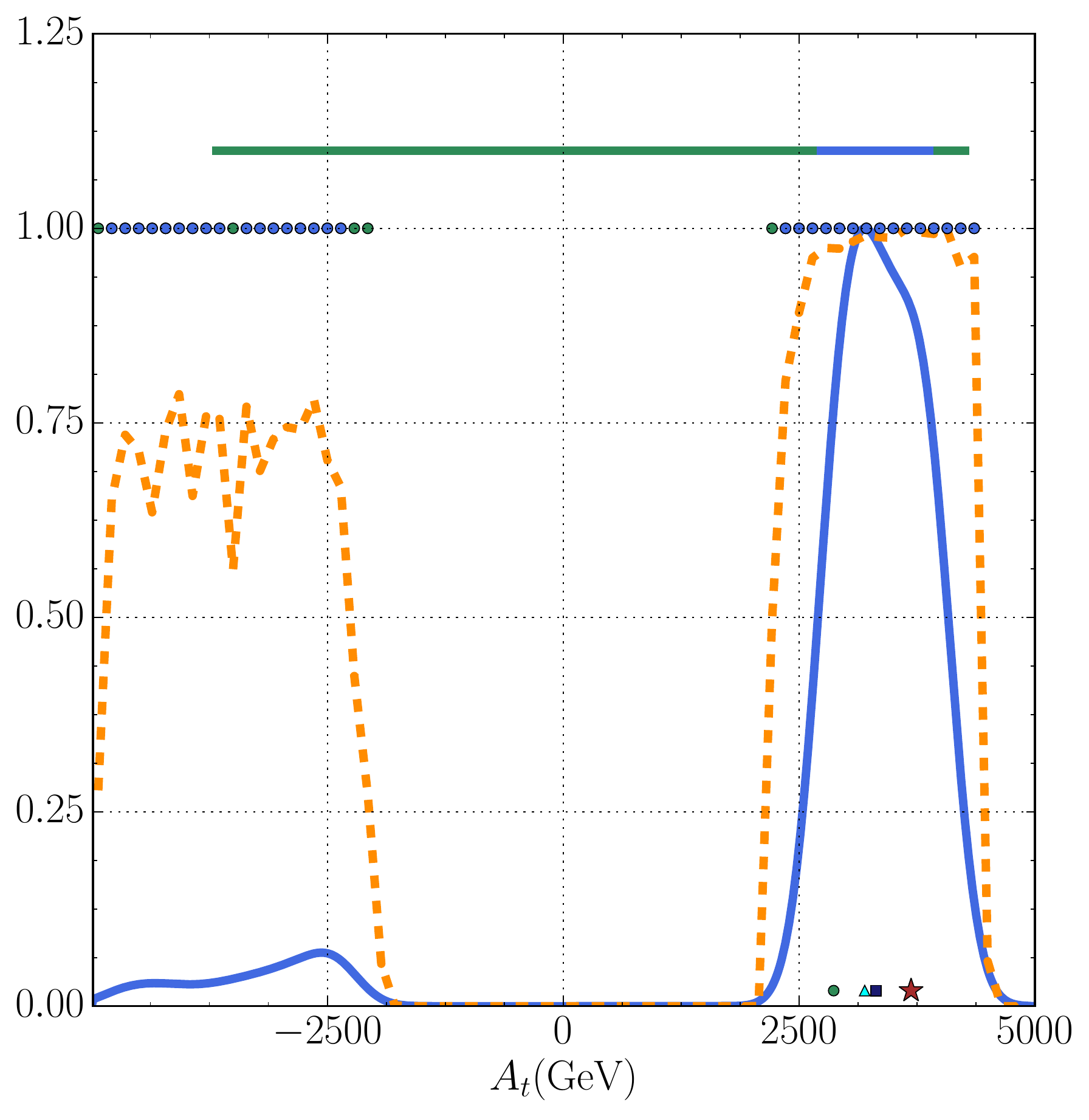}
		}
		\resizebox{0.7\textwidth}{!}{
		\includegraphics{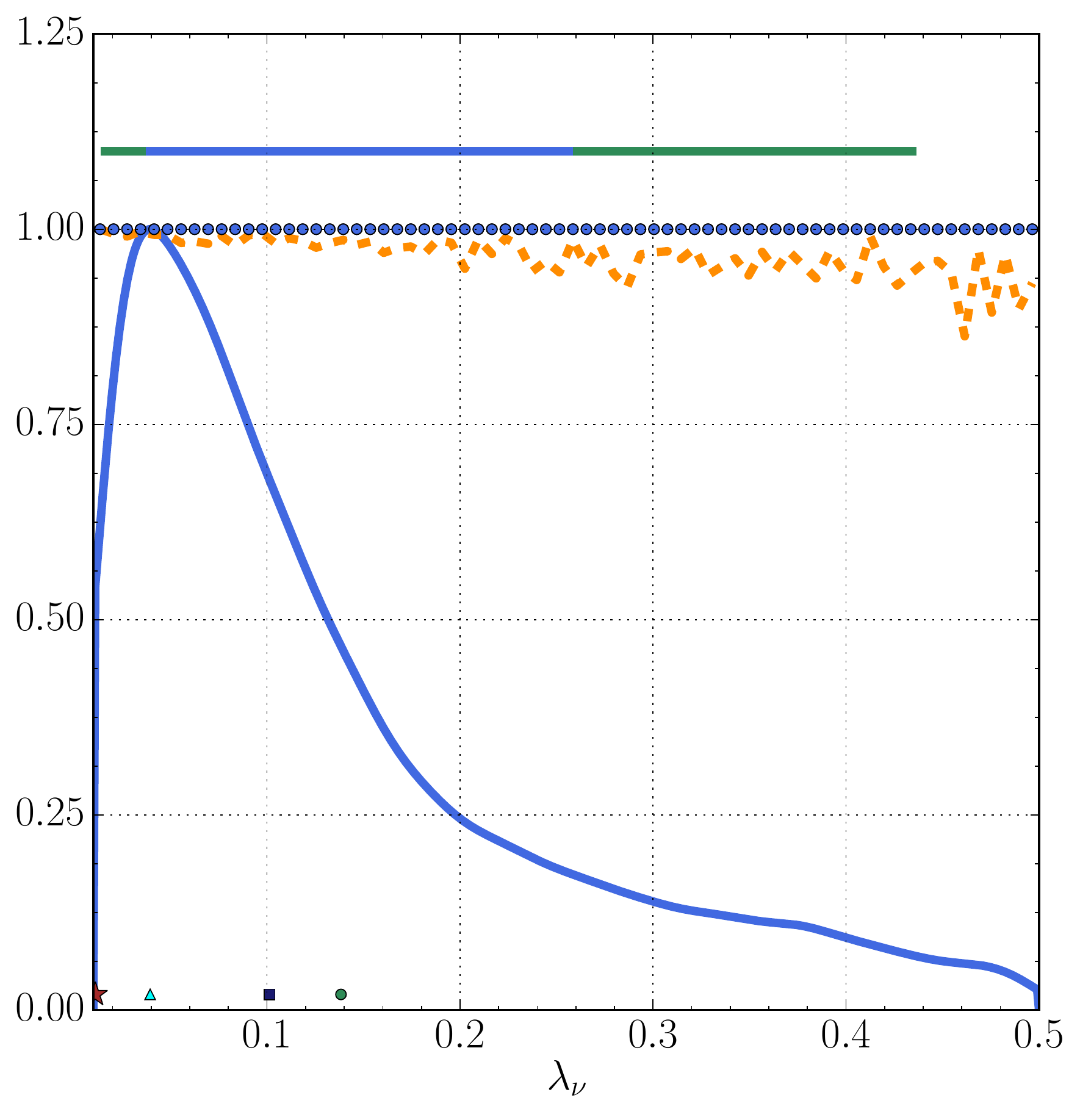}
		\includegraphics{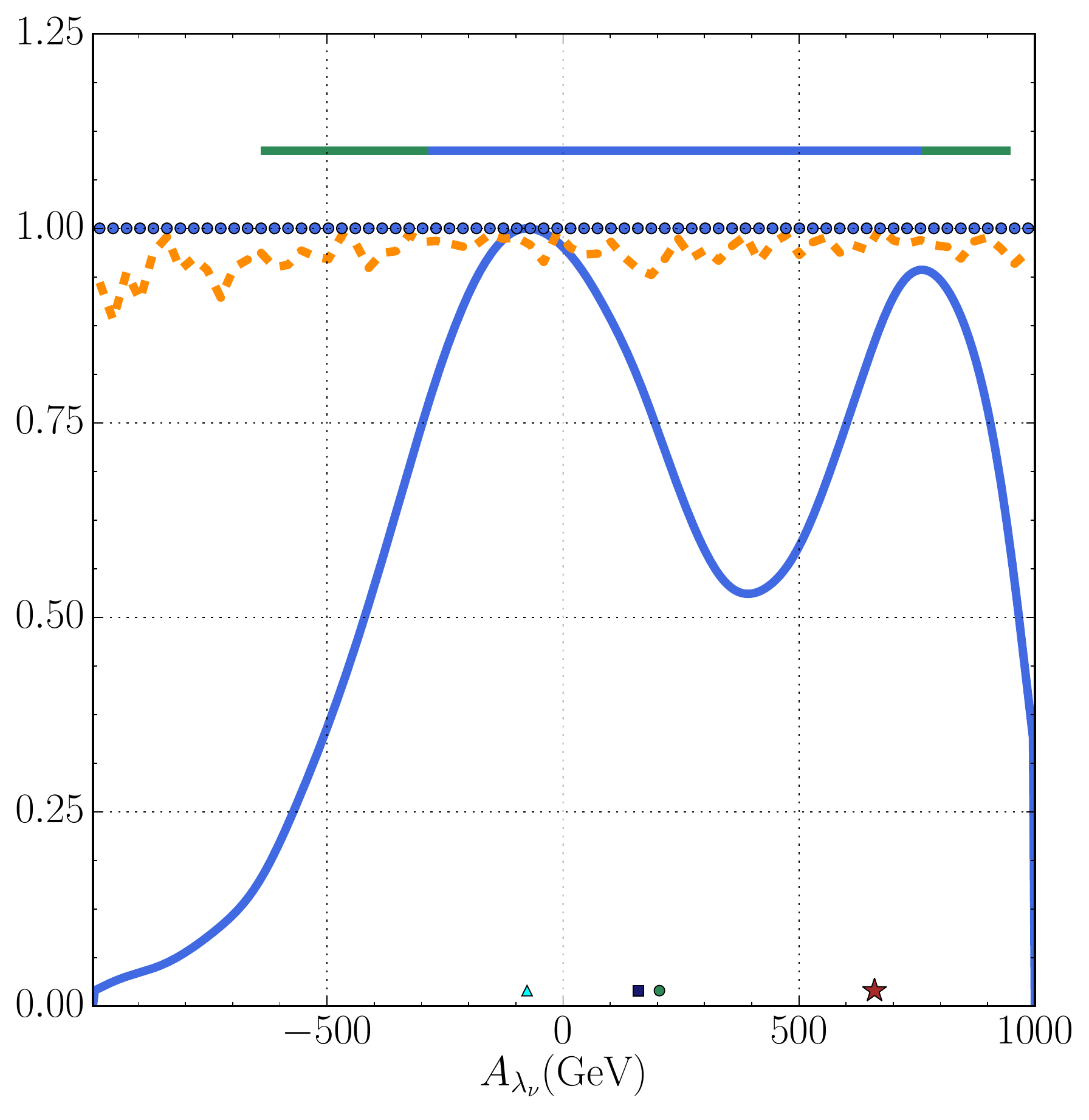}
		\includegraphics{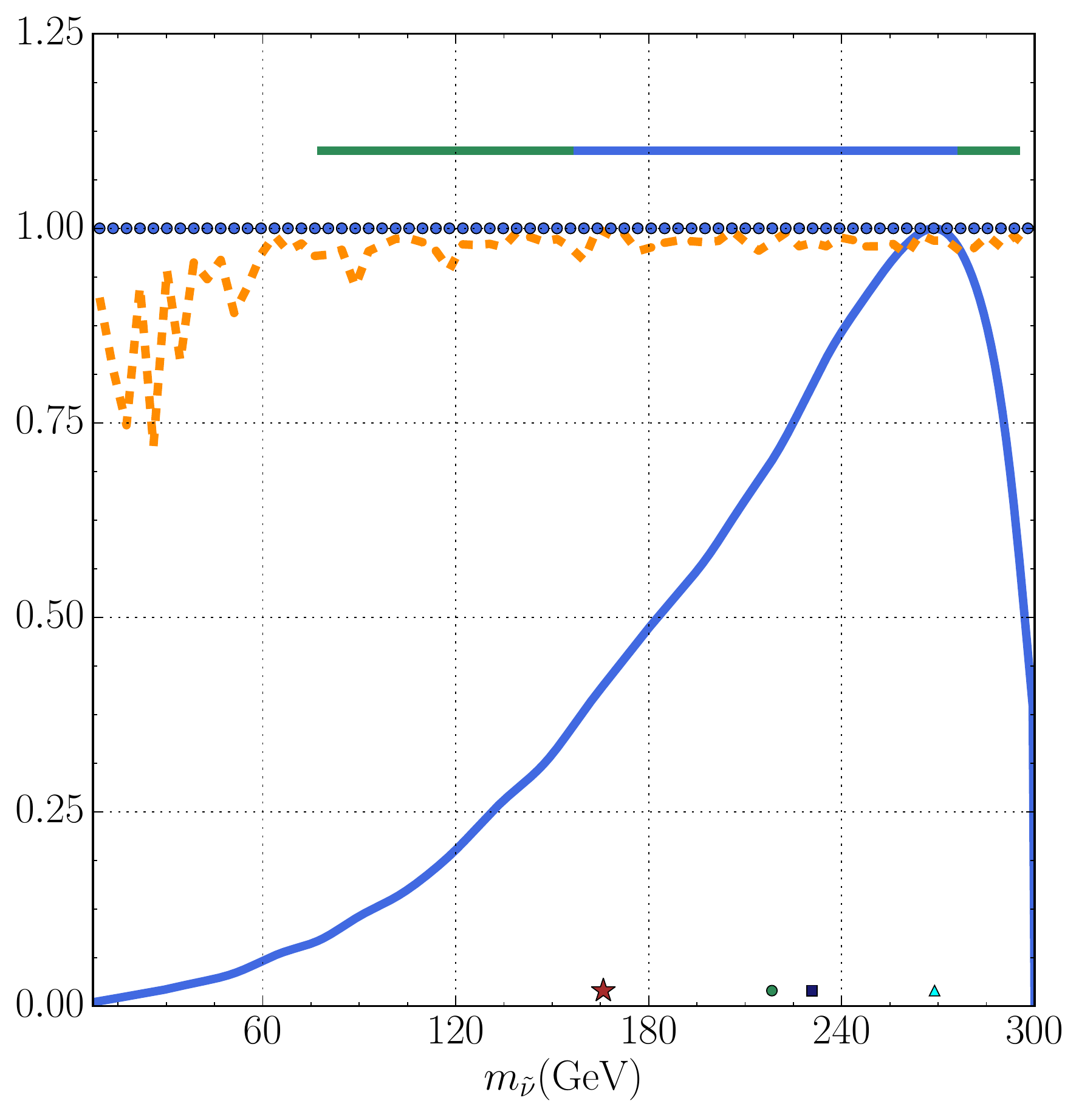}
		}

       \vspace{-0.4cm}

		\caption{Same as FIG.\ref{fig1}, but for the case that the DM candidate $\tilde{\nu}_1$ is CP-odd.  \label{fig7} }
	\end{figure*}	

	\begin{figure*}[htbp]
		\centering
		\resizebox{0.8\textwidth}{!}{
		\includegraphics{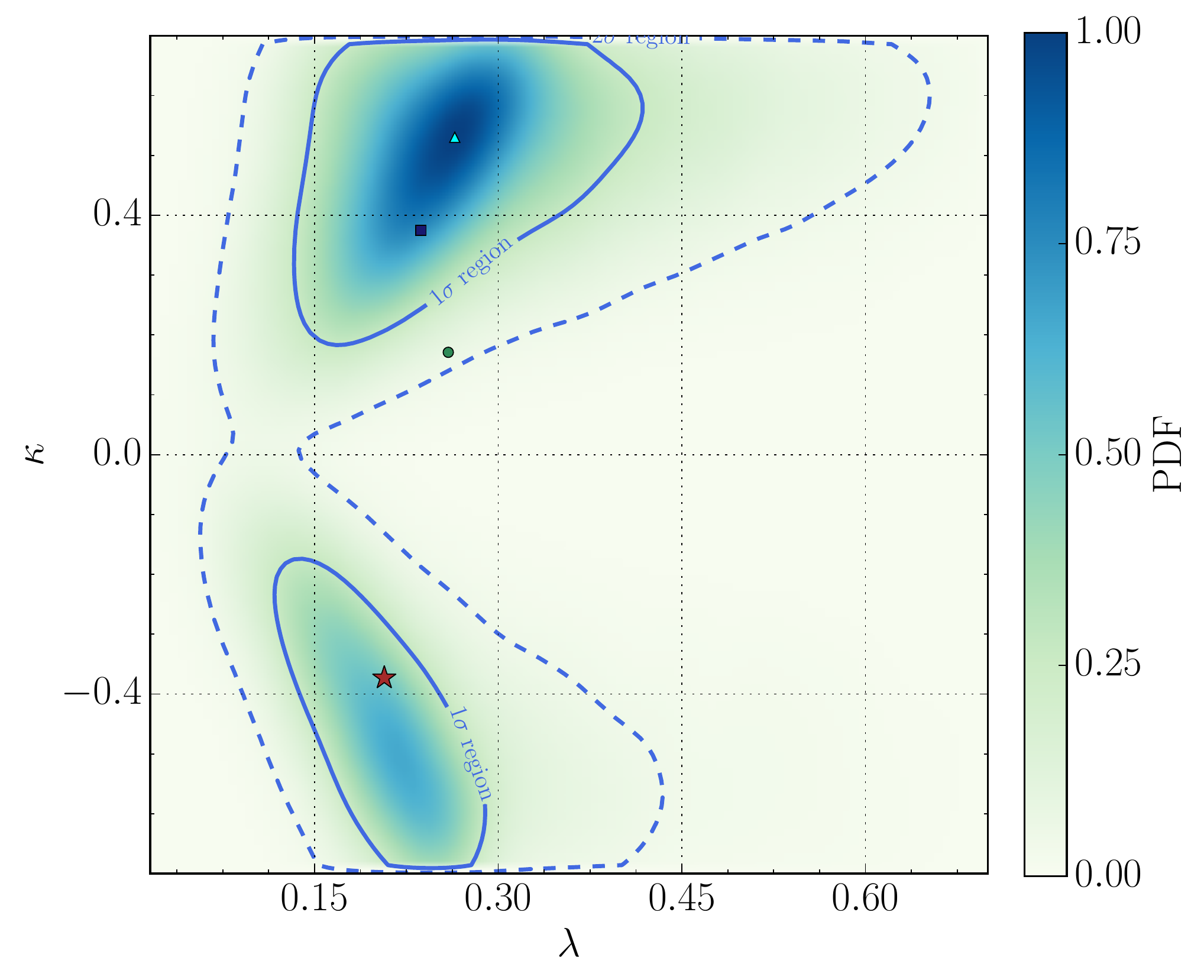}
		\includegraphics{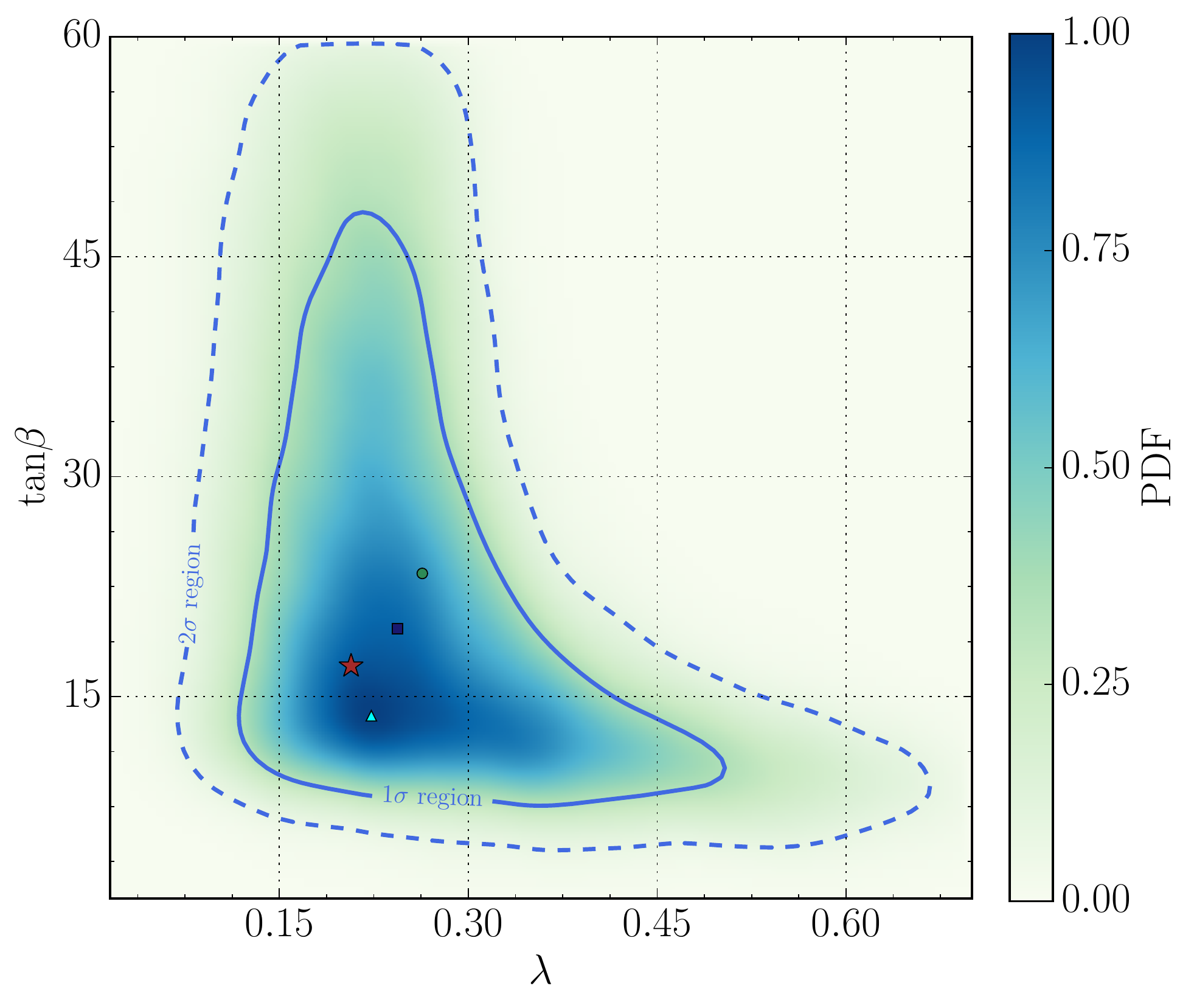}
		\includegraphics{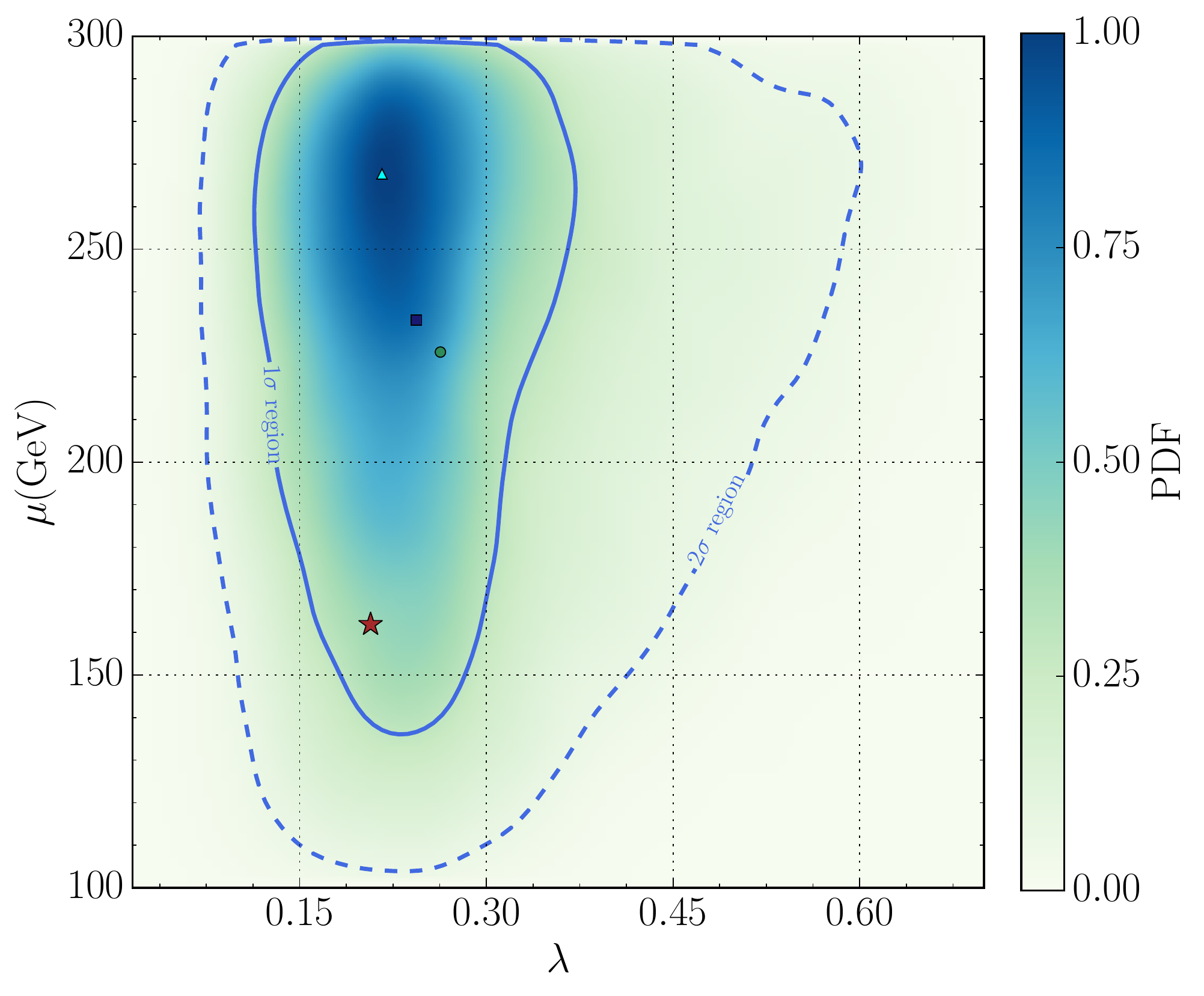}
		}
		\resizebox{0.8\textwidth}{!}{
		\includegraphics{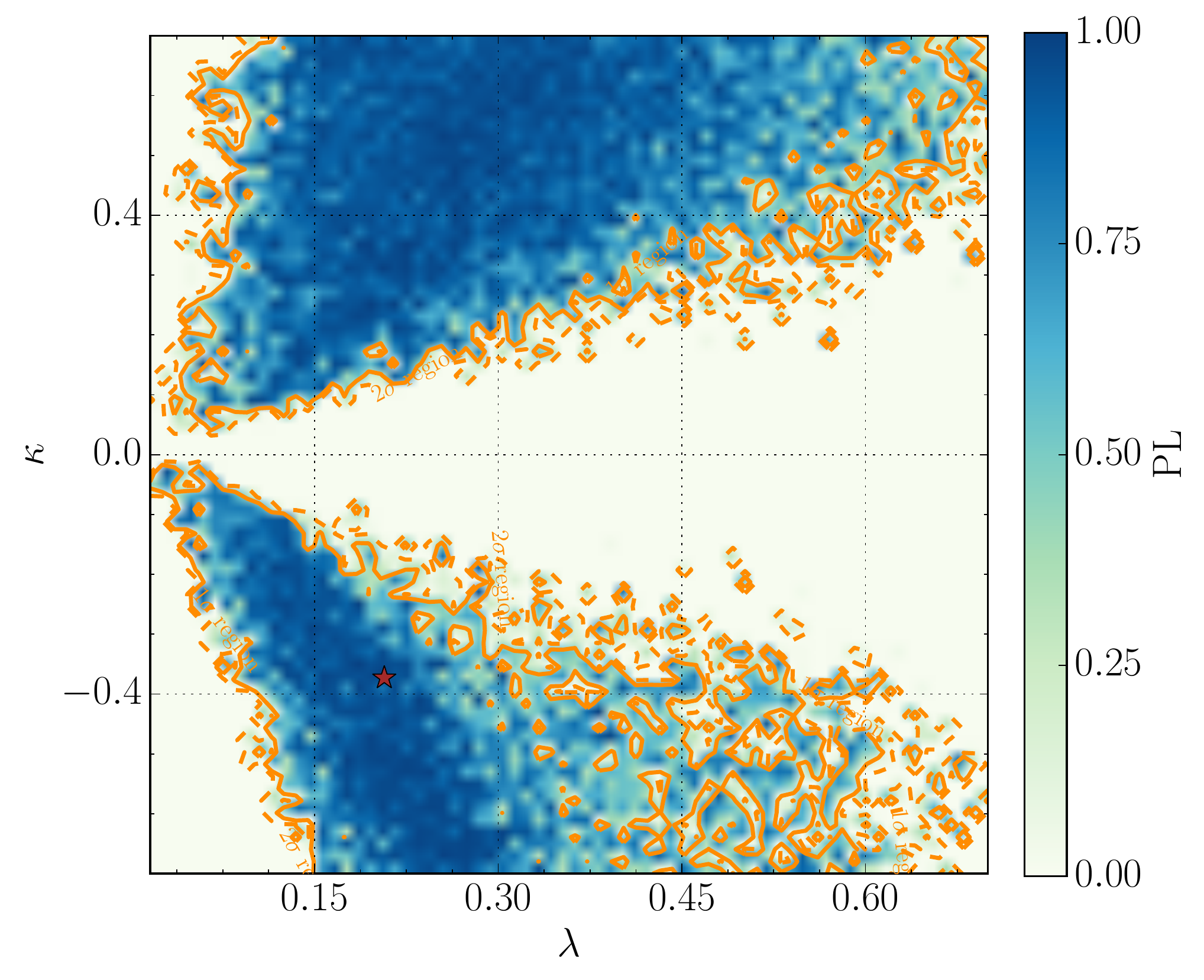}
		\includegraphics{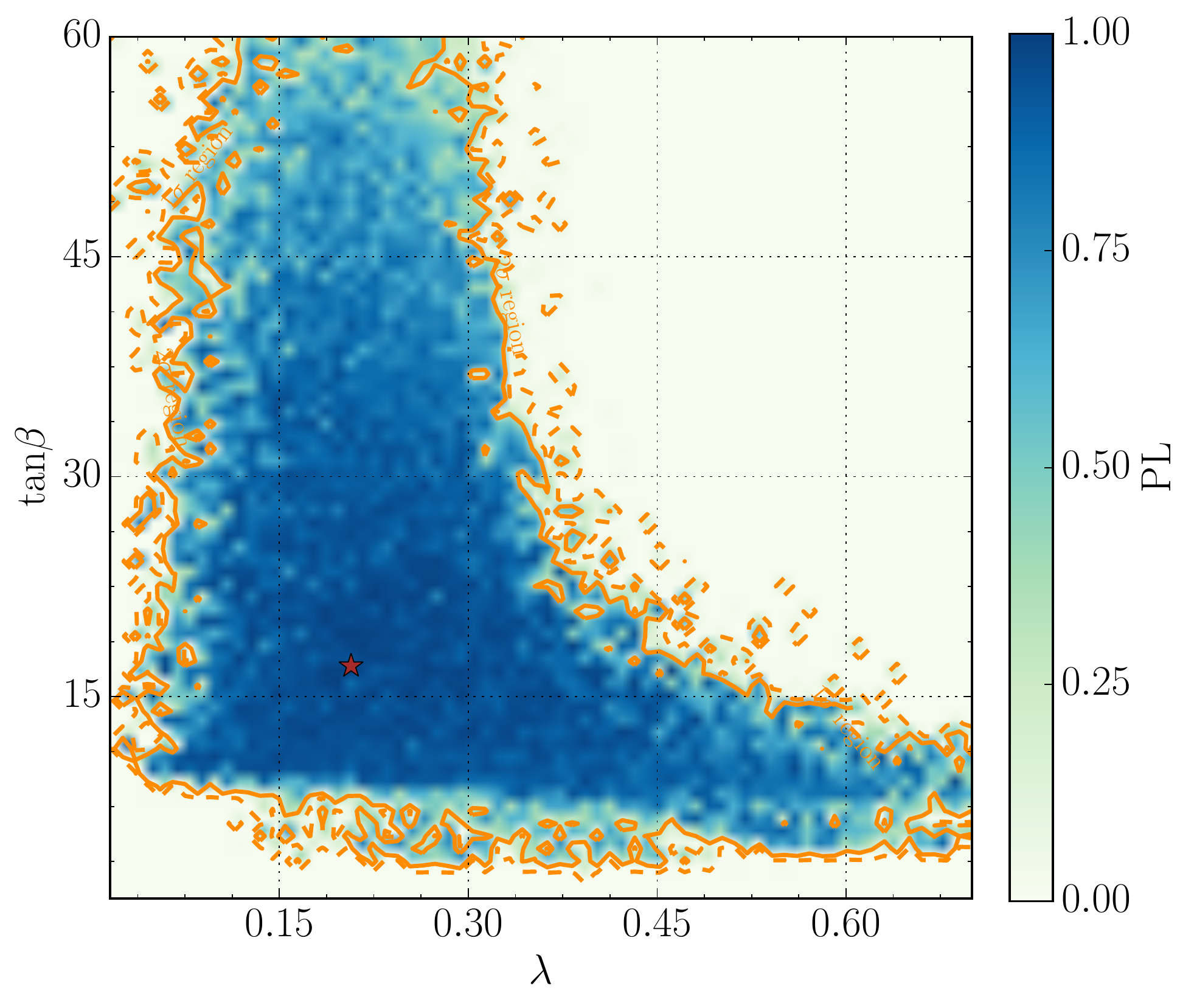}
		\includegraphics{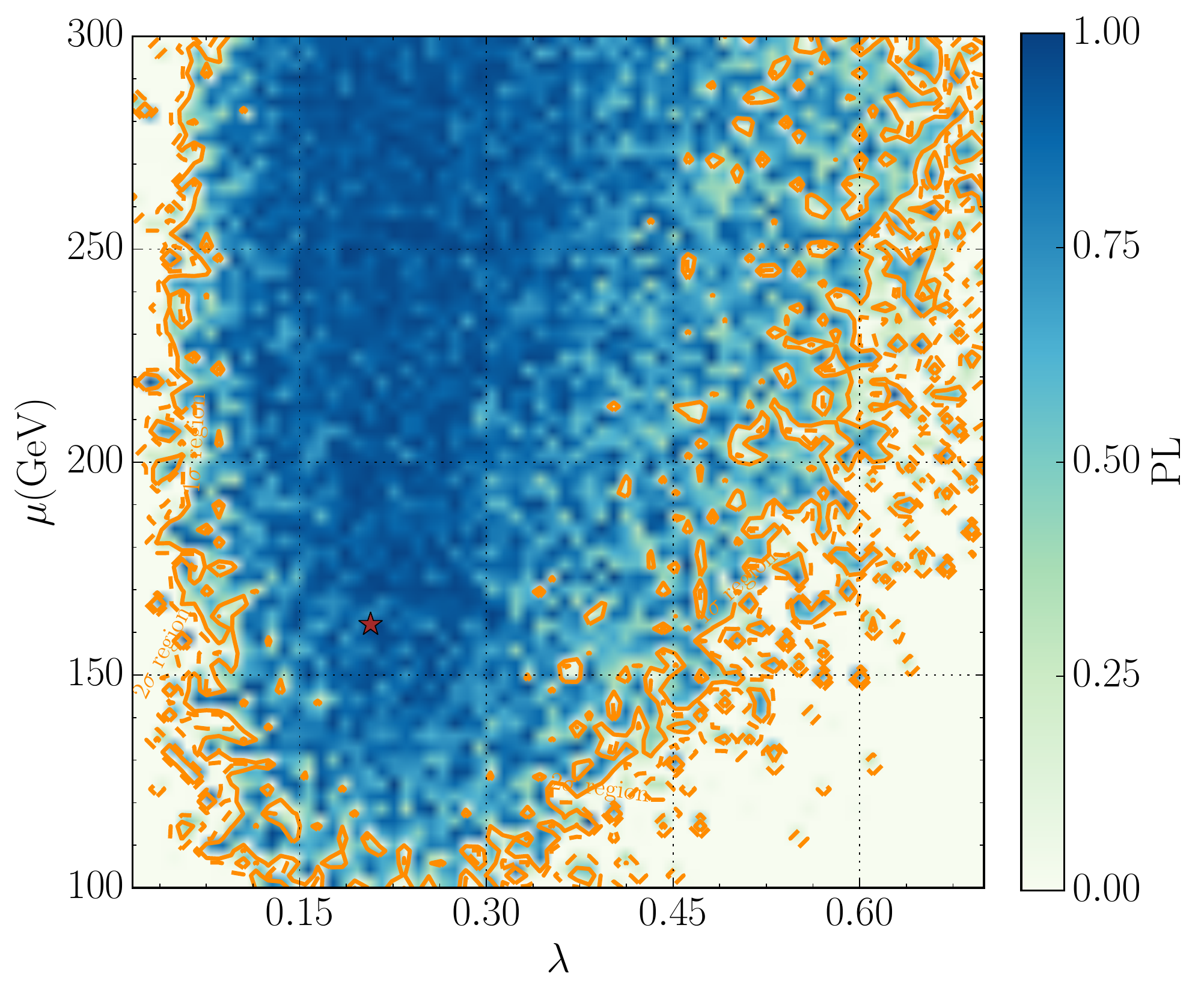}
		}

       \vspace{-0.4cm}

		\caption{Similar to FIG.\ref{fig2}, but for the case that the DM candidate $\tilde{\nu}_1$ is CP-odd. \label{fig8}}
	\end{figure*}

	\subsection{DM annihilation mechanisms}

	In this subsection, we study the annihilation mechanisms of the CP-even sneutrino DM. For this purpose, we project the posterior PDF and PL on $m_{A_1}-m_{\tilde{\nu}_1}$,  $m_{h_2}-m_{\tilde{\nu}_1}$ and $\mu-m_{\tilde{\nu}_1}$ planes in FIG.\ref{fig3}, and use the red dashed lines to denote the cases of $m_{A_1} = m_{\tilde{\nu}_1}$,  $m_{h_2} = 2 m_{\tilde{\nu}_1}$ and  $\mu = m_{\tilde{\nu}_1}$, respectively. We also show in FIG.\ref{fig4}  the distribution of the mass splitting $\Delta m$ on $ m_{\tilde{\nu}_{1, I}} - m_{\tilde{\nu}_1}$ plane, where we have split the plane into $70 \times 70 $ equal boxes (i.e. we divide each dimension of the plane by 70 equal bins), and define $\Delta m$ in each box as $\Delta m \equiv {\bf{min}} ( | m_{A_1} - m_{\tilde{\nu}_1} - m_{\tilde{\nu}_{1, I}} |)$ for the samples allocated in the box. Different CRs and IRs are also plotted for the left and right panel of this figure, respectively, with the red dashed lines corresponding to the case of $ m_{\tilde{\nu}_{1, I}} = m_{\tilde{\nu}_1}$. From these figures, one can learn following facts:
	\begin{itemize}
       \item For the samples in $1 \sigma$ CRs, the singlet dominated $A_1$ and $h_2$ are heavier than about $300 {\rm GeV}$.

		\item A small portion of samples satisfy $m_{A_1} \lesssim m_{\tilde{\nu}_1}$, which implies that the DM may annihilate dominantly into $A_1 A_1$ final state. Since this annihilation channel is mainly a s-wave process, $\langle \sigma v \rangle_0 \simeq \langle \sigma v \rangle_{FO} \simeq 3 \times 10^{-26} cm^3 s^{-1}$ with $\langle \sigma v \rangle_0$ denoting current DM annihilation rate and $\langle \sigma v \rangle_{FO}$ representing the rate at freeze-out temperature, and consequently the annihilation is tightly limited by the Fermi-LAT data from drawf galaxies~\cite{Ackermann:2015zua}. As was pointed in our previous work~\cite{Cao:2017cjf}, such a constraint can be avoided by the forbidden annihilation proposed in~\cite{Griest:1990kh} for $m_{\tilde{\nu}_1} \lesssim 100 {\rm GeV}$\footnote{Note that this case is rare since it requires specific kinematics, i.e. $A_1$ should be slightly heavier than $\tilde{\nu}_1$.} or simply by requiring $m_{\tilde{\nu}_1} > 100 {\rm GeV}$ where the constraints become loose~\cite{Ackermann:2015zua}. Our results verify this point.

Note that for the case considered here, the sneutrino DM and the singlet-dominated $A_1$ compose a self-contained DM sector, which interacts with the SM sector only via the small Higgs-doublet component of $A_1$, and the corresponding likelihood function may be quite large, which is shown by the PL distribution on $m_{A_1}-m_{\tilde{\nu}_1}$ plane.  This is a typical hidden or secluded DM scenario~\cite{Pospelov:2007mp}.
		\item Quite a few samples predict $m_{h_2} \simeq 2 m_{\tilde{\nu}_1}$. In this case, the DM annihilation mainly proceeds via a s-channel resonant $h_2$ funnel. Note that most samples in the $1\sigma$ CR on the $m_{h_2}-m_{\tilde{\nu}_1}$ plane satisfy $m_{h_2} > 2 m_{\tilde{\nu}_1}$, which implies $ \langle \sigma v \rangle_{FO} > \langle \sigma v \rangle_0 $ if the $h_2$-mediated contribution is dominant in the annihilation~\cite{Griest:1990kh}.
		\item For all samples in the $1 \sigma$ CR on the $\mu-m_{\tilde{\nu}_1}$ plane, $\mu$ and $m_{\tilde{\nu}_1}$ are approximately equal. In this case, the DM and Higgsinos can be in thermal equilibrium in early Universe before DM freeze-out due to the transition between the DM pair and the Higgsino pair, which is mediated by the singlet-dominated Higgs bosons. As a result, $\tilde{\nu}_1$  can reach its correct relic density by coannihilating with the Higgsino-dominated neutralinos,
 although its couplings with the SM particles are tiny. This situation is quite similar to the sneutrino DM scenario in the inverse seesaw extension of the NMSSM~\cite{Cao:2017cjf}. Note that a heavy $\tilde{\nu}_1$ corresponds to a larger parameter space for the annihilation, so the posterior PDF is maximized at $\mu \simeq 300 {\rm GeV}$. In Appendix C of this work, we will further discuss the situation of the marginal PDF for $\mu > 300 {\rm GeV}$ case.
		\item The lightest CP-odd sneutrino $\tilde{\nu}_{1,I}$ may roughly degenerate in mass with $\tilde{\nu}_1$. In this case, $\tilde{\nu}_1$ can coannihilate with $\tilde{\nu}_{1,I}$ to reach the right relic density by exchanging the CP-odd scalar $A_1$. In particular, the relation $m_{A_1} \simeq m_{\tilde{\nu}_1} + m_{\tilde{\nu}_{1,I}}$ may hold when $55 {\rm GeV} \leq \tilde{\nu}_1 \leq 190 {\rm GeV}$, which means that the coannihilation is mediated by a resonant $A_1$.
    \item From the PL distributions in FIG.\ref{fig3} and FIG.\ref{fig4}, one can infer that the DM may also annihilate by a resonant $h_1$, but the probability of the occurrence is small.
    \item We emphasize that these results are based on flat prior PDF of the input parameters, which has no bias on these parameters before the scan. If one focuses on low $|A_\kappa|$ region, e.g. by choosing a log prior PDF of $|A_\kappa|$, a light $A_1$ becomes preferred, and consequently more samples below the red line $m_{A_1} = m_{\tilde{\nu}_1}$ can be obtained. We will discuss this issue in Appendix B.
	\end{itemize}

	In summary, FIG.\ref{fig3} and FIG.\ref{fig4} reveal the fact that the singlet dominated Higgs boson may either serve as the DM annihilation product, or mediate the DM annihilations in the Type-I extension of the NMSSM, and consequently, the lightest sneutrino in the theory can act as a viable DM candidate. This situation is quite different from the Type-I extension of the MSSM. Moreover, we note that the $\tilde{\nu}_1$-Higgsino coannihilation was neglected in previous studies, but from our study it is actually the most important annihilation mechanism from both the marginal PDF and PL distributions.

	\subsection{DM-nucleon scattering}

In Section \ref{Section-Model}, we have shown by analytic formulae the natural suppression of the cross section for $\tilde{\nu}_1$-nucleon scattering. In this subsection, we present relevant numerical results. In FIG.\ref{fig5}, we plot the marginal posterior PDFs and PLs on $\xi_2 - \xi_1$ planes (top panels) and $\sigma_{\tilde{\nu}_1-p}-m_{\tilde{\nu}_1}$  planes (bottom panels). The top panels indicate that the typical magnitudes of $\xi_i$ (i=1,2) are $10^{-11} {\rm GeV^{-3}}$, and both the CRs and CIs are not symmetric under the sign exchange of $\xi_i$. Correspondingly, the cross section for $\tilde{\nu}_1$-nucleon scattering is usually less than $10^{-46} {\rm cm^2}$ with its marginal PDF maximized around $4 \times 10^{-47} {\rm cm^2}$, and may be as low as  $10^{-49} {\rm cm^2}$ for a significant part of samples (see bottom panels).
	
	From Eq.(\ref{Csnn}) and Eq.(\ref{SI-simplify1}) in Section II and the favored parameter space in FIG.\ref{fig2}, one can learn that the scattering rate is sensitive to the parameters $\lambda_\nu$ and $A_{\lambda_\nu}$, so we concentrate on these two parameters in following discussion. In the upper panels of FIG.\ref{fig6}, we show the posterior PDF and PL on $A_{\lambda_\nu}-\lambda_\nu$ plane, and in the lower panels we plot the distribution of the fine tuning parameter $\Delta_{FT}$ defined in Eq.(\ref{Fine-tuning}) of Section II on the same plane. The upper panels indicate that both the $1 \sigma$ posterior PDF and $1 \sigma$ PL spread a broad region on the plane, reflecting that the theory can readily accommodate the tight constraints of the XENON1T-2018 experiment. The lower panels, on the other hand, show that the theory does not need any fine tuning to survive the tight constraints.

	\begin{figure*}[htbp]
		\centering
		\resizebox{0.8\textwidth}{!}{
		\includegraphics{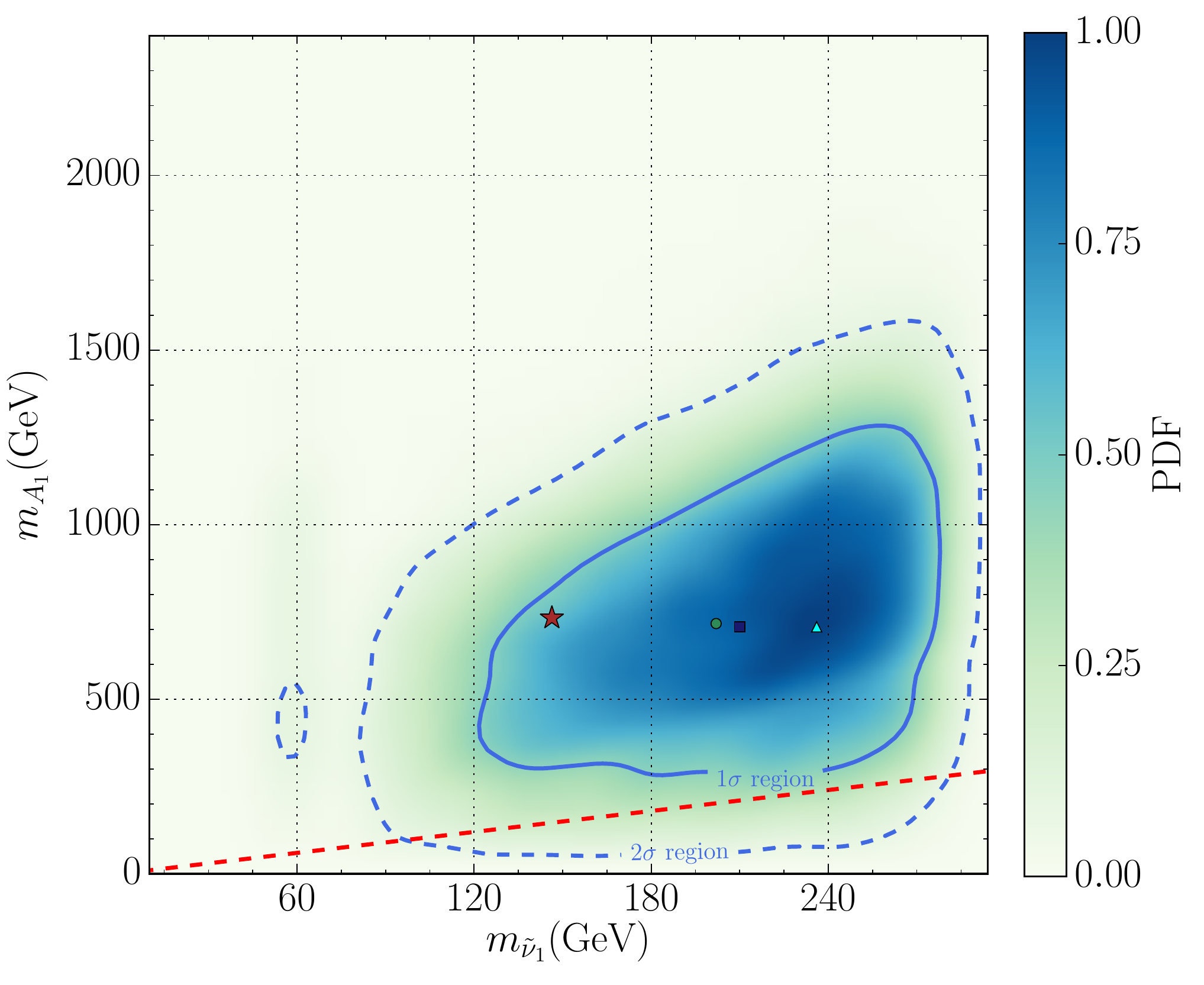}
		\includegraphics{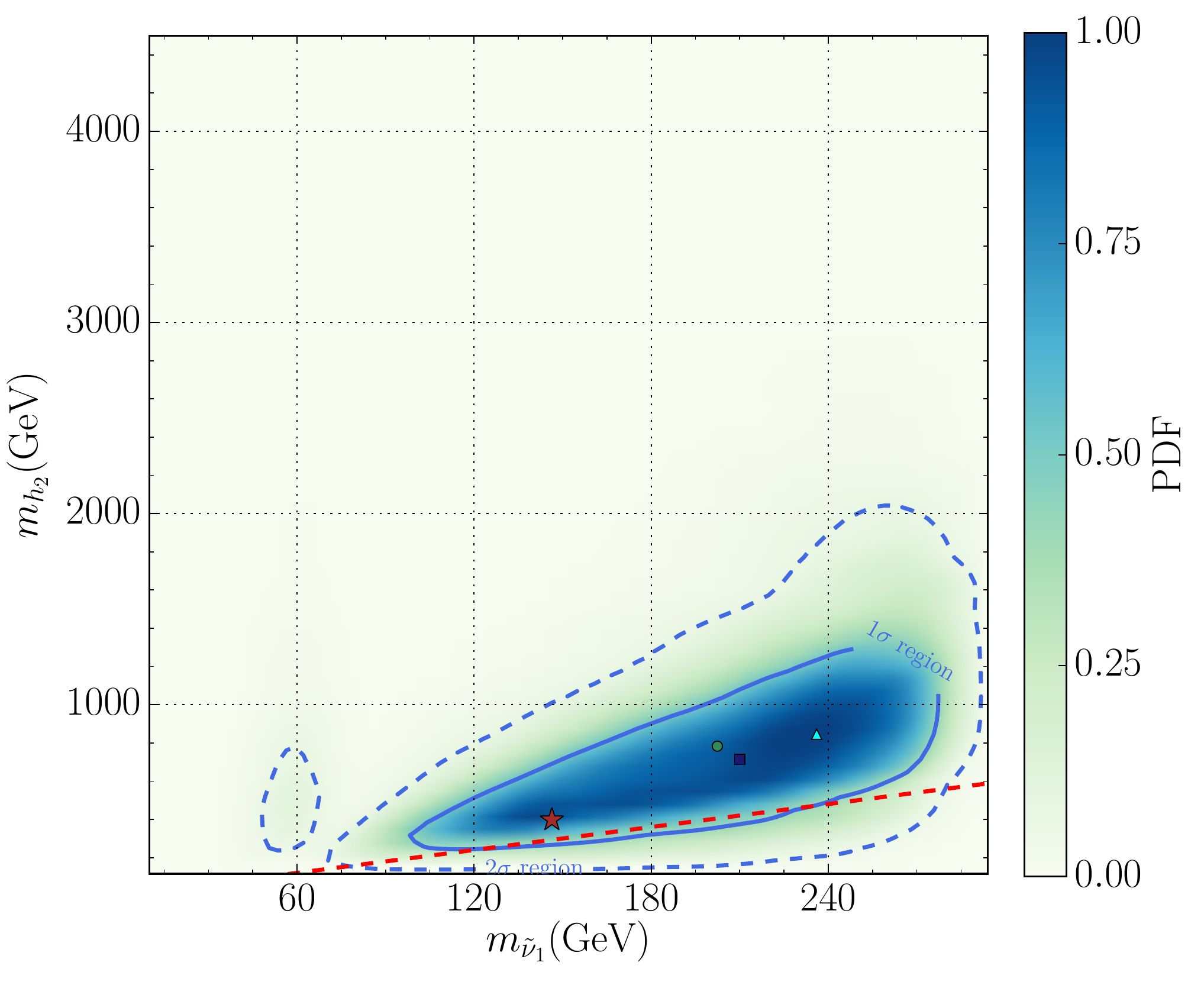}
		\includegraphics{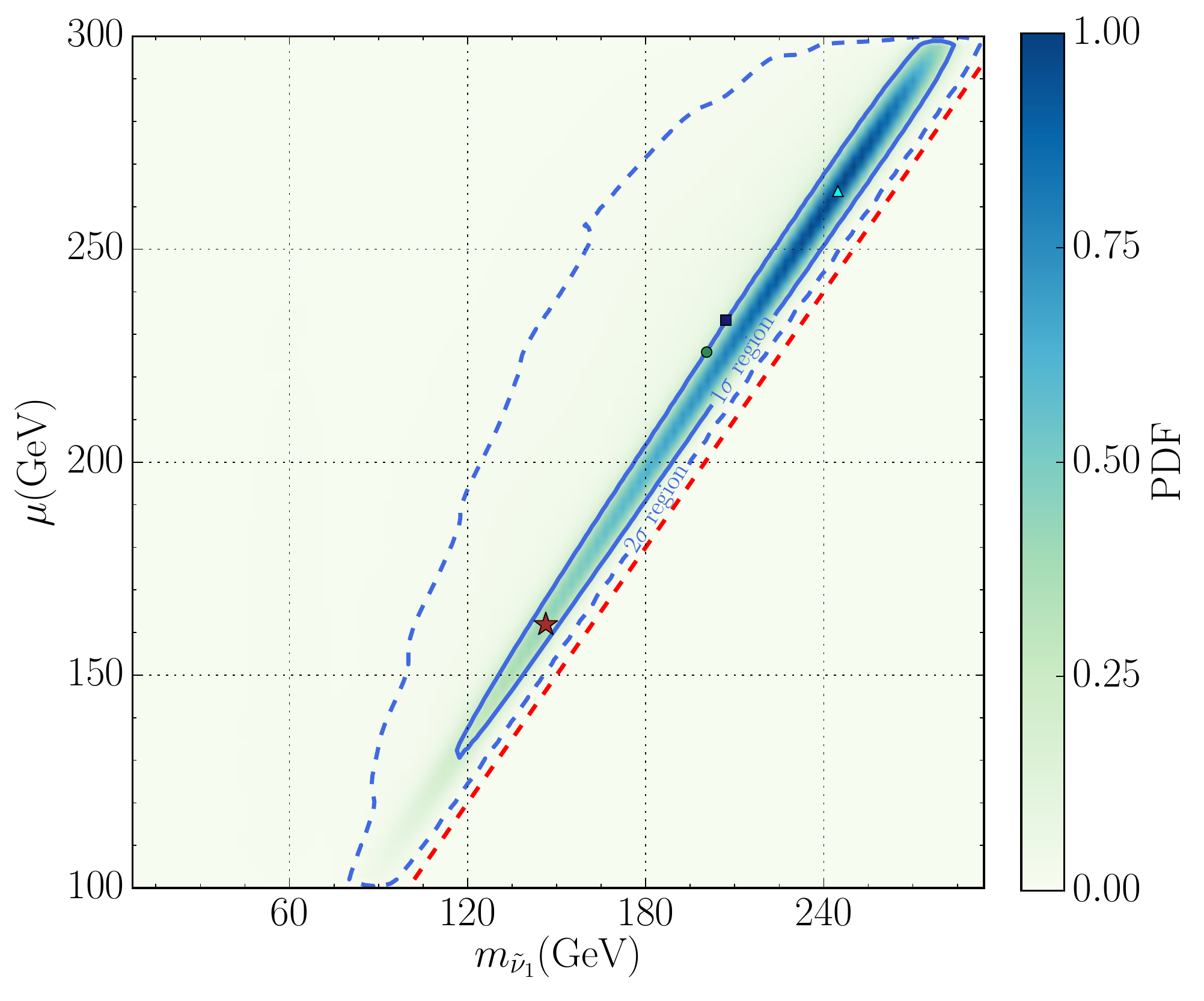}
		}
		\resizebox{0.8\textwidth}{!}{
		\includegraphics{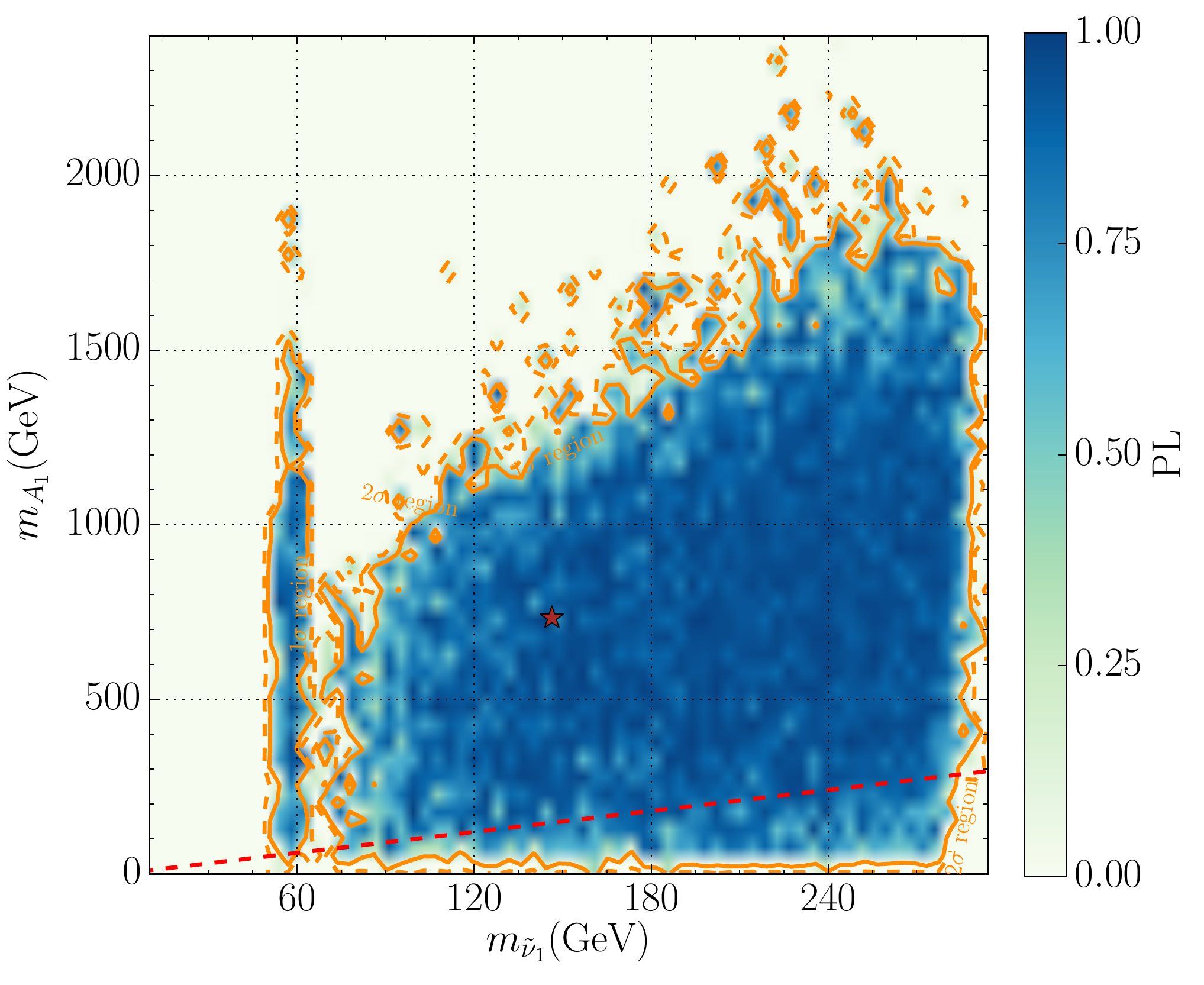}
		\includegraphics{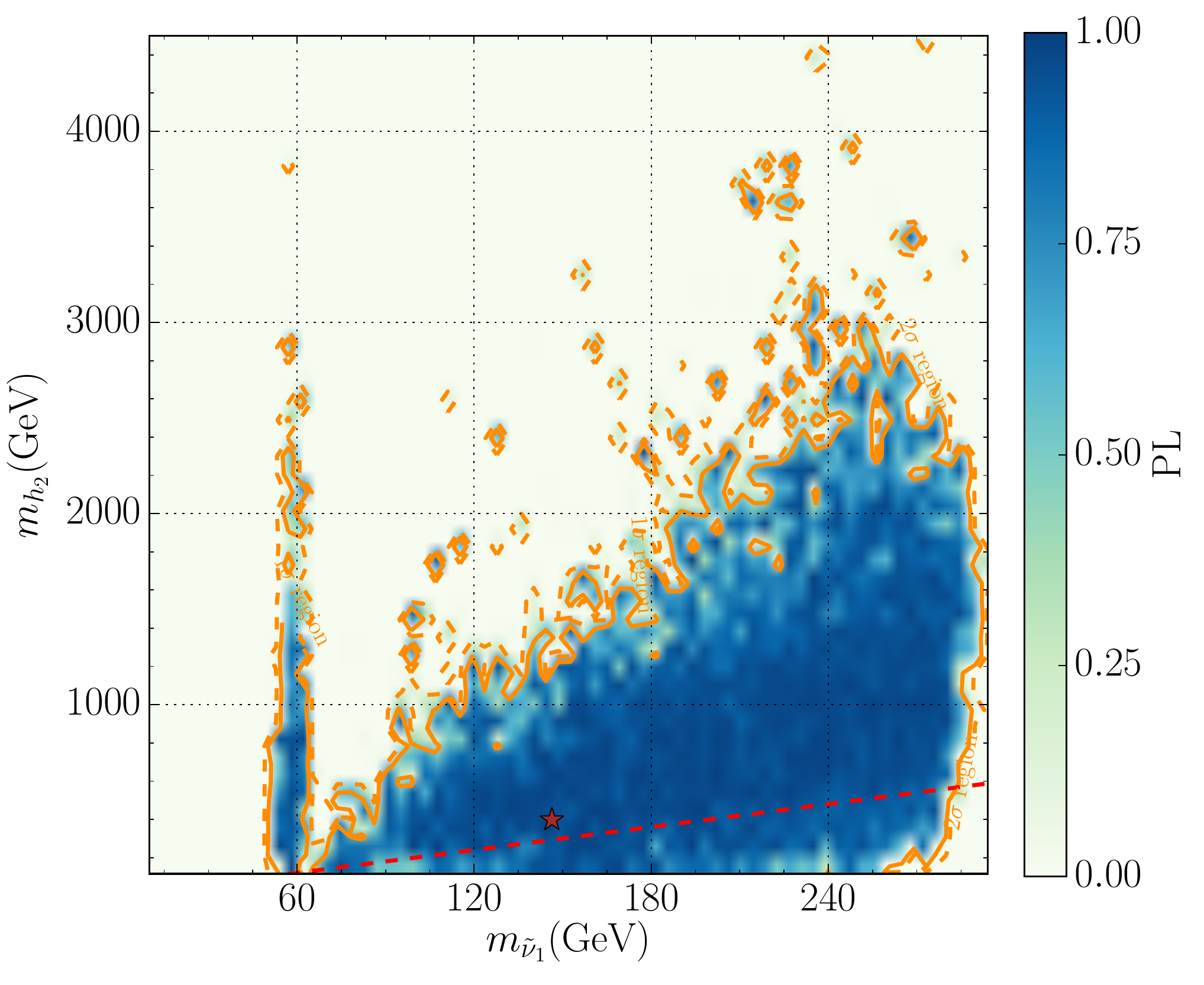}
		\includegraphics{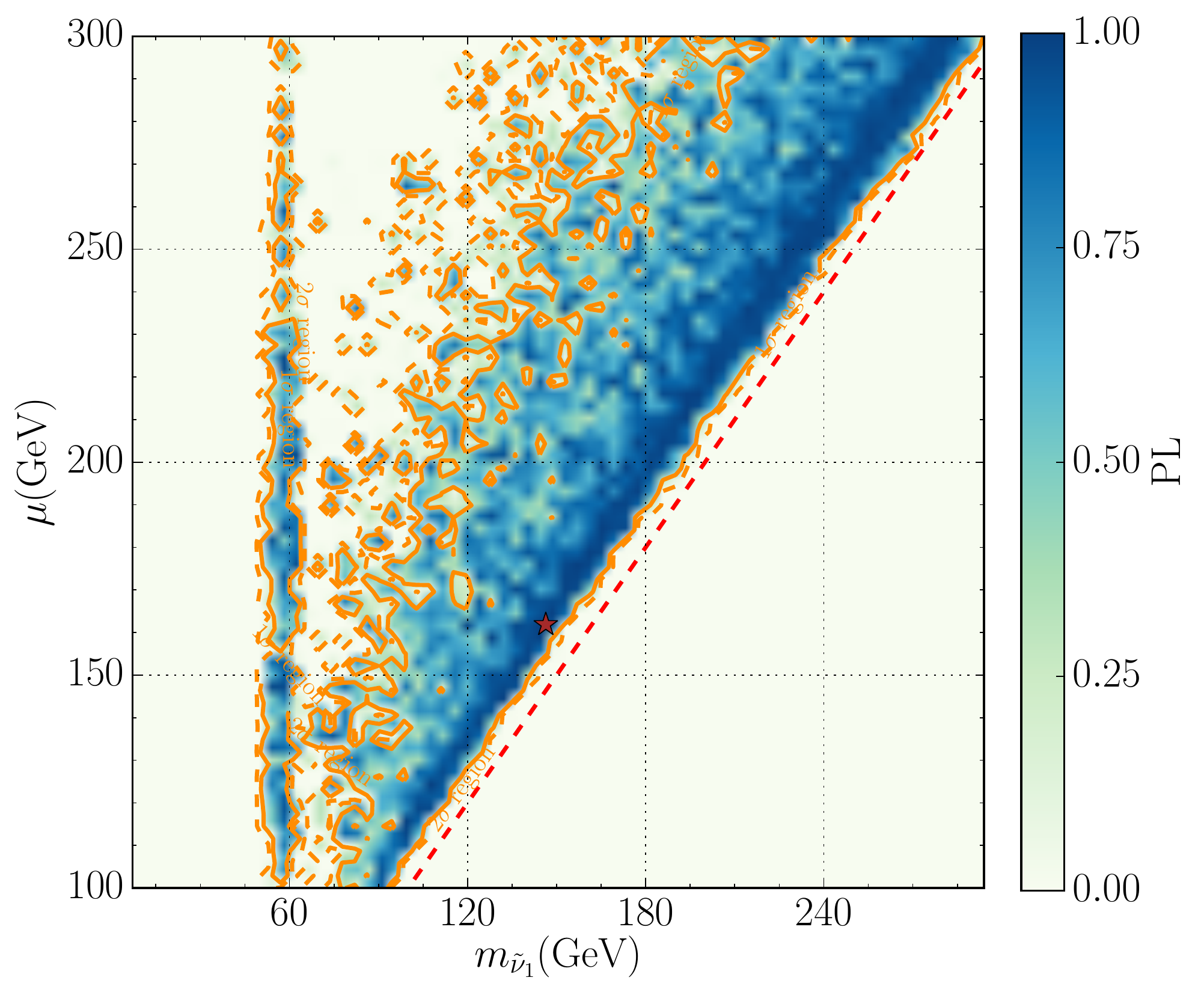}
		}

       \vspace{-0.4cm}

		\caption{Similar to FIG.\ref{fig3}, but for the case that the DM candidate $\tilde{\nu}_1$ is CP-odd.  \label{fig9}}
	\end{figure*}

	\begin{figure*}[htbp]
		\centering
		\resizebox{0.58\textwidth}{!}{
		\includegraphics{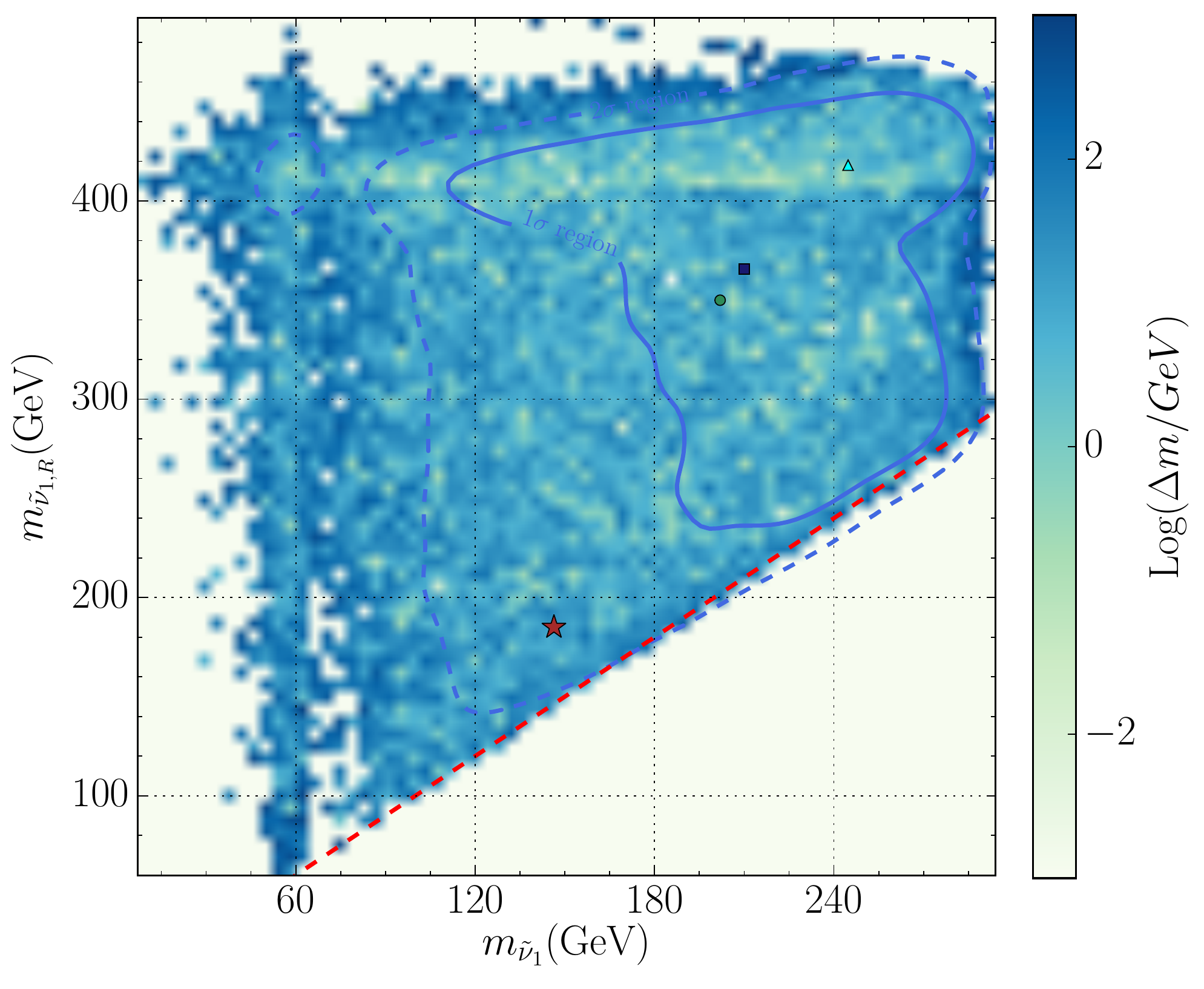}
		\includegraphics{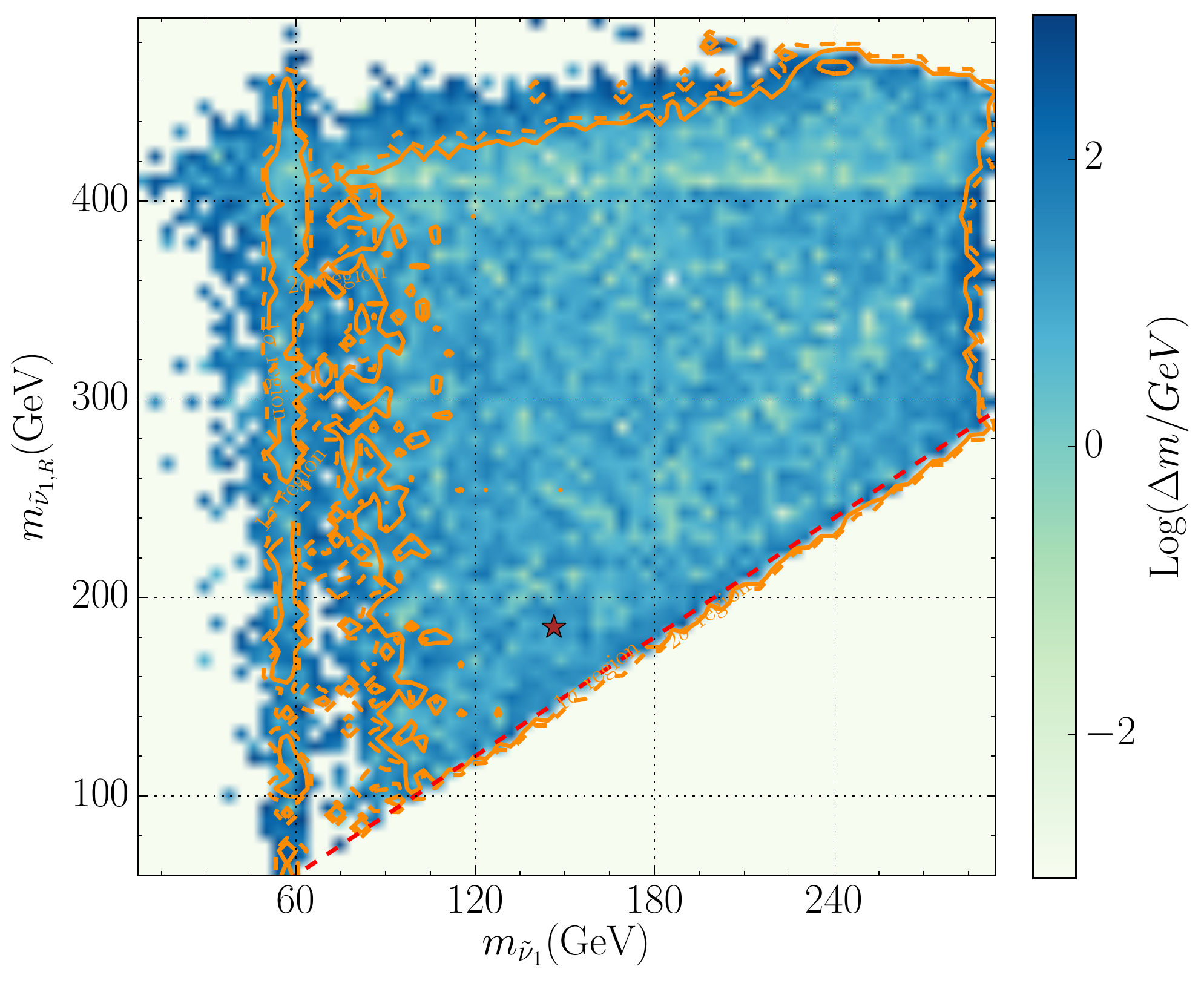}
		}

      \vspace{-0.4cm}

		\caption{Similar to FIG.\ref{fig4}, but for the case that the DM candidate $\tilde{\nu}_1$ is CP-odd.  \label{fig10}}
	\end{figure*}

	\begin{figure*}[htbp]
		\centering
		\resizebox{0.58\textwidth}{!}{
		\includegraphics{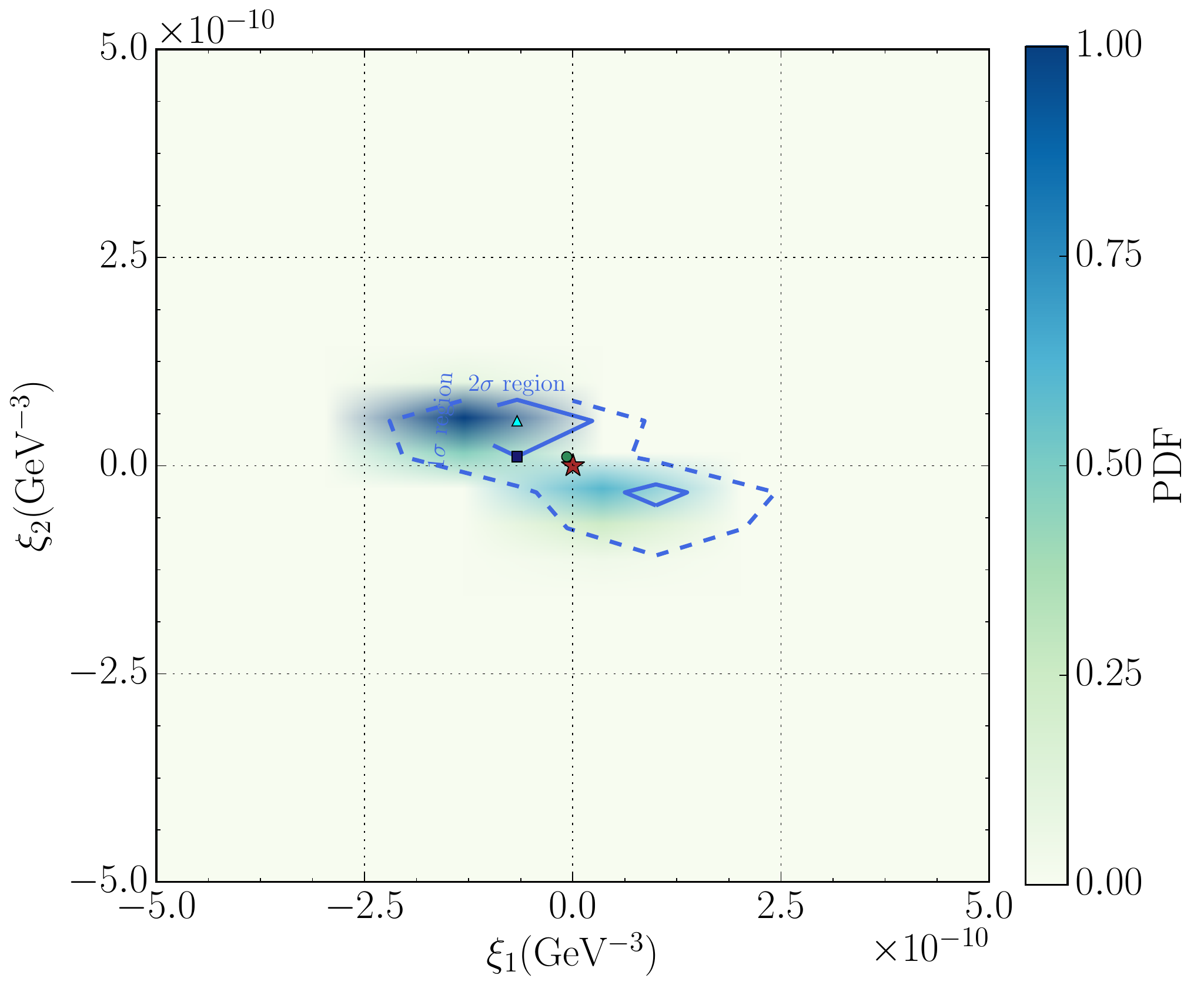}
		\includegraphics{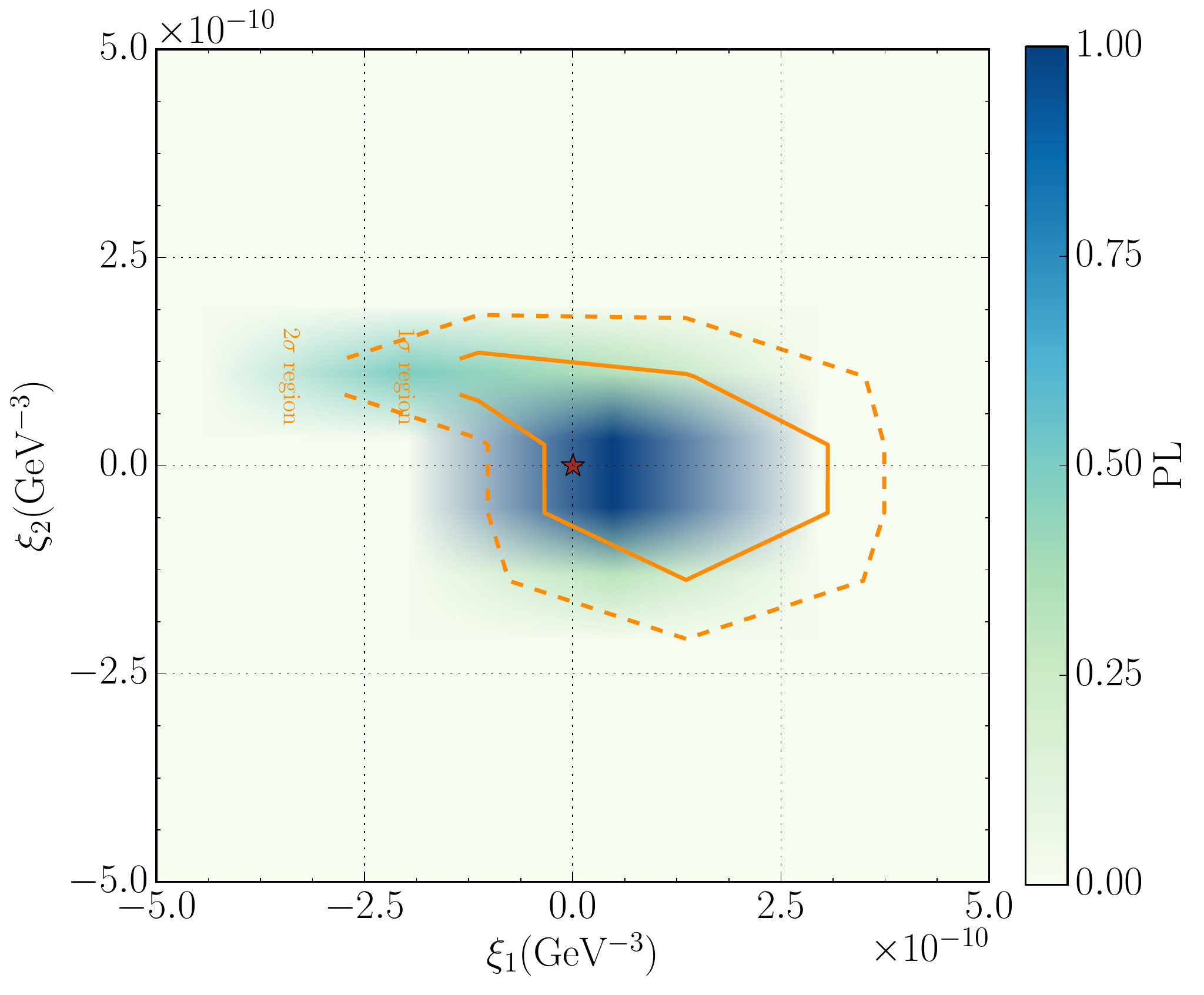}
		}
		\resizebox{0.58\textwidth}{!}{
		\includegraphics{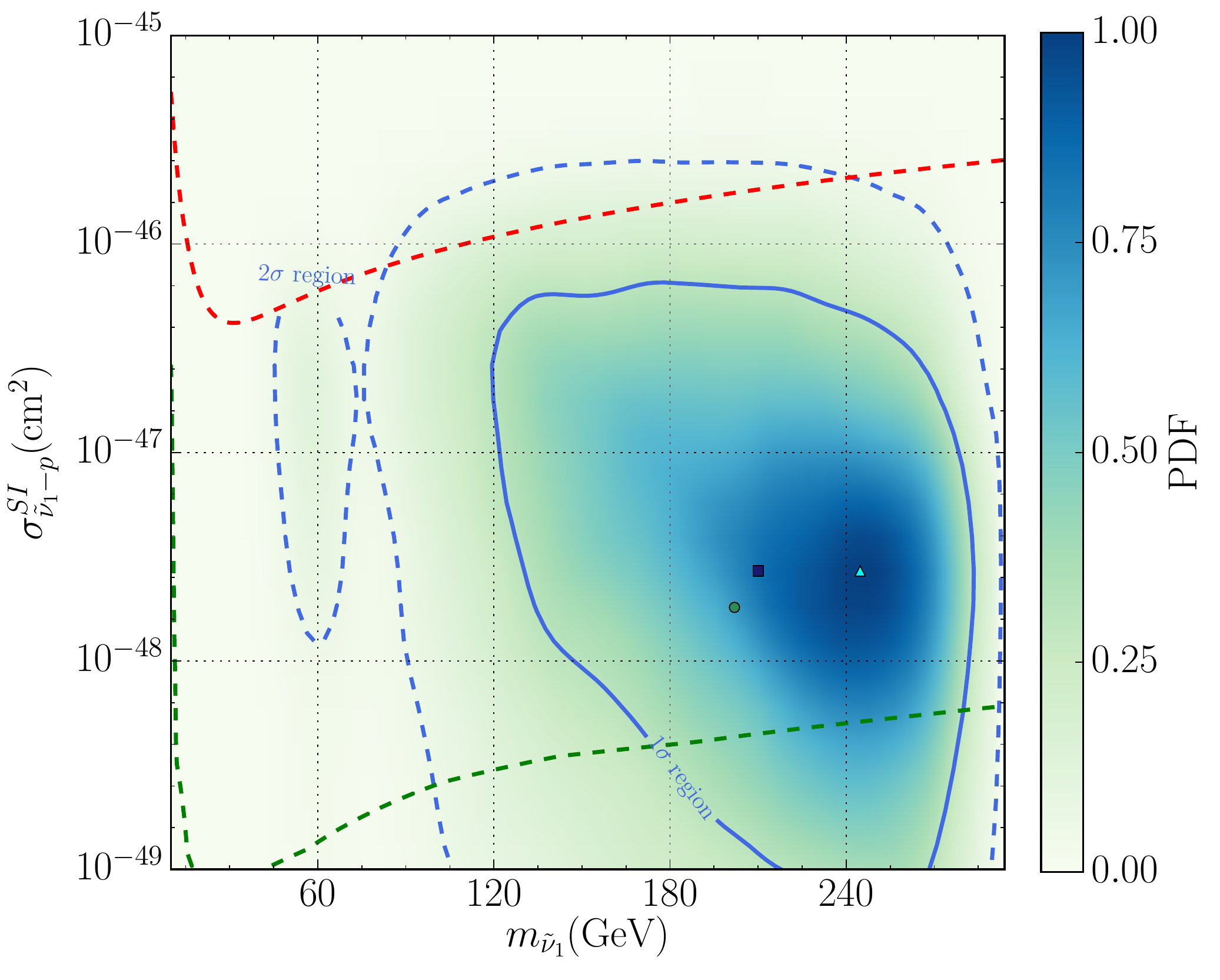}
		\includegraphics{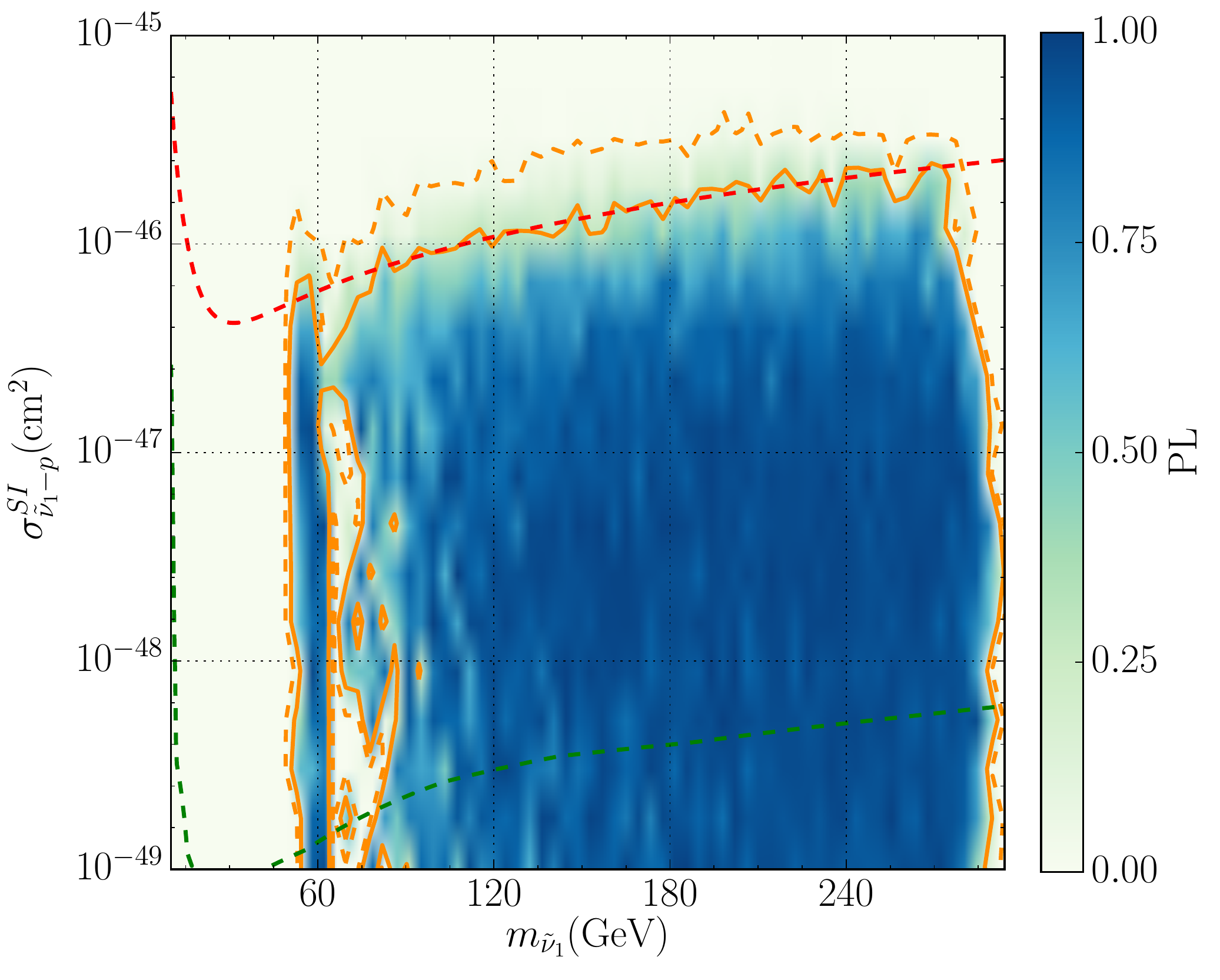}
		}

       \vspace{-0.4cm}

      \caption{Similar to FIG.\ref{fig5}, but for the case that the DM candidate $\tilde{\nu}_1$ is CP-odd.
			\label{fig11}}
	\end{figure*}
	
	\begin{figure*}[htbp]
		\centering
		\resizebox{0.58\textwidth}{!}{
		\includegraphics{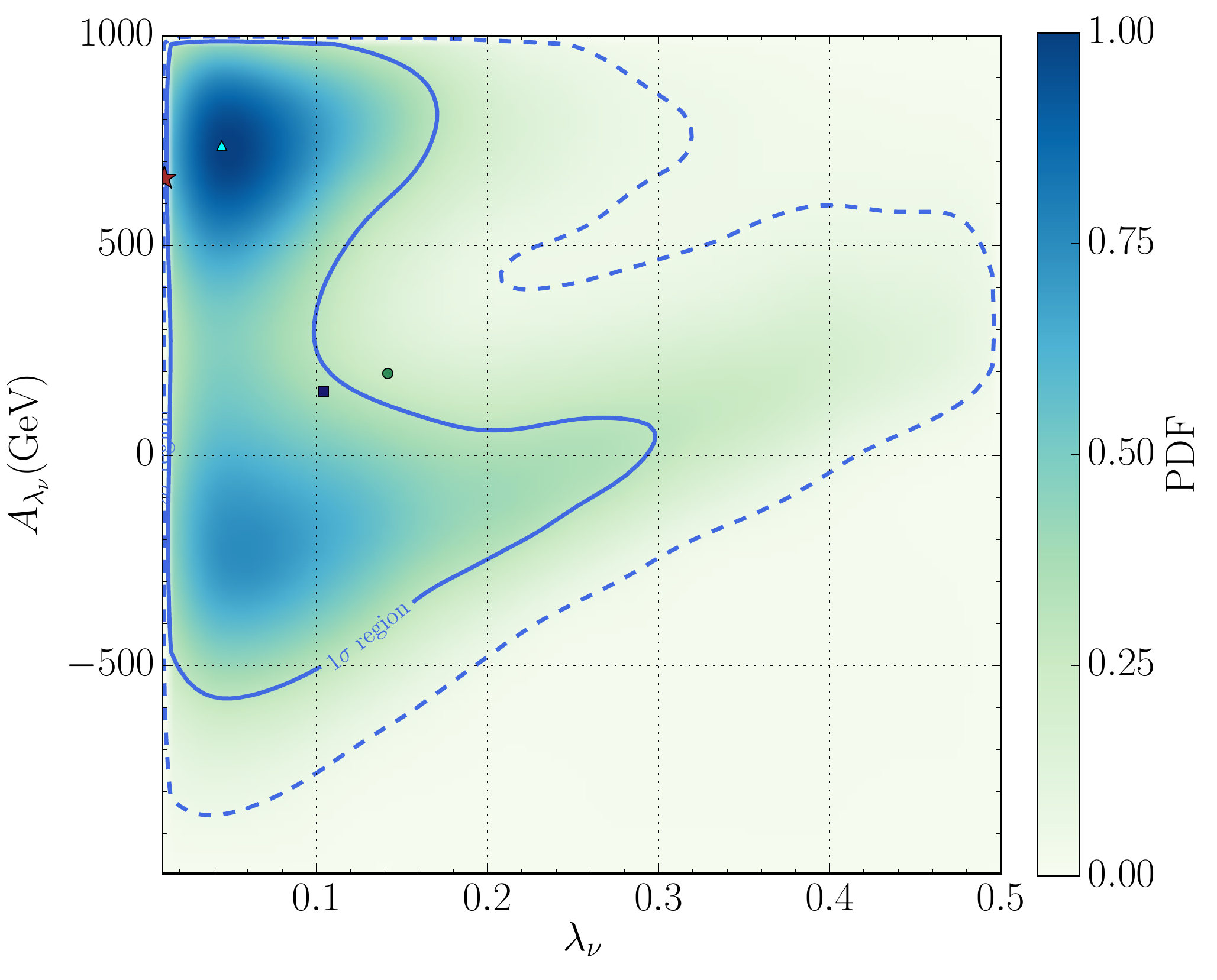}
		\includegraphics{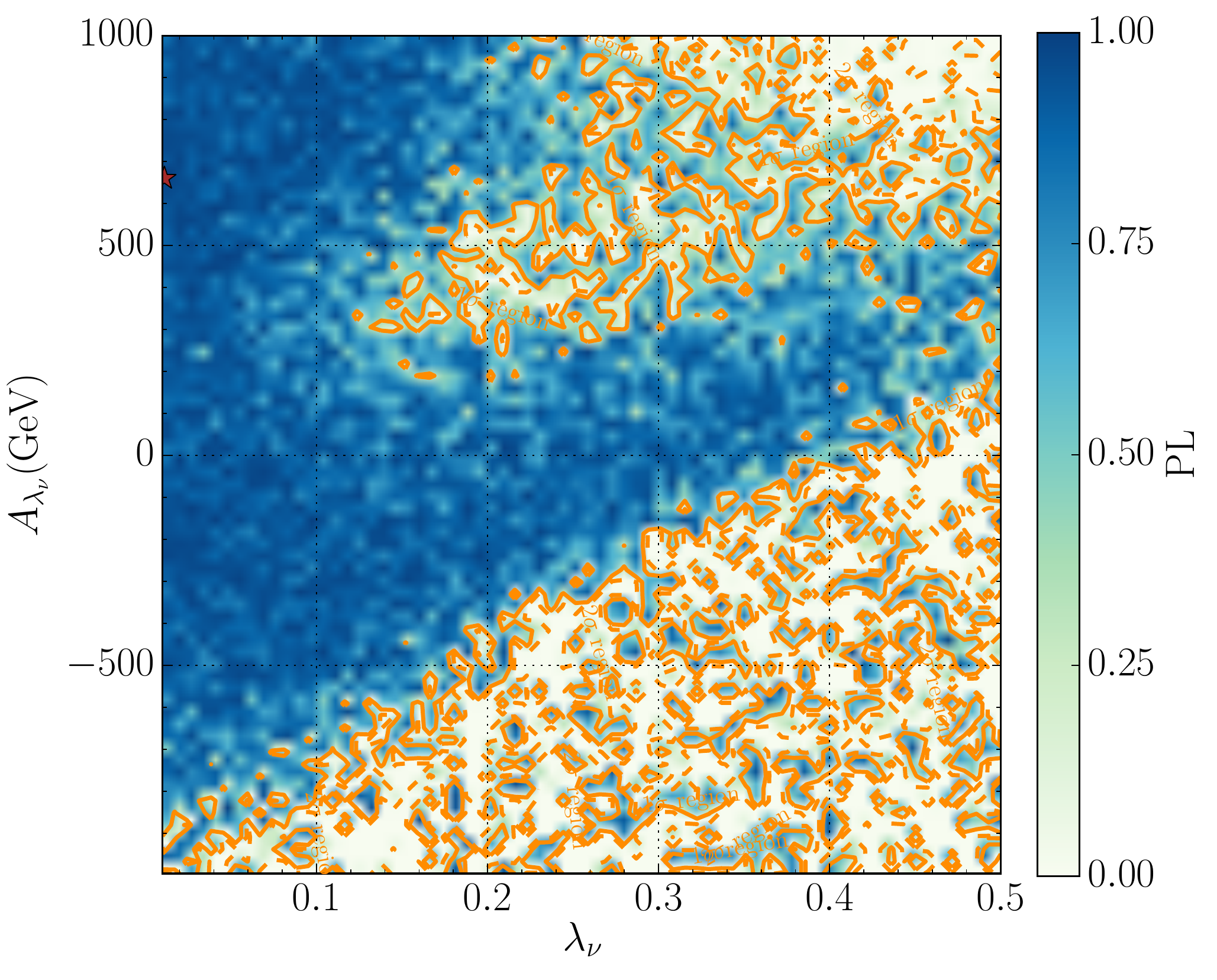}
		}
		\resizebox{0.58\textwidth}{!}{
		\includegraphics{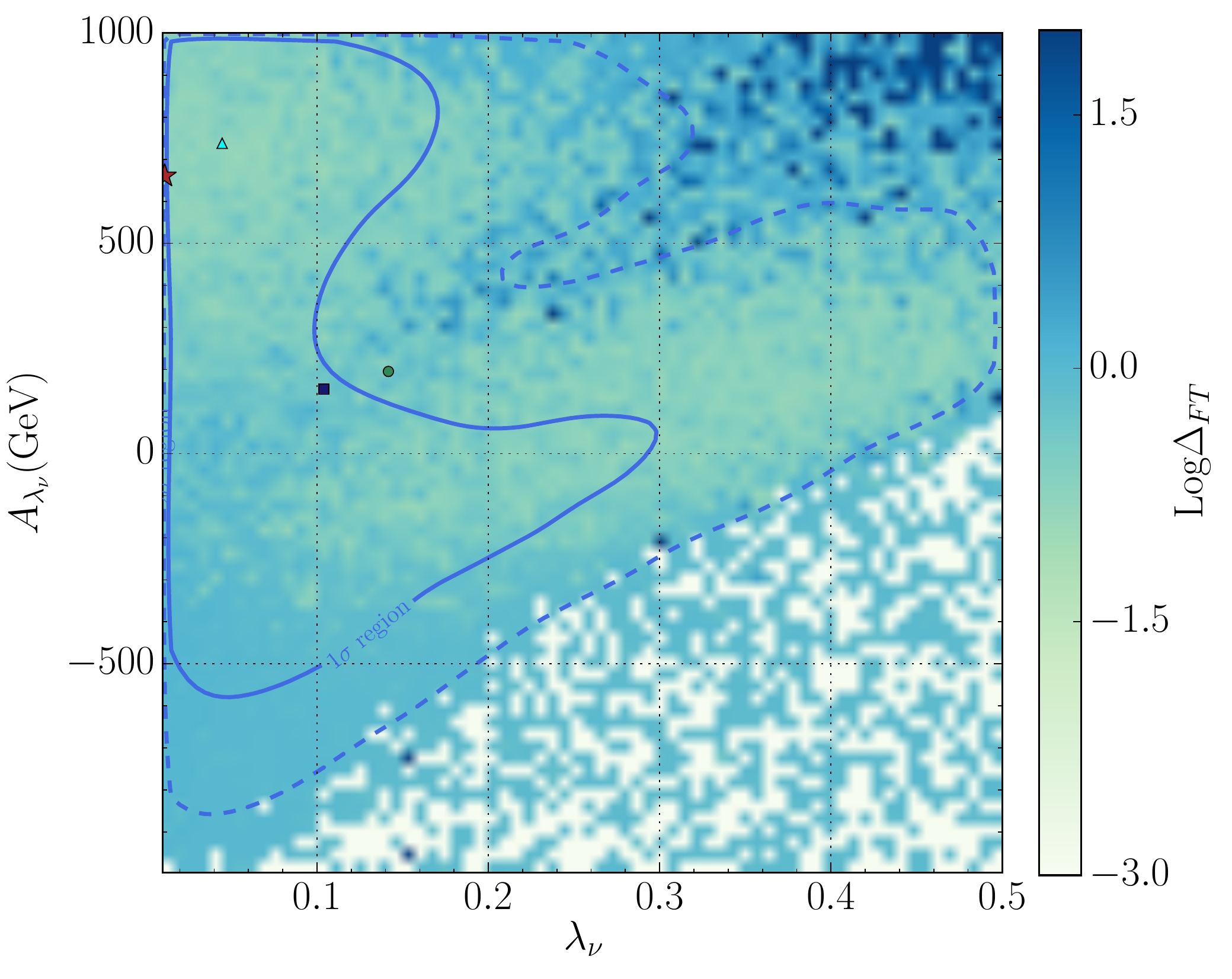}
		\includegraphics{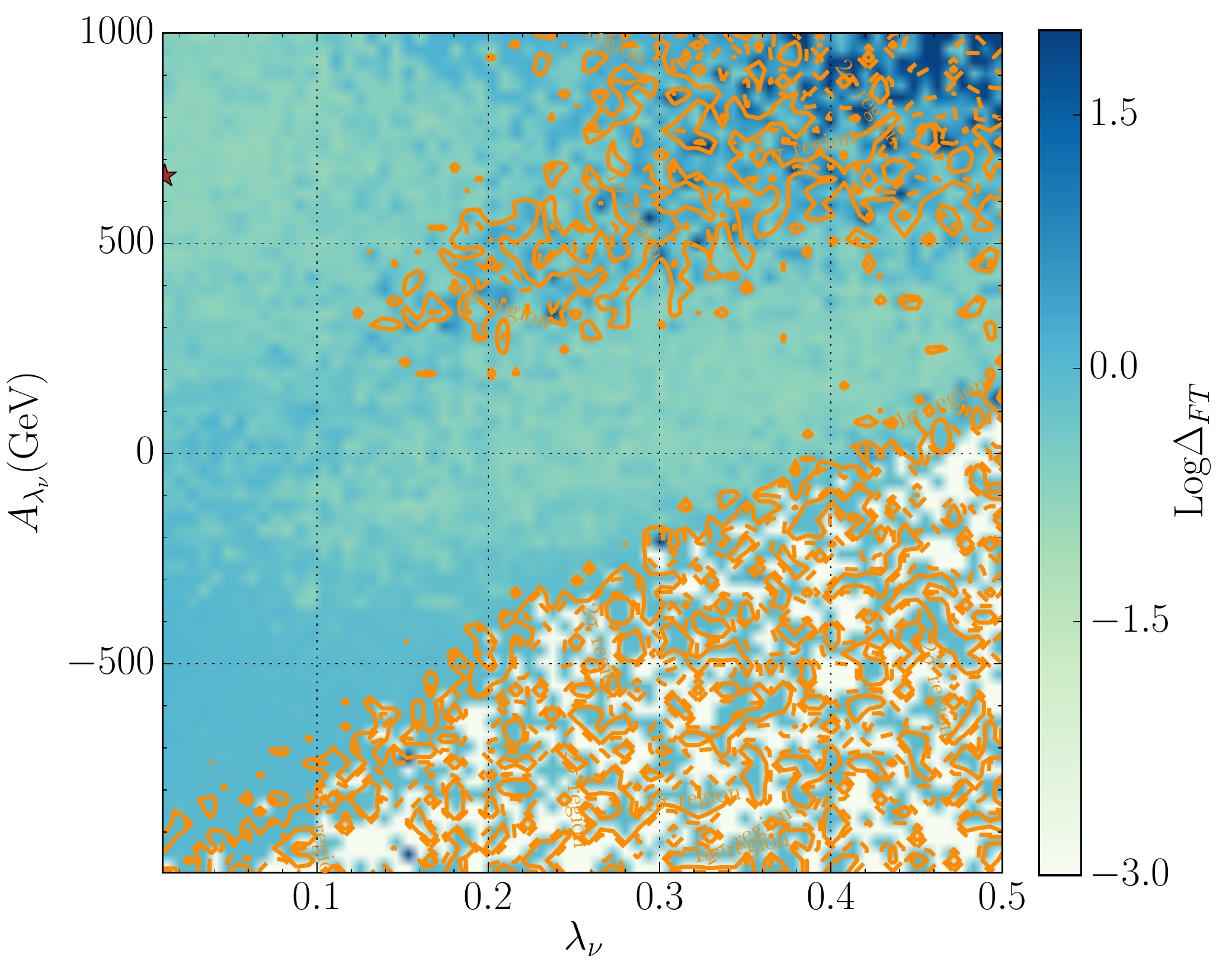}
		}

       \vspace{-0.4cm}

		\caption{Similar to FIG.\ref{fig6}, but for the case that the DM candidate $\tilde{\nu}_1$ is CP-odd.   \label{fig12}}
	\end{figure*}	

\begin{small}
	\begin{table*}[htbp]
		\renewcommand\arraystretch{1.5}
		\resizebox{0.7\textwidth}{!}{
			\begin{tabular}{|c|c|c|c|c|c|c|c|}
				\hline
				\multicolumn{4}{|c|}{CP-even Sneutrino DM} & \multicolumn{4}{c|}{CP-odd Sneutrino DM} \\ \hline
				$\lambda$ 					& 0.28 			& $\kappa$ 									& -0.60 				& $\lambda$ 					& 0.21 		& $\kappa$ 				& -0.37      \\ 
				$\tan\beta$ 					& 23.7 			& $\lambda_\nu$ 							& 0.17 				& $\tan\beta$ 					& 17.1  		& $\lambda_\nu$ 		& 0.01   \\ 
				$\mu$ 						& 109 \ GeV 		& $A_{\lambda_\nu}$ 						& -193 \ GeV 			& $\mu$ 						& 162 \ GeV 	& $A_{\lambda_\nu}$     & 660 \  GeV    \\ 
				$A_\kappa$ 					& 427 \ GeV 		& $A_t$ 									& 3133 \ GeV 			&  $A_\kappa$ 					& 623 \ GeV 	& $A_t$                 & 3688 \ GeV     \\ 
				$m_{\tilde\nu}$ 			& 206 \ GeV 		& $m_{\tilde\nu_1}$							& 54.3 \ GeV 			& $m_{\tilde\nu}$ 			& 166 \ GeV 	& $m_{\tilde\nu_1}$     & 146.4 \ GeV         \\ 
				$m_{h_1}$						& 125.2 \ GeV 		& $m_{h_2}$ 								& 338 \ GeV 			& $m_{h_1}$ 					& 125.1 \ GeV 	& $m_{h_2}$ 			& 398 \  GeV     \\ 
				$m_{h_3}$ 					& 2188.8 \ GeV 		& $m_{A_1}$ 								& 535 \ GeV 			& $m_{h_3}$ 					& 2231.5 \ GeV 	& $m_{A_1}$ 			& 733.0 \ GeV     \\ 
				$m_{A_2}$ 					& 2189.3 \ GeV 		& $m_{\tilde\chi^0_1}$ 						& 107.6 \ GeV 			& $m_{A_2}$ 					& 2231.8 \ GeV 	& $m_{\tilde\chi^0_1}$ 	& 159.1 \ GeV      \\ 
				$m_{\tilde\chi^0_2}$ 			& -114.6 \ GeV 		& $m_{\tilde\chi^\pm_1}$ 					& 111.9 \ GeV 			& $m_{\tilde\chi^0_2}$ 			& -169.6 \ GeV 	& $m_{\tilde\chi^\pm_1}$& 165.1 \  GeV       \\ 
				$\sigma^{SI}_{\tilde\nu_1-p}$ & $1.0 \times 10^{-48} \ cm^2$ 	& ${\left\langle \sigma v\right\rangle}_0$ 	& $7.0 \times 10^{-29} \ cm^3 s^{-1}$   & $\sigma^{SI}_{\tilde\nu_1-p}$ 	& $1.0 \times 10^{-49} cm^2$ & ${\left\langle \sigma v\right\rangle}_0$ & $2.4 \times 10^{-30} \ cm^3 s^{-1}$       \\ 
				$\Omega h^2$ 					& 0.119             & 	 $\chi^2$						& 	74.5						& $\Omega h^2$						 & 0.118 		  &  $\chi^2$      & 74.6  \\ \hline
	  \end{tabular}}

       \vspace{-0.2cm}

		\caption{Detailed information of the best point for the CP-even and CP-odd sneutrino DM cases.}
		\label{table3}
	\end{table*}
     \end{small}

     	\begin{figure*}[tbp]
		\centering
		\resizebox{0.7\textwidth}{!}{
		\includegraphics{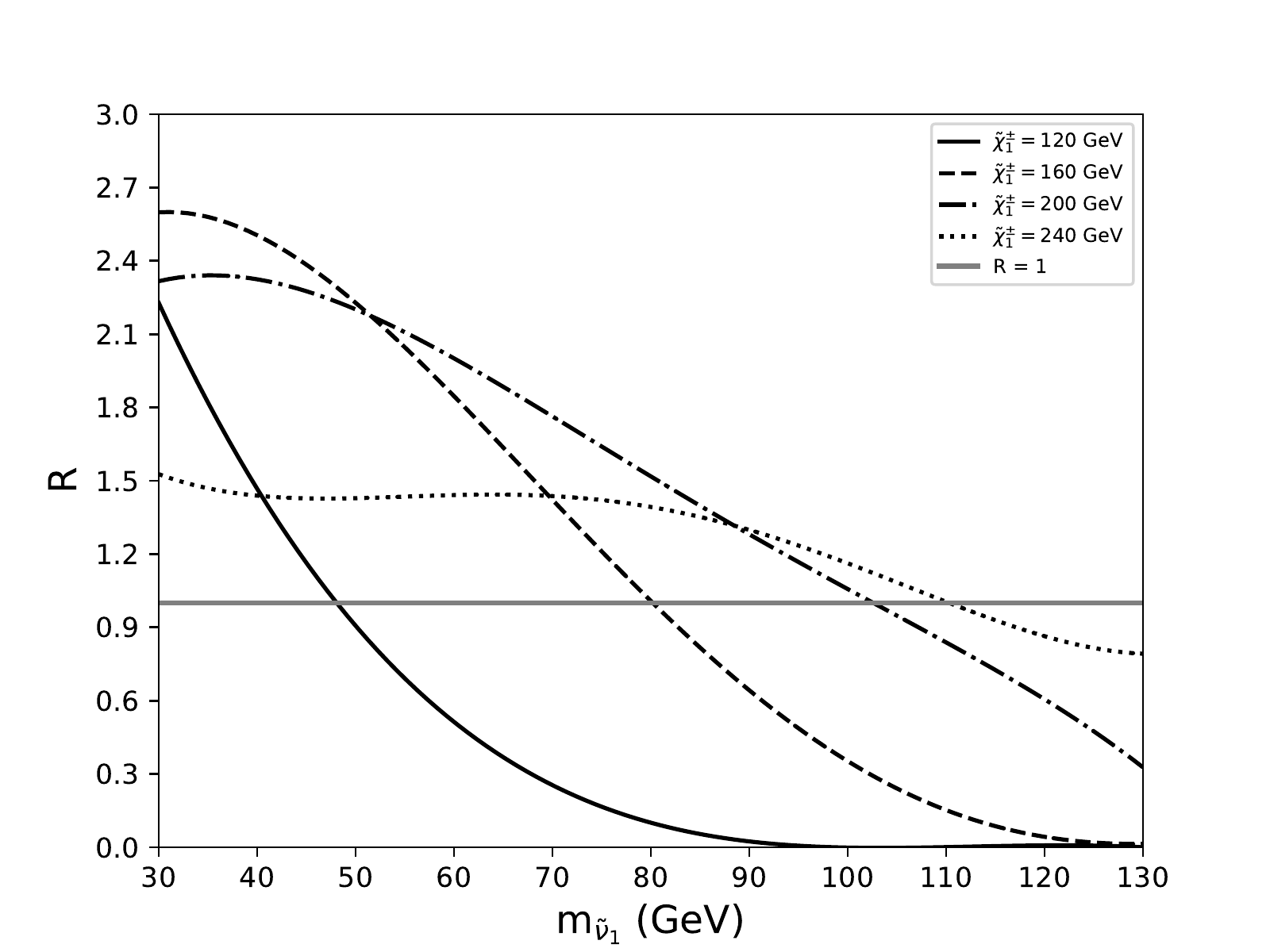}
		\vspace{-0.4cm}
        \includegraphics{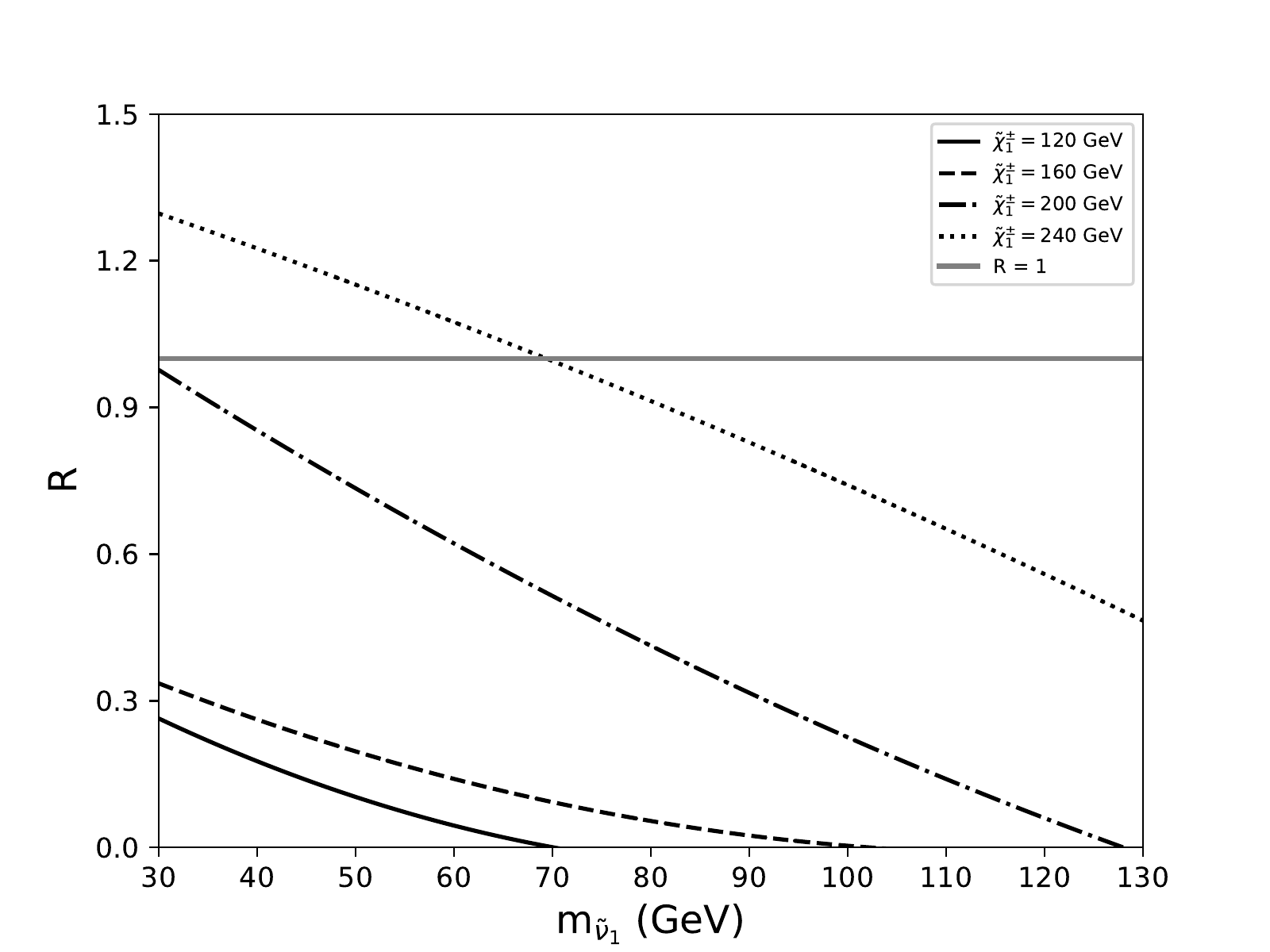}
		}

       \vspace{-0.4cm}

		\caption{The dependence of $R\equiv  S/S_{95}^{OBS}$ on the sneutrino DM mass for the analysis of the $2 \tau + E_{T}^{miss}$ signal at 8TeV LHC (left panel) and at 13TeV
          LHC (right panel). In the Type-I seesaw extension of the NMSSM, this signal comes from the process $p p \to \tilde{\chi}_1^\pm \tilde{\chi}_1^\mp$. In getting this figure, we choose $m_{\tilde{\chi}^\pm} = 120 {\rm GeV}, 160 {\rm GeV}, 200 {\rm GeV}, 240 {\rm GeV}$ as benchmark points, and calculate the cross sections of the process at next-to-leading order by the code Prospino\cite{prospino}.  We also assume $Br(\tilde{\chi}_1^\pm \to \tilde{\nu}_1 \tau) = 100\%$.
			If the assumption is not satisfied, the $R$ value should be rescaled by the factor $Br^2(\tilde{\chi}_1^\pm \to \tilde{\nu}_1 \tau)$. } \label{fig13}
	\end{figure*}
		
In a similar way, one may discuss the physics about a CP-odd sneutrino DM. The results are presented in FIG.\ref{fig7} for 1D posterior PDFs and PLs of the input parameters, and in FIG.\ref{fig8}-\ref{fig12} for 2D distributions which are analogy with FIG.\ref{fig2}-\ref{fig6}. By comparing each panel for the CP-odd case with its corresponding one for the CP-even case, one can learn that the two cases are somewhat similar except that the former favors positive $\kappa$ and $A_{\lambda_\nu}$ and a negative $A_\kappa$,
while the latter prefers opposite signs. Such a difference can be understood from following facts:
\begin{itemize}
\item As indicated by the expression of sneutrino mass in Eq.(\ref{sneutrino_matrix}) of Section II, positive $\kappa$ and $A_{\lambda_\nu}$ can predict a CP-odd sneutrino state which is lighter than its corresponding CP-even state.
\item From the discussion about the annihilation of the sneutrino DM, one can learn that its couplings with the singlet Higgs fields play an important role in determining the favored parameter space. Eq.(\ref{Csnn}) and Eq.(\ref{CsnnAA}) of section II show that the couplings of the CP-odd state are related with those of the CP-even state by the substitution $\kappa \to -\kappa$ and $A_{\lambda_\nu} \to -A_{\lambda_\nu}$. Moreover, the analysis in Appendix A reveals that the Higgs spectrum is invariant under the transformation $\kappa \to -\kappa$ and $A_\kappa \to -A_\kappa$ if $A_\lambda \gg \kappa v_s$. These intrinsic relations imply that, as a rough approximation, the results of the CP-odd case can be obtained from those of the CP-even case by the replacement $\kappa \to -\kappa$, $A_\kappa \to -A_\kappa$ and $A_{\lambda_\nu} \to -A_{\lambda_\nu}$, which is justified by our calculation.
\item In any case, a negative $\kappa A_\kappa$ is favored by the positiveness of the squared mass for the CP-odd singlet field. Related discussions about the parameters $\kappa$ and $A_\kappa$ can be found in Appendix A and B.
\end{itemize}
Besides, we compute the Bayesian evidence for the CP-odd case, and find $\ln Z = - 50.12 \pm 0.08$,
which is slightly larger than $\ln Z = -50.39 \pm 0.06$ for the CP-even case. Since the Jeffreys' scale defined in~\cite{Jeffreys} is 0.27 and small, we infer that the data has no significant preference of the CP-odd case over the CP-even case~\footnote{Given two scenarios to be compared beside one another, the Jeffreys' scale presents a calibrated spectrum of significance for the relative strength between the Bayesian evidences of the scenarios~\cite{Jeffreys}. For the application of Jeffreys' scale in particle physics, see for example~\cite{Feroz:2008wr}. }.
We also show the detailed information of the best fit point for the two cases in Table \ref{table3}.
We find that the $\chi^2$s of the best points are roughly equal, $\chi^2_{min} \simeq 74.6$, with most of its contribution coming from the Higgs sector, which contains $74$ observables in the peak-centered method of the package HiggsSignal~\cite{HiggsSignal}.

\section{\label{LHC-search} Constraints from the LHC experiments}

	In supersymmetric theories, moderately light Higgsinos are favored by naturalness. So they are expected to be copiously produced at the LHC, and the detection of their signals can significantly limit the theory~\cite{Arina:2013zca,Guo:2013asa,Chatterjee:2017nyx,Cao:2015efs}. In the NMSSM with Type-I seesaw mechanism, the neutral Higgsinos mix with the Singlino, the fermionic component field of the singlet superfield $\hat{S}$, to form mass eigenstates called neutralinos. Due to the tininess of the Yukawa coupling $Y_\nu$, the Higgsino-dominated neutralinos may have a much stronger coupling with the $\tilde{\nu}_1 \nu_R$ state, which is induced by the $\lambda_\nu \hat{s} \hat{\nu} \hat{\nu}$ term in the superpotential, than with the $\tilde{\nu}_1 \nu_L$ state, which comes from the $Y_\nu \hat{l}\cdot \hat{H}_u \hat{\nu}$ interaction. The decay product of the neutralinos is then complicated by the decay chain $\tilde{\chi}^0 \to  \tilde{\nu}_1 \nu_R $ with $\nu_R \to W^{(\ast)} l, Z^{(\ast)} \nu_L, h^{(\ast)} \nu_L$\cite{Cerdeno:2013oya}\footnote{If the neutralino decay $\tilde{\chi}^0 \to  \tilde{\nu}_1 \nu_R $ is kinematically forbidden, $\tilde{\chi}^0$ has to decay into $\tilde{\nu}_1 \nu_L $ assuming no mass splitting among the Higgsino-dominated particles. In this case, the experimental constraints on the Mono-jet signal from the neutralino pair production is rather weak, which has been shown in our previous work \cite{Cao:2017cjf}.}. On the other hand, this situation is simplified greatly for the Higgsino-dominated chargino, which may decay mainly by the channel $\tilde{\chi}^\pm \to \tilde{\nu}_1 \tau^\pm$ if the LSP $\tilde{\nu}_1$ carries $\tau$ flavor, and meanwhile the magnitude of the 33 element in the matrix $Y_\nu$ is much larger than that of the other elements~\footnote{We remind that although the effect of the neutrino Yukawa coupling $Y_\nu$ on sneutrino mass can be safely neglected, the non-diagonality of $Y_\nu$ can induce flavor changing decays $\tilde{\chi}_1^\pm \to \tilde{\nu}_1 e, \tilde{\nu}_1 \mu$, which has been tightly limited by the latest CMS search for sleptons if the mass splitting between $\tilde{\chi}_1^\pm$ and $\tilde{\nu}_1$ is sizable~\cite{Sirunyan:2018nwe}. From theoretical point of view, these decays can be suppressed greatly if the $33$ element of $Y_\nu$ is much larger than the other elements.  In practice, one can easily obtain the hierarchy structure by adopting the Casas-Ibarra parameterization of $Y_\nu$ and scanning randomly the angles of the orthogonal matrix $R$ and the right-handed neutrino masses involved in the parameterization \cite{Casas:2001sr}. }. So in this subsection we investigate the constraints of the ATLAS analysis at 8TeV LHC on the signal of two hadronic $\tau$s plus $E^{miss}_T$\cite{tausLHC1}, which arises from the process $p p \to \tilde{\chi}_1^\pm \tilde{\chi}_1^\mp \to 2 \tau + E_T^{miss}$ in the Type-I extension. We note that recently both the ATLAS and CMS collaboration updated their analysis on the $2 \tau + E_T^{miss}$ signal with the data collected at 13TeV LHC~\cite{Aaboud:2017nhr,Sirunyan:2018nwe,CMS:2017wox,CMS:2017rio}. As a comparison, we also study
the constraints from the renewed ATLAS analysis~\cite{Aaboud:2017nhr}.

In our calculation, we use the simulation tools MadGraph/MadEvent~\cite{mad-1,mad-2} to generate the parton level events of the processes, Pythia6~\cite{pythia} for parton fragmentation and hadronization, Delphes~\cite{delphes} for fast simulation of the performance of the ATLAS detector, and CheckMATE~\cite{cmate-1,cmate-2,cmate-3} to implement
the cut selections of the analysis. The validation on the implementation of the analysis in CheckMATE was provided in our work~\cite{Cao:2017cjf,Cao:2018rix}.
	
The procedure to get the constraints is as follows: we first determine the signal region (SR) with the largest expected sensitivity
for a given parameter point $(m_{\tilde{\nu}_1}, m_{\tilde{\chi}_1^\pm})$ (see footnote 6 in~\cite{Cao:2017cjf} for more details),
then we calculate its $R$ value defined by $R \equiv S/S_{95}^{OBS}$, where $S$ stands for the number of signal events in the SR with the
statistical uncertainty considered and $S_{95}^{OBS}$ denotes the observed limit at 95\% confidence level for the SR.  Obviously, $R$ represents the capability
of the LHC in exploring the point. $R > 1$ implies that the point is excluded, or else it is allowed. In FIG.\ref{fig13}, we present our results of
$R$ as a function of $m_{\tilde{\nu}_1}$ for four choices of $m_{\tilde{\chi}_1^\pm}$, $m_{\tilde{\chi}_1^\pm} = 120 {\rm GeV}, 160 {\rm GeV}, 200 {\rm GeV}, 240 {\rm GeV}$.
The left and right panels are for 8TeV results and 13TeV results respectively. From the left panel, one can learn that
$R$ increases monotonously with the enlarged mass splitting $\Delta m \equiv m_{\tilde{\chi}_1^\pm} - m_{\tilde{\nu}_1}$ for fixed
$m_{\tilde{\chi}_1^\pm}$, and $R=1$ corresponds to $\Delta m = 65 {\rm GeV}, 70 {\rm GeV}, 85 {\rm GeV}, 150 {\rm GeV}$, respectively,
for the $m_{\tilde{\chi}_1^\pm}$s. The underlying physics is that a large mass splitting tends to enhance the cut efficiency, and the number of the signal event $S$ depends on both the cut efficiency and the production rate $\sigma (p p \to \tilde{\chi}_1^\pm \tilde{\chi}_1^\mp)$, which decreases as the chargino becomes heavy~\cite{prospino}.
The right panel shows similar features, except that the $R$ values are significantly smaller than corresponding ones at 8TeV LHC.
This reflects that the analysis at the $8{\rm TeV}$ LHC is more powerful in limiting the model than that at the $13 {\rm TeV}$ LHC when $m_{\tilde{\chi}_1^\pm} \lesssim 260{\rm GeV}$. The difference comes from the fact that the updated analysis focuses on heavy chargino case, which requires more energetic jets and larger missing energy than the original analysis.
	
With respect to the results in FIG.\ref{fig13}, two points should be noted. One is that we have assumed $Br(\tilde{\chi}_1^\pm \to \tilde{\nu}_1 \tau) = 100\%$ in getting the figure. But in practice, $\tilde{\nu}_1$ may be either $\tilde{\nu}_{1,R}$ or $\tilde{\nu}_{1,I}$, and the two channels $\tilde{\chi}_1^\pm \to \tilde{\nu}_{1,R}  \tau, \tilde{\nu}_{1,I}  \tau$ can be kinematically accessible simultaneously. Moreover, the chargino may decay first into $\tilde{\chi}_1^0 j j^\prime/\pi^\pm$ after considering the mass splitting among the Higgsino-dominated particles\cite{Chatterjee:2017nyx}.  So in case of $Br(\tilde{\chi}_1^\pm \to \tilde{\nu}_1 \tau) < 1$, the $R$ value in FIG.\ref{fig13} should be rescaled by the factor of $Br^2 (\tilde{\chi}_1^\pm \to \tilde{\nu}_1 \tau)$, which can weaken the constraint. The other is that, from the plots on the $\mu-\tilde{\nu}_1$ plane in FIG.\ref{fig3} and FIG.\ref{fig9}, the mass splitting $\Delta m = \mu - m_{\tilde{\nu}_1}$ is of ${\cal{O}}(10 {\rm GeV})$ for most samples. In this case, the LHC searches for the $2 \tau$ signal actually have no exclusion ability.

	\section{\label{Section-Conclusion}Conclusions}

	Given the strong tension between the naturalness for $Z$ boson mass and the DD experiments for customary neutralino DM candidate in minimal supersymmetric theories, it is essential to explore the DM physics in any extension of the MSSM or the NMSSM. In this work, we augment the NMSSM with Type-I seesaw mechanism, which is the simplest extension to reconcile the neutrino non-zero masses and neutrinos oscillation experiment, and carry out a comprehensive study on sneutrino DM physics. The highlight of the theory is that the singlet Higgs field plays an important role in various aspects, including generating the Higgsino mass and the heavy neutrino masses dynamically, mediating the transition between the DM pair and Higgsino pair to keep them in thermal bath in early Universe, acting as DM annihilation final state or mediating DM annihilations, as well as contributing to DM-nucleon scattering rate. Moreover, since sneutrino DM does not interact with SM particles directly, the scattering of the sneutrino DM with nucleon can be suppressed in a natural way and by several mechanisms so that the tension is alleviated greatly even after considering the latest XENON1T results. In order to illustrate these features, we carry out a sophisticated scan over the vast parameter space by Nested Sampling method, and adopt both Bayesian and frequentist statistical quantities to analyze the favored parameter space of different scenarios, the DM annihilation mechanism as well as the behavior of the DM-nucleon scattering confronted with the tight DD constraints. To get the statistical inference, we construct the likelihood function by considering Higgs data, B-physics measurements, DM relic density as well as DM direct detection and indirect detection limits. We obtain following key conclusions:
	\begin{itemize}
		\item[(1)] The model provides a viable sneutrino DM candidate over a broad parameter space. In particular, moderately light Higgsinos, $\mu \lesssim 250 {\rm GeV}$, are
allowed for a large portion of the parameter space, which, as a theoretical advantage, can predict $Z$ boson mass in a natural way.  To the best of our knowledge, the model may be the minimal framework to accommodate sneutrino as WIMP DM in supersymmetric theory.
		\item[(2)] The DM and the singlet-dominated Higgs bosons can compose a realistic DM sector, which communicates with the SM sector only by the small doublet component of the bosons. This is a typical feature of hidden or secluded DM scenario.
		\item[(3)] In most cases, the DM co-annihilated with the Higgsinos to get its right relic density, which was omitted in previous studies.
		\item[(4)] The sneutrino DM scenarios can satisfy the tight constraints from the recent XENON1T experiment on DM-nucleon scattering without any fine tuning.
		\item[(5)] The sneutrino DM scenarios can naturally survive the constraints from the direct searches for electroweakinos with the final state of $2\tau + E^T_{miss}$ signal
		at the LHC.
	\end{itemize}

Finally, we have more explanations about this work:
\begin{itemize}
\item Throughout our discussion, we do not consider the scenario that the second lightest CP-even Higgs $h_2$ as the discovered Higgs boson. Compared with this scenario, the situation for $h_1$ as the SM-like Higgs boson with a CP-even sneutrino DM is much more strongly supported by the experimental data. Numerically speaking, we find that the Jeffreys' scale is about 6.0 (5.9) for the $h_2$ scenario with a CP-even sneutrino DM (a CP-odd sneutrino DM). There are at least two reasons for the suppression of the evidence for the $h_2$ scenario. One is that the parameter space to predict $m_{h_2} \simeq 125 {\rm GeV}$ is relatively narrow~\cite{Cao:2012fz}. The other is that a light $h_1$ can enhance the scattering of the sneutrino DM with nucleon, and in order to satisfy the constraint from the XENON-1T experiment, the DM usually annihilated by a resonant $h_1$ or $h_2$ to get its measured relic density. The tuning to get the right density is usually more than 100~\cite{Cao:2019aam}. Since the $h_2$ scenario becomes relatively unimportant after considering the latest XENON-1T experiment, we will present its features elsewhere.

\item Although we have optimized our computer code for the calculation, it still took us about 0.4 million core-hours for Intel
	I9 7900X CPU to finish the scans. This is challenging for our computer system.

\item The conclusions listed above may be applied to the inverse seesaw extension of the NMSSM proposed in~\cite{Cao:2017cjf} due to the similarities of the two frameworks, and a careful study of the model is necessary.
\end{itemize}

\section*{Acknowledgement}
	
	We thank Dr. Yangle He, Xiaofei Guo, Yang Zhang and Pengxuan Zhu for helpful discussion about dark matter indirect detection experiments. This work is supported by the National Natural Science Foundation of China (NNSFC) under grant No. 11575053.

	\begin{figure*}[htbp]
		\centering
		\resizebox{0.7 \textwidth}{!}{
		\includegraphics{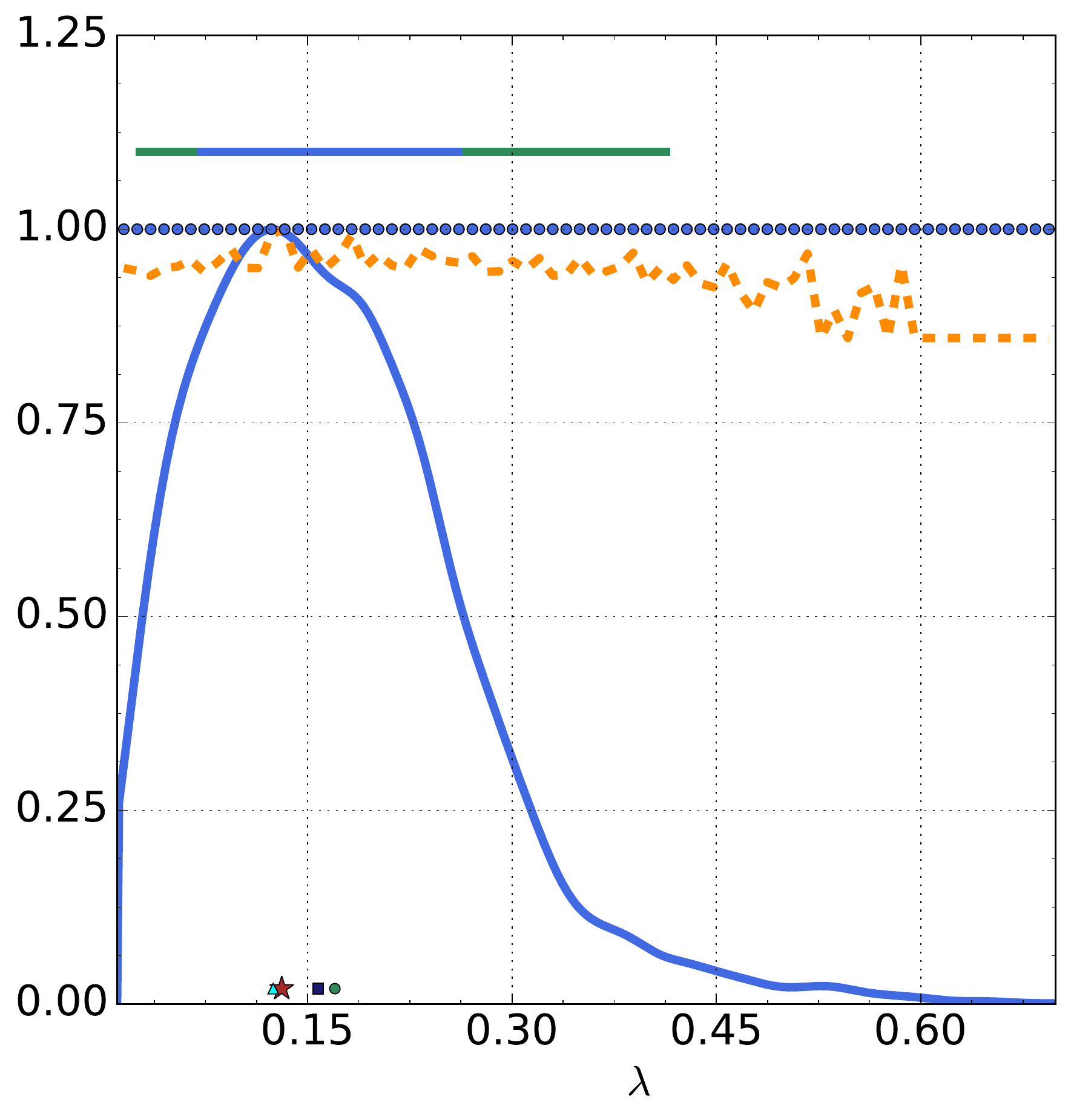}
		\includegraphics{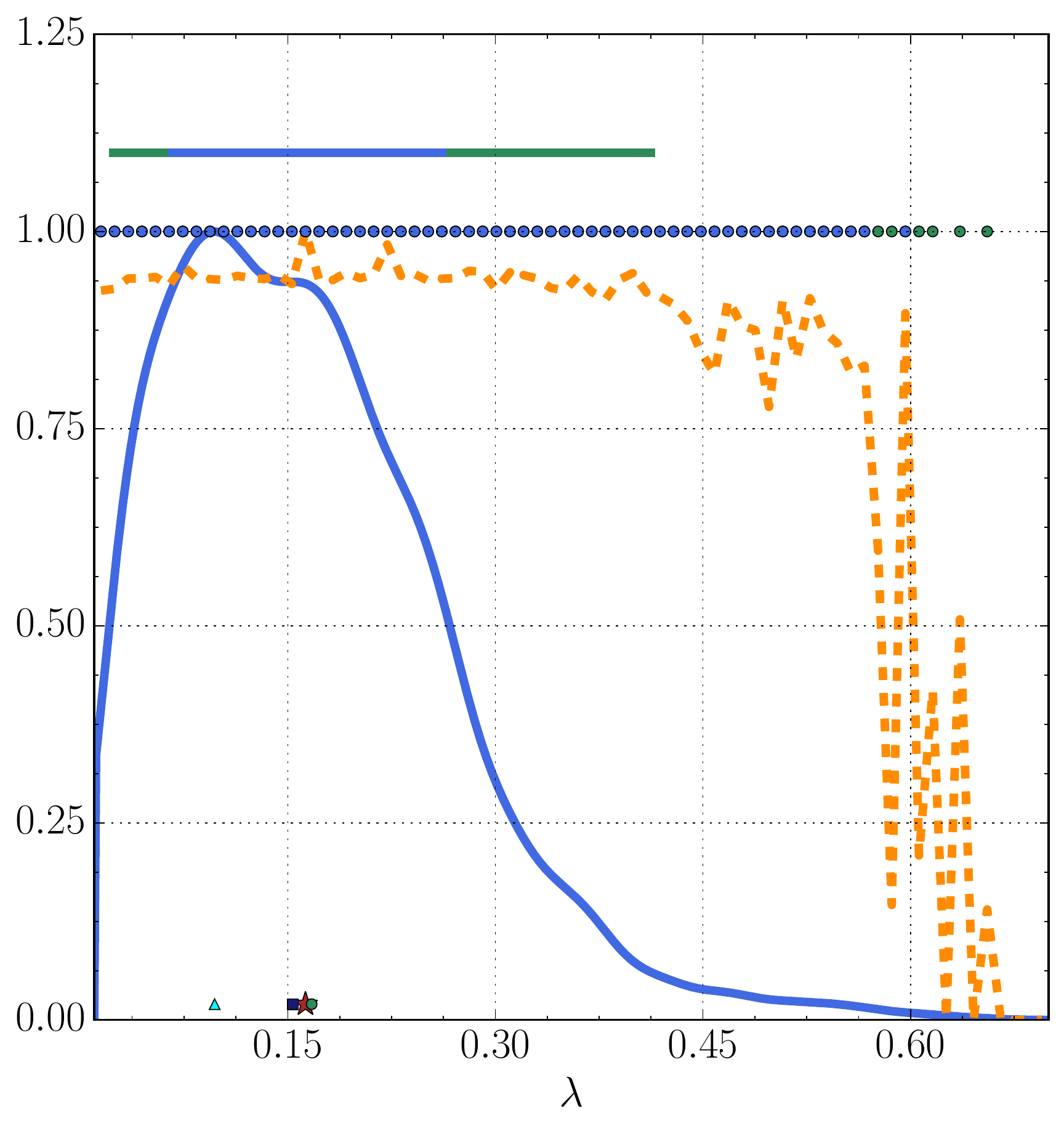}
		\includegraphics{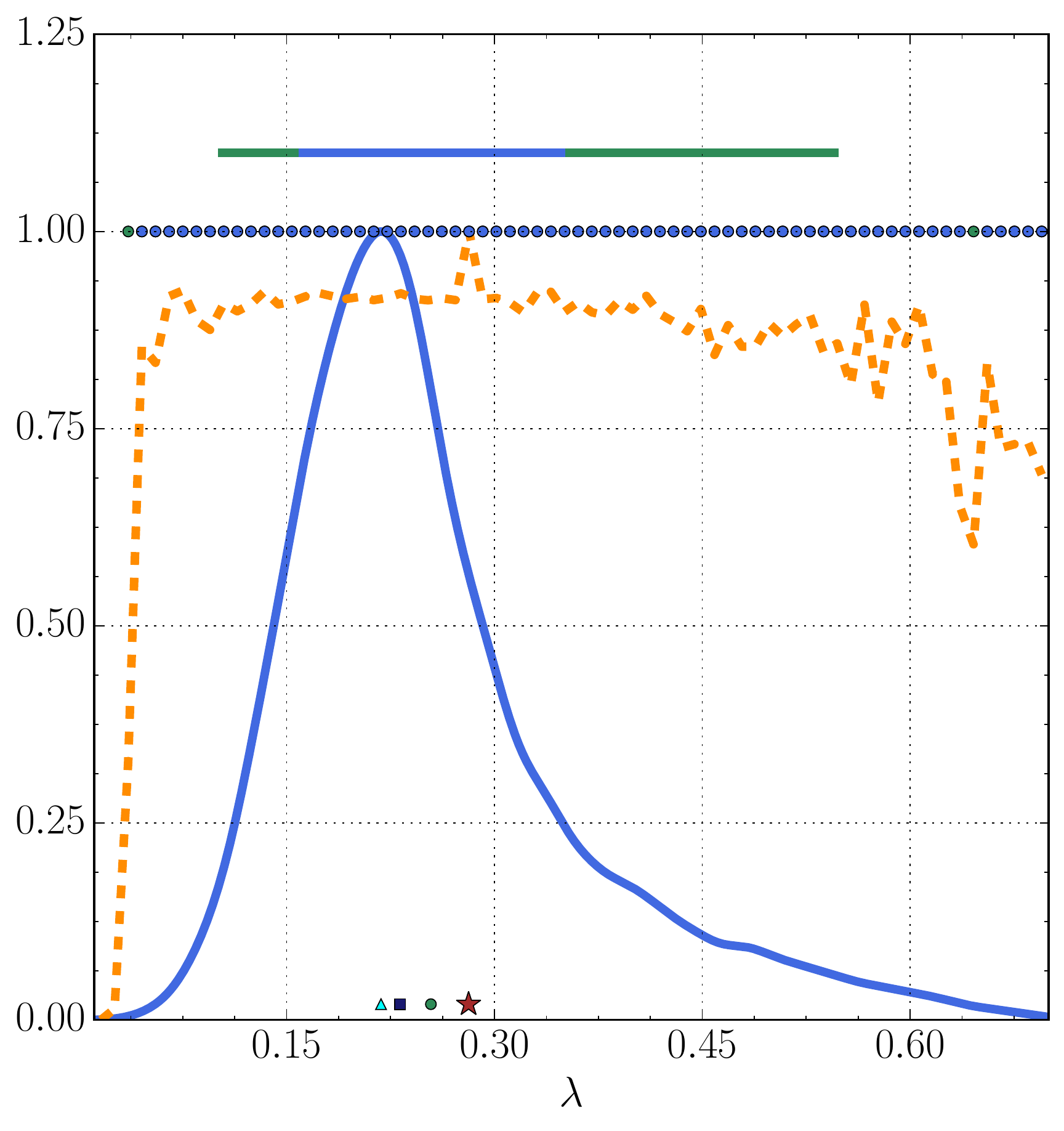}
		}
		\resizebox{0.7\textwidth}{!}{
		\includegraphics{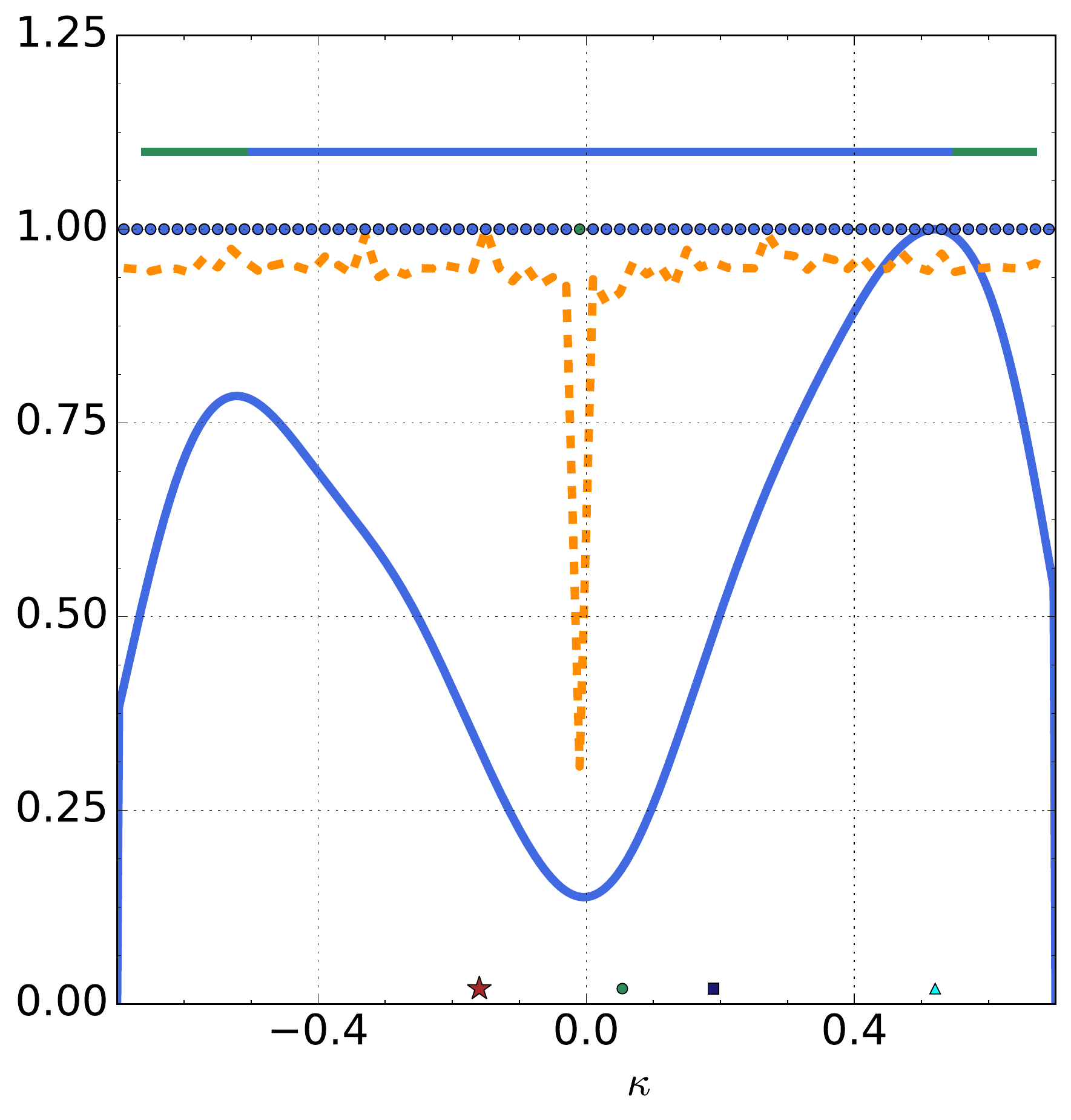}
		\includegraphics{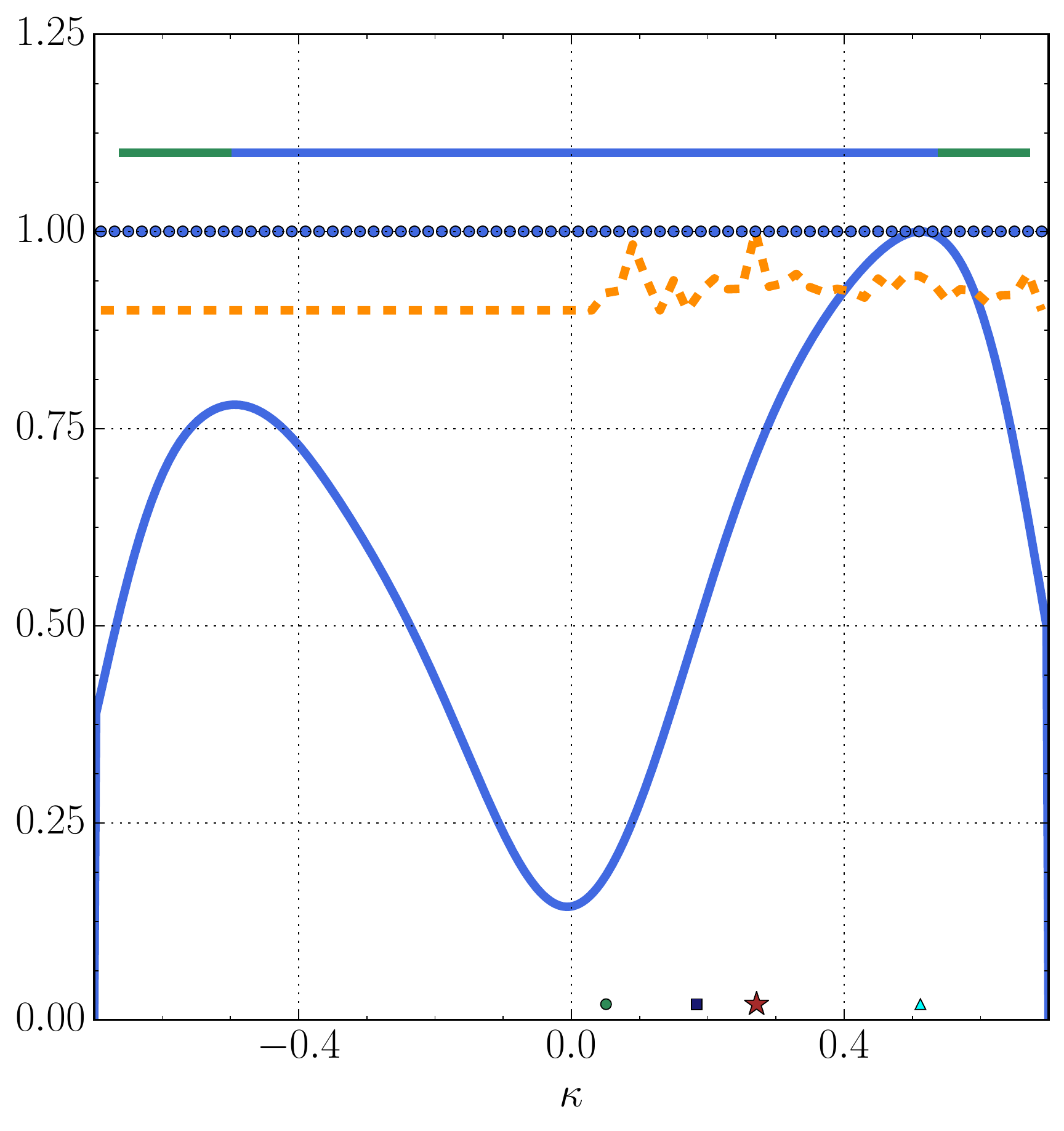}
		\includegraphics{Fig1-2.pdf}
		}
		\resizebox{0.7 \textwidth}{!}{
		\includegraphics{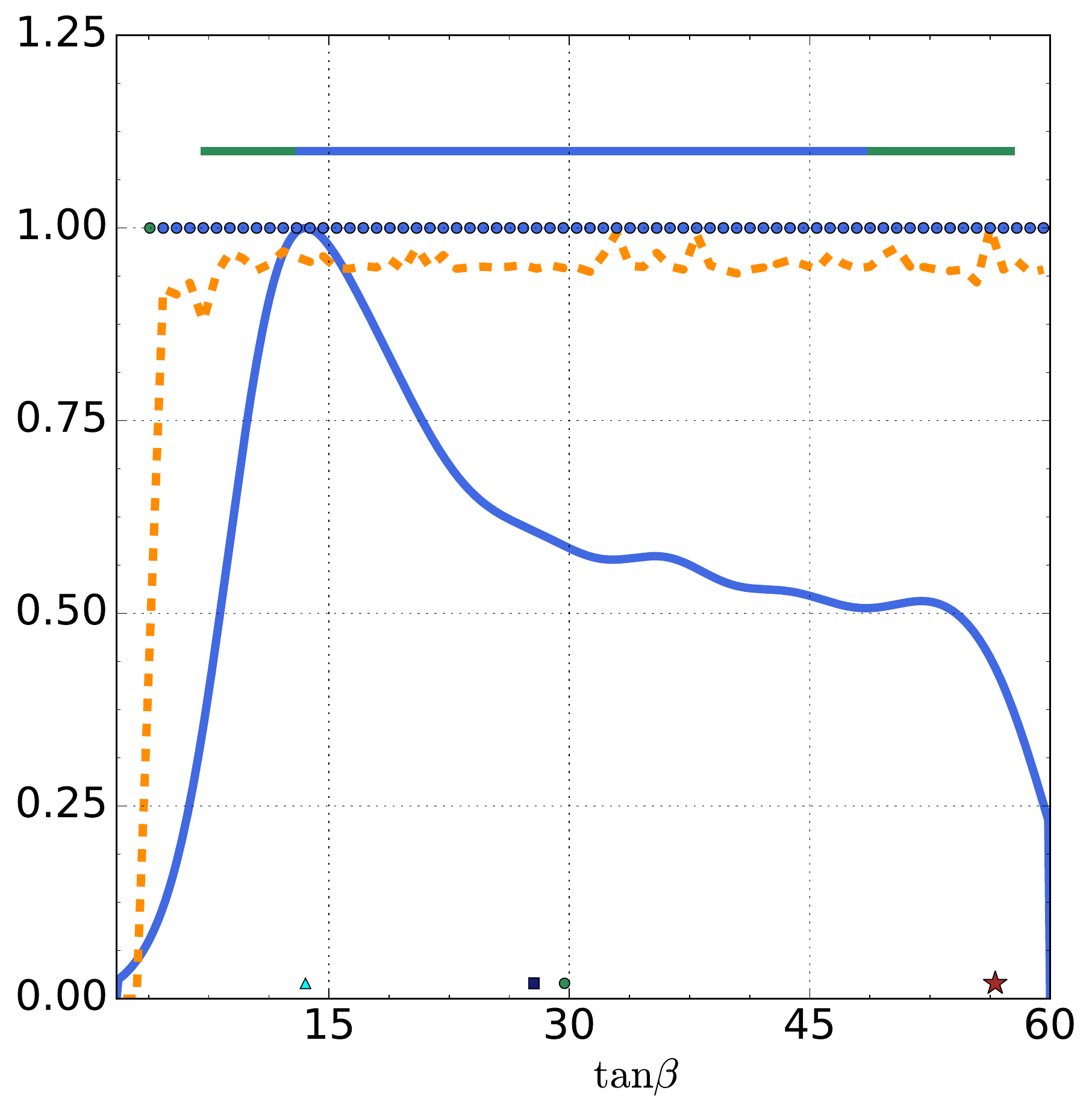}
		\includegraphics{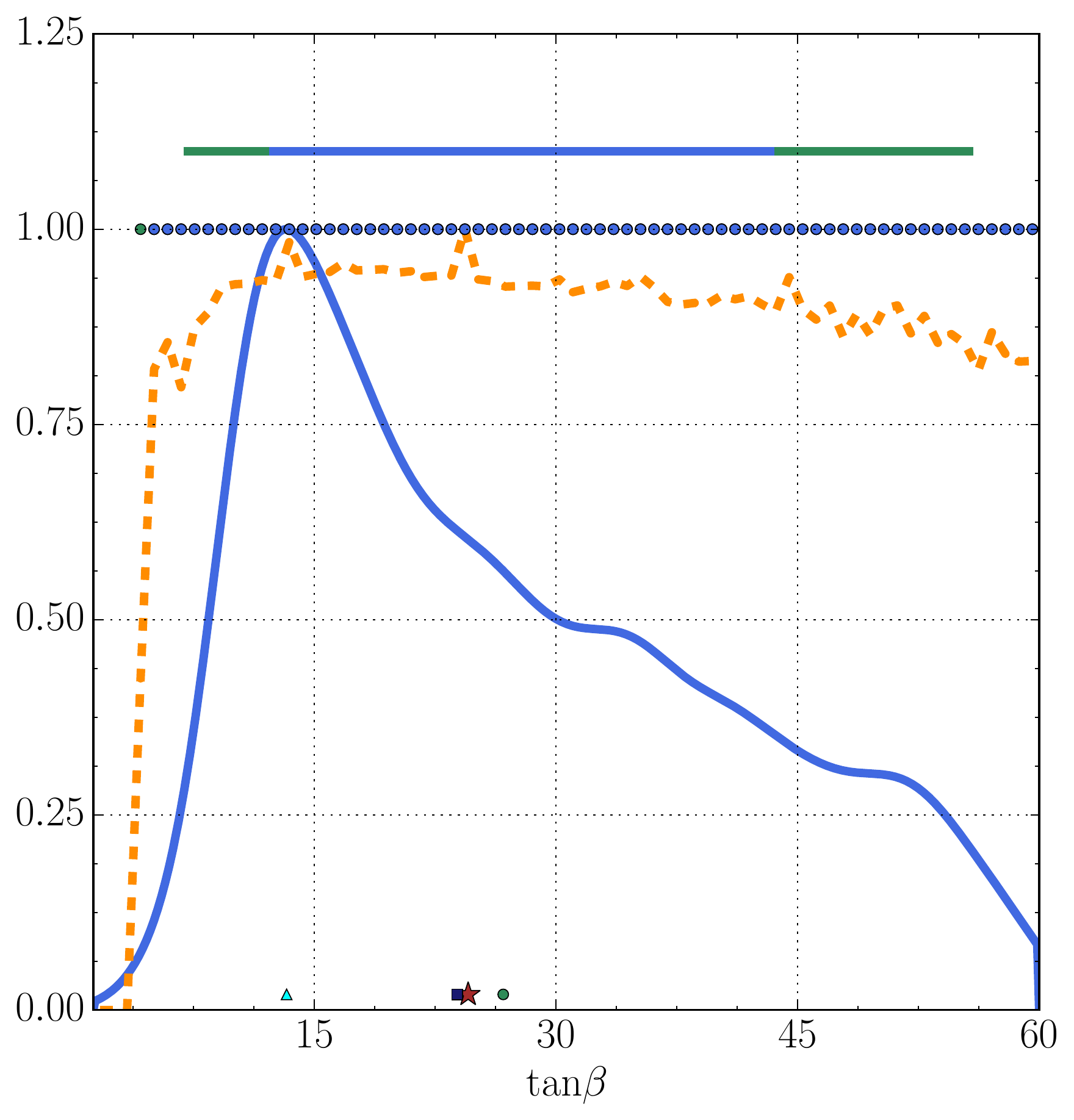}
		\includegraphics{Fig1-3.pdf}
		}

       \vspace{-0.4cm}

	\caption{Impact of different experimental measurements on the PDFs and PLs of the parameters $\lambda$, $\kappa$ and $\tan \beta$. The left, middle and right panels are obtained by the likelihood function from the Higgs data, Higgs + B data and Higgs + B + DM data introduced in the text respectively with the log prior PDF. To get this figure, we set {\it nlive}=2000 for the left and middle panels, and take the right panel from FIG.\ref{fig1} directly. \label{fig14}}
	\end{figure*}

	\begin{figure*}[htbp]
		\centering
		\resizebox{0.7\textwidth}{!}{
		\includegraphics{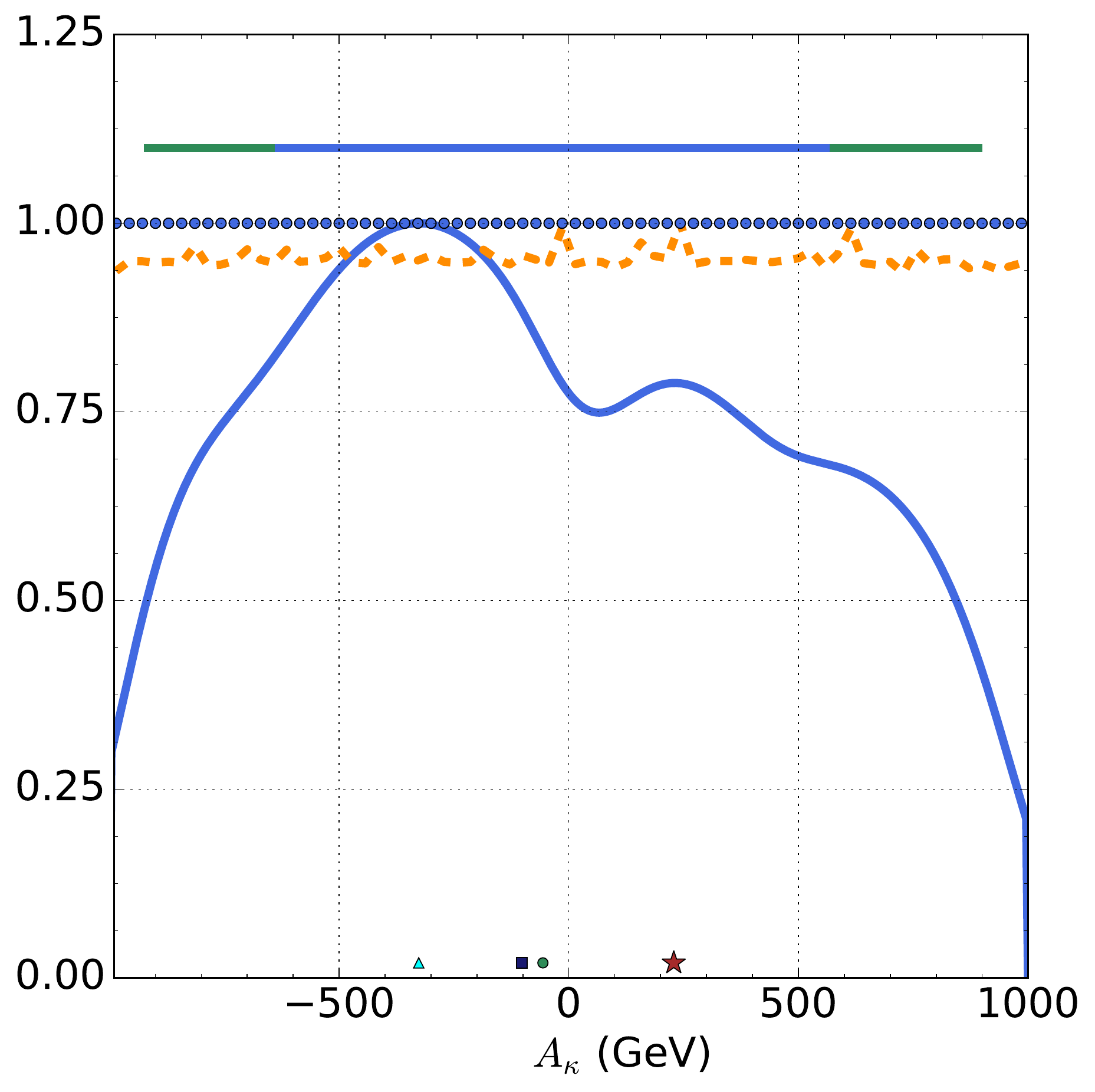}
		\includegraphics{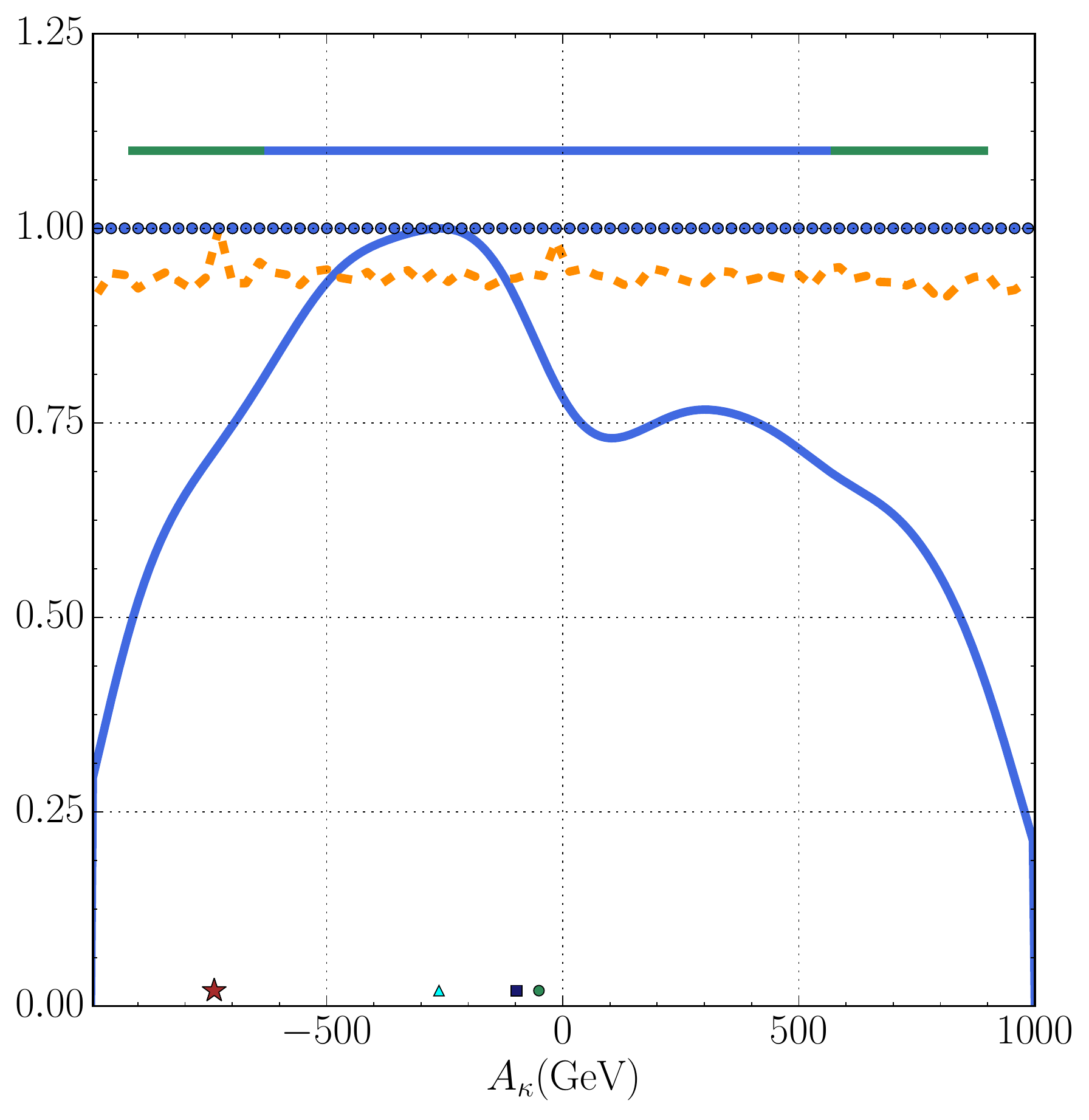}
		\includegraphics{Fig1-4.pdf}
		}
		\resizebox{0.7\textwidth}{!}{
		\includegraphics{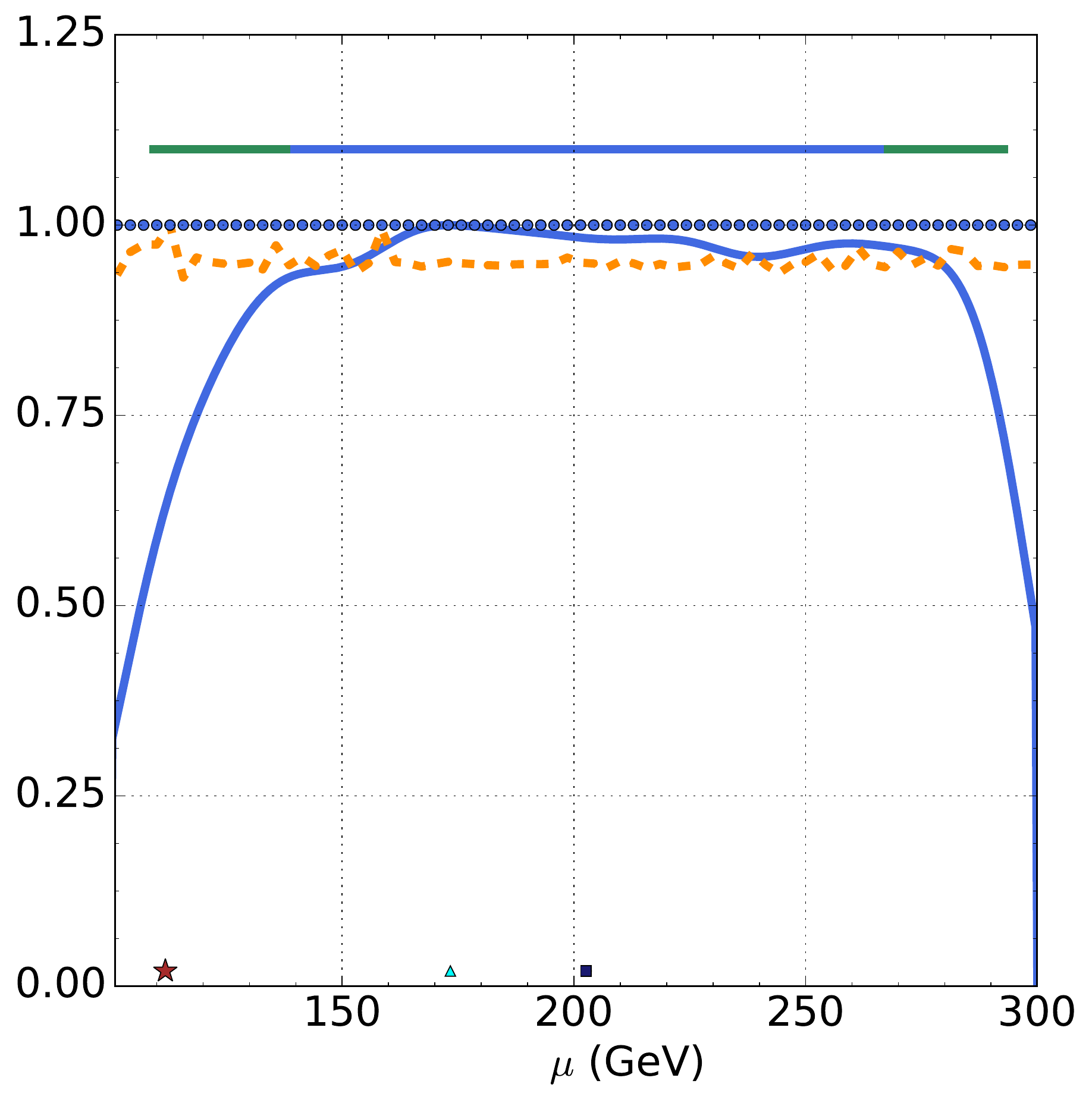}
		\includegraphics{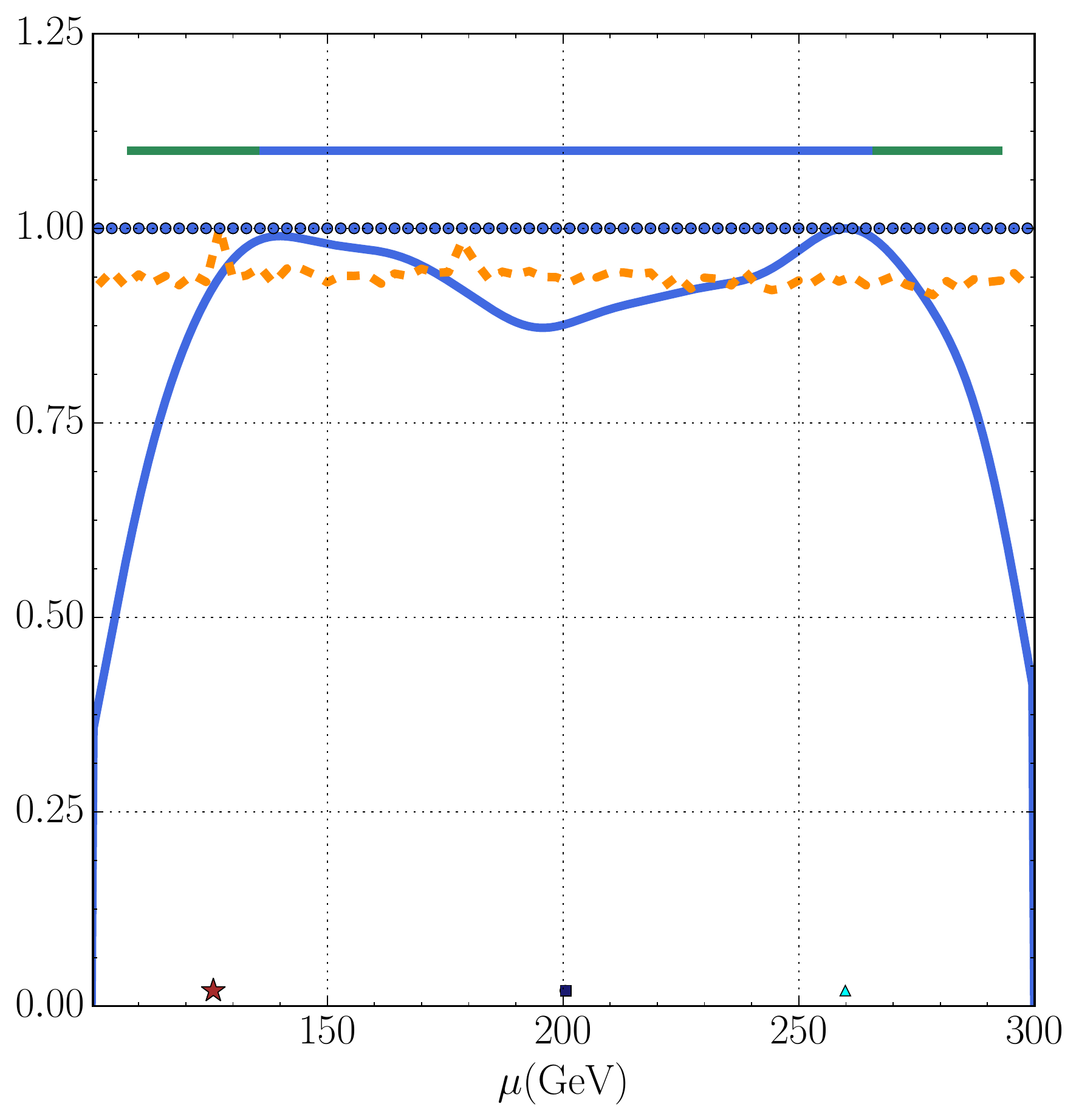}
		\includegraphics{Fig1-5.pdf}
		}
		\resizebox{0.7\textwidth}{!}{
		\includegraphics{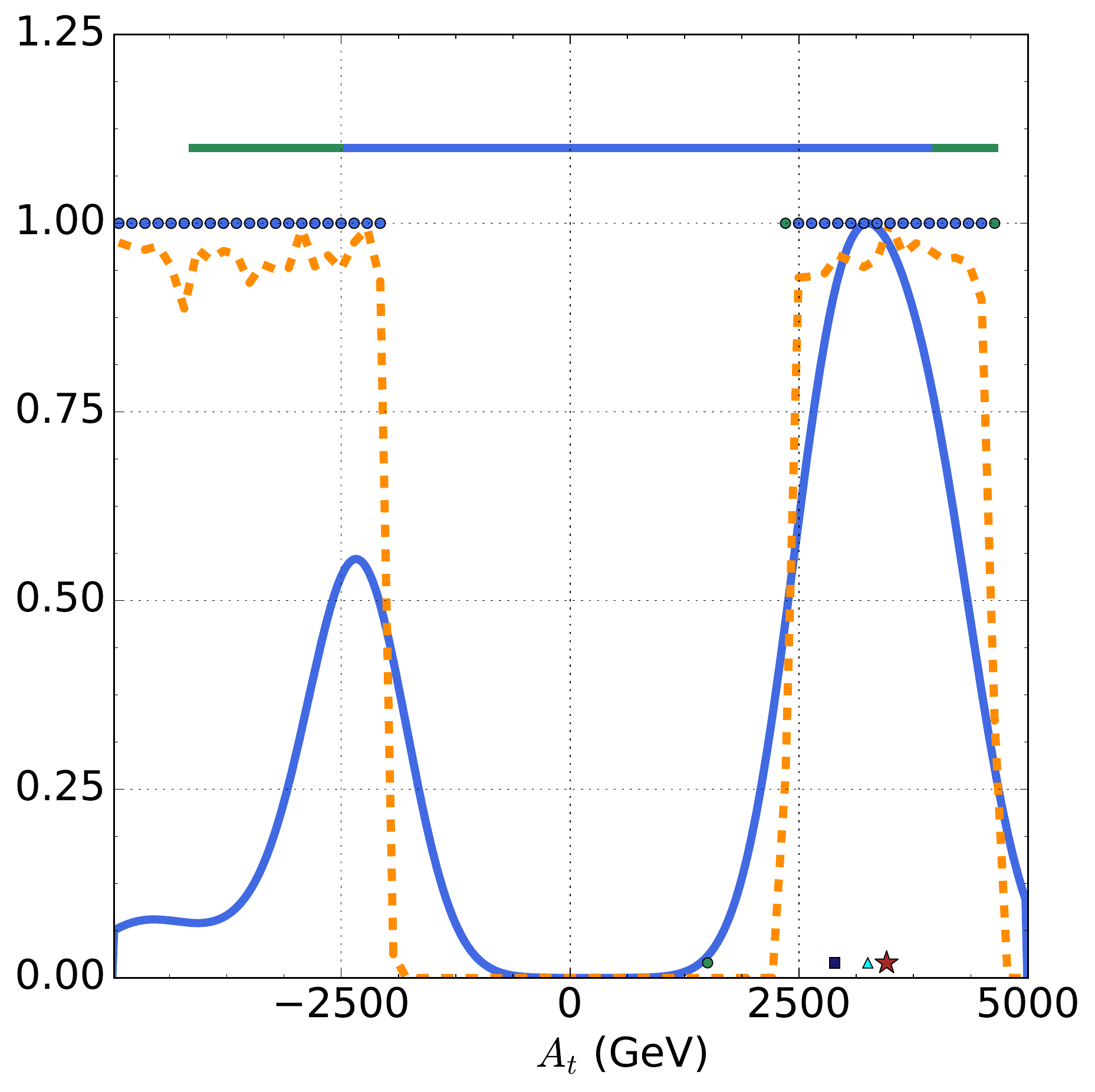}
		\includegraphics{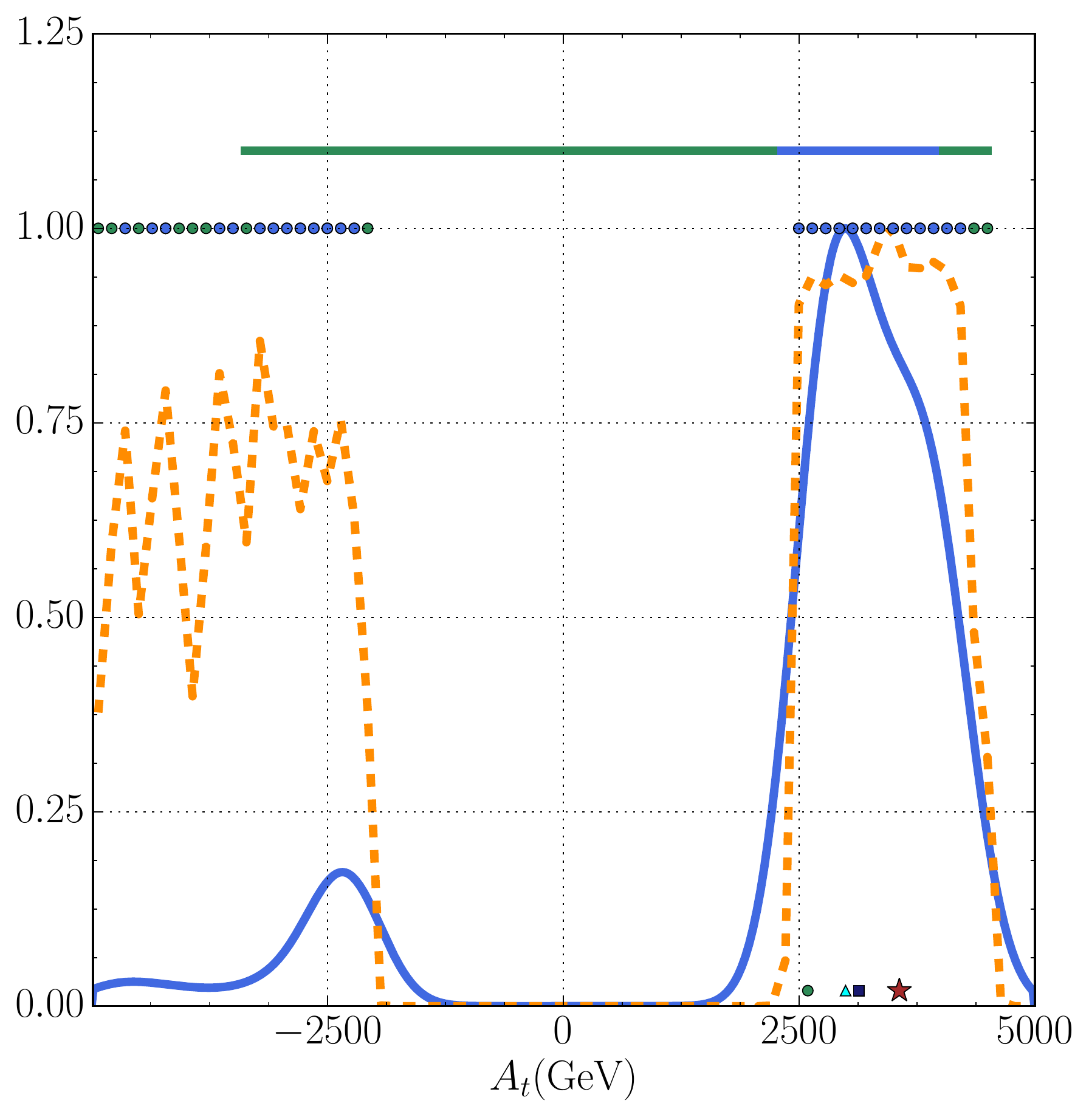}
		\includegraphics{Fig1-6.pdf}
		}

       \vspace{-0.4cm}

	\caption{Same as FIG.\ref{fig14}, but for the results of the parameters $A_\kappa$, $\mu$ and $A_t$. \label{fig15}}
	\end{figure*}

\appendix

\section{Impacts of different measurements on posterior PDFs}

In this Appendix, we study the impacts of different experimental measurements on the marginal PDFs of the theoretical input parameters in the $h_1$ scenario with DM being a CP-even sneutrino. We note that in the Type-I seesaw extended NMSSM, the neutrino and sneutrino sectors have little effect on the Higgs and B physics observables considered in the text. So in order to study the influence of the observables on the parameters, we first fix $\bar{m}_{\tilde{\nu}} = 0.4 {\rm TeV} \times \bf{1}$, $\bar{\lambda}_\nu = 0.3 \times \bf{1}$ and $\bar{A}_{\lambda_\nu} = 0 $ with $\bf{1}$ denoting unit matrix in flavor space, and take the other unimportant parameters in the NMSSM except $A_\lambda$ from Table \ref{benchmark}, then we explore following parameter space
\begin{eqnarray}
&& 0 < \lambda \leq 0.7, |\kappa| \leq 0.7, 1 \leq tan\beta \leq 60, |A_\kappa| \leq 1 {\rm TeV}, \nonumber
\\
&&  |A_t| \leq 5 {\rm TeV}, \quad 0.1 {\rm TeV} \leq \mu \leq 0.3{\rm TeV}, \label{scan-ranges1}
\end{eqnarray}
by MultiNest algorithm with flat prior PDF for the input parameters.

\subsection{Preference of Higgs physics on input parameters}

\begin{figure*}[htbp]
		\centering
		\resizebox{0.7\textwidth}{!}{
		\includegraphics{Fig14-A-1.pdf}
		\includegraphics{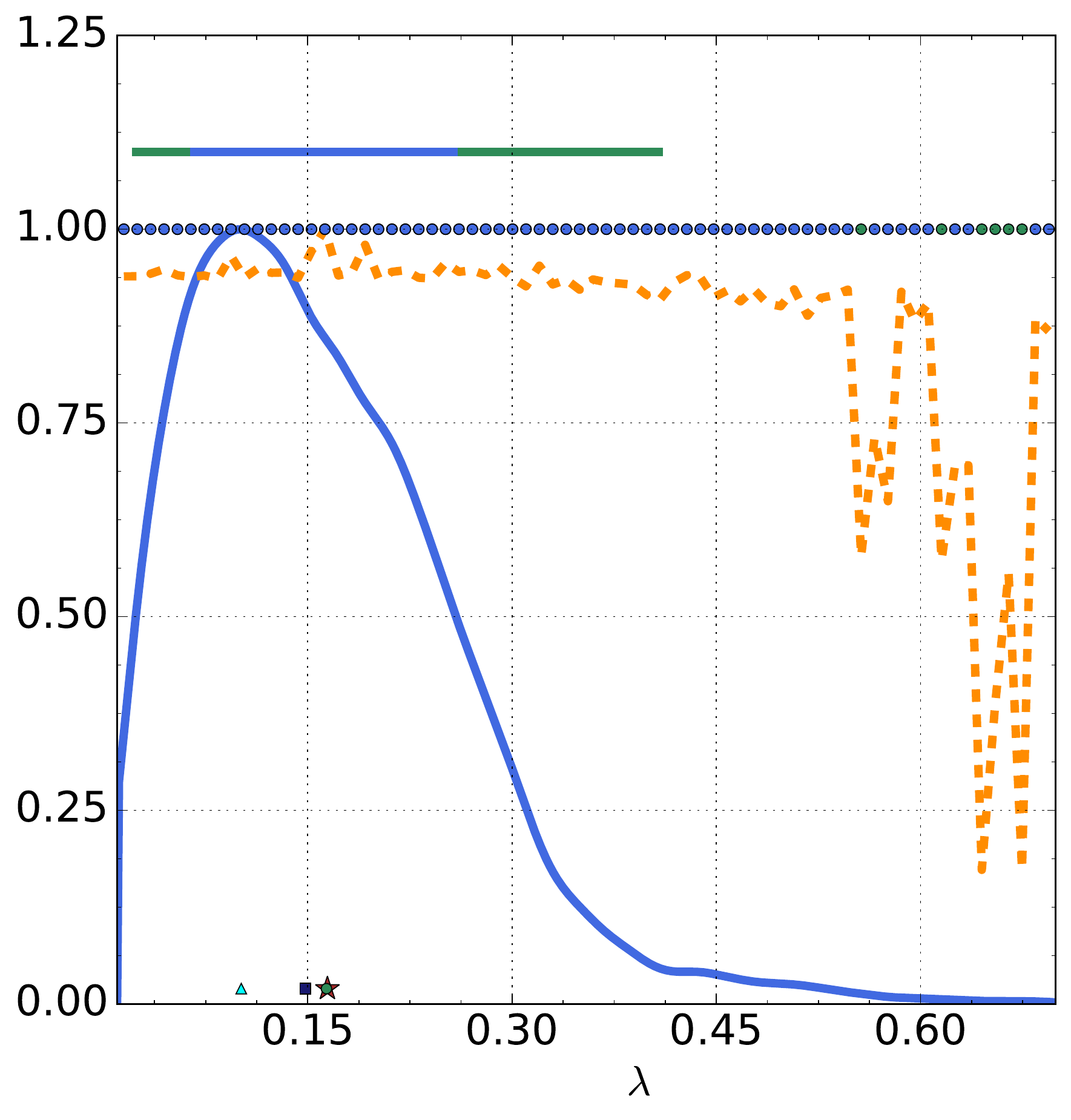}
		\includegraphics{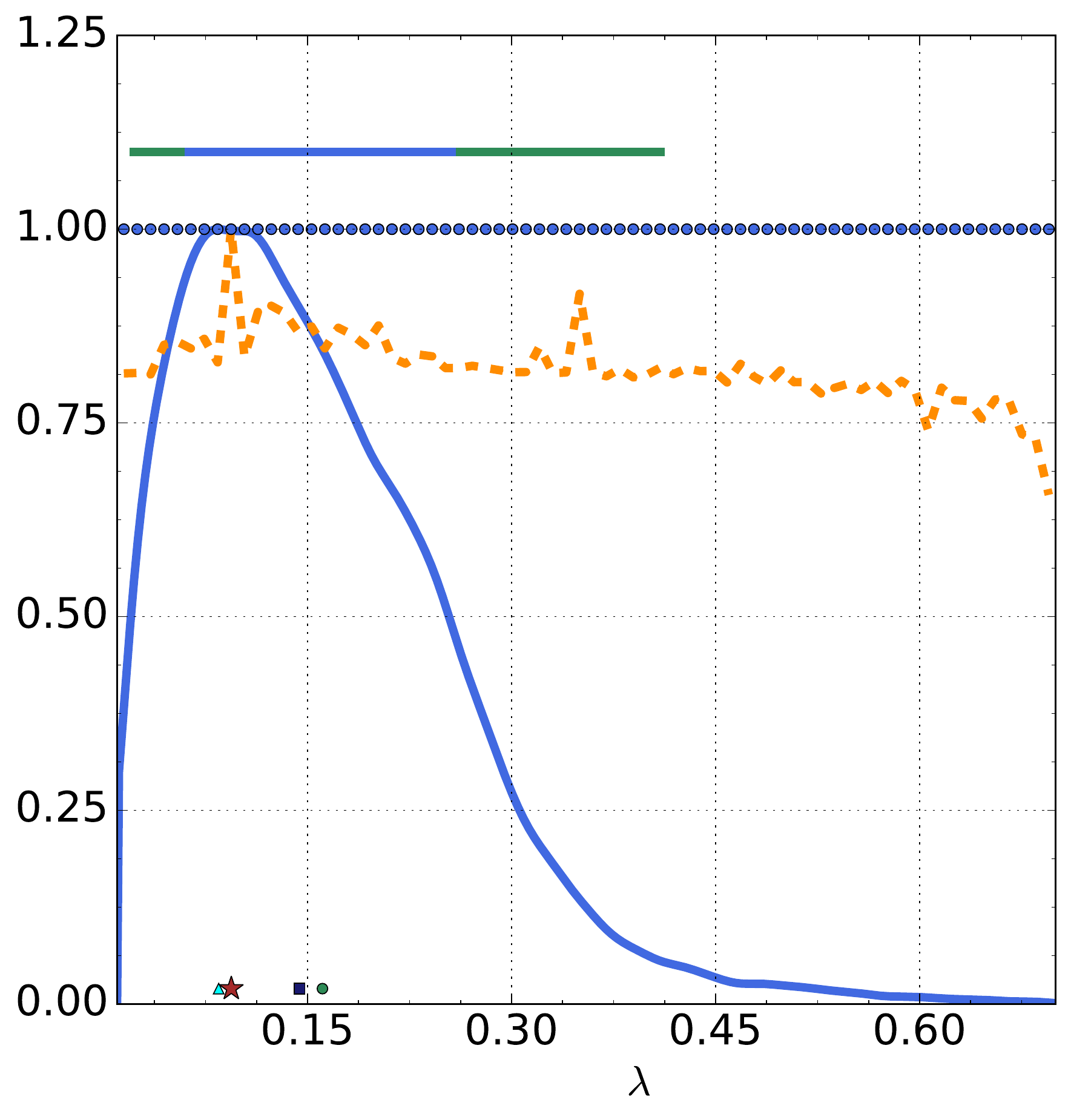}
		}
		\resizebox{0.7\textwidth}{!}{
		\includegraphics{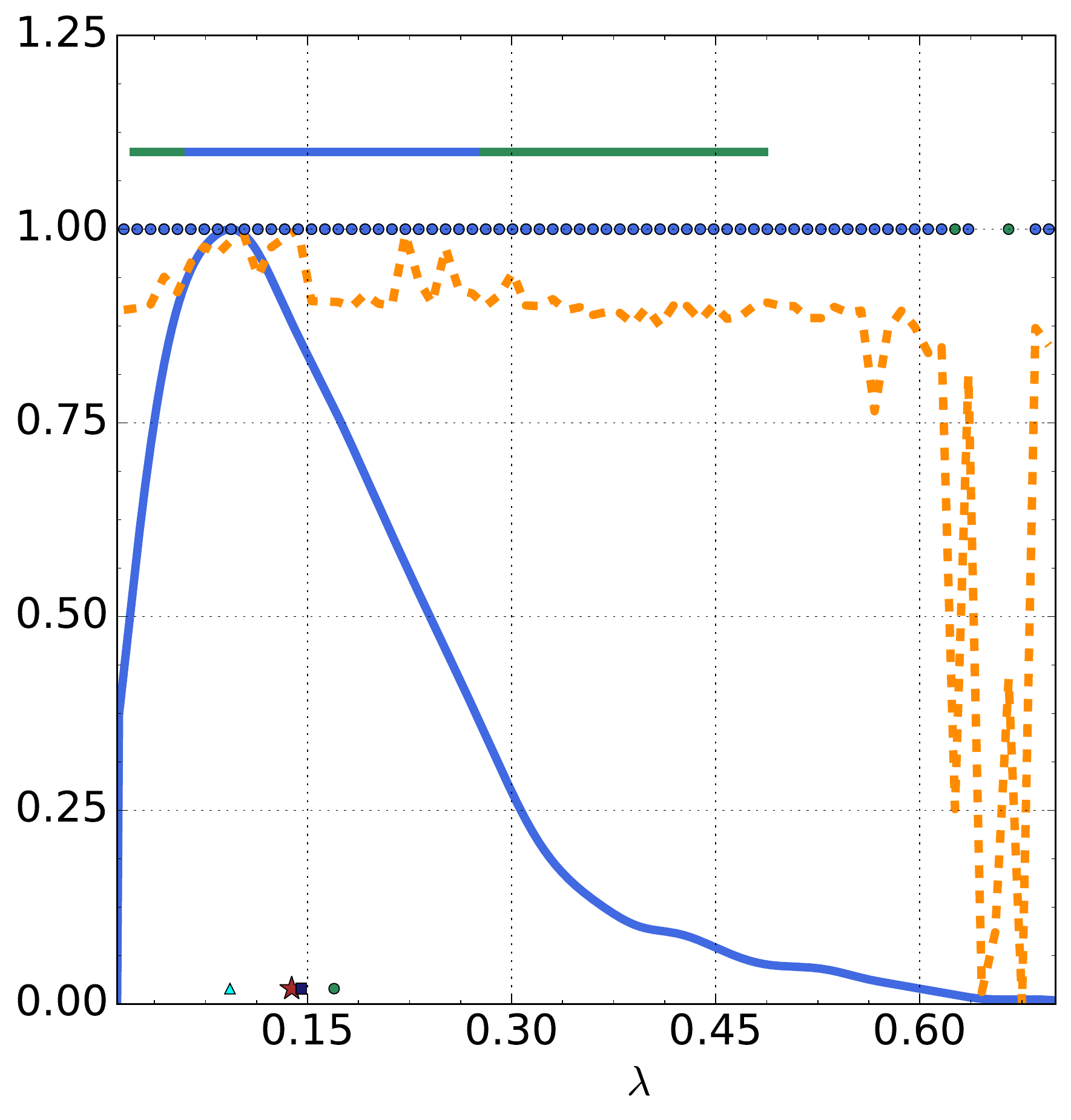}
		\includegraphics{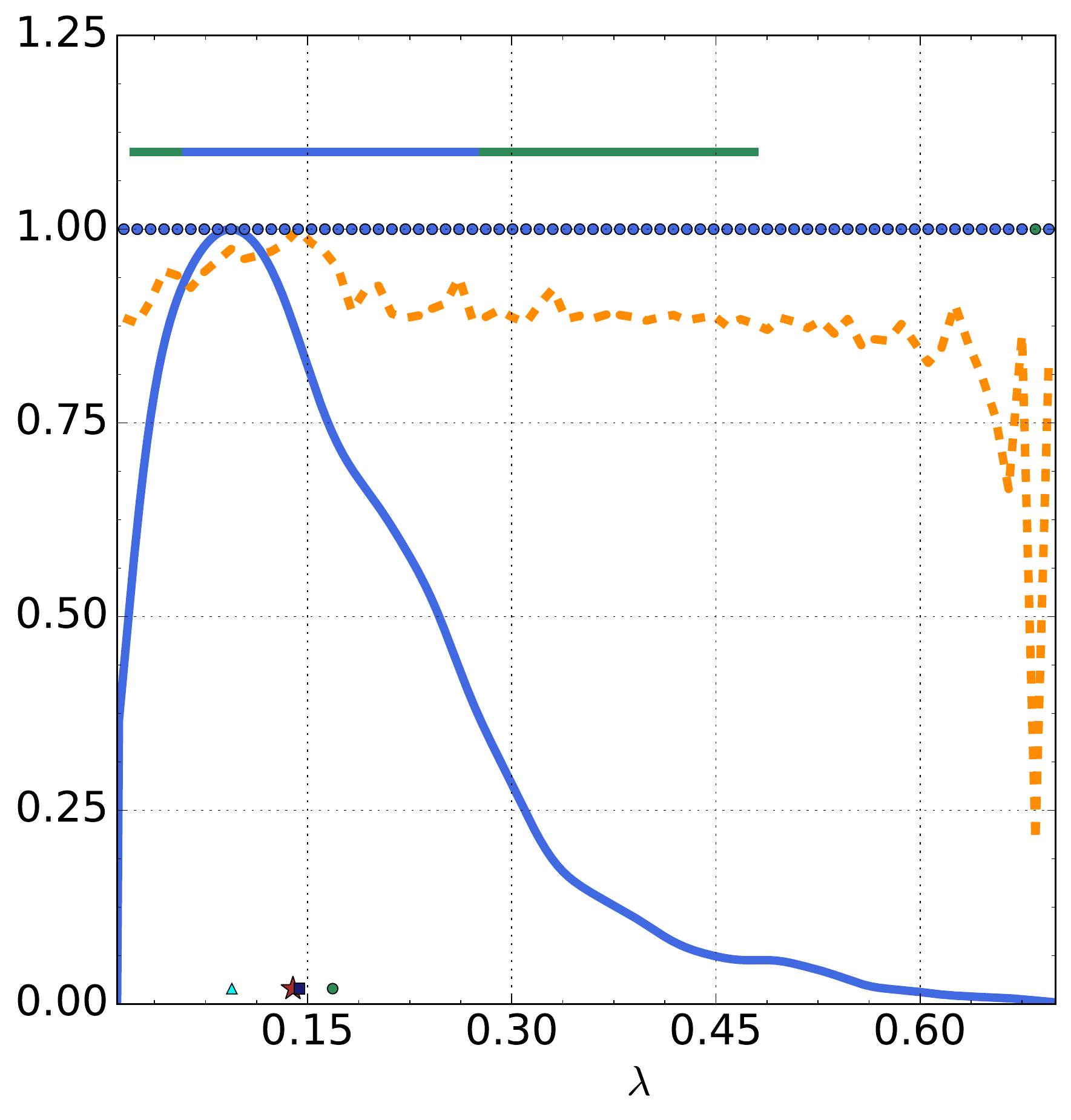}
		\includegraphics{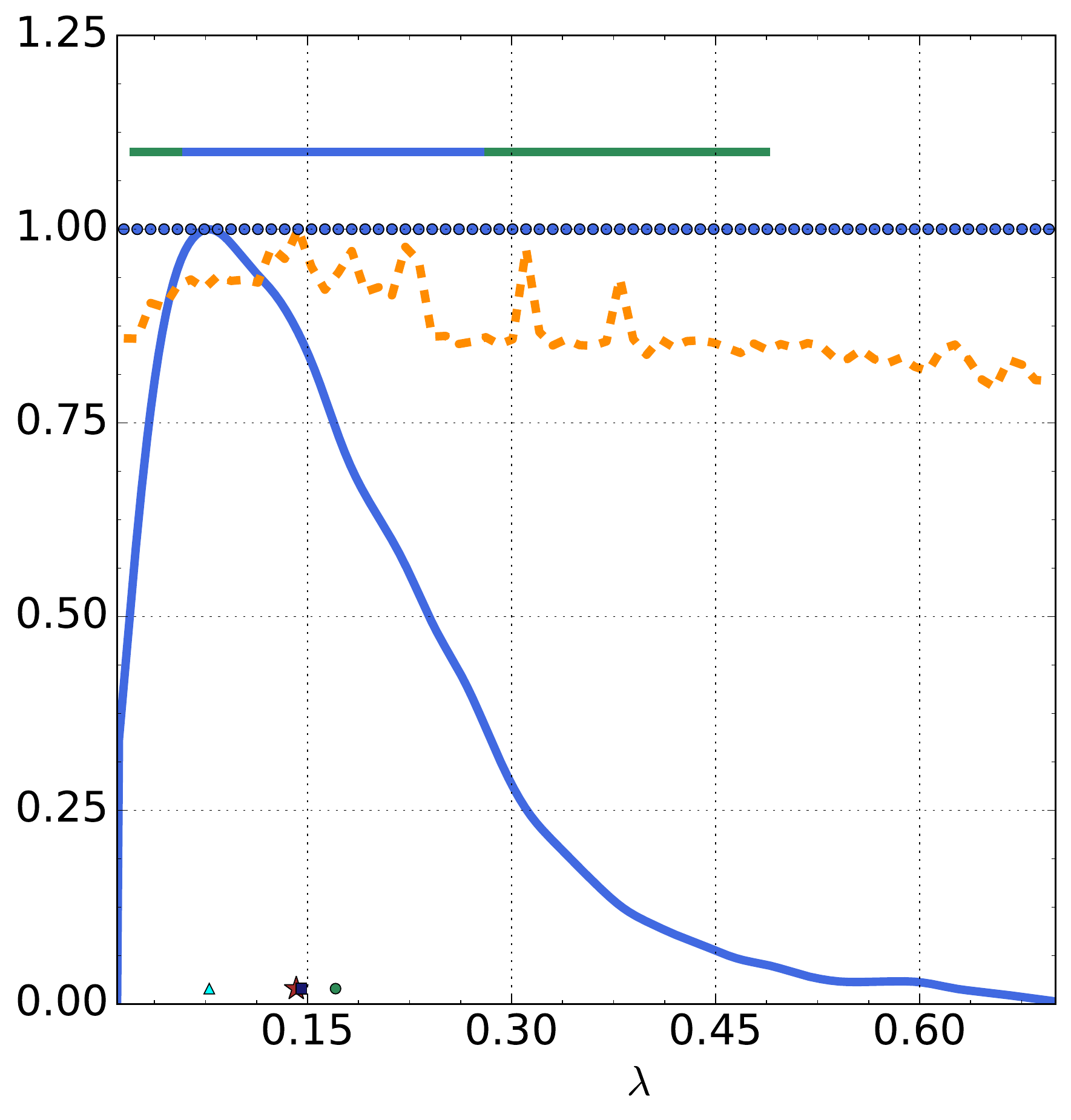}
		}
		\resizebox{0.7\textwidth}{!}{
		\includegraphics{Fig14-A-2.pdf}
		\includegraphics{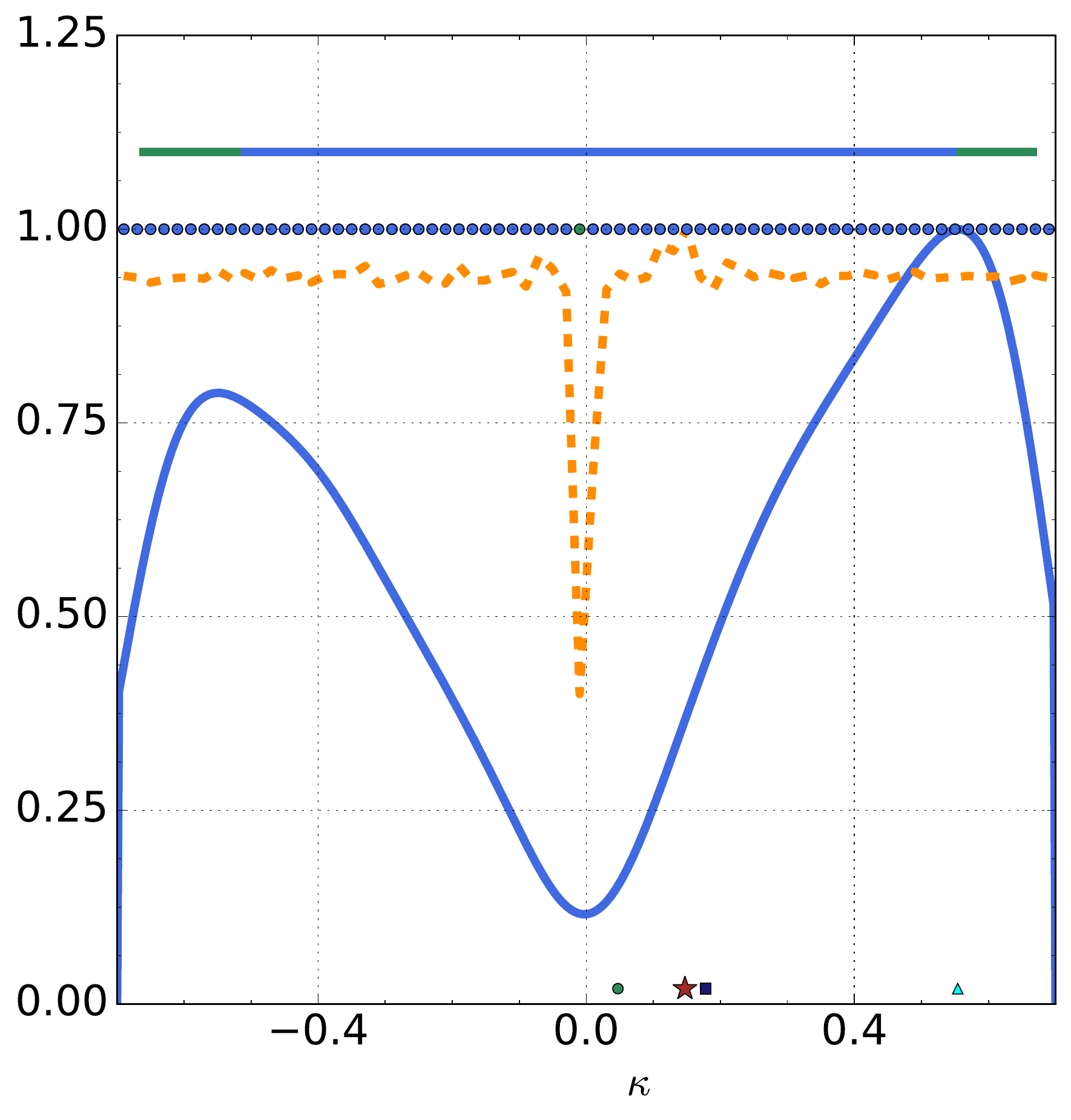}
		\includegraphics{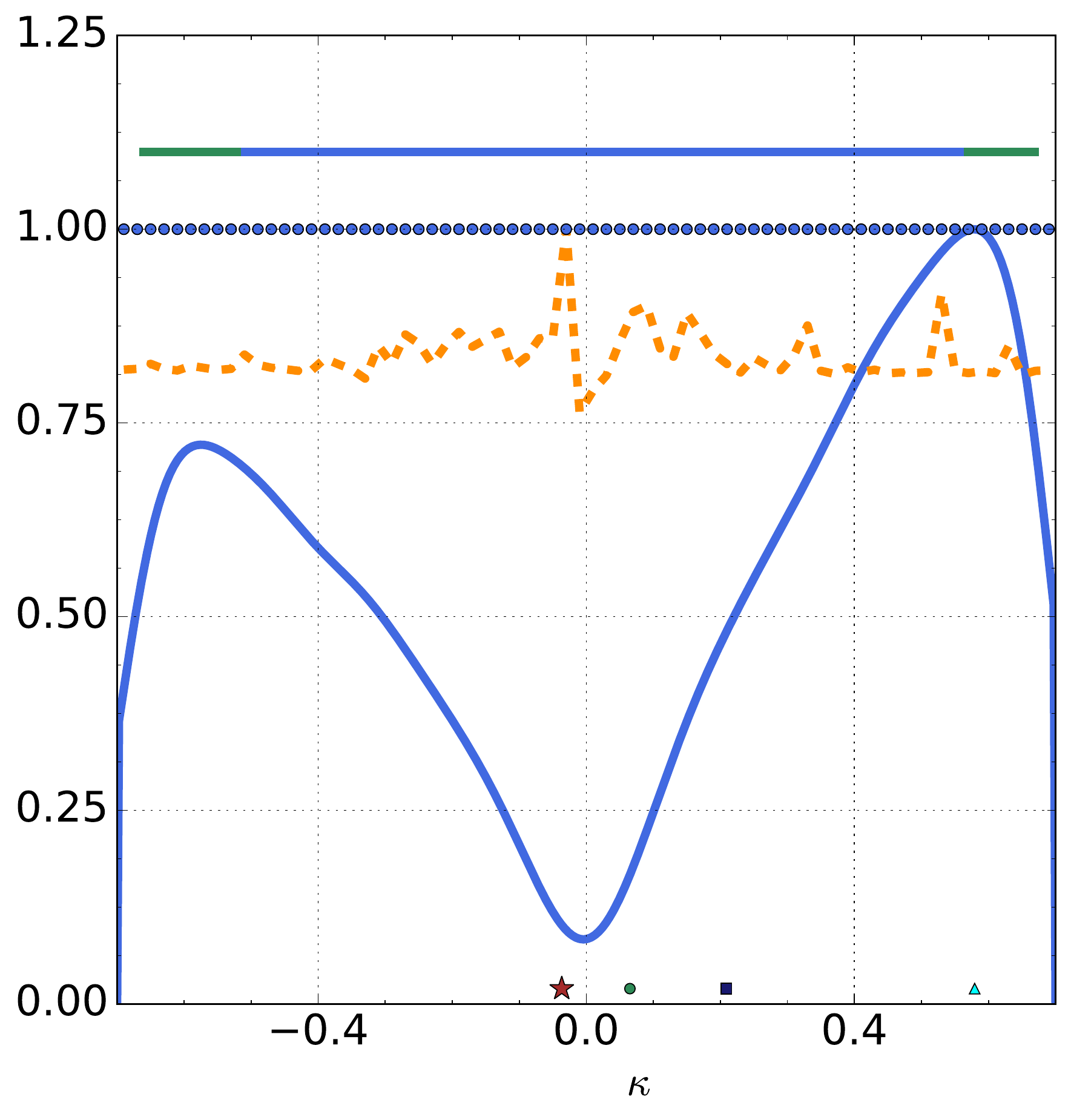}
		}
		\resizebox{0.7\textwidth}{!}{
		\includegraphics{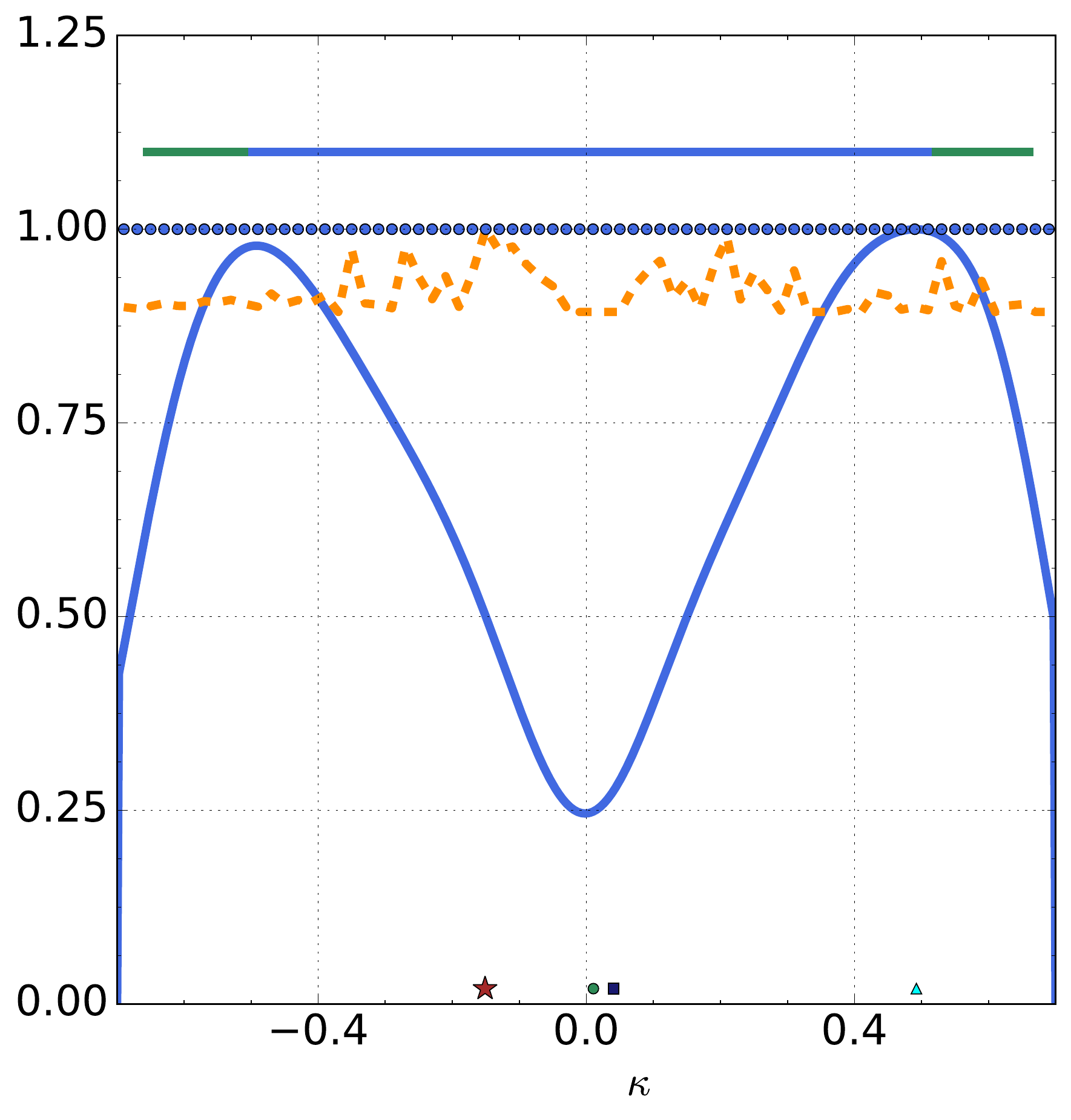}
		\includegraphics{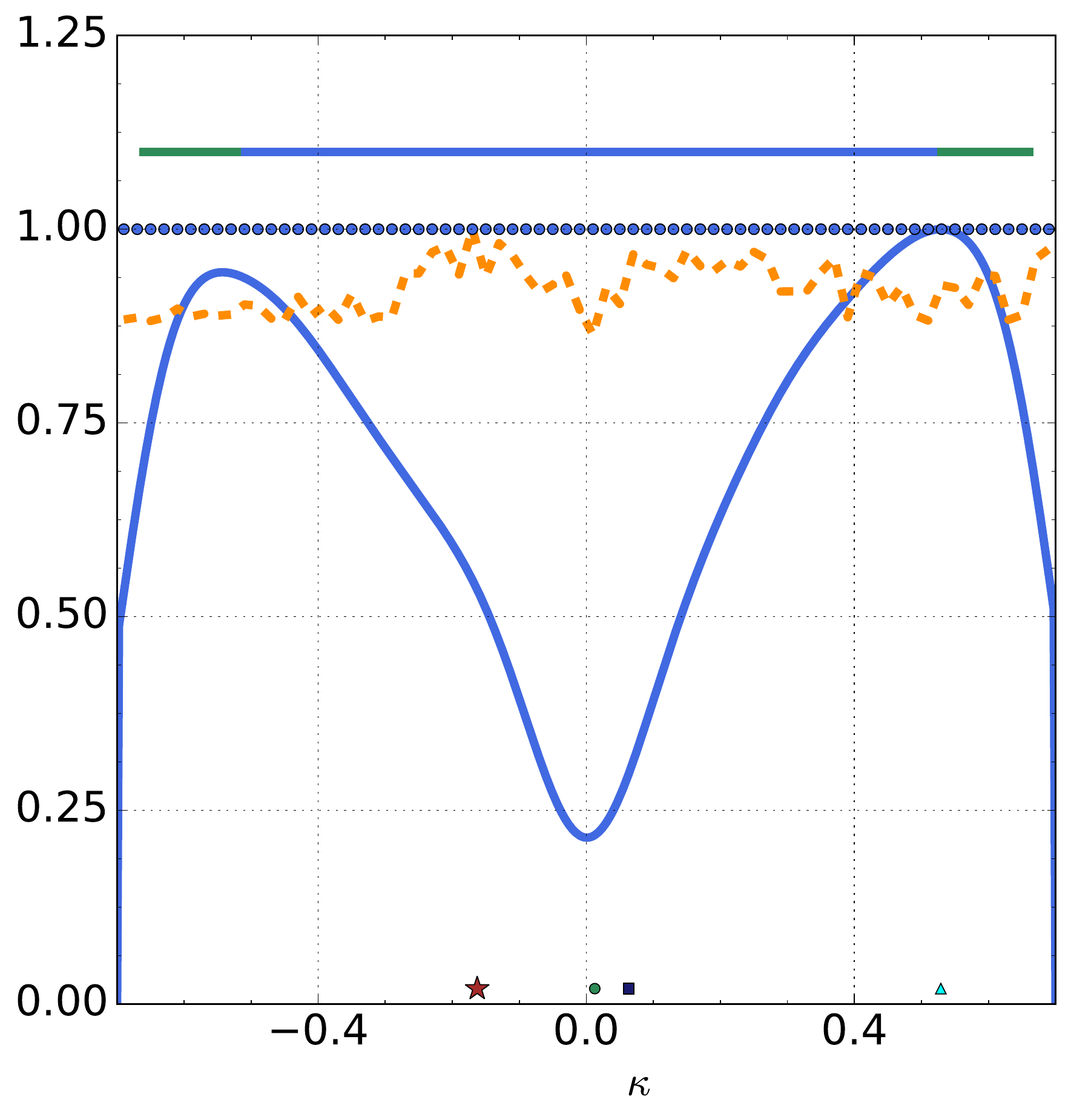}
		\includegraphics{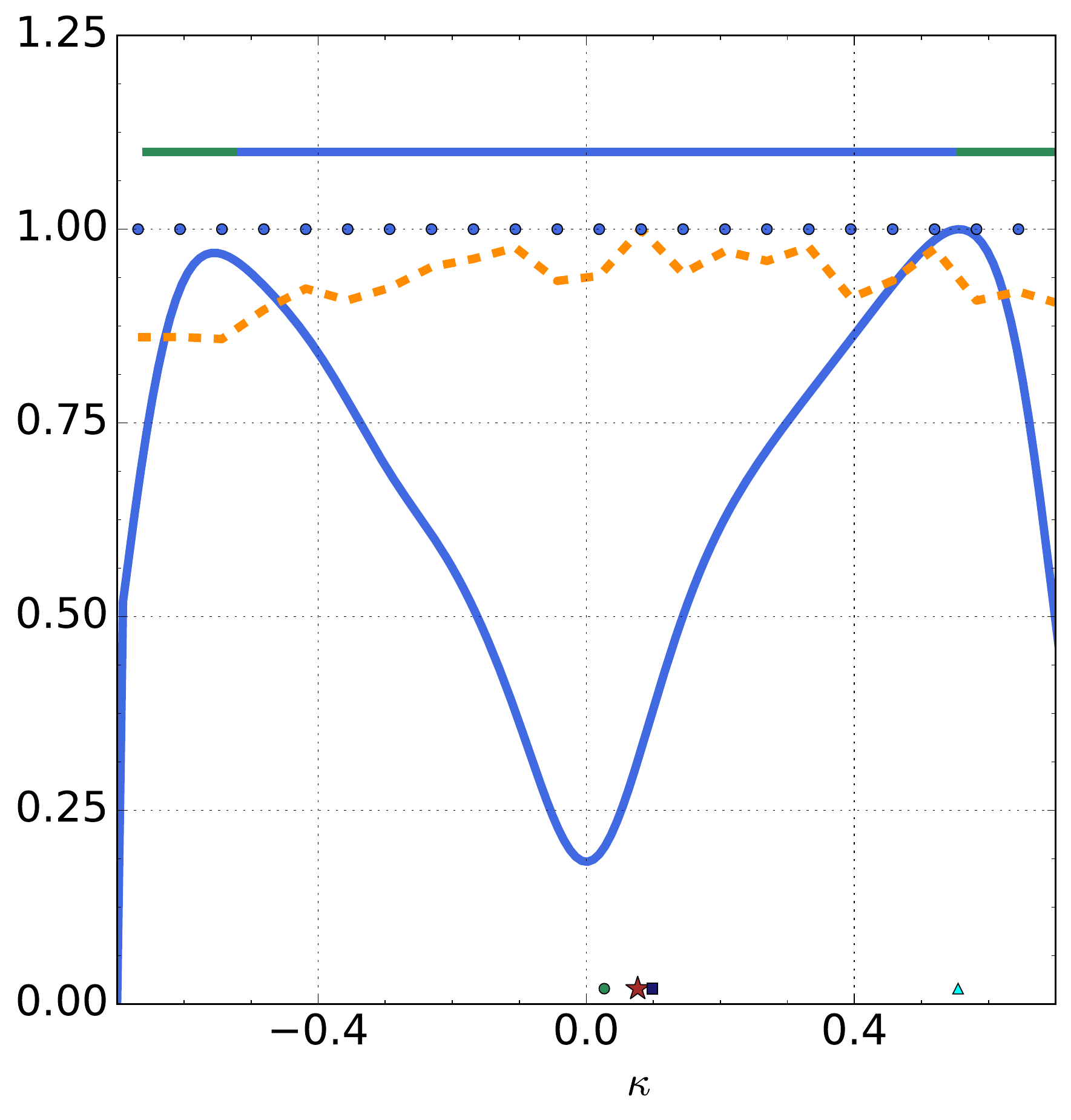}
		}

       \vspace{-0.4cm}

	\caption{The marginal PDFs and PLs of the parameters $\lambda$ (the top two rows) and $\kappa$ (the last two rows) obtained with the likelihood function $\mathcal{L} = \mathcal{L}_{Higgs} $. The panels in the first and third rows are obtained by the flat prior PDF with {\it nlive} = 2000 (left panel), 6000 (middle panel) and 24000 (right panel). The rest panels are obtained in a similar way, but by the log prior PDF.   \label{fig16}}
   \end{figure*}

\begin{figure*}[htbp]
		\centering
		\resizebox{0.7\textwidth}{!}{
		\includegraphics{Fig14-A-3.pdf}
		\includegraphics{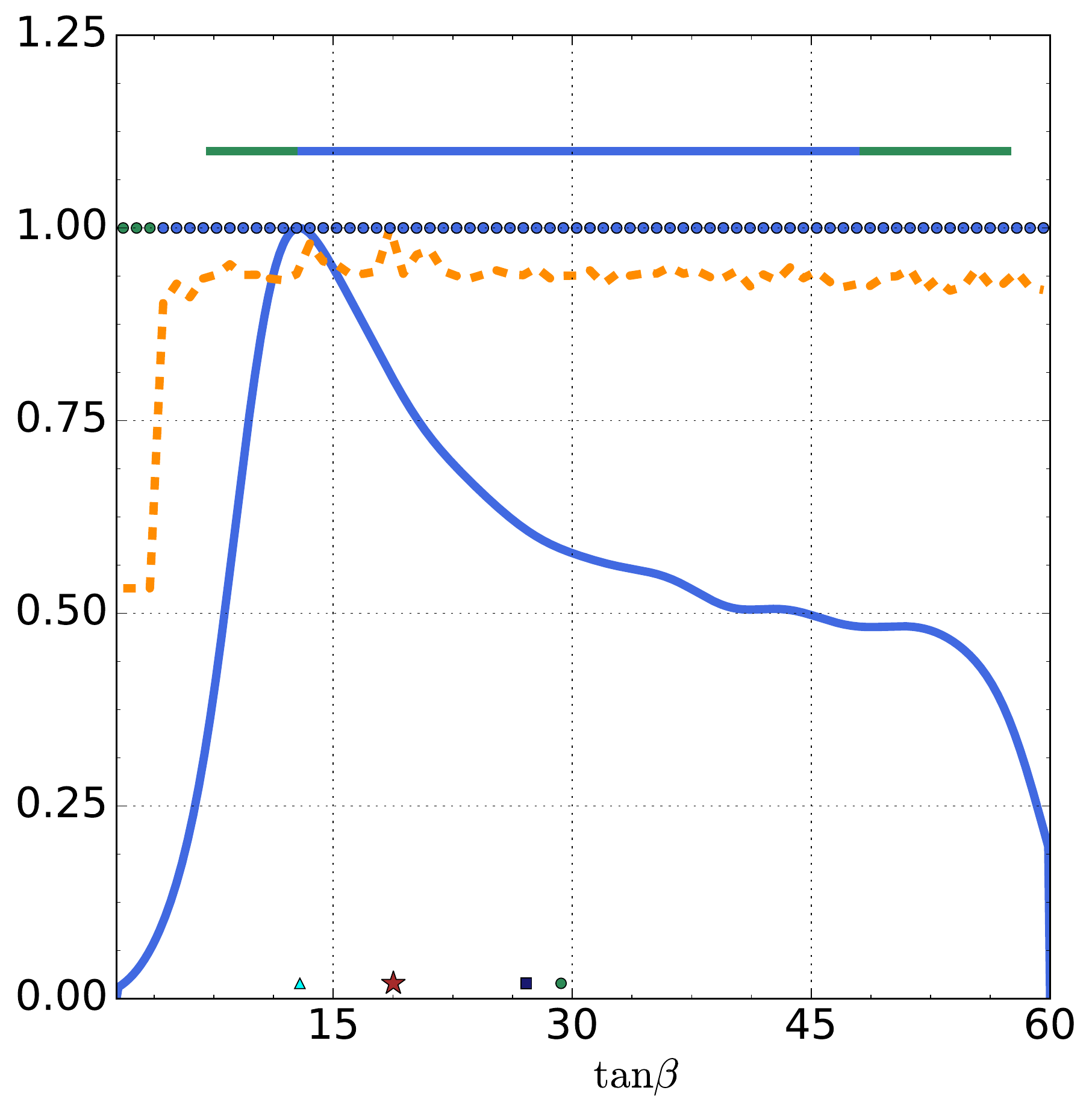}
		\includegraphics{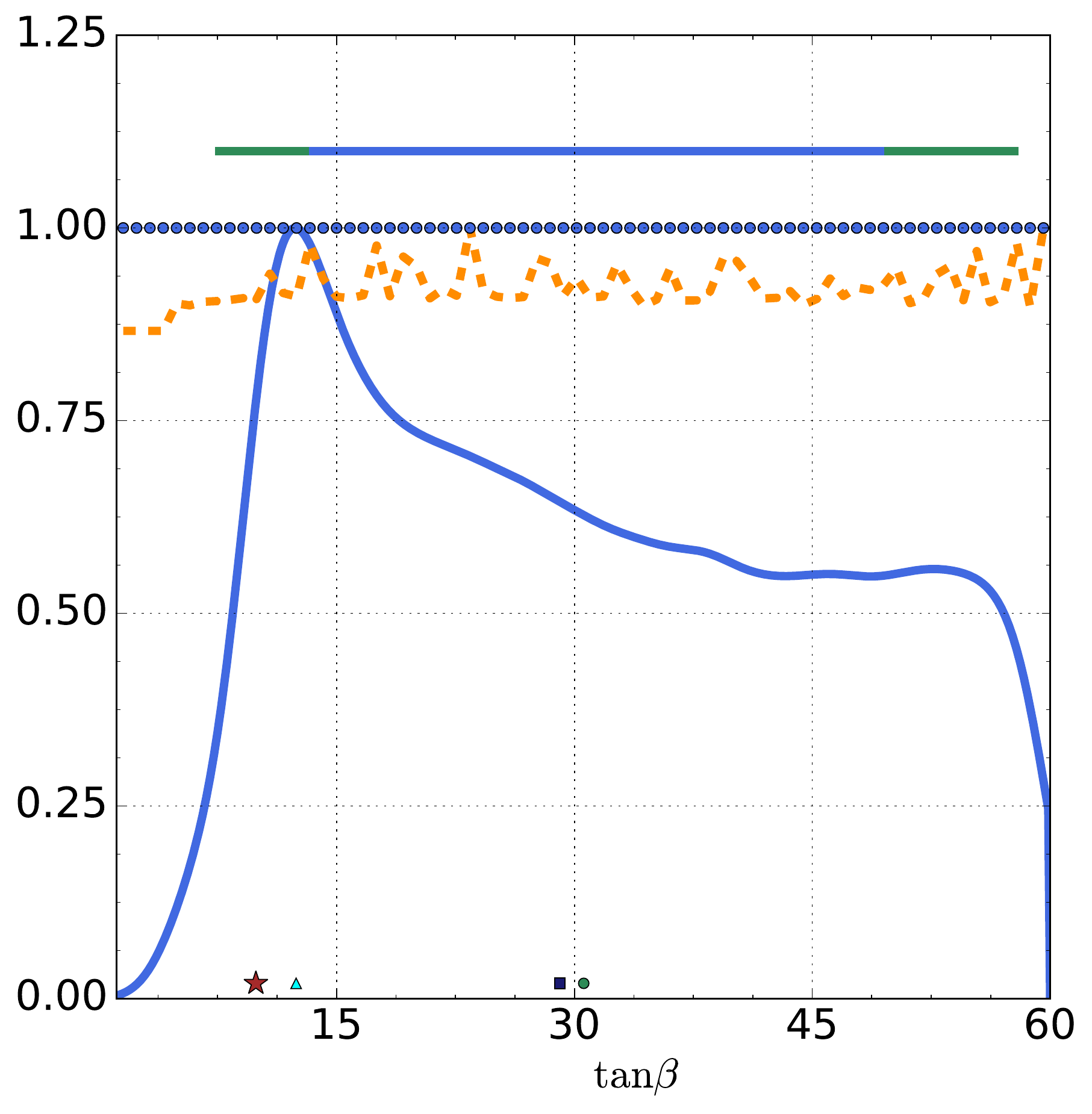}
		}
		\resizebox{0.7\textwidth}{!}{
		\includegraphics{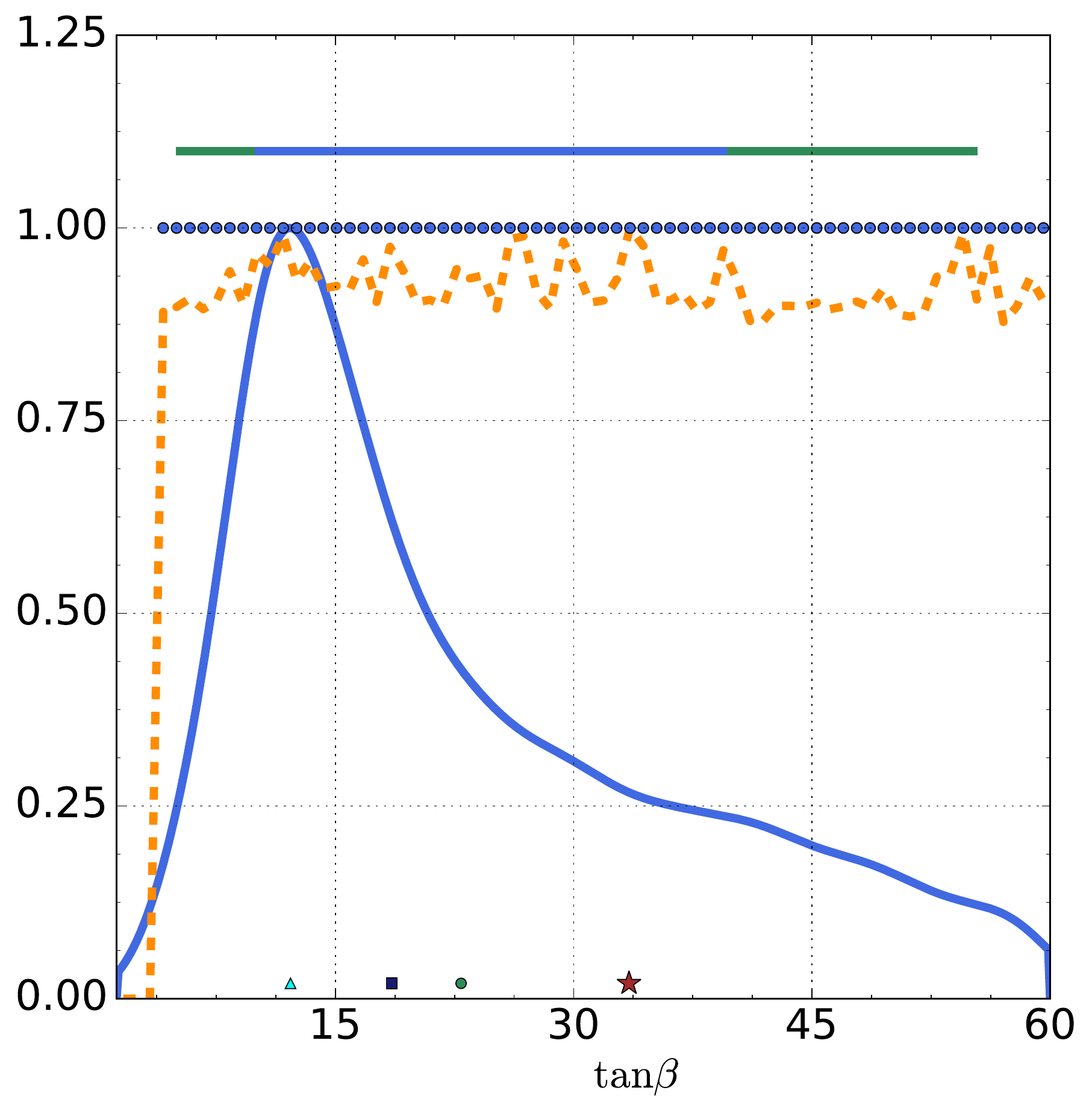}
		\includegraphics{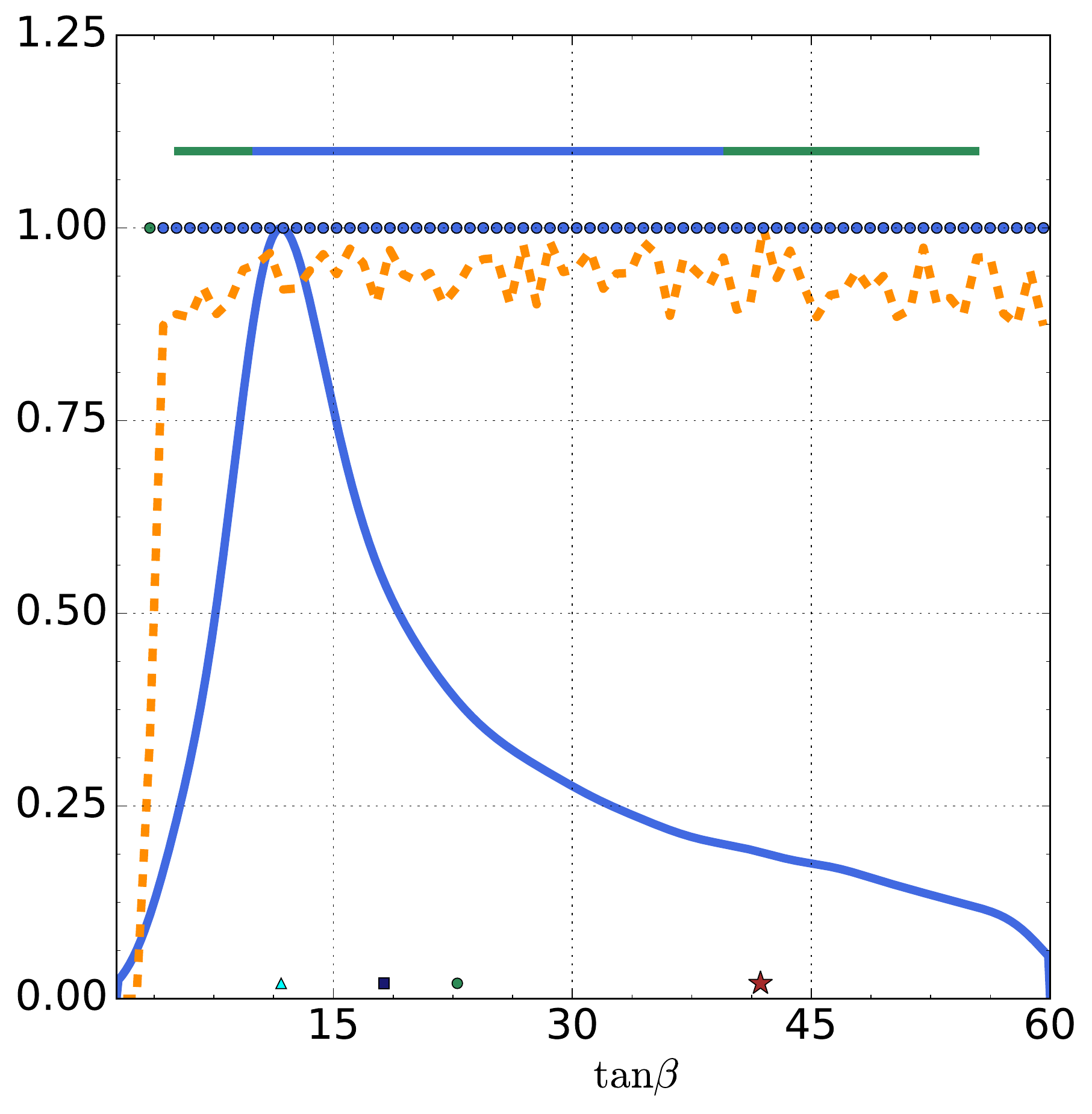}
		\includegraphics{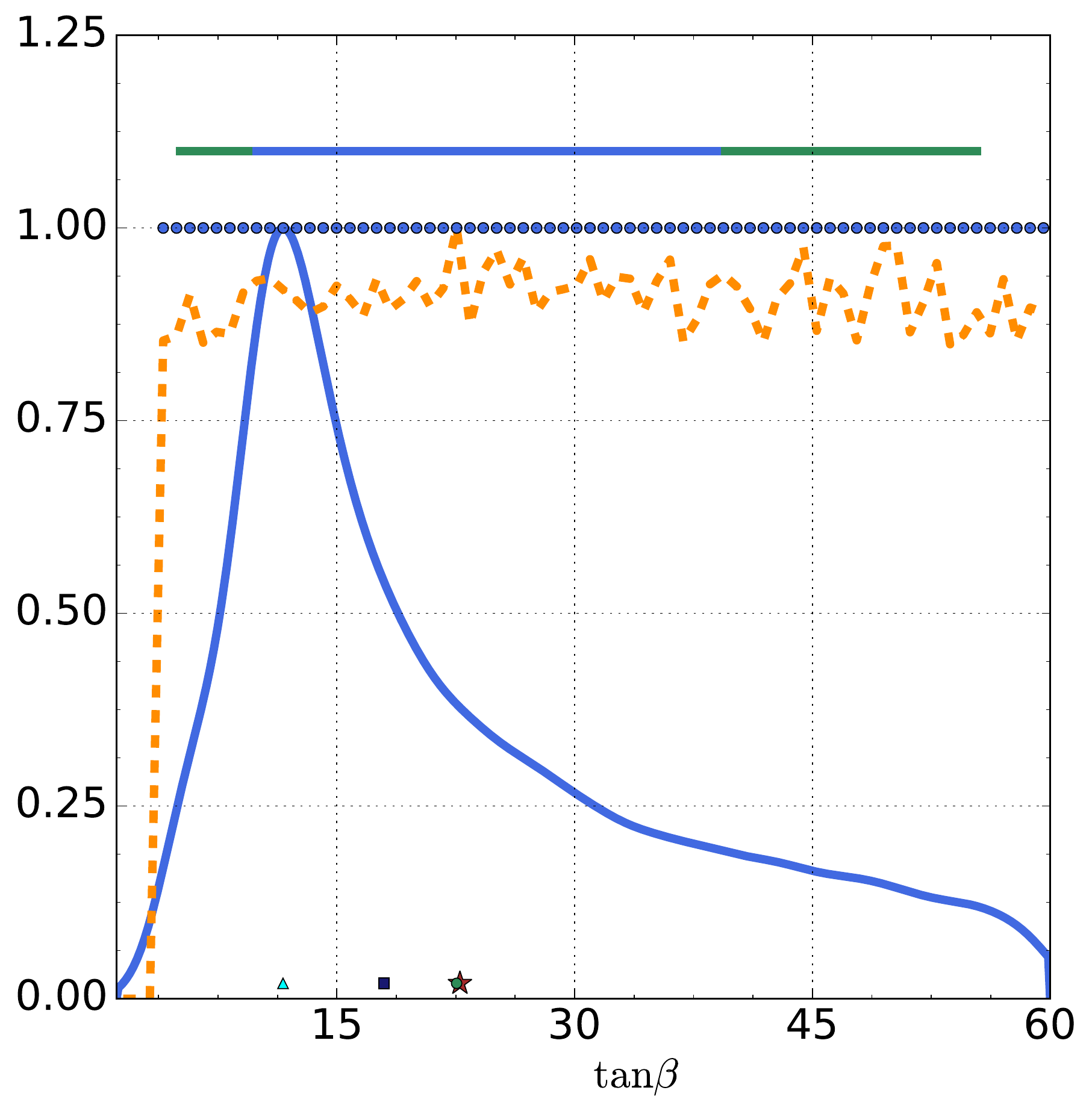}
		}
		\resizebox{0.7\textwidth}{!}{
		\includegraphics{Fig15-A-1.pdf}
		\includegraphics{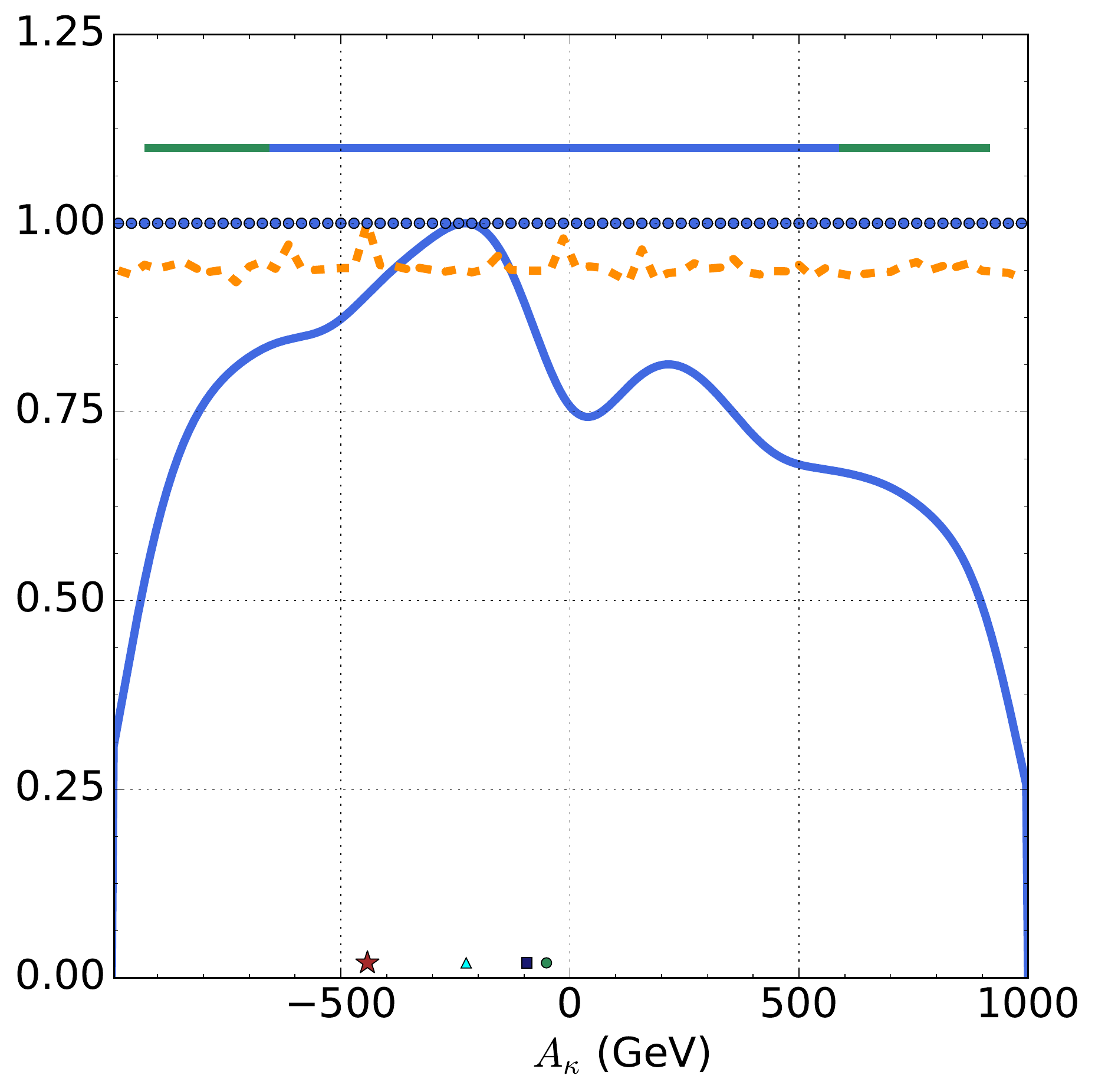}
		\includegraphics{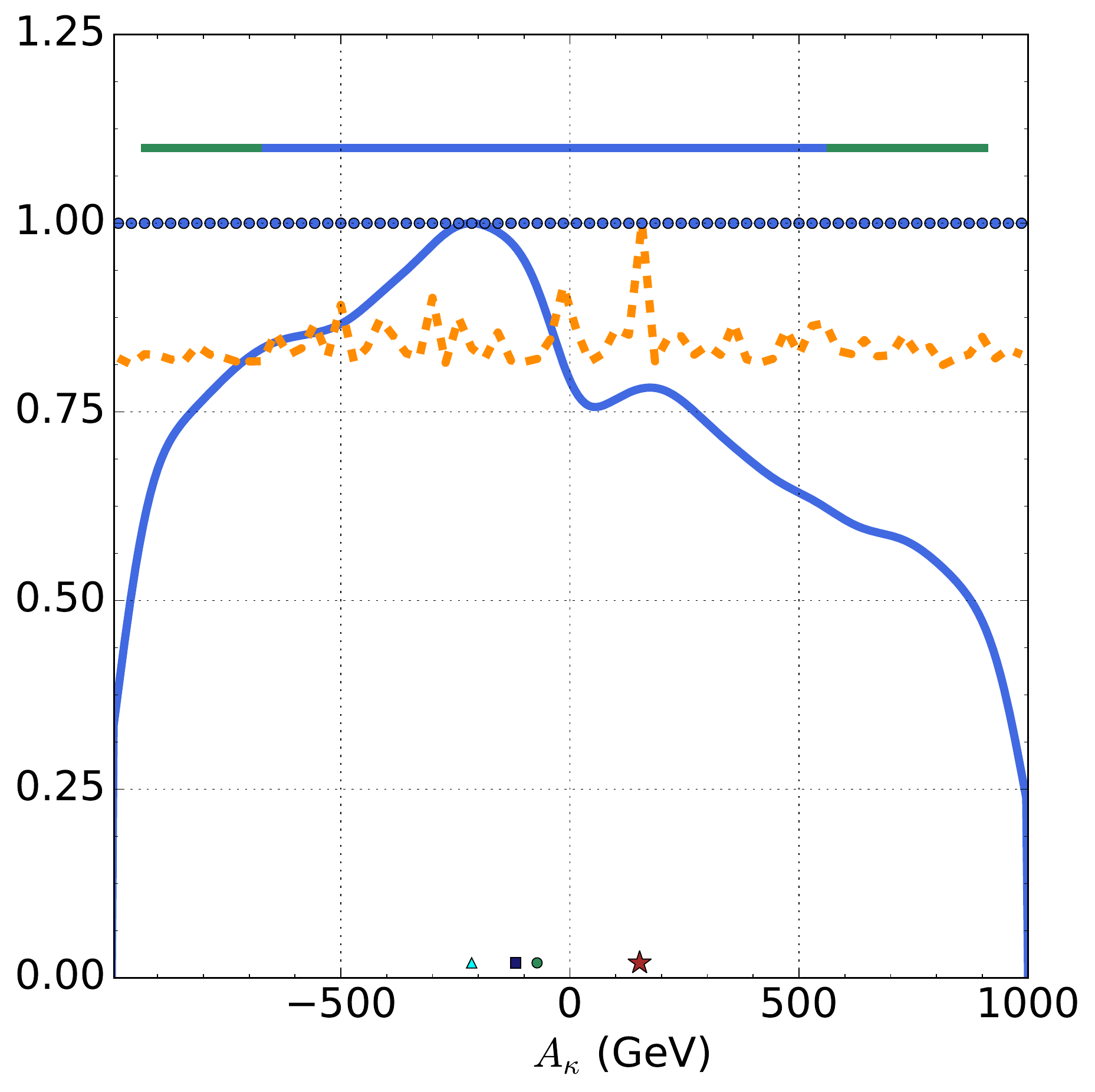}
		}
		\resizebox{0.7\textwidth}{!}{
		\includegraphics{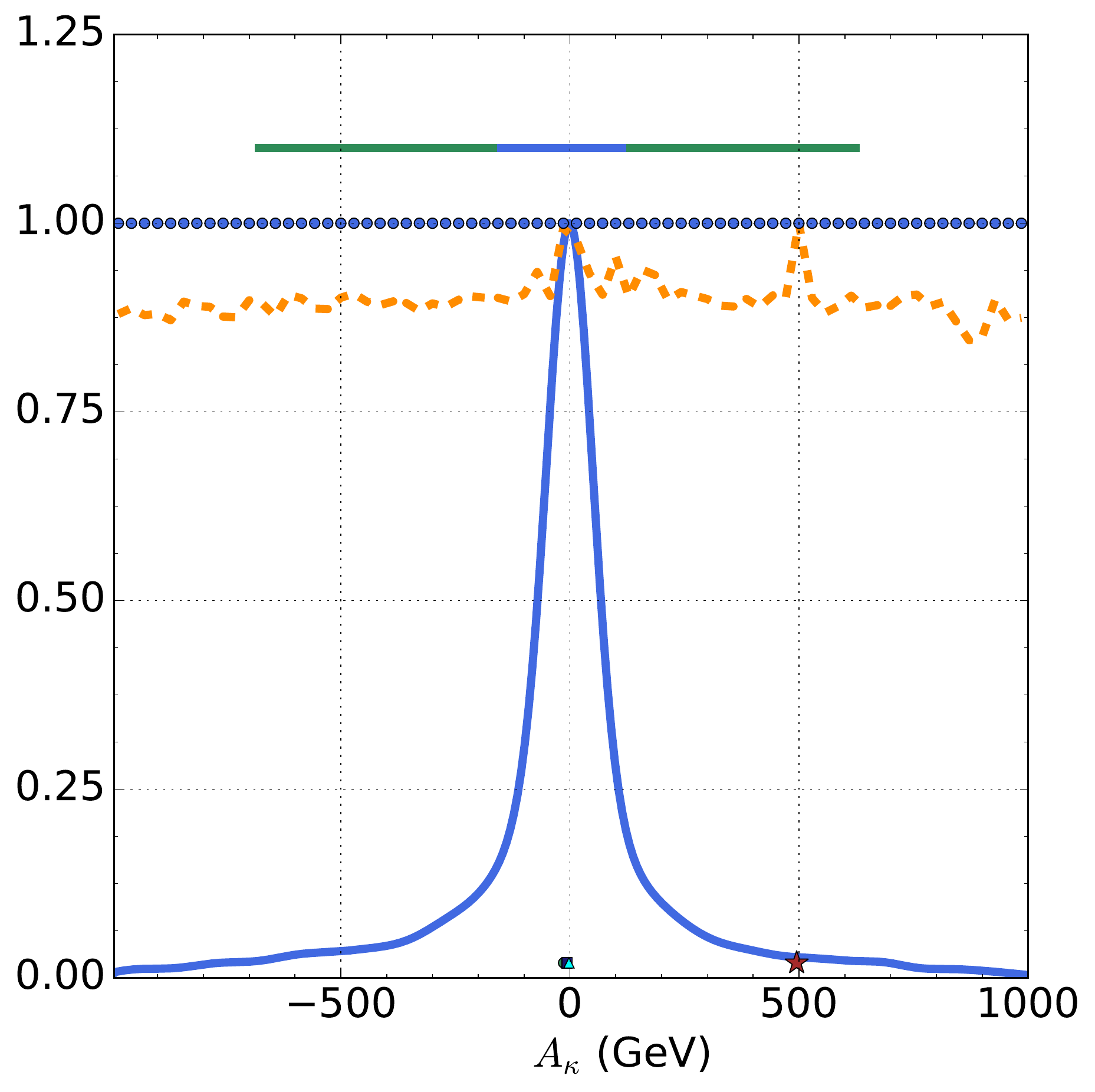}
		\includegraphics{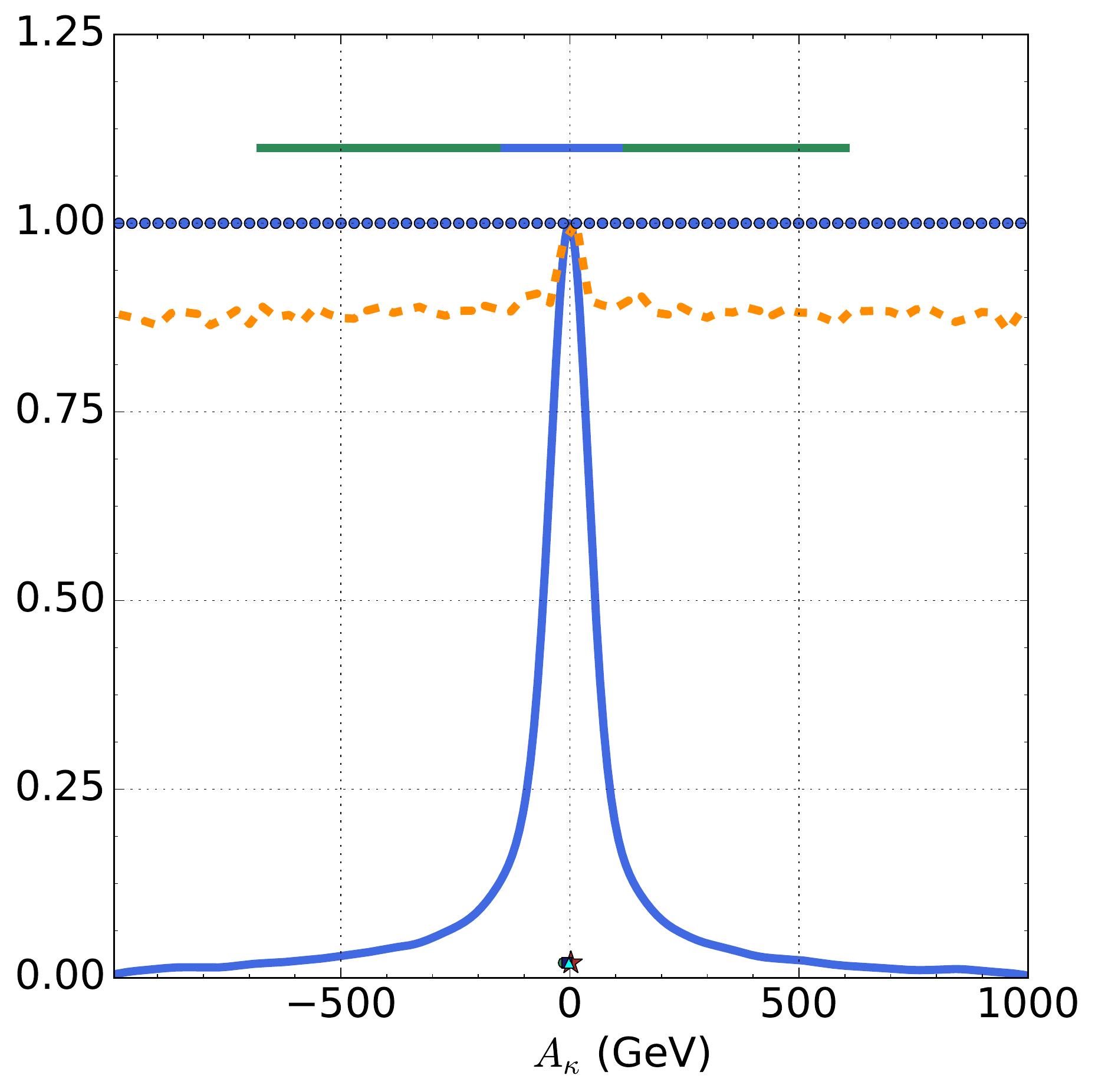}
		\includegraphics{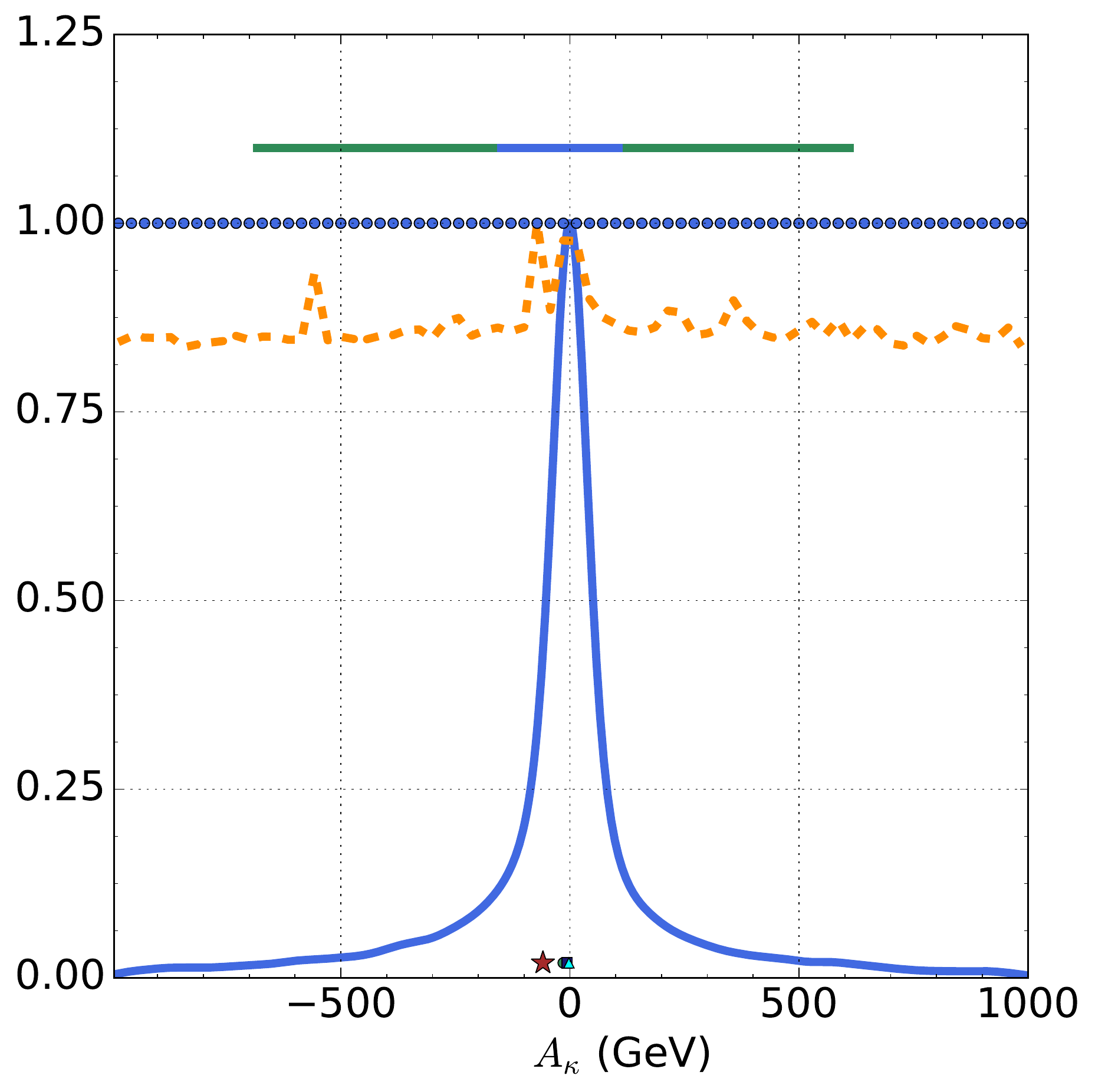}
		}
       \vspace{-0.4cm}

	\caption{Same as FIG.\ref{fig16}, but for the distributions of the parameters $\tan \beta$ and $A_\kappa$. \label{fig17}}
    \end{figure*}

\begin{figure*}[htbp]
		\centering
		\resizebox{0.7\textwidth}{!}{
		\includegraphics{Fig15-A-2.pdf}
		\includegraphics{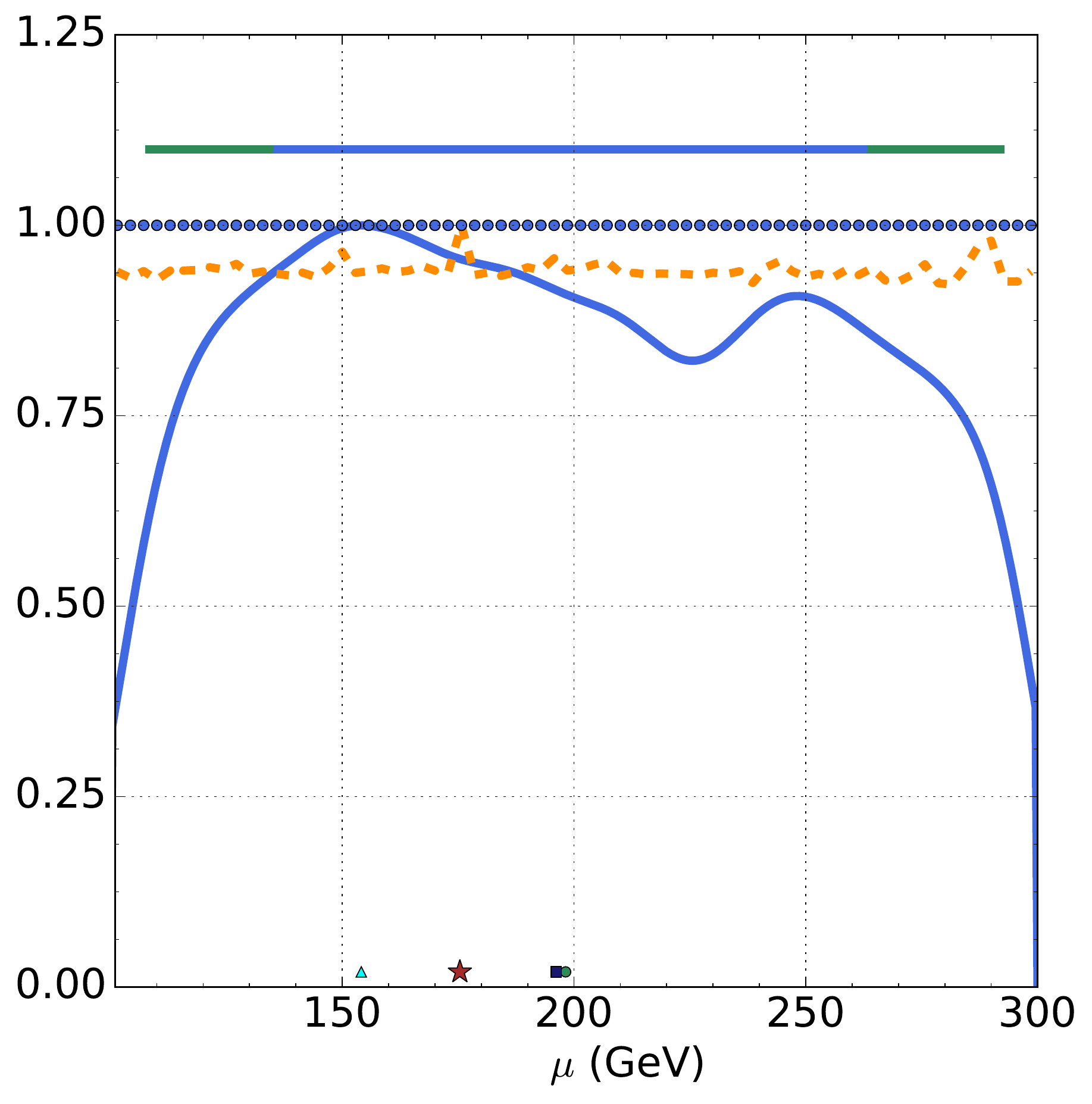}
		\includegraphics{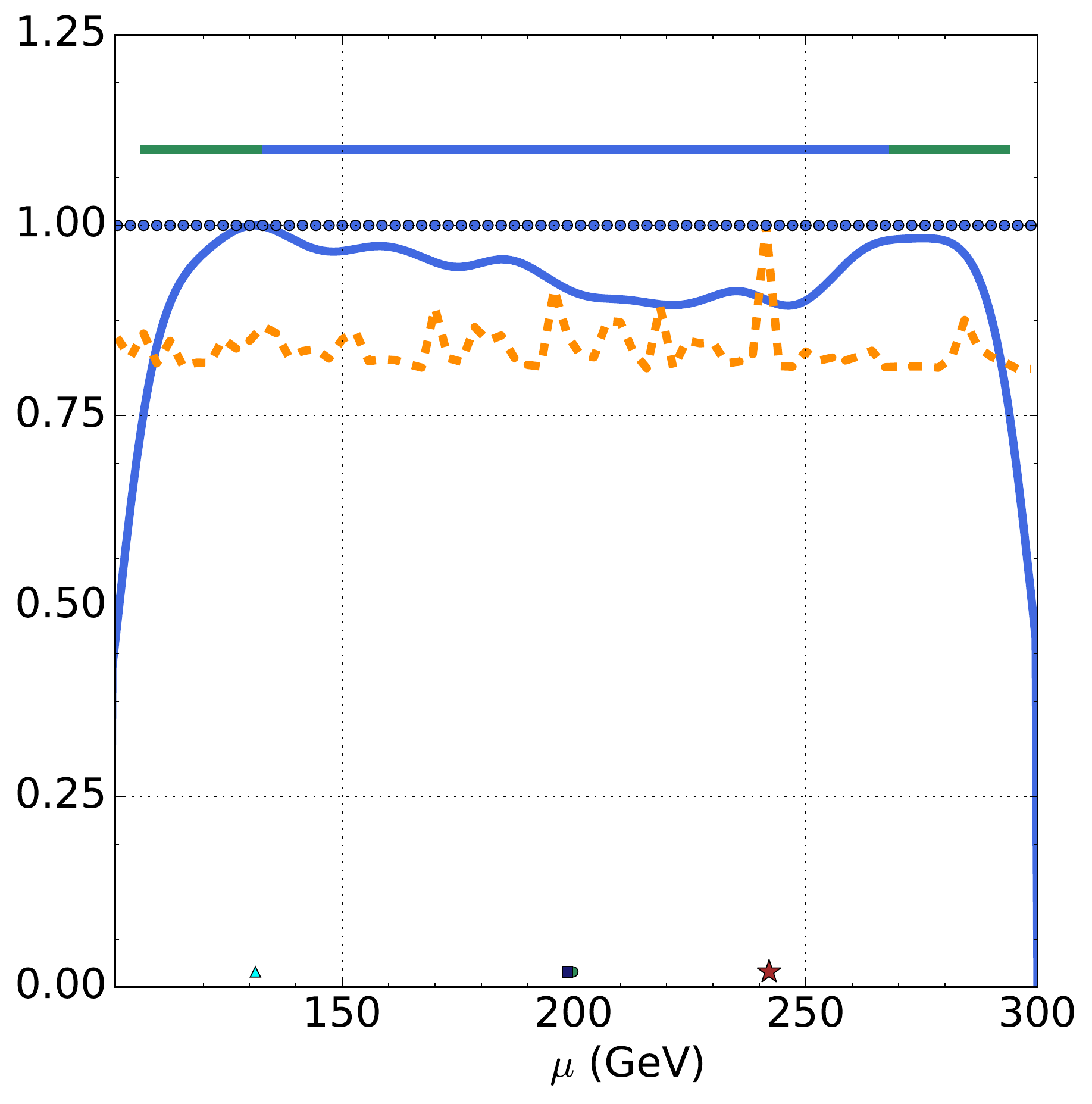}
		}
		\resizebox{0.7\textwidth}{!}{
		\includegraphics{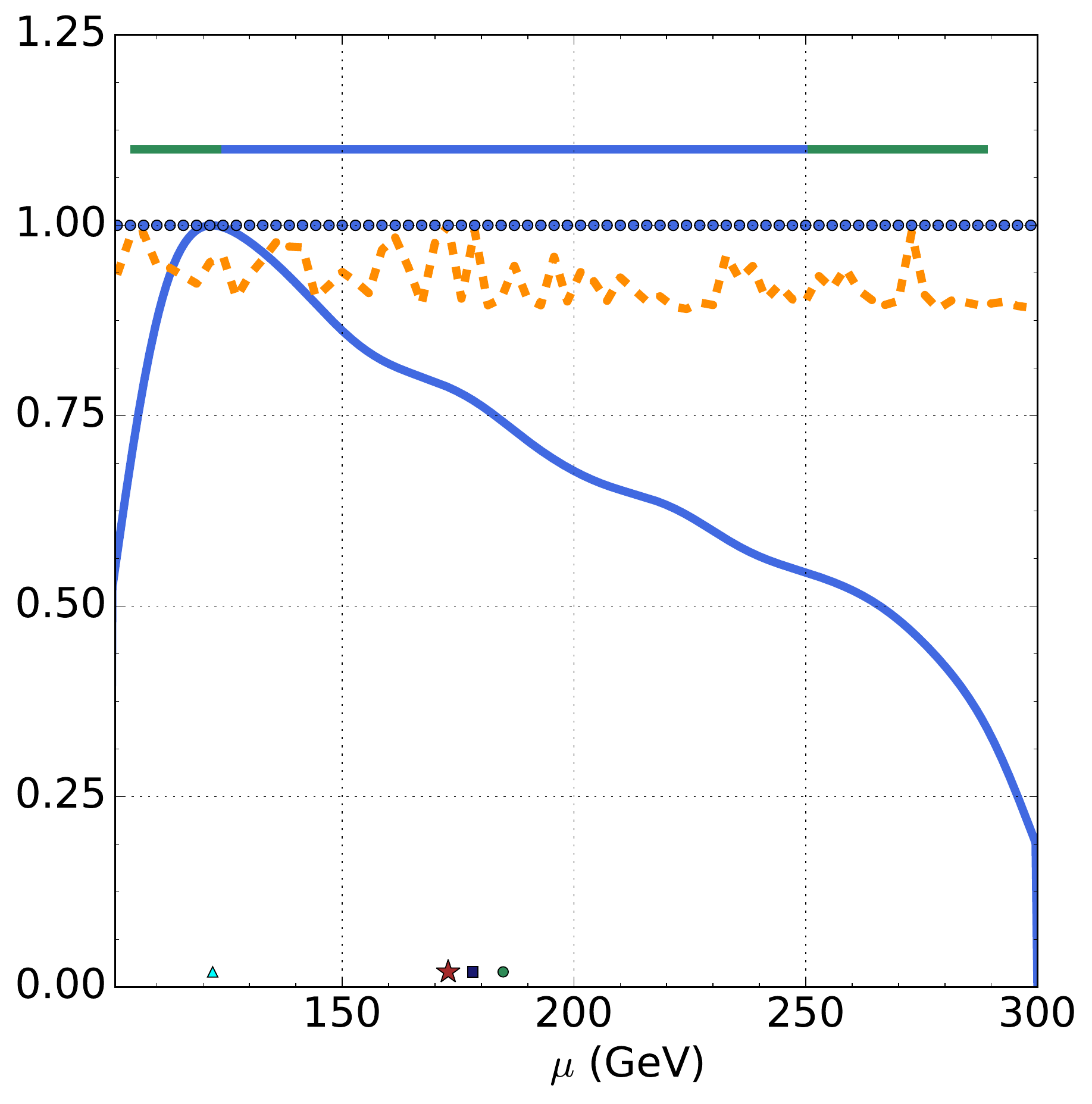}
		\includegraphics{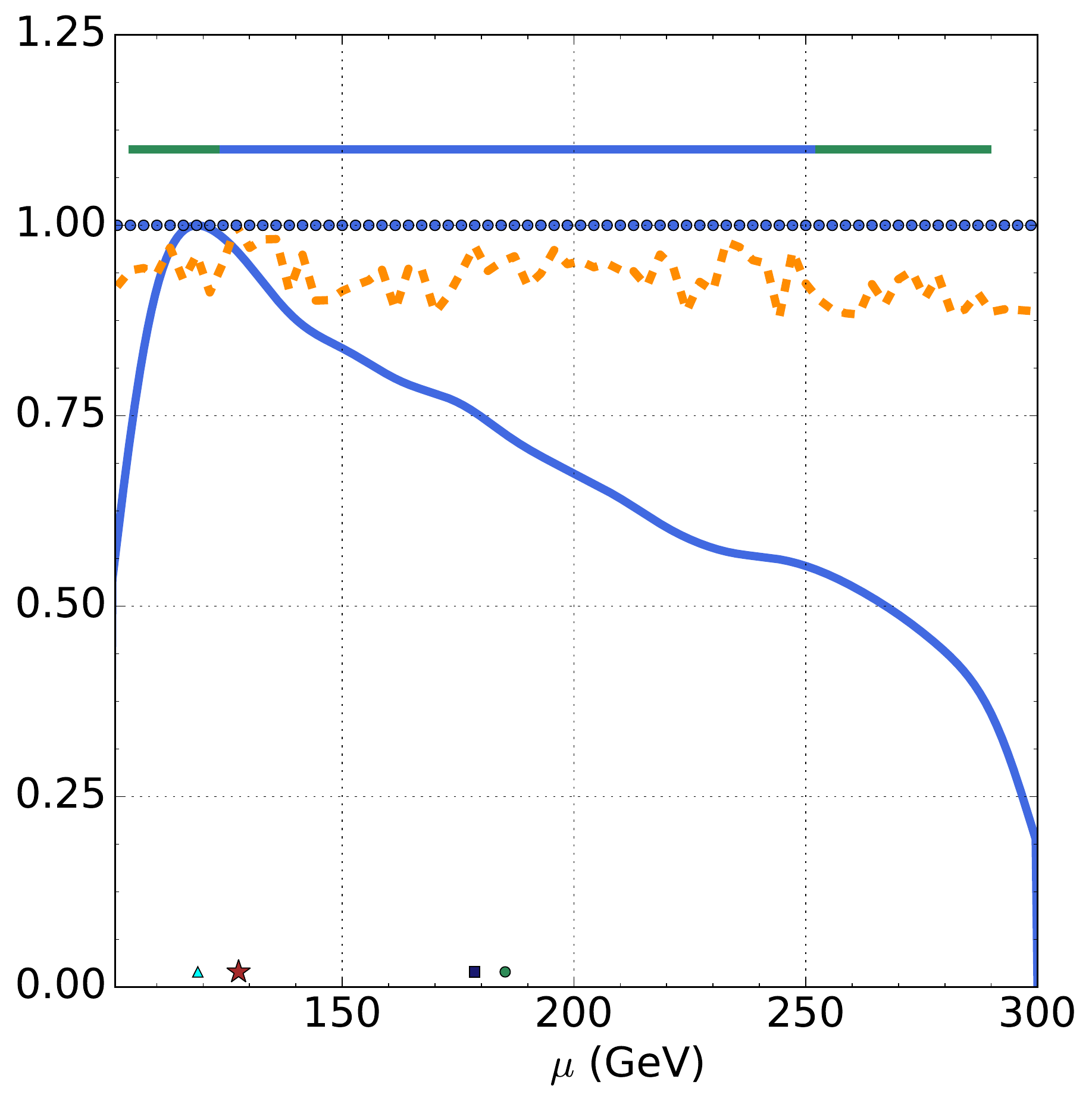}
		\includegraphics{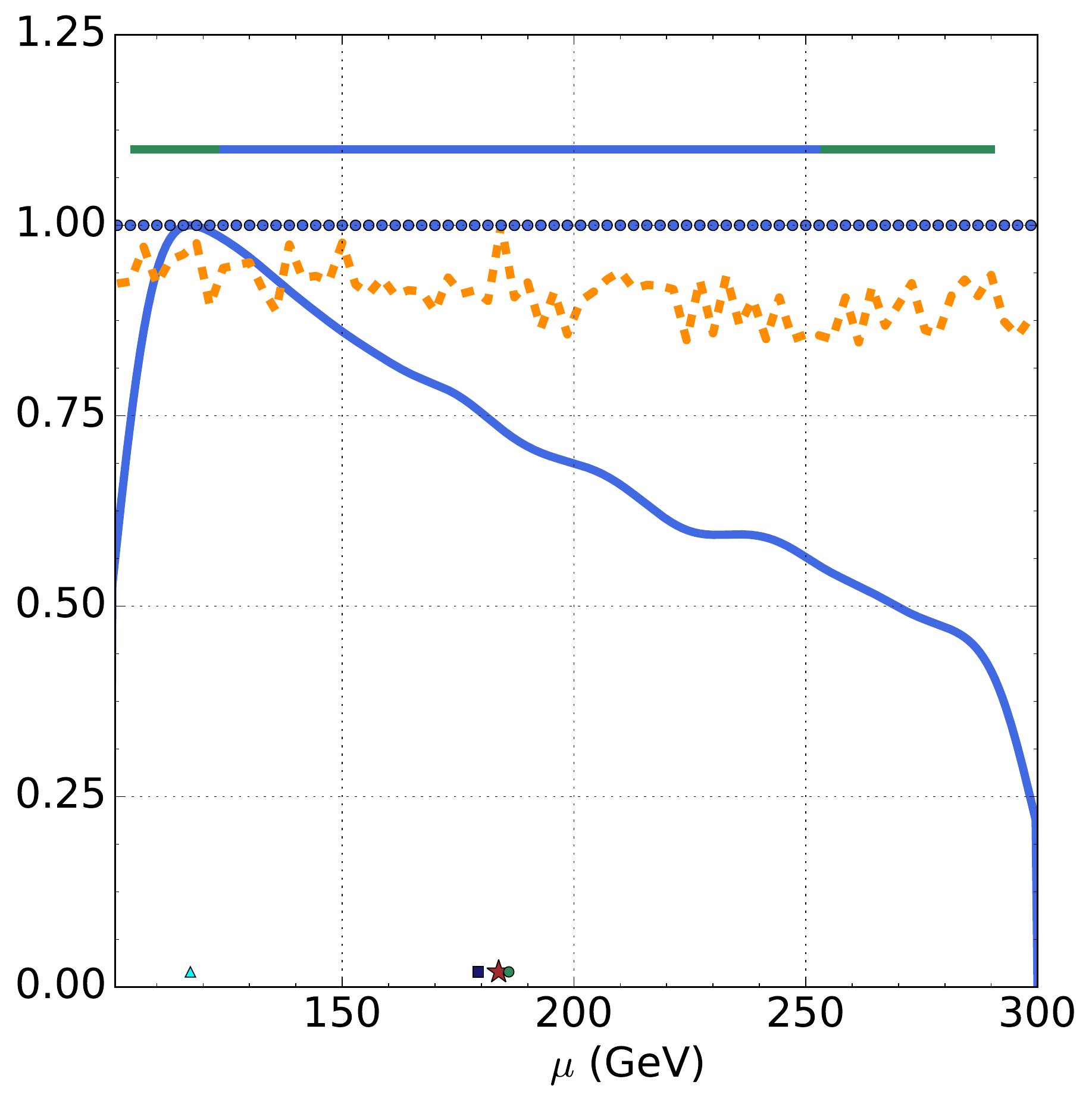}
		}
		\resizebox{0.7\textwidth}{!}{
		\includegraphics{Fig15-A-3.pdf}
		\includegraphics{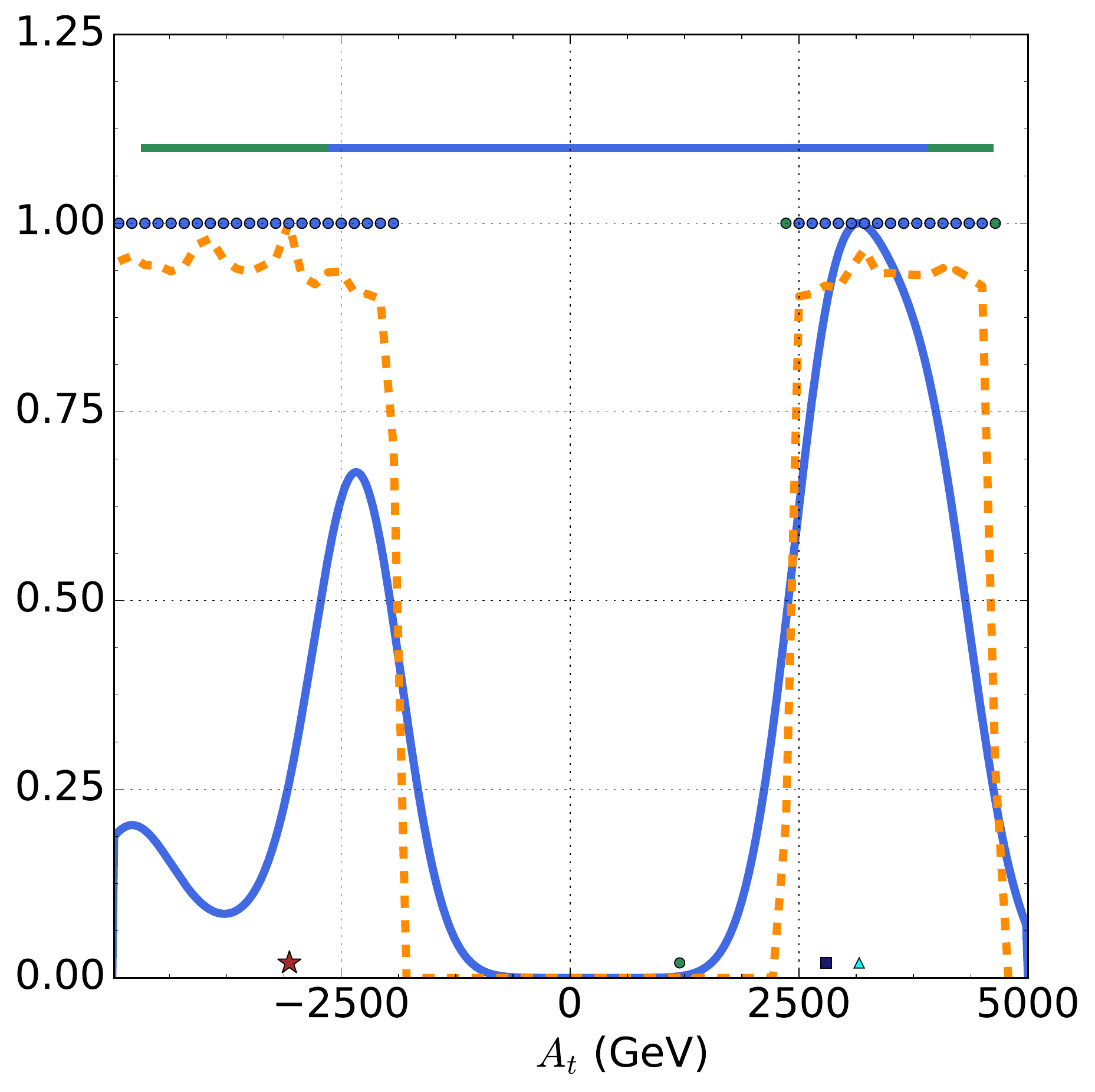}
		\includegraphics{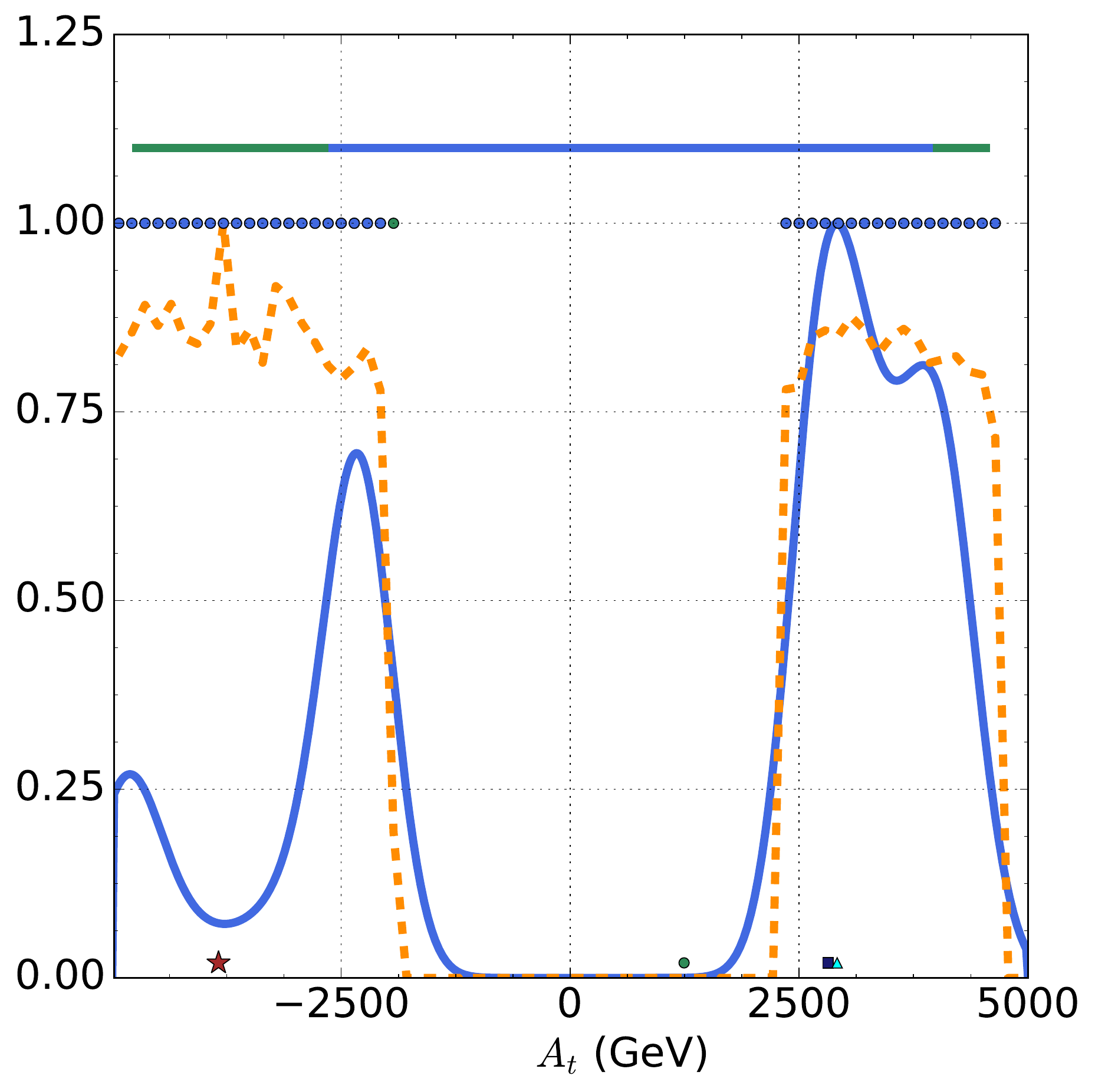}
		}
		\resizebox{0.7\textwidth}{!}{
		\includegraphics{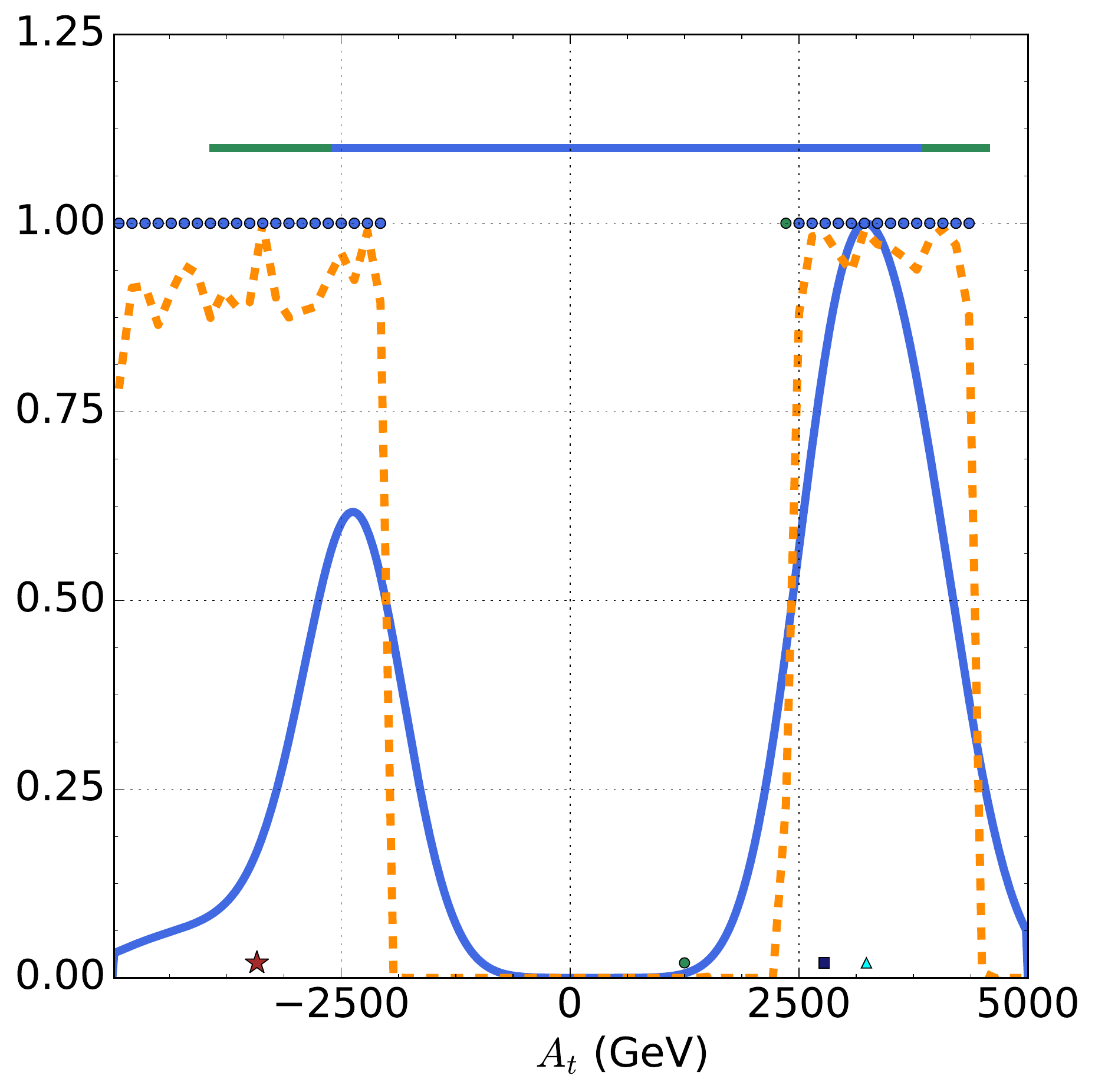}
		\includegraphics{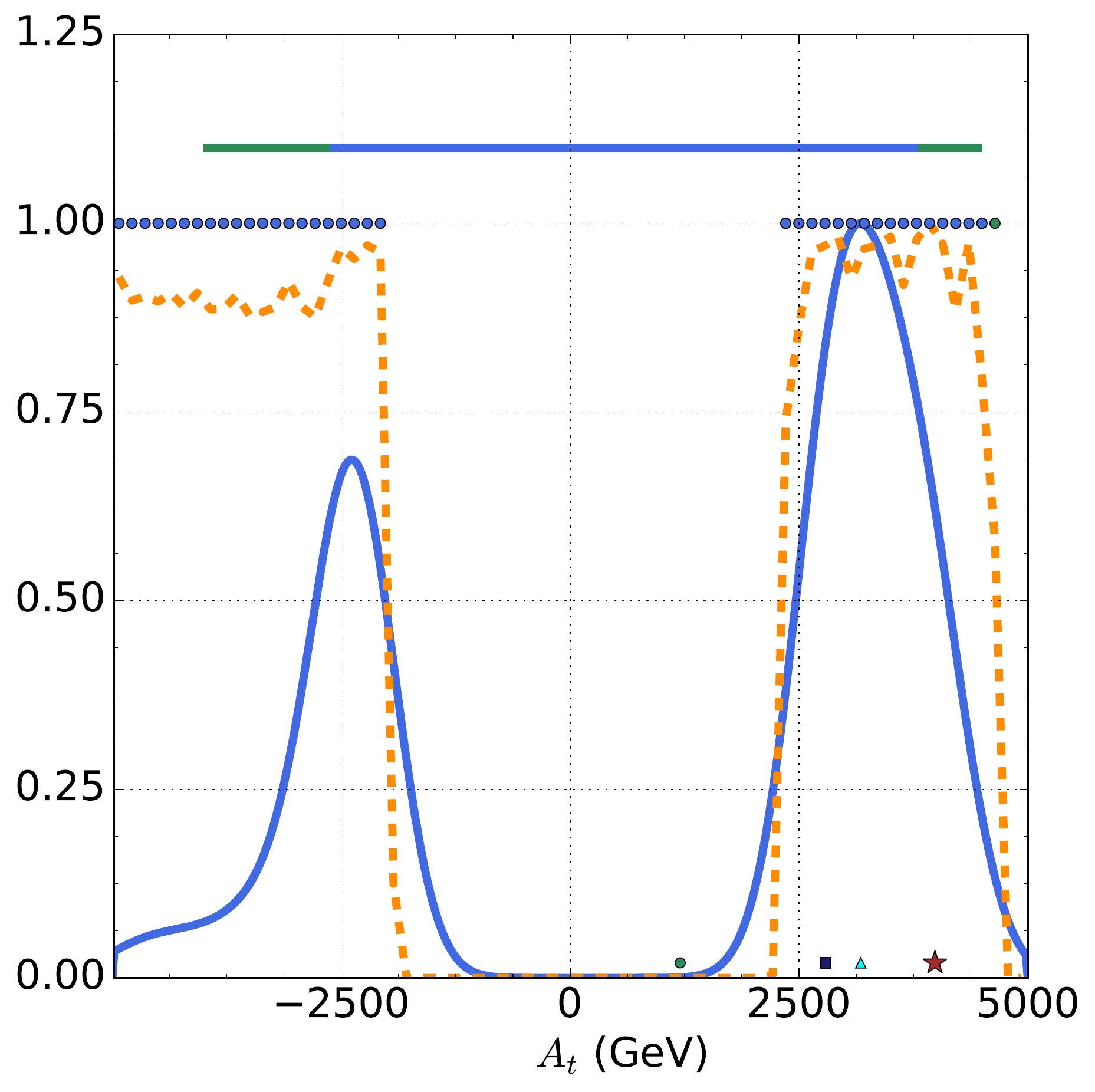}
		\includegraphics{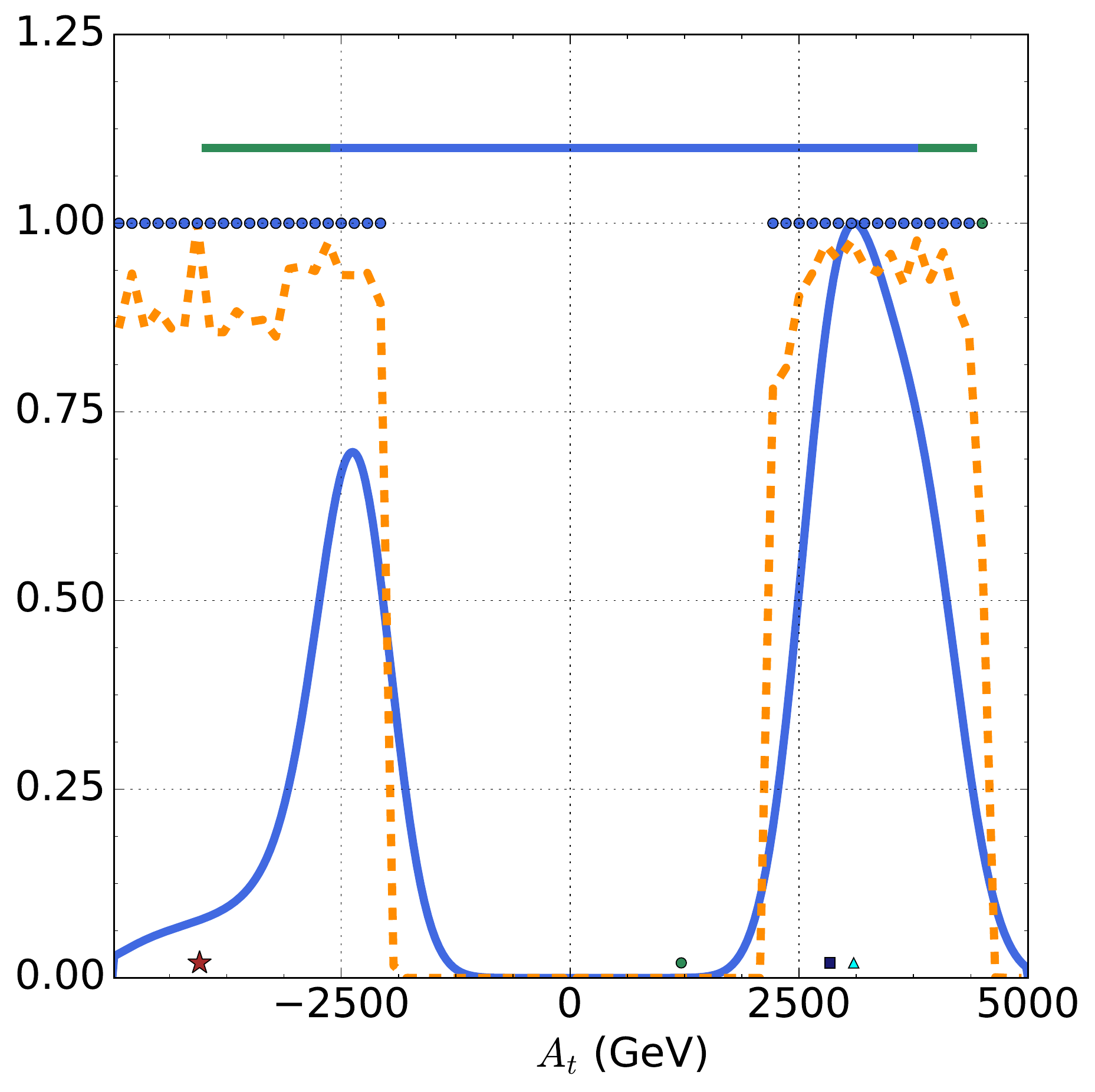}
		}
       \vspace{-0.4cm}

	\caption{Same as FIG.\ref{fig16}, but for the distributions of the parameters $\mu$ and $A_t$. \label{fig18}}
    \end{figure*}

In order to get the key features of the parameter space favored by Higgs data, we take $\mathcal{L}=\mathcal{L}_{Higgs} $ in the scan, and adopt a large {\it nlive}, {\it nlive} = 24000,  since the involved calculation is relatively fast and we want to
get the marginal PDFs as precise as possible. We realize that varying $A_\lambda$ may be helpful to improve our understanding on the features, so we choose $A_\lambda = 2 {\rm TeV}$, $4 {\rm TeV}$, $6 {\rm TeV}$ and $10 {\rm TeV}$ as four benchmark cases in the study. Part of our results, i.e. the posterior PDFs of the parameters $\lambda$, $\kappa$, $\tan \beta$, $A_\kappa$, $\mu$ and $A_t$ for $A_\lambda = 2 {\rm TeV}$, are presented in some of the right panels in FIG.\ref{fig16}-\ref{fig18}.  From these results, we have following observations:
\begin{enumerate}
\item With $A_\lambda$ becoming larger, the most favored value of $\lambda$ decreases monotonously from about 0.09 for $A_\lambda = 2 {\rm TeV}$ to about 0.035 for $A_\lambda = 10 {\rm TeV}$. Correspondingly, the most favored value of the product $\lambda A_\lambda$ increases from about $180 {\rm GeV}$ to about $350 {\rm GeV}$.

\item The marginal PDF of $\kappa$ ($A_\kappa$) is maximized at $\kappa \simeq 0.6 $ ($A_\kappa \simeq -200 {\rm GeV}$) for $A_{\lambda} = 2 {\rm TeV}$, $4 {\rm TeV}$ cases and $\kappa \simeq -0.6 $ ($A_\kappa \simeq 200 {\rm GeV}$) for the other cases, and both the PDFs (especially that for $\kappa$) exhibit a rough reflection symmetry (see the right panels in the third row of FIG.\ref{fig16} and FIG.\ref{fig17}).

    The product $\kappa A_\kappa$ is always negative. Its distribution is peaked at $\kappa A_\kappa \simeq -40 {\rm GeV}$, and extends to about $- 400 {\rm GeV}$ for $1 \sigma$ CR.  The shape of the distribution is insensitive to the choice of $A_\lambda$.

\item With the increase of $A_\lambda$, the most favored value of $\tan \beta$ moves upward from about 13 for $A_\lambda = 2 {\rm TeV}$ to about 22 for $A_\lambda = 10 {\rm TeV}$.

\item For $A_\lambda = 2 {\rm TeV}$, the marginal PDF of $\mu$ is evenly distributed in the range from $120 {\rm GeV}$ to $300 {\rm GeV}$ (see the right panel in the first row of FIG.\ref{fig18}), but with the increase of $A_\lambda$, the preference of a moderately low $\mu$ gradually emerges (see the right panel of FIG.\ref{fig23}).

\item The marginal PDFs of the ratio $\mu/\lambda \equiv v_s/\sqrt{2}$ and $|\kappa \mu|/\lambda$ are peaked at $950 {\rm GeV}$ and $400 {\rm GeV}$ respectively for $A_\lambda = 2 {\rm TeV}$, and these most favored values move to  $1200 {\rm GeV}$ and $600 {\rm GeV}$ respectively for $A_\lambda = 10 {\rm TeV}$. This reflects the general conclusion that the ratios tend to increase with $A_\lambda$.


\item As for the ratio
   \begin{eqnarray}
   R_{23} \equiv \frac{(\lambda A_\lambda + 2 \kappa \mu)  \sin 2 \beta}{2 \lambda \mu}  \nonumber
   \end{eqnarray}
   which parameterizes the degree of cancellation between different terms in 23 element of the CP-even Higgs mass matrix (see discussion below), our results indicate that its marginal PDF is peaked around 0.4 and its $1\sigma$ CR corresponds to $ 0.2 \lesssim R_{23} \lesssim 1 $ for $A_\lambda = 2 {\rm TeV}$.  With the increase of $A_\lambda$ to $10{\rm TeV}$, these quantities shift to about 1.8 and $ 1 \lesssim R_{23} \lesssim 4$ respectively. Our results also indicate that a small $\lambda$ usually allows $R_{23}$ to deviate from 1 significantly, e.g.  $R_{23} > 3 $ is possible only when $\lambda \lesssim 0.15$ for $A_\lambda = 10 {\rm TeV}$, which is shown by the 2D marginal PDF on $R_{23}-\lambda$ plane. We will explain these behaviors later.

  With the definition of $R_{23}$, one can calculate its derivatives to the parameters $\lambda$, $\kappa$, $\tan \beta$, $\mu$ and $A_\lambda$. The expression of these derivatives indicates that $R_{23}$ is very sensitive to $\tan \beta$ for the favored values of the parameters given in items 1-4.

\item The marginal PDF of $A_t$ is insensitive to $A_\lambda$, and its shape is quite similar to that in FIG.\ref{fig18} for any benchmark cases of $A_{\lambda}$.
\end{enumerate}

\begin{table*}[thbp]
	\begin{center}
\begin{tabular}{|c|c|c|c|c|c|c|}
\hline
\multirow{2}{*}{} & \multicolumn{3}{c|}{\bf{1D marginal PDF}} & \multicolumn{3}{c|}{\quad \quad \quad \quad \quad \quad \quad \bf{1D PL}} \\ \cline{2-7}
 & Higgs Data & B Physics Data & DM Data & Higgs Data & B Physics Data & DM Data \\ \hline
$\lambda$ & $\surd$   & $X$  &  $\surd$  & $\surd \hspace{-0.15cm} \bf{\smallsetminus}$  & $X$  & {$\surd$}\  (for lower end) \\ \hline
$\kappa$ & $\surd$  & $X$  & {$\surd$}   &   $\surd \hspace{-0.15cm} \bf{\smallsetminus}$ &  $X$  &  $\surd$   \\ \hline
$\tan \beta$ & $\surd$  & {$\surd$}\  (for upper end)  & $\surd$ \ (for upper end) &  $\surd \hspace{-0.15cm} \bf{\smallsetminus}$  & {$\surd$} \  (for upper end)  & $\surd$\  (for upper end)  \\ \hline
$A_\kappa$ & $\surd$  & $X$   & {$\surd$}   & $X$   & $X$  & $X$ \\ \hline
$\mu$ & $X$   & $\surd \hspace{-0.15cm} \bf{\smallsetminus}$  & {$\surd$}   & $X$   & $X$   & $X$  \\ \hline
$A_t$ & $\surd$  & {$\surd$} \  (for $A_t < 0$)  & {$\surd$} \   (for $A_t < 0$)  & {$\surd$}   & {$\surd$}\  (for $A_t < 0$)  & {$X$}  \\ \hline
\end{tabular}
\caption{Dependence of the 1D marginal PDFs and PLs on different experimental measurements for $A_\lambda = 2 {\rm TeV}$. The symbols $\surd$, $\surd \hspace{-0.15cm} \bf{\smallsetminus}$ and $X$ represent strong dependence, mildly strong dependence and weak dependence respectively. }
		\label{table4}
\end{center}
\end{table*}

At this stage, we remind that the favored values can be regarded as the typical size of corresponding quantities. Given the expressions of Higgs mass matrix presented in Eq.(\ref{Mass-CP-even-Higgs}) and Eq.(\ref{Mass-CP-odd-Higgs}) of Section II, they are helpful
to understand some features of the marginal PDFs. In the following, we will illustrate this point.
\begin{itemize}
\item The facts in item 1, 2 and 3 indicate that
\begin{eqnarray}
\frac{\lambda A_\lambda \sin 2 \beta}{4 \mu} \ll 1, \quad  \frac{(\lambda A_\lambda + 3 \kappa \mu) \sin 2 \beta}{4 \mu} \ll 1.  \nonumber
\end{eqnarray}
So one can conclude that
\begin{eqnarray}
{\cal M}^2_{33} &\simeq &  \frac{\mu}{\lambda} (\kappa A_\kappa +  \frac{4 \kappa^2 \mu}{\lambda} ) \sim \frac{4 \kappa^2 \mu^2}{\lambda^2}, \nonumber \\
{\cal M}^2_{P,22} &\simeq & - \frac{3 \mu}{\lambda} \kappa A_\kappa.
\end{eqnarray}
 Without a strong mixing of the singlet field with the other fields, the right sides of the approximations denote the squared masses of the singlet states. Then the positivity of ${\cal M}^2_{P,22}$ requires a negative $\kappa A_\kappa$. Moreover, from item 2 and 5 one can estimate that the masses of the singlet dominated scalars are usually
 below $1 {\rm TeV}$ for $A_\lambda = 2 {\rm TeV}$.

\item  The square of the mass scale for the heavy doublet field,
   \begin{eqnarray}
   M_A^2 \equiv {\cal M}_{P,11}^2 =  \frac{2 \mu (\lambda A_\lambda + \kappa \mu)}{\lambda \sin 2 \beta},  \nonumber
   \end{eqnarray}
   is enhanced by a factor $1/(\lambda \sin 2 \beta)$. With the increase of $A_\lambda$, this factor tends to increase and consequently the doublet scalars become heavier. We checked that $M_A \gtrsim 2 {\rm TeV}$ ($M_A \gtrsim 4.5 {\rm TeV}$) in its $1 \sigma$ CR for $A_\lambda = 2 {\rm TeV}$ ($A_\lambda = 10 {\rm TeV}$).

\item In the decoupling limit of the heavy doublet field, the effective Higgs potential at electroweak scale contains only the SM Higgs field and the singlet field. In the basis ($S_2$, $S_3$), the effective squared mass matrix is
\begin{equation}
{\cal M}^2_{eff} = \begin{pmatrix}
{\cal M}^2_{22} \quad \quad
&  {\cal M}^2_{23}    \\
{\cal M}^2_{23} \quad \quad
&  {\cal M}^2_{33} - ({\cal M}^2_{13})^2/{\cal M}^2_{11}     \nonumber
\end{pmatrix}
\end{equation}
when  $|{\cal M}^2_{12}| \ll |{\cal M}^2_{13}| $. Then the mixing angle between the two fields satisfies
   \begin{eqnarray}
   \tan 2 \theta & \simeq & \frac{2 {\cal M}^2_{23}}{{\cal M}^2_{22} - {\cal M}^2_{33} + ({\cal M}^2_{13})^2/({\cal M}^2_{11})} \nonumber \\
     & \simeq & \frac{- \lambda^2 v \left [ 2 \lambda \mu - (\lambda A_\lambda + 2 \kappa \mu ) \sin 2 \beta \right ] }{2 \sqrt{2} \kappa^2 \mu^2}.  \nonumber
     \end{eqnarray}
if ${\cal M}^2_{33} \gg (100 {\rm GeV})^2$. This approximation indicates that the mixing angle can be suppressed either by a small $\lambda/\kappa$, or by the cancellation between $ 2 \lambda \mu $ and $(\lambda A_\lambda + 2 \kappa \mu ) \sin 2 \beta$ if $R_{23} \sim 1$. Such a suppression is favored by the $125 {\rm GeV}$ Higgs data of the LHC. From the results in item 1, 2 and 3 or 5 and 6, one can infer that the mixing is small and it is very sensitive to $\tan \beta$ instead of to $\mu$.

 Moreover, the natural size of $\mu$ can be inferred by the minimization condition of the Higgs potential with respect to the field $H_u^0$ as follows~\cite{Ellwanger:2009dp}: \footnote{Note that this formula is based on the assumption of $\mu^2 \gg |m_{H_u}^2|, m_Z^2$, and it can be used only in estimating the magnitude of $\mu$.}
   \begin{eqnarray}
   \mu \sim \frac{A_\lambda}{\tan \beta - \kappa/\lambda}.  \nonumber
   \end{eqnarray}
   In fact, we studied the marginal PDF of the expression $\frac{A_\lambda}{\tan \beta - \kappa/\lambda}$. We found that it is peaked around $300 {\rm GeV}$ for $A_\lambda = 10 {\rm TeV}$, and with $A_\lambda$ diminishing, the value usually becomes smaller.

\item From the expressions of ${\cal M}^2_{23}$ and $({\cal M}^2_{13})^2/{\cal M}^2_{11}$ in the effective mass matrix, one can learn that terms proportional to $\kappa$ appear either in the form $\lambda \kappa v^2 \sin 2 \beta$ or in the form $\kappa^2 \mu^2 v^2/M_A^2$, and obviously all of them are suppressed.  As a result, the Higgs sector is approximately invariant under the transformation $\kappa \to - \kappa$ and $A_\kappa \to - A_\kappa$~\cite{Cao:2012fz}, and that is why the marginal posterior PDFs of $\kappa$ and $A_\kappa$ have a rough reflection symmetry.

\end{itemize}

        \begin{figure*}[htbp]
		\centering
		\resizebox{0.62\textwidth}{!}{
		\includegraphics{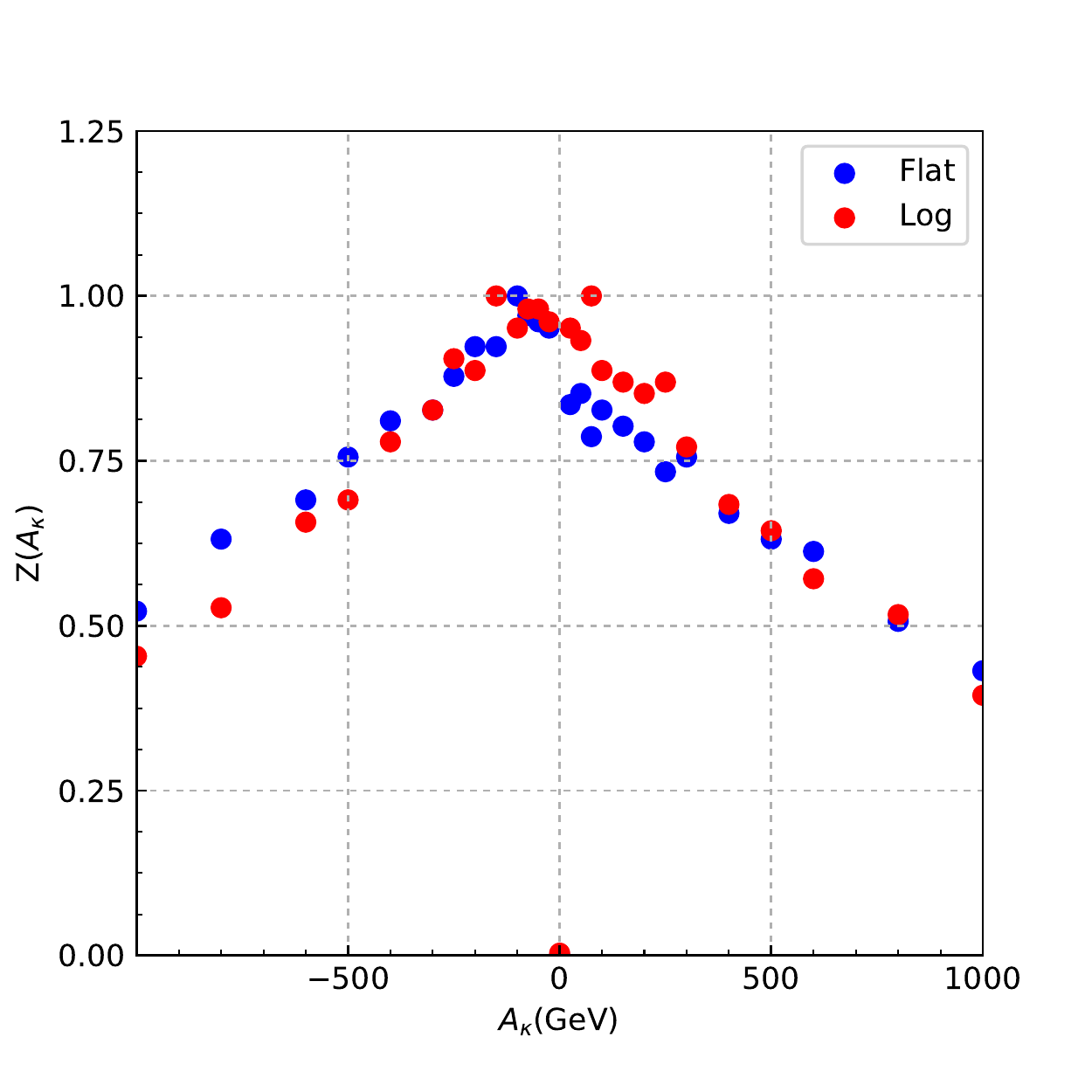}
		\includegraphics{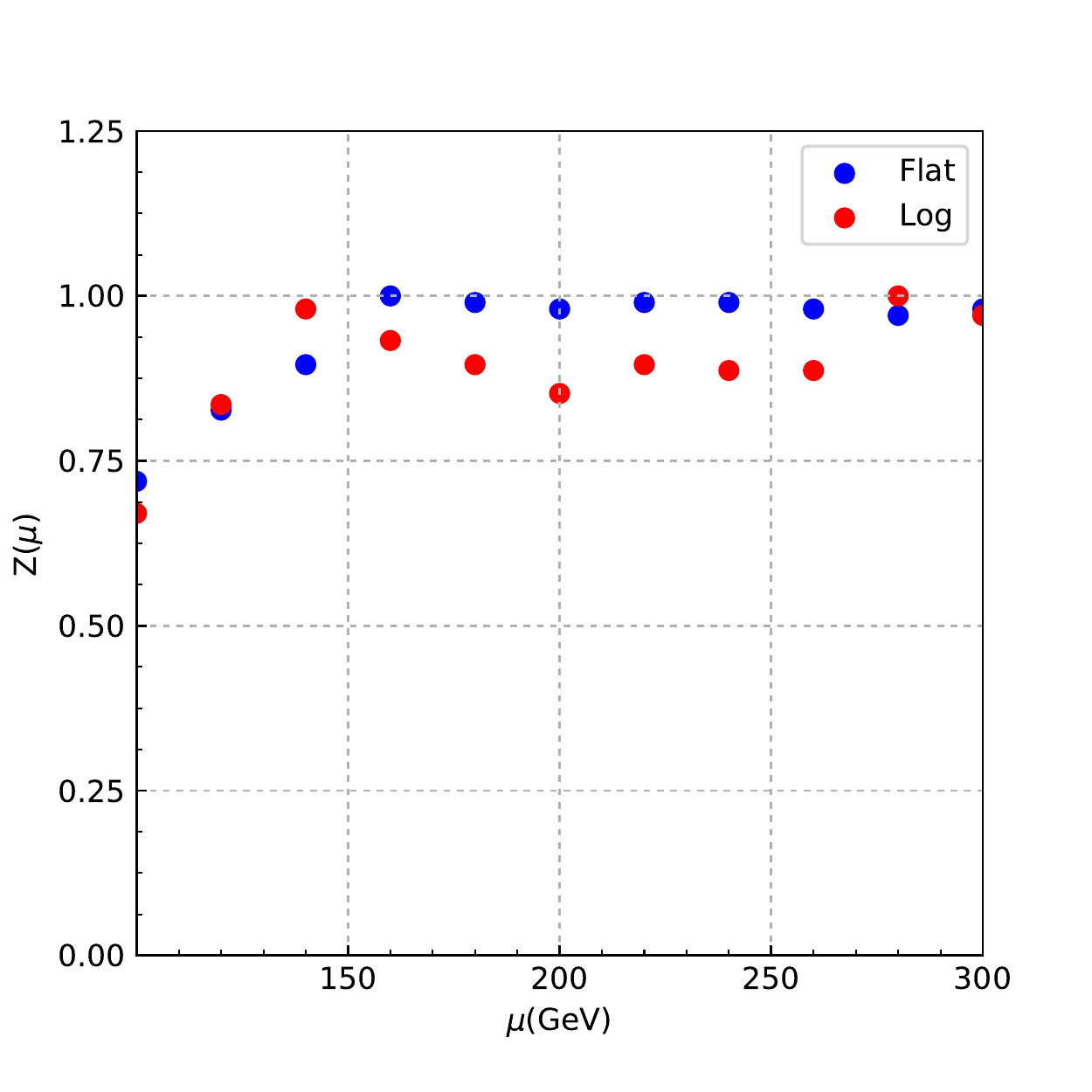}
		}

       \vspace{-0.4cm}

	\caption{Left panel: normalized distribution of $Z(A_\kappa)$ obtained by the scan in Eq.(\ref{scan-ranges1}) of Appendix A with the parameter $A_\kappa$ fixed for each point in the panel. The procedure to get $Z(A_\kappa)$ is depicted below Eq.(\ref{Alternative}) in Appendix B. The blue points and red points correspond to the results from the flat prior PDF and log prior PDF, respectively. Right panel: normalized distribution of $Z(\mu)$, which is obtained in a similar way to $Z(A_\kappa)$. \label{fig19} }
	\end{figure*}	
	
   \begin{figure*}[htbp]
		\centering
		\resizebox{0.58\textwidth}{!}{
		\includegraphics{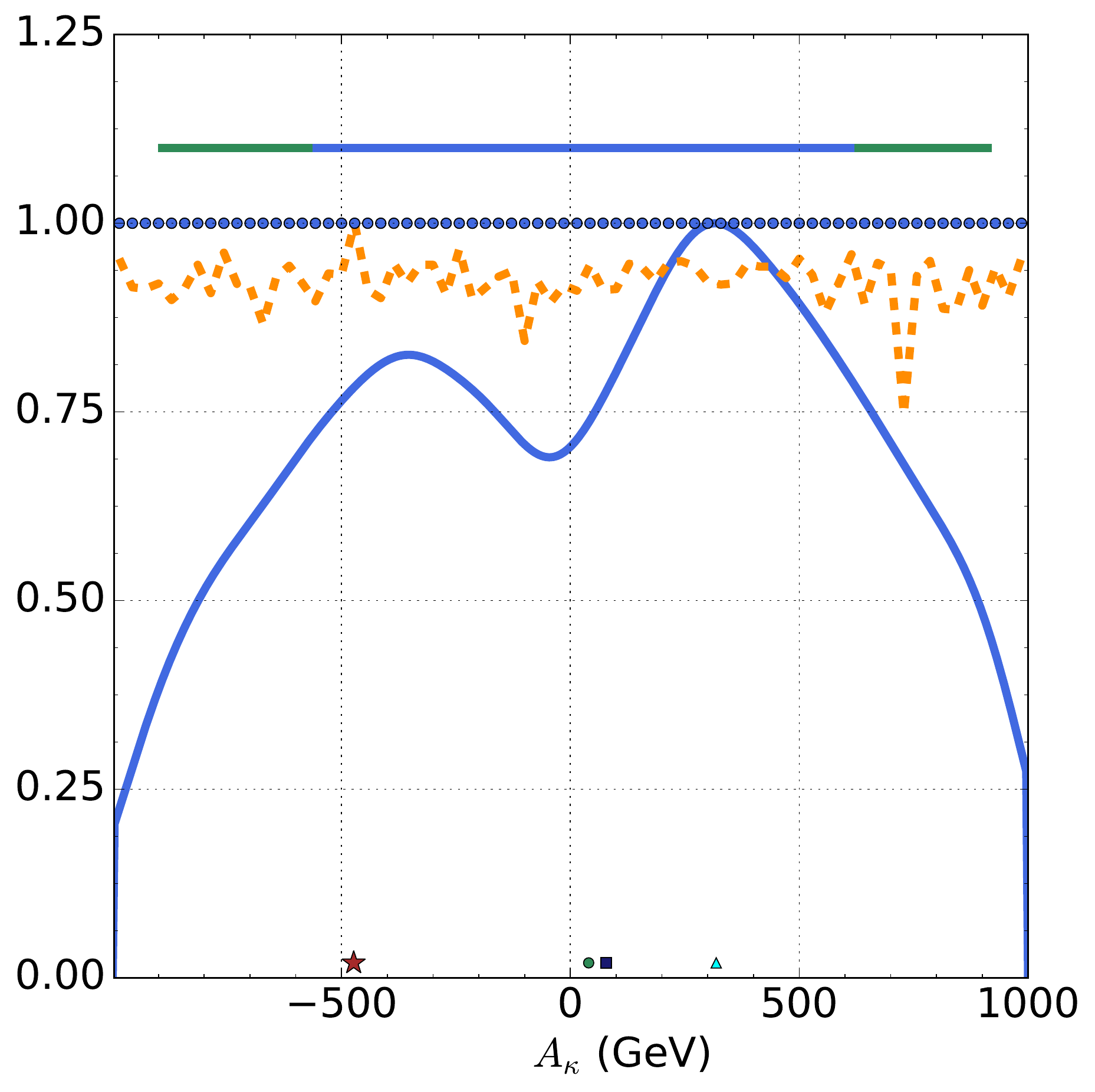}
		\includegraphics{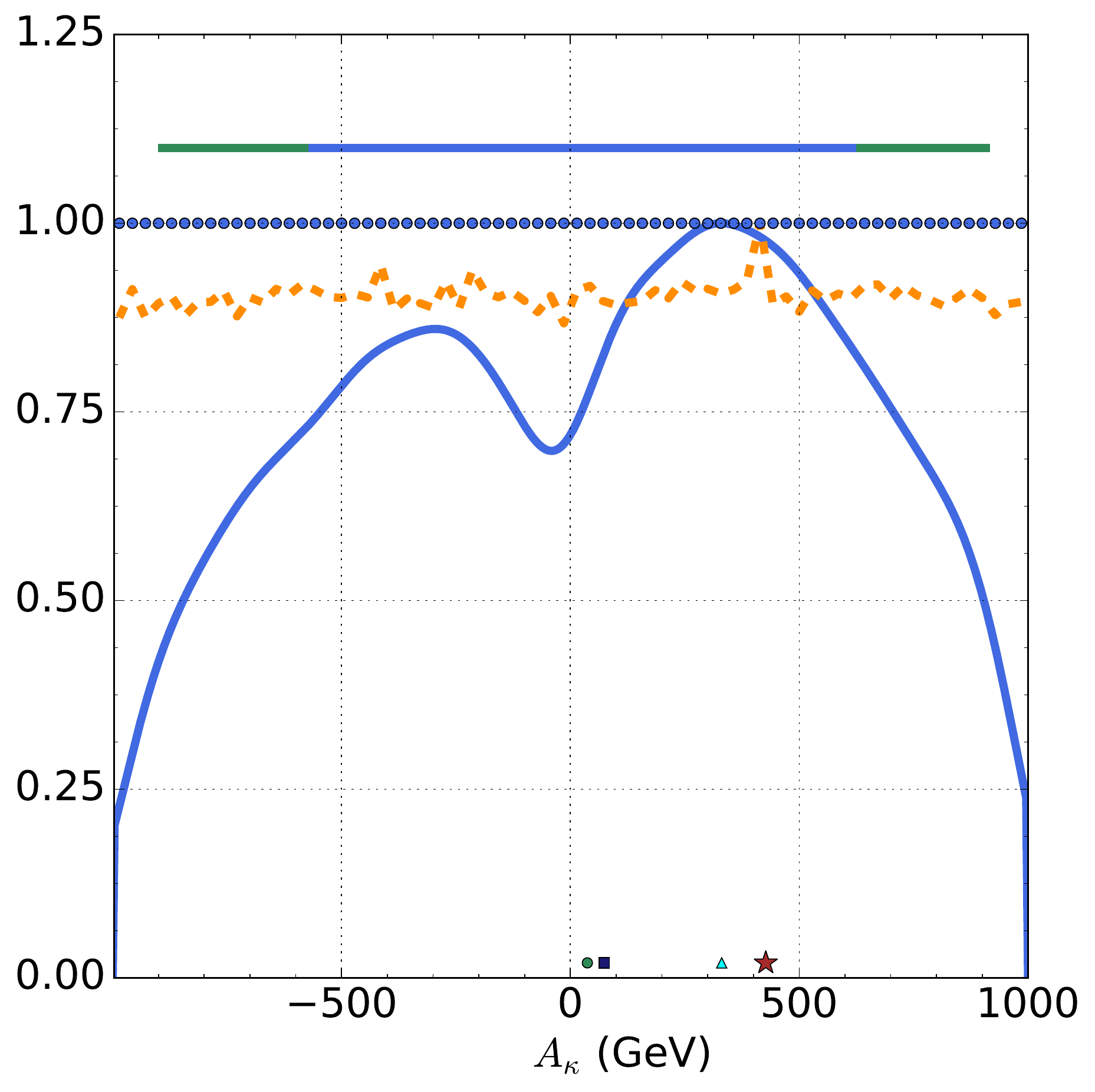}
		}
		\resizebox{0.58\textwidth}{!}{
		\includegraphics{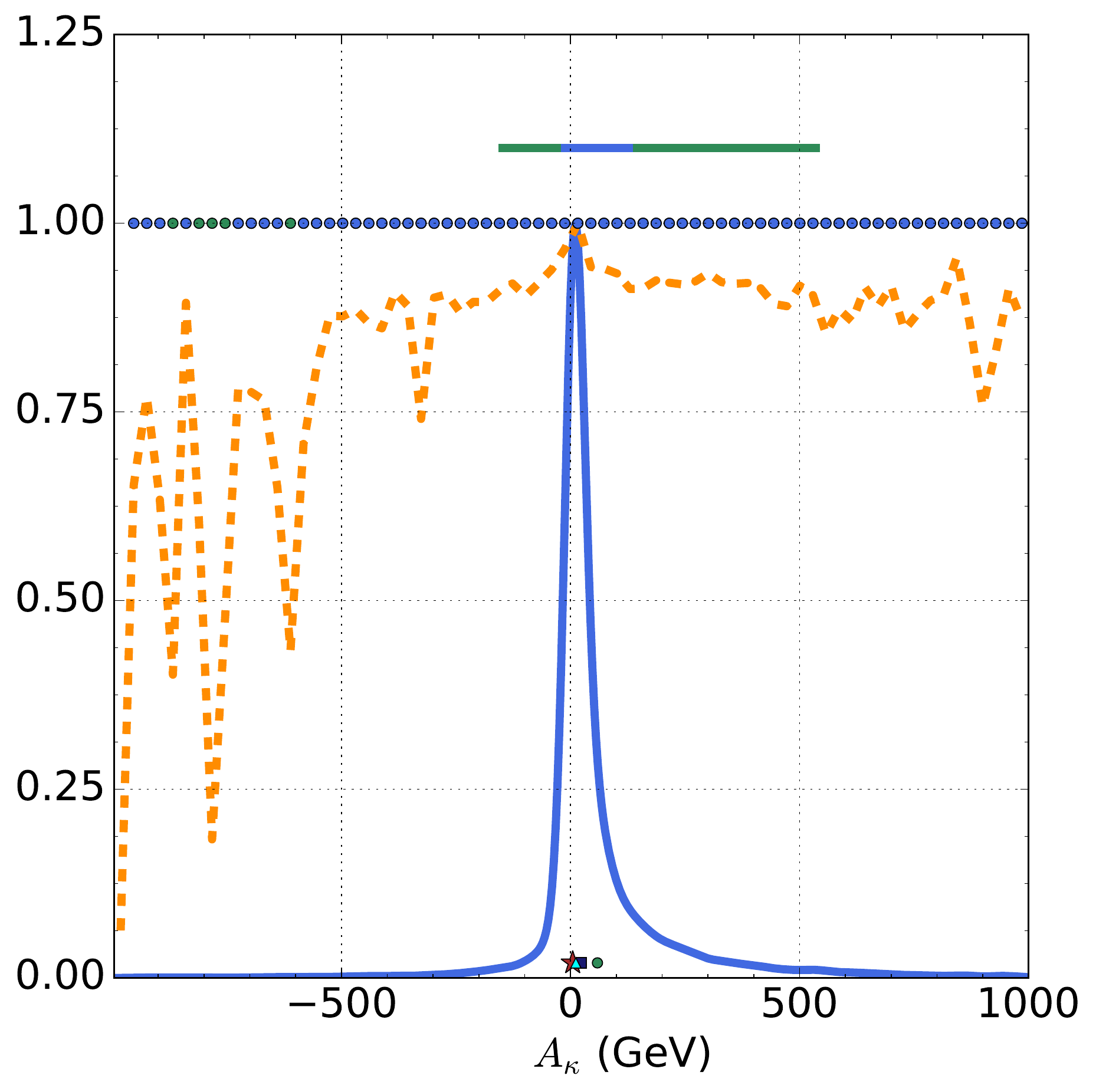}
		\includegraphics{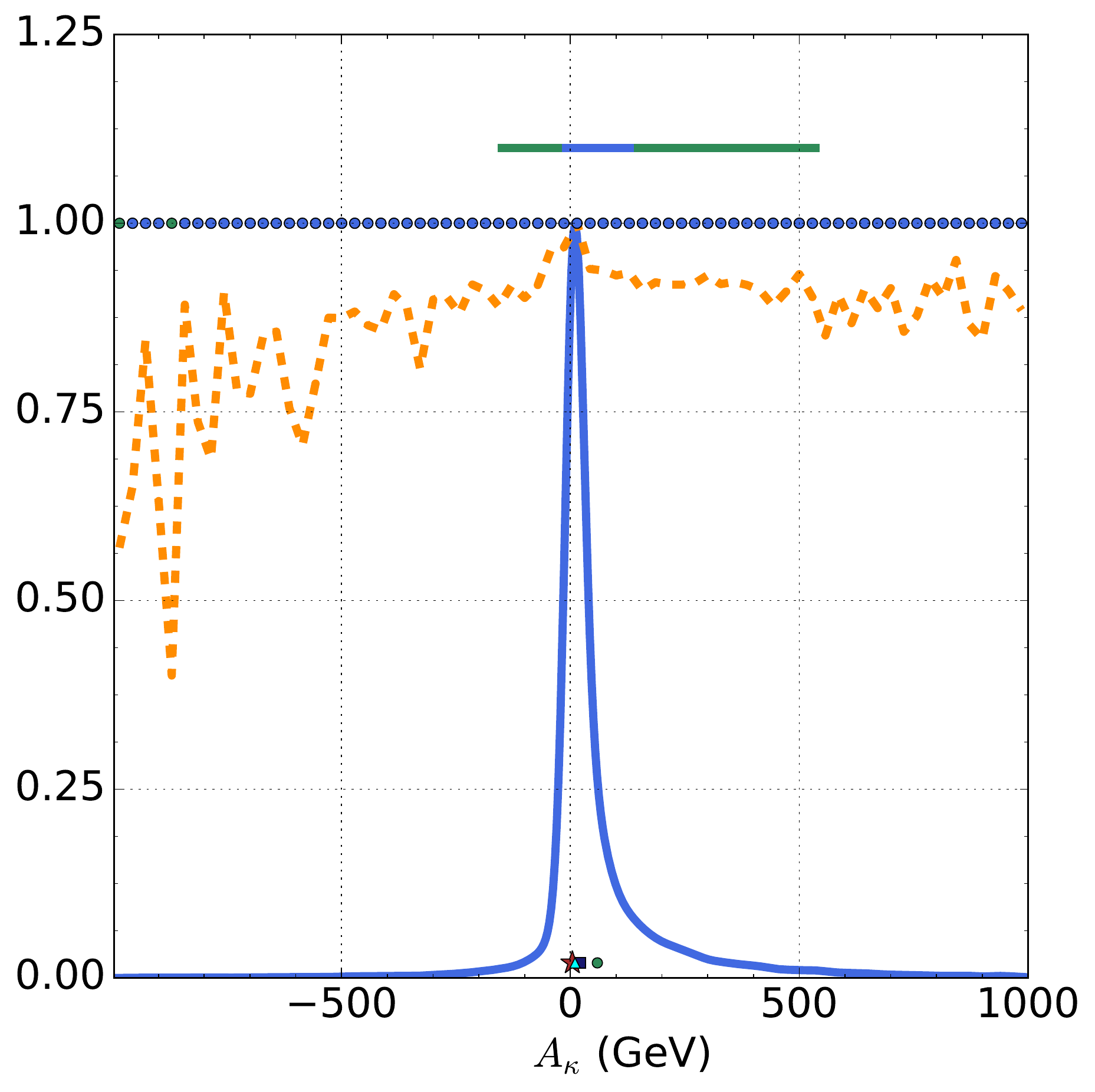}
		}
       \vspace{-0.4cm}

	\caption{The posterior PDF and PL of $A_\kappa$, which are obtained by the flat prior PDF (top panels) and the log prior PDF (bottom panels) with the total likelihood function given in Eq.(\ref{Likelihood}) of Section III. The left panels correspond to the setting {\it nlive} =2000, and the right panels correspond to {\it nlive} =6000.  \label{fig20}}
	\end{figure*}

\subsection{Cumulative effect of various measurements \\ on final results}

In this subsection, we perform two independent scans over the parameter space with the likelihood function given by
$\mathcal{L}=\mathcal{L}_{Higgs} $ and $\mathcal{L} =\mathcal{L}_{Higgs} \times  \mathcal{L}_{Br(B_s\rightarrow \mu^+\mu^-)} \times \mathcal{L}_{Br(B_s \rightarrow X_s\gamma)}$ respectively, and compare their predictions on the marginal PDFs of the input parameters with those presented in FIG.\ref{fig1}.

Our results are presented in FIG.\ref{fig14} and FIG.\ref{fig15} for the 1D marginal posterior PDFs and PLs of the parameter $\lambda$, $\kappa$, $\tan \beta$,
$A_\kappa$, $\mu$ and $A_t$. Since the features of the PDFs and PLs for the parameters $\lambda$, $\tan \beta$, $\mu$ and $A_t$ have been depicted in the text,
we in the following concentrate on those of $\kappa$ and $A_\kappa$. The main conclusions are
\begin{itemize}
\item As we mentioned before, the marginal distribution of $\kappa$ is roughly symmetric under the substitution $\kappa \to -\kappa$ if only the Higgs measurements are
considered. Especially when $\kappa$ approaches zero from either direction, the symmetry becomes more and more accurate (see the left panel of FIG.\ref{fig14}) since the terms which violate the symmetry are proportional to $\kappa$. Similarly, the marginal posterior PDF of $A_\kappa$ also exhibits a rough reflection symmetry if only the Higgs observables are considered (see the left panel in the first row of FIG.\ref{fig15}).

We point out that a moderately large $\kappa$ can suppress the mixing between the SM Higgs field and the singlet field,
which may explain why the $\kappa$ distribution is maximized at $\kappa \simeq 0.6$.

\item B physics measurements have little impact on the PDFs and PLs of $\kappa$ and $A_\kappa$. The reason is that the parameters $\kappa$ and $A_\kappa$ mainly affect the property of the singlet field, whose effects on $B \to X_s \gamma$ and $B_s \to \mu^+ \mu^-$ are usually negligible.

\item The marginal PDFs of $\kappa$ and $A_\kappa$ are modified significantly by the DM physics, and a negative $\kappa$ is now favored (see the right panels in the second row of FIG.\ref{fig14} and the first row of FIG.\ref{fig15}). This can be understood from the expression of the sneutrino mass in Section II, which shows that it depends on the parameter $\kappa$, and the CP-even state as the lightest sneutrino prefers a negative $\kappa$.
\end{itemize}

In Table \ref{table4}, we summarize the dependence of these 1D marginal PDFs and PLs on different experimental measurements. This table indicates that all the PDFs except that of $\mu$ are affected greatly by the Higgs data for $A_\lambda = 2 {\rm TeV}$, while the PLs (except that of $A_t$) usually have a weak dependence on the Higgs data.

\section{Impact of different prior PDFs \\ on posterior PDFs}

In this section, we study the difference of the posterior PDFs induced by the choices of the flat prior PDF adopted in the text and the log prior PDF where $\lambda$, $\kappa$ and $\lambda_\nu$ are flat distributed and the other input parameters are log distributed. From the discussion in previous section, we learn that the Higgs data play an important role in determining most of the posterior PDFs. Noting that the calculation of the $\mathcal{L}_{Higgs} $ is much faster than that of the total likelihood function in Eq.(\ref{Likelihood}) of Section III, we first consider the scan in Eq.(\ref{scan-ranges1}) of Appendix A with $\mathcal{L} = \mathcal{L}_{Higgs} $ as an example to illustrate the features of the difference. We remind that even after the simplification, the sampling process is still time consuming for our computer clusters when we take a large value of the {\it nlive}, which is needed to test the stability of the results when more and more samplings are considered. We then compare the results calculated by the total likelihood function with different choices of the prior PDFs and also with different settings of  {\it nlive}.

\subsection{Posterior PDFs from different prior PDFs under Higgs data: a comparative study}

   \begin{figure*}[htbp]
		\centering
		\resizebox{0.58\textwidth}{!}{
		\includegraphics{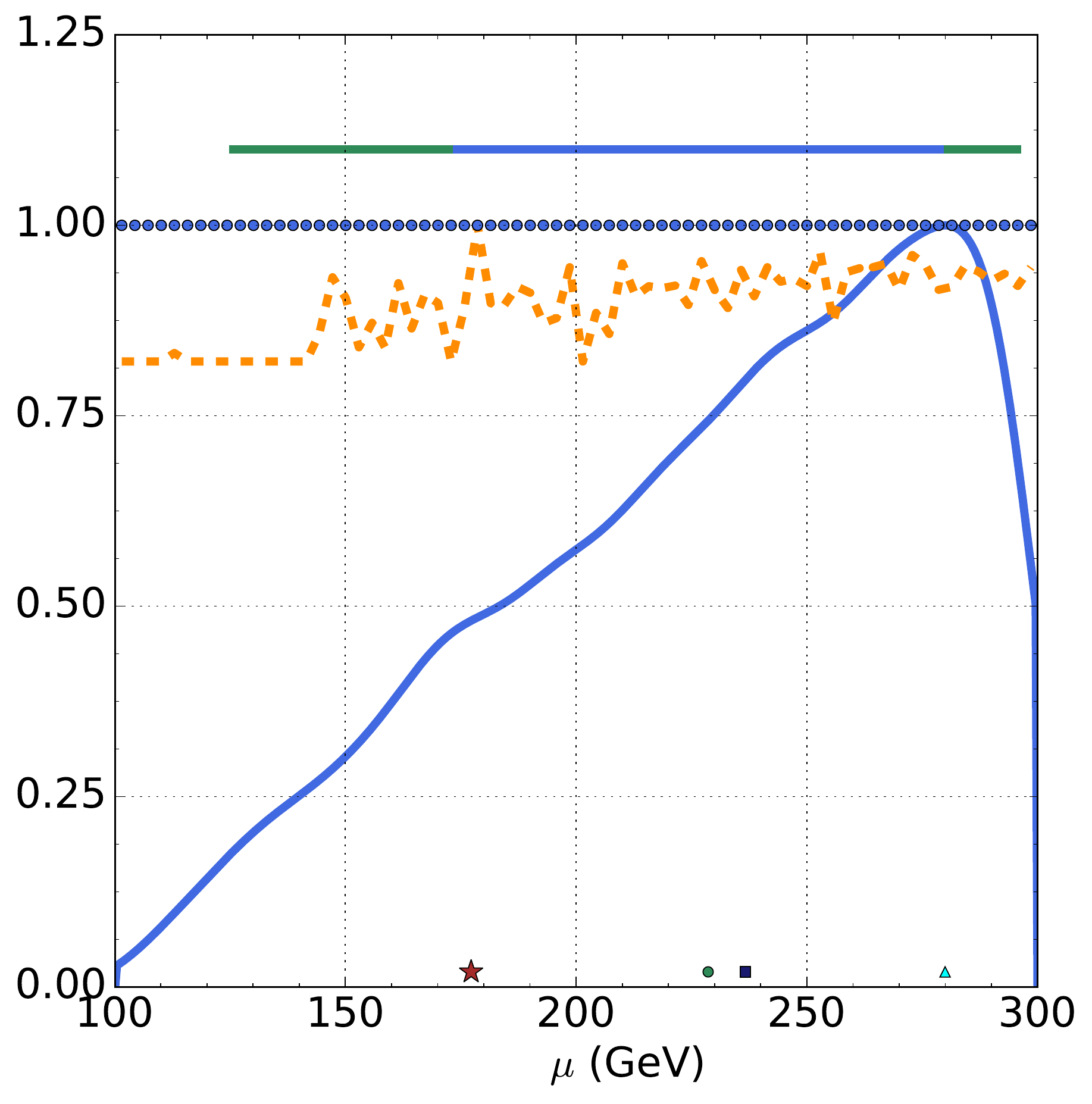}
		\includegraphics{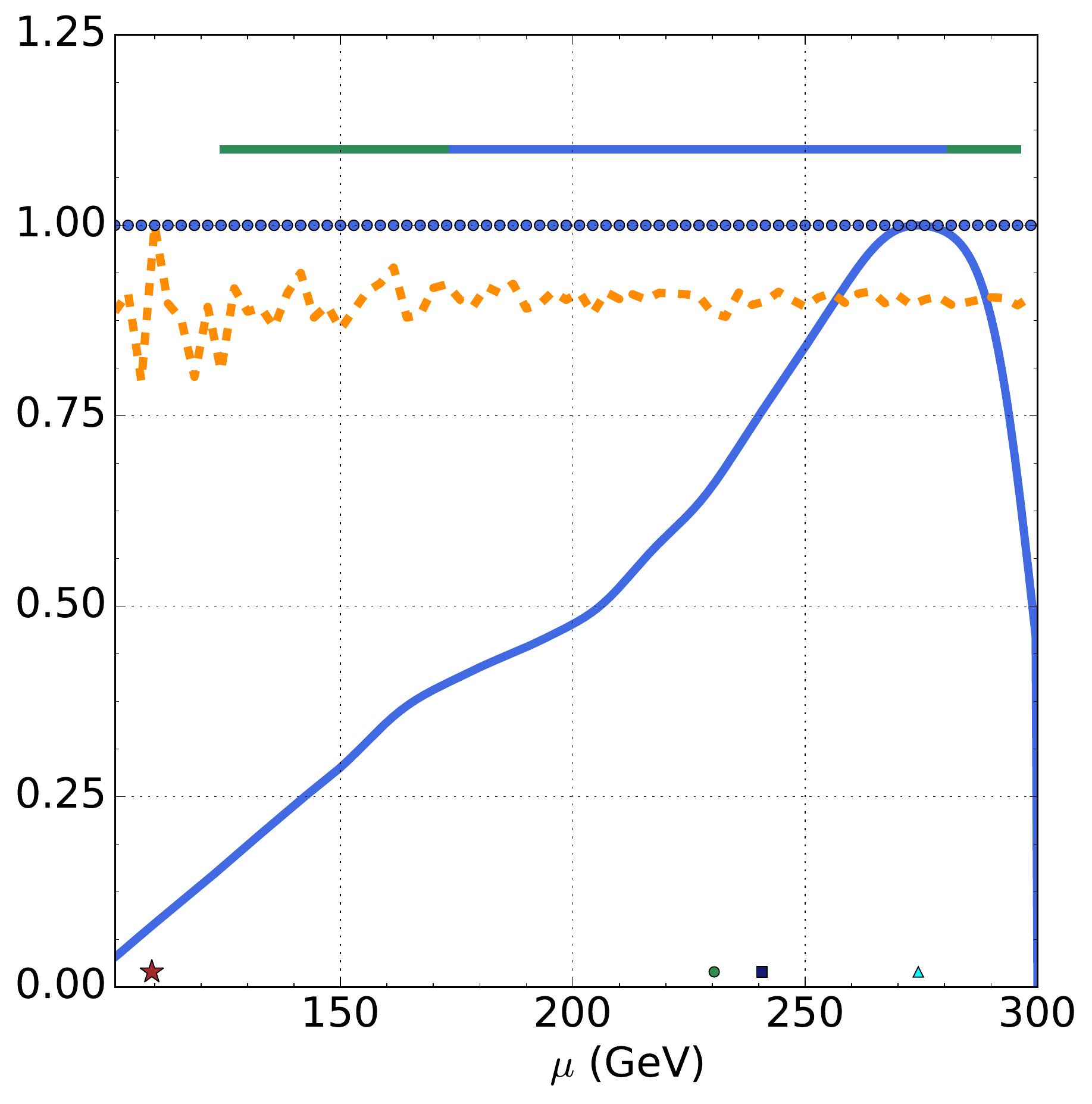}
		}
		\resizebox{0.58\textwidth}{!}{
		\includegraphics{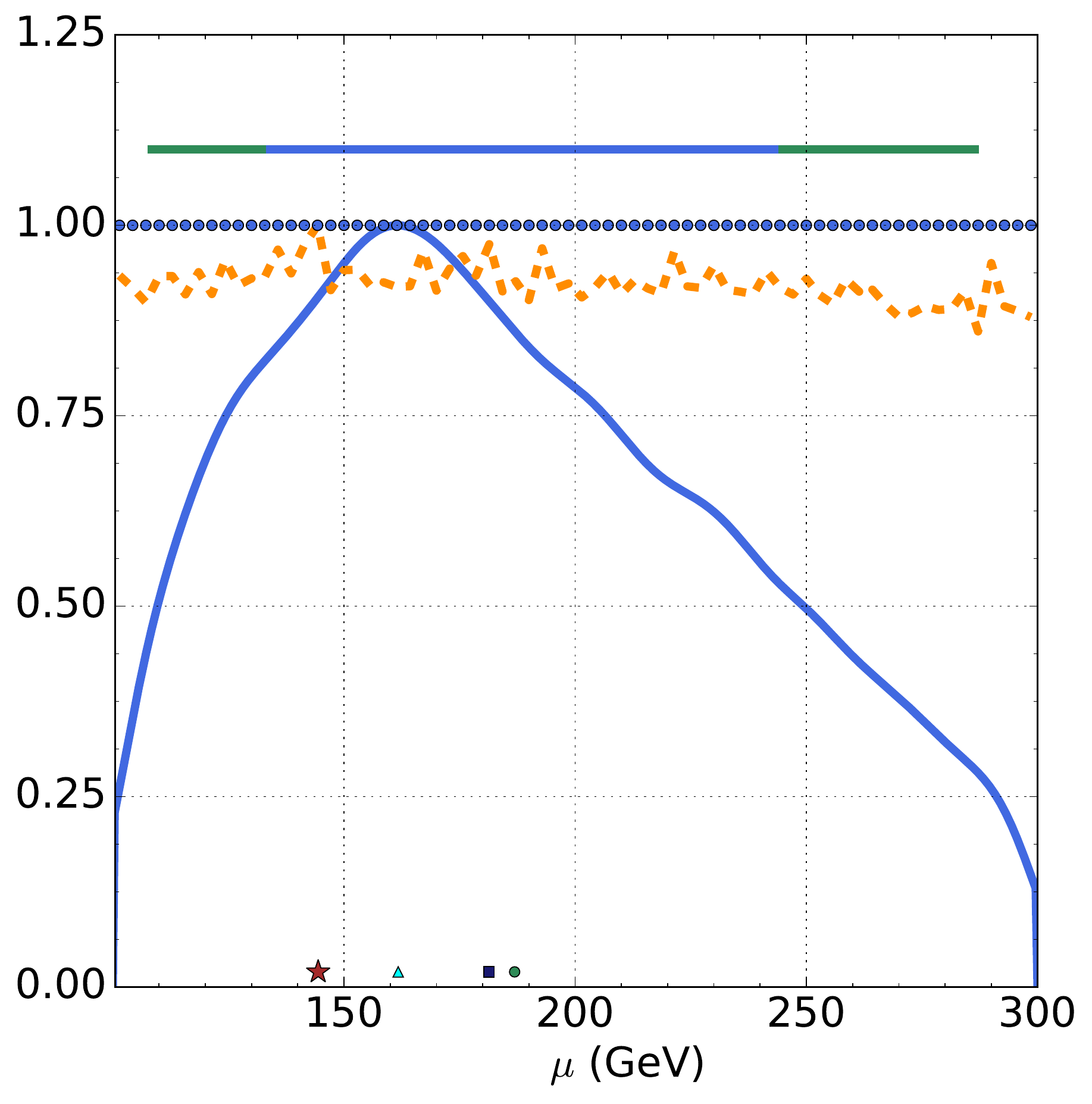}
		\includegraphics{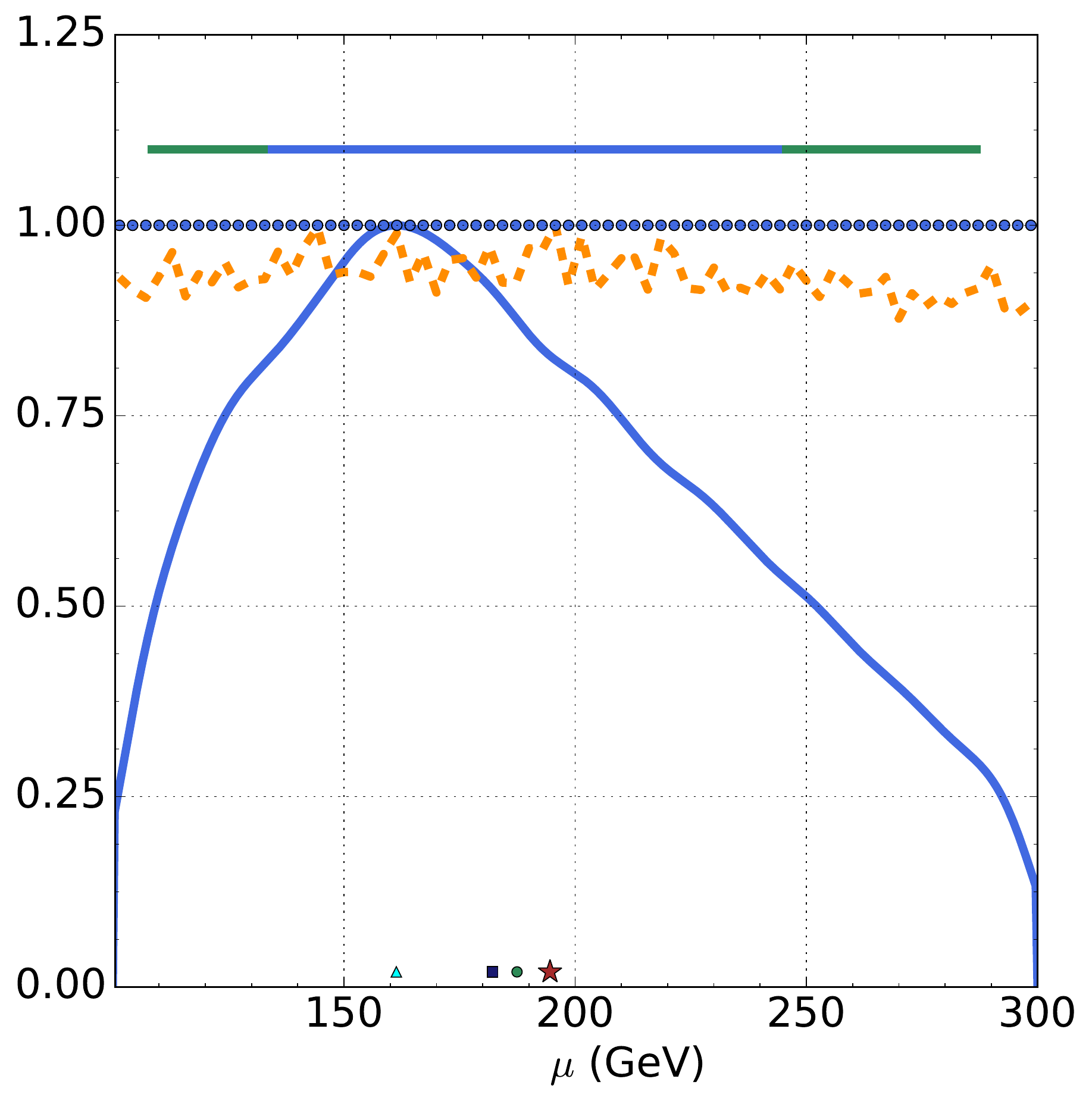}
		}
       \vspace{-0.4cm}

	\caption{Same as FIG.\ref{fig20}, but for the distribution of $\mu$. \label{fig21}}
	\end{figure*}	

   \begin{figure*}[htbp]
		\centering
		\resizebox{0.7\textwidth}{!}{
		\includegraphics{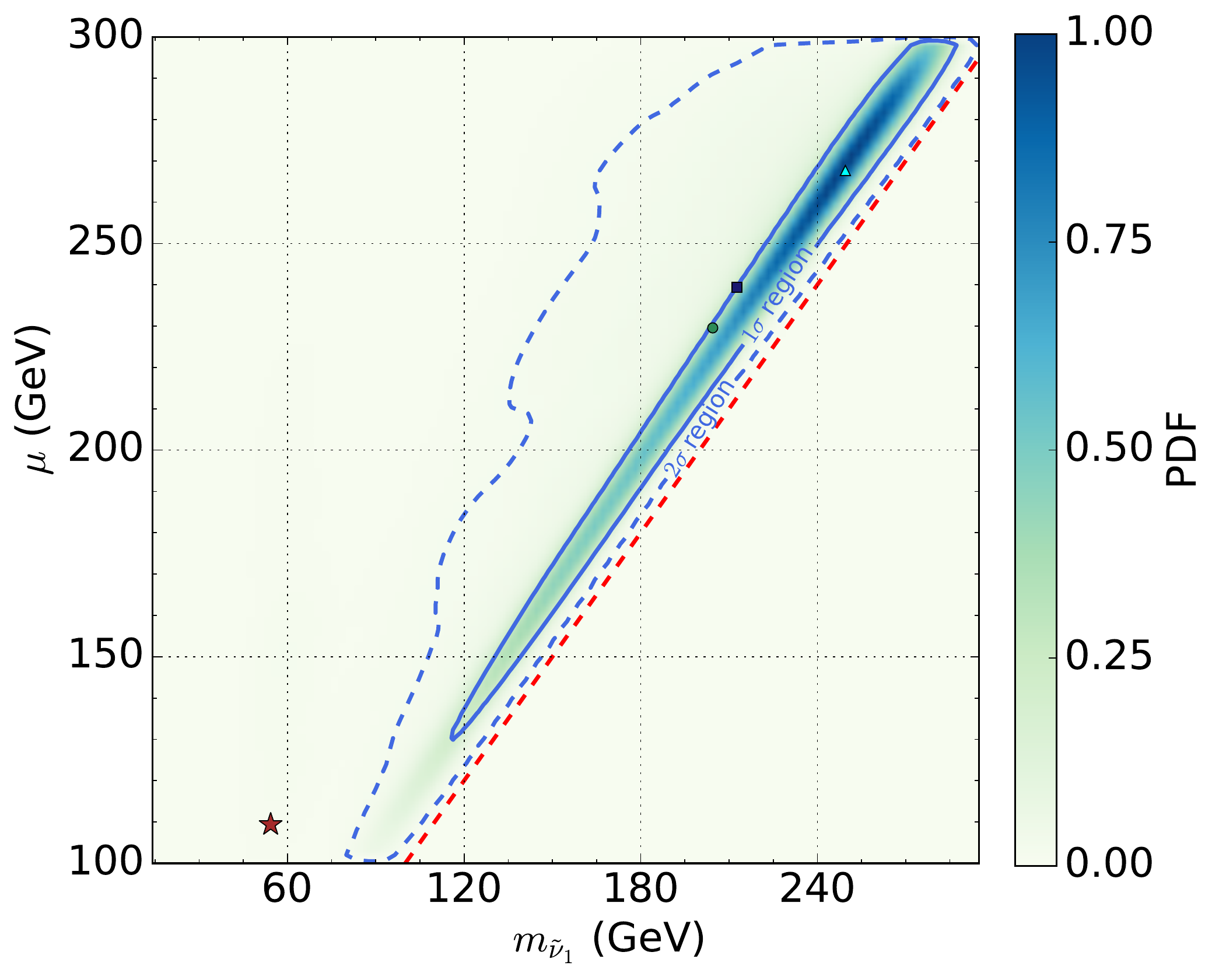}
		\includegraphics{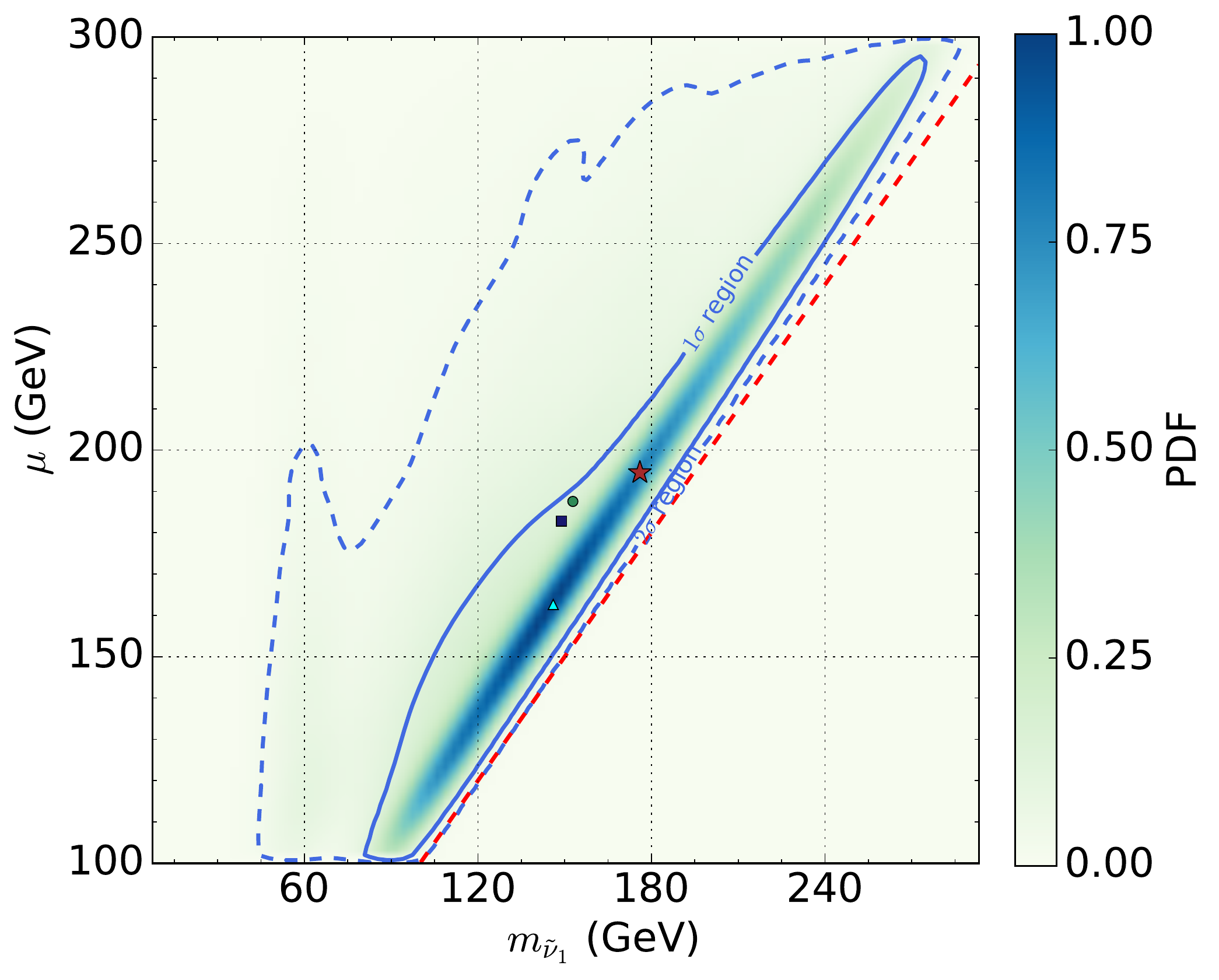}
		}
       \vspace{-0.4cm}

	\caption{Comparison of the marginal posterior PDF on $\mu-m_{\tilde{\nu}_1}$ plane obtained by the flat prior PDF (left panel) and the log prior PDF (right) with the total likelihood function. We have set {\it nlive} =6000 to get these panels.  \label{fig22}}
	\end{figure*}

This subsection is devoted to explore the difference of the posterior PDFs obtained by the flat and log prior distributions. We will show that the log prior PDF overemphasizes the region characterized by a low value of $|A_\kappa|$ and consequently distorts
the marginal PDFs of $A_\kappa$ and $\mu$. Except those that we clarify explicitly, all the results presented in this subsection are based on the scan in Eq.(\ref{scan-ranges1}) of Appendix A with $\mathcal{L} = \mathcal{L}_{Higgs} $.

From FIG.\ref{fig16} to Fig.\ref{fig18}, we present the posterior PDFs of the parameter $\lambda$, $\kappa$, $\tan \beta$, $A_\kappa$, $\mu$ and $A_t$ respectively.
For each parameter, the three panels in the upper row are obtained by the flat prior PDF with ${\it nlive} = 2000, 6000, 24000$ respectively, and those in the lower row are obtained by the log prior distribution. From these figures, one can get following conclusions:
\begin{itemize}
\item The results obtained by the log distribution are rather stable when one alters the settings of the {\it nlive}, and by contrast those obtained by the flat distribution change sizably with the increase of  the {\it nlive}.

\item The largest difference induced by the two choices of the prior PDFs comes from the marginal posterior PDFs of $A_\kappa$ in the third and fourth rows of  FIG.\ref{fig17}. Explicitly speaking, the prediction of the log prior distribution is sharply peaked in low $|A_\kappa|$ region, while the prediction of the flat distribution shows a roughly uniform distribution over a broad region, and it is moderately unstable for insufficient samplings, which affects the stability of the other marginal PDFs with the change of {\it nlive} since the PDFs are correlated. Moreover, with the increase of {\it nlive}, the peak of the $A_\kappa$ posterior distribution moves in a slow way to lower $|A_\kappa|$ region for the flat prior PDF, which seems to coincide with the log prediction.

\item The marginal posterior PDFs of the parameter $\mu$ shown in the first and second rows of FIG.\ref{fig18} are also significantly different. For that obtained by the log prior PDF, a low value of $\mu$ is significantly preferred, and the peak locates at $\mu \simeq 120 {\rm GeV}$. By contrast the marginal PDF from the flat prior PDF is more or less a constant over a broad range of $\mu$, and with the increase of {\it nlive}, it is enhanced in low $\mu$ region.
\item Both the prior PDFs predict  the peak of the marginal PDF for $\tan \beta$ at $\tan \beta \simeq 13$, which is shown in the first two rows in FIG.\ref{fig17}, but for the region of $\tan \beta > 16$, the difference of the marginal posterior PDFs becomes apparent.
\item The $\lambda$ distributions predicted by the two prior distributions agree well with each other (see the first two rows in FIG.\ref{fig16}), and so is the $A_t$ distributions (see the last two rows in FIG.\ref{fig18}). Especially, the marginal posterior PDFs obtained by the flat PDF approach steadily to corresponding predictions of the log prior PDF with the increase of {\it nlive}.

\item For the marginal PDF of $\kappa$ shown in the third and fourth rows of FIG.\ref{fig16}, both the prior PDFs can produce the symmetry mentioned in Appendix A. Especially
   the symmetry becomes more accurate in low $|\kappa|$ region.
\end{itemize}

   \begin{figure*}[thbp]
		\centering
		\resizebox{0.58\textwidth}{!}{
		\includegraphics{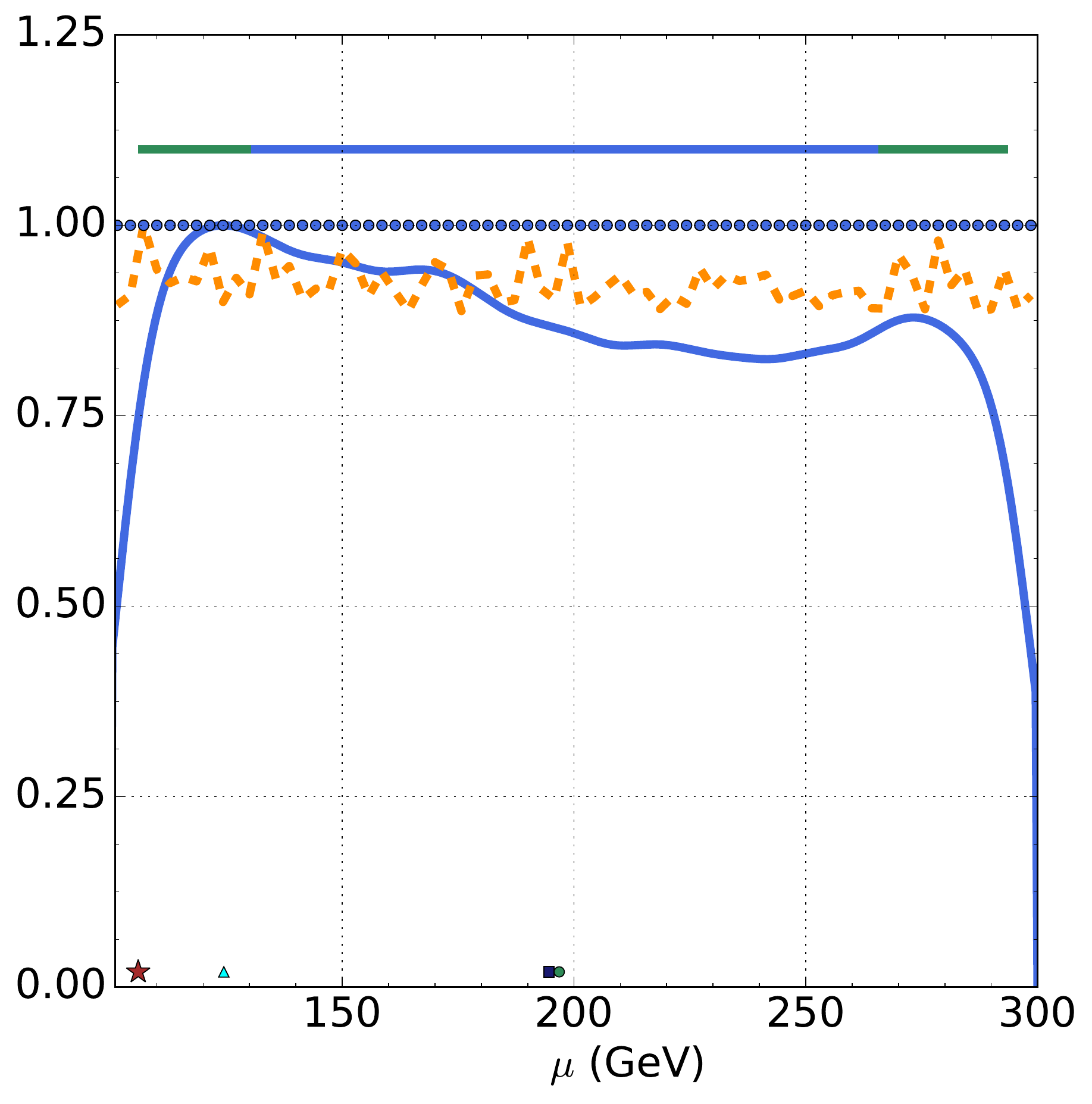}
		\includegraphics{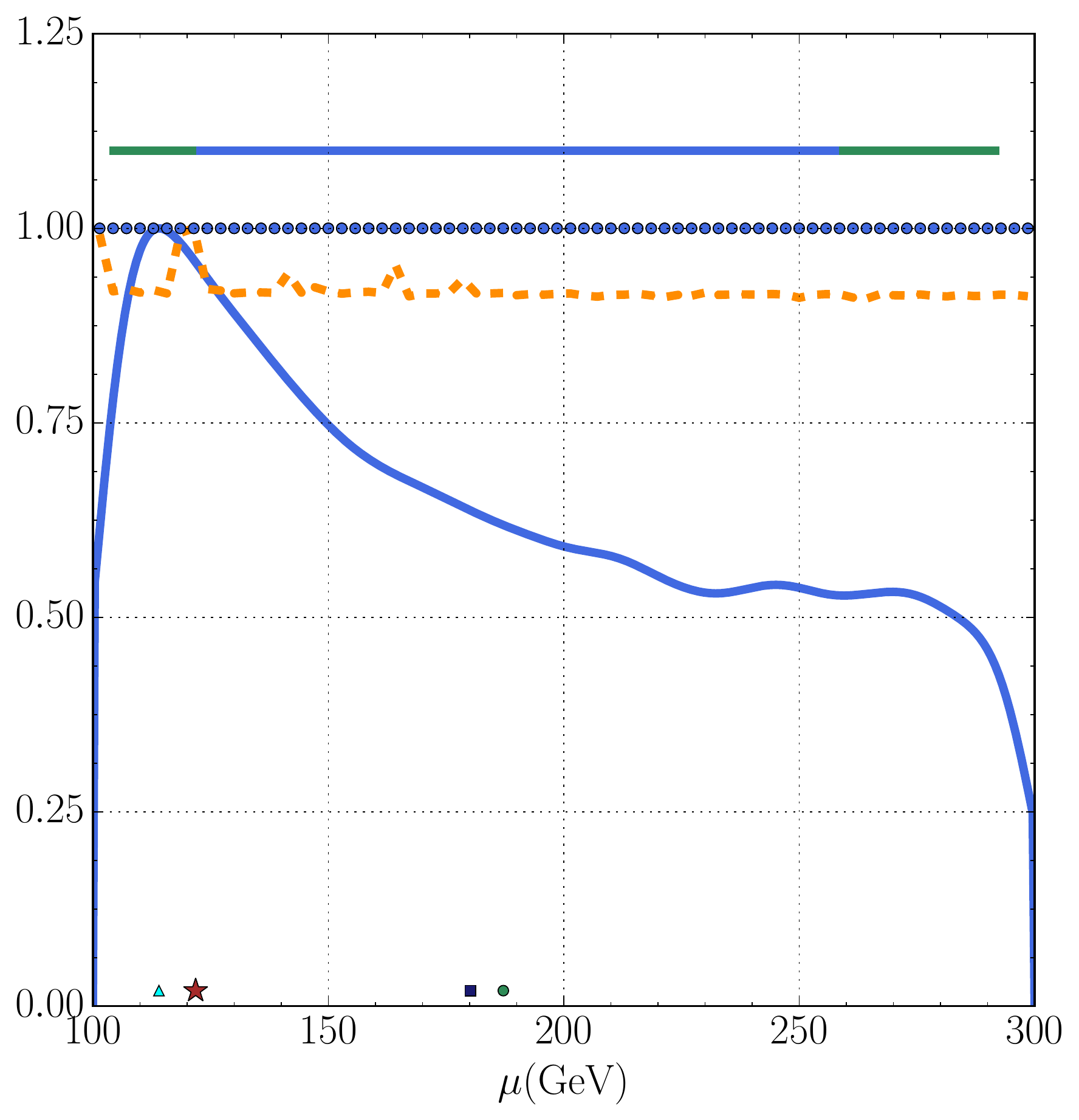}
		}

       \vspace{-0.4cm}

	\caption{The marginal PDF and PL of the parameter $\mu$ obtained by the flat prior PDF, the likelihood function $\mathcal{L} = \mathcal{L}_{Higgs} $ and {\it nlive} = 24000. The left panel (the right panel) is based on $A_\lambda = 2 {\rm TeV}$ ($A_\lambda = 10 {\rm TeV}$) and the parameter space presented in Eq.(\ref{scan-ranges1}) of Appendix A with (without) $-200 {\rm GeV} \leq A_\kappa \leq 200 {\rm GeV}$ further required.  \label{fig23}}
	\end{figure*}	

   \begin{figure*}[thbp]
		\includegraphics[height=5cm]{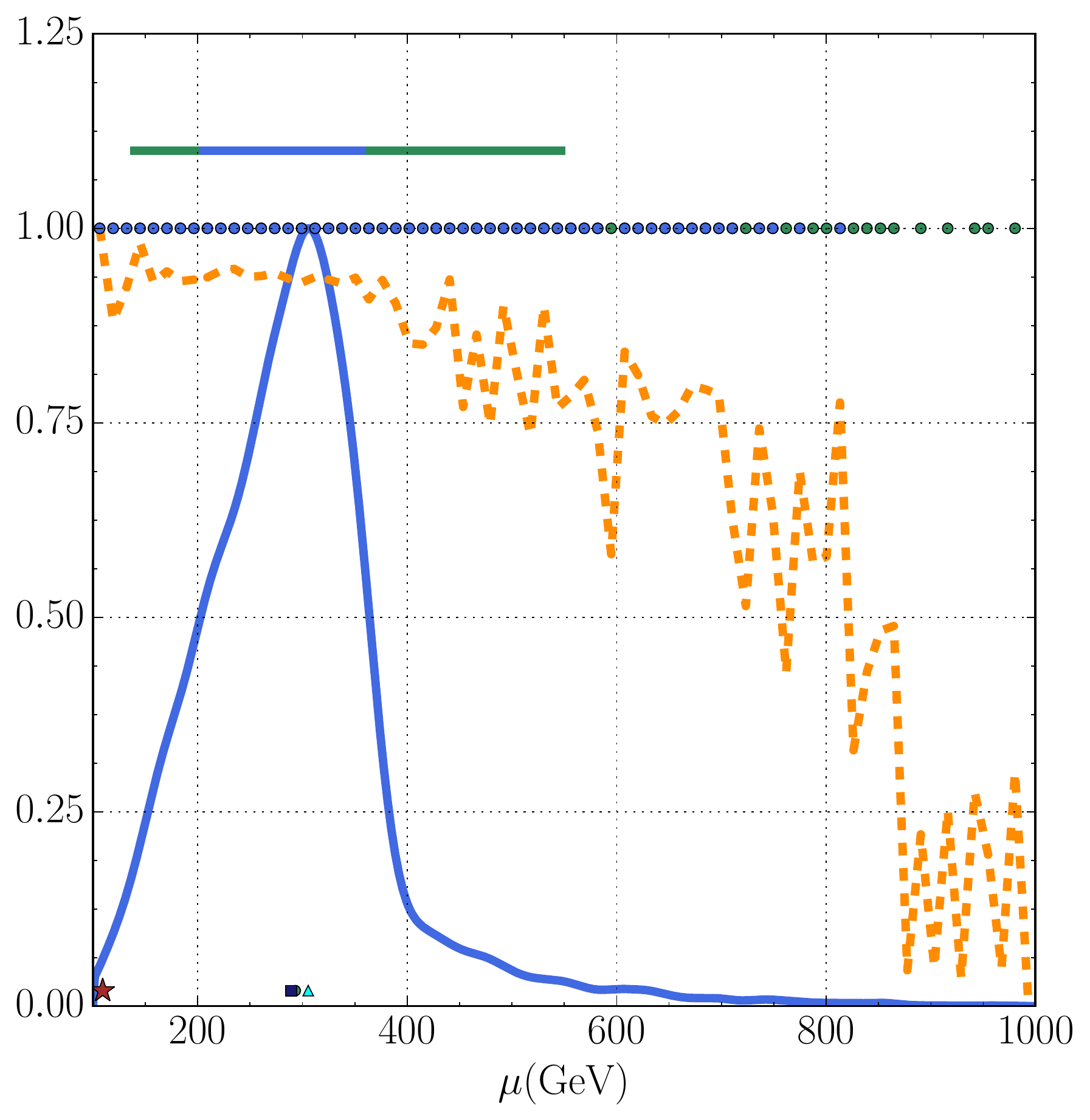}
		\includegraphics[height=5cm]{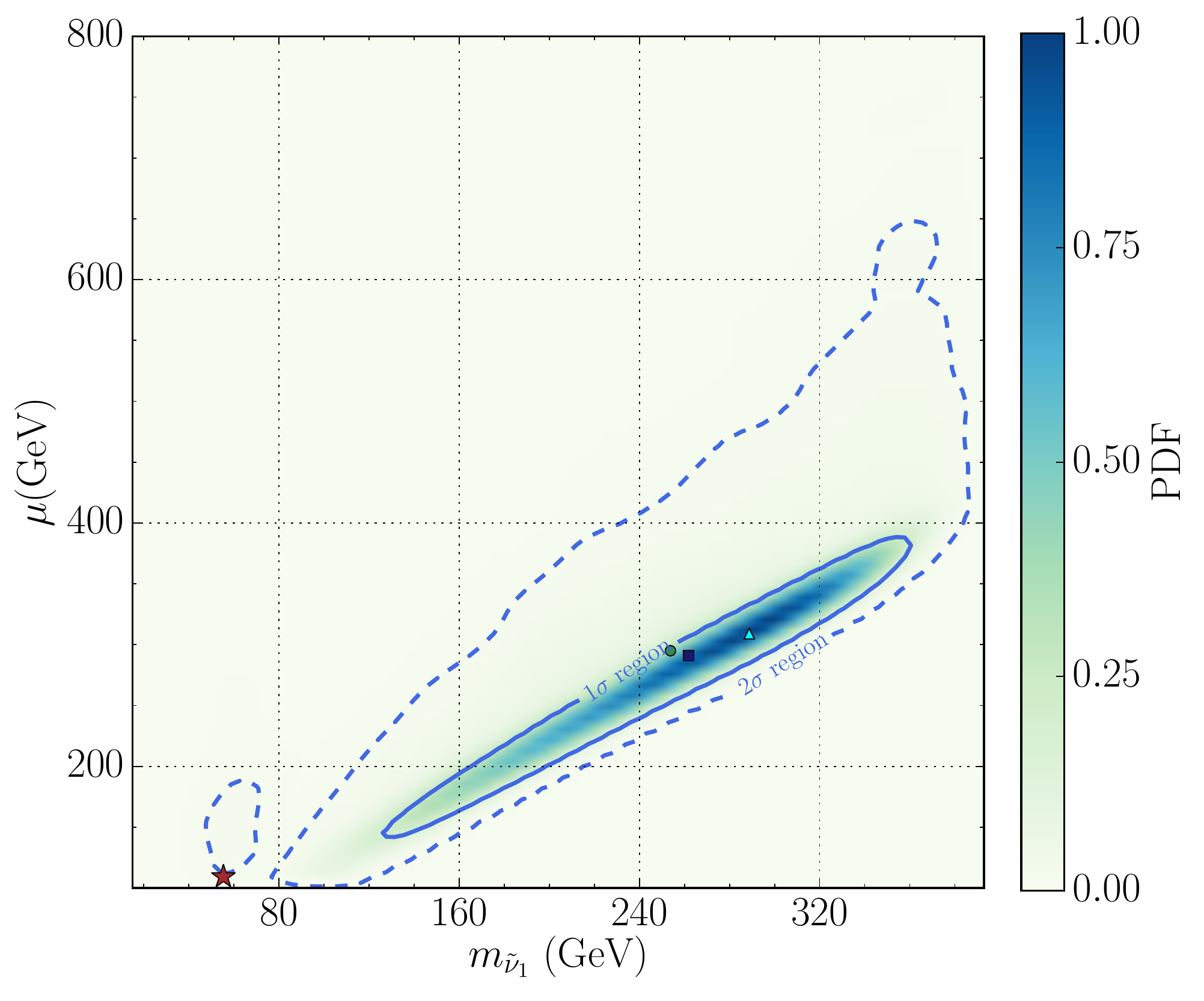}

       \vspace{-0.4cm}

	\caption{The 1D marginal PDF and PL of the parameter $\mu$ (left panel) and 2D marginal PDF on $\mu-\tilde{\nu}_1$ plane (right panel) obtained by the flat prior PDF, the total likelihood function, and {\it nlive} =6000. In getting this figure, we extend the upper bound of $\mu$ in Eq.(\ref{scan-ranges}) of Section III.A to $1 {\rm TeV}$.  \label{fig24}}
	\end{figure*}	

        \begin{figure*}[htbp]
		\centering
		\resizebox{0.8\textwidth}{!}{
		\includegraphics{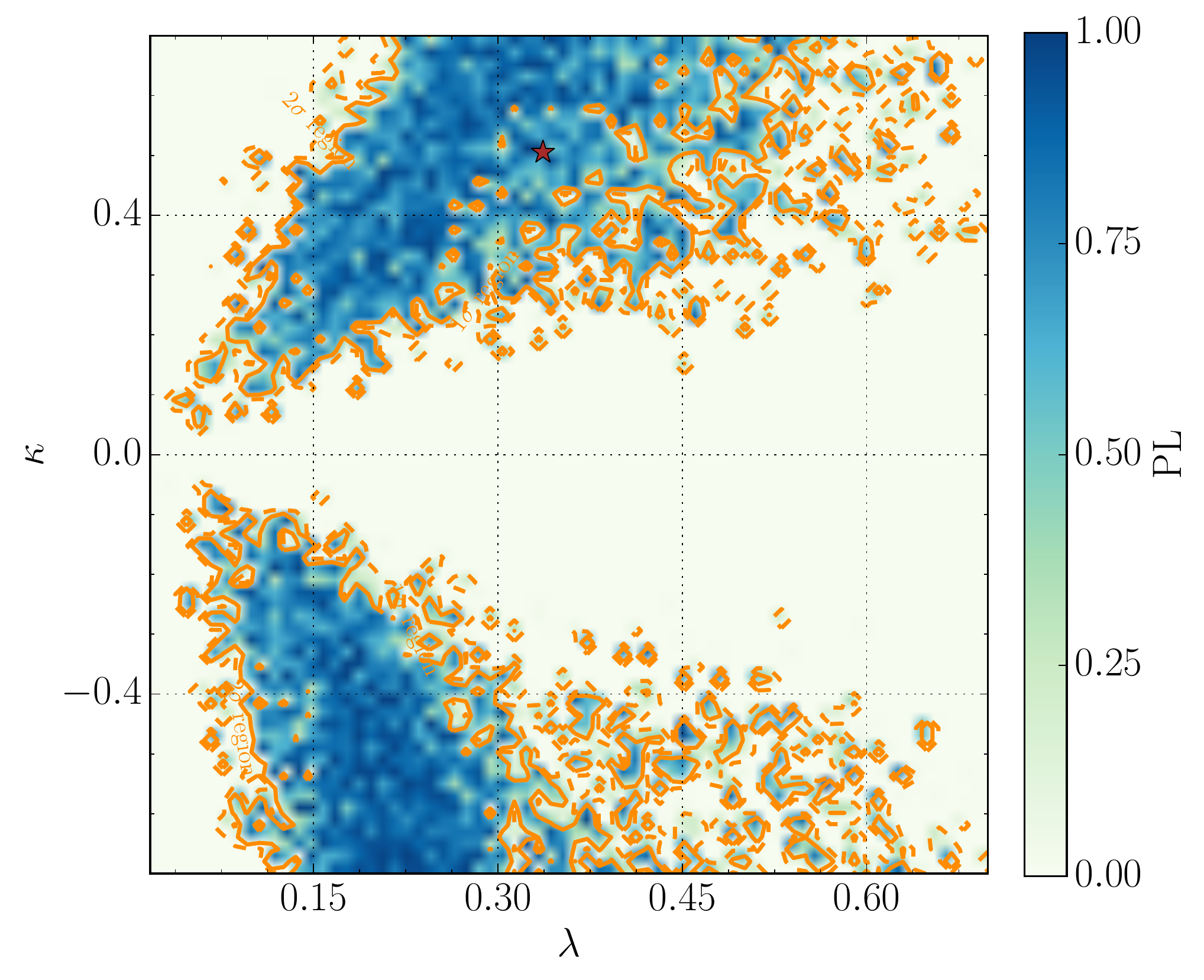}
		\includegraphics{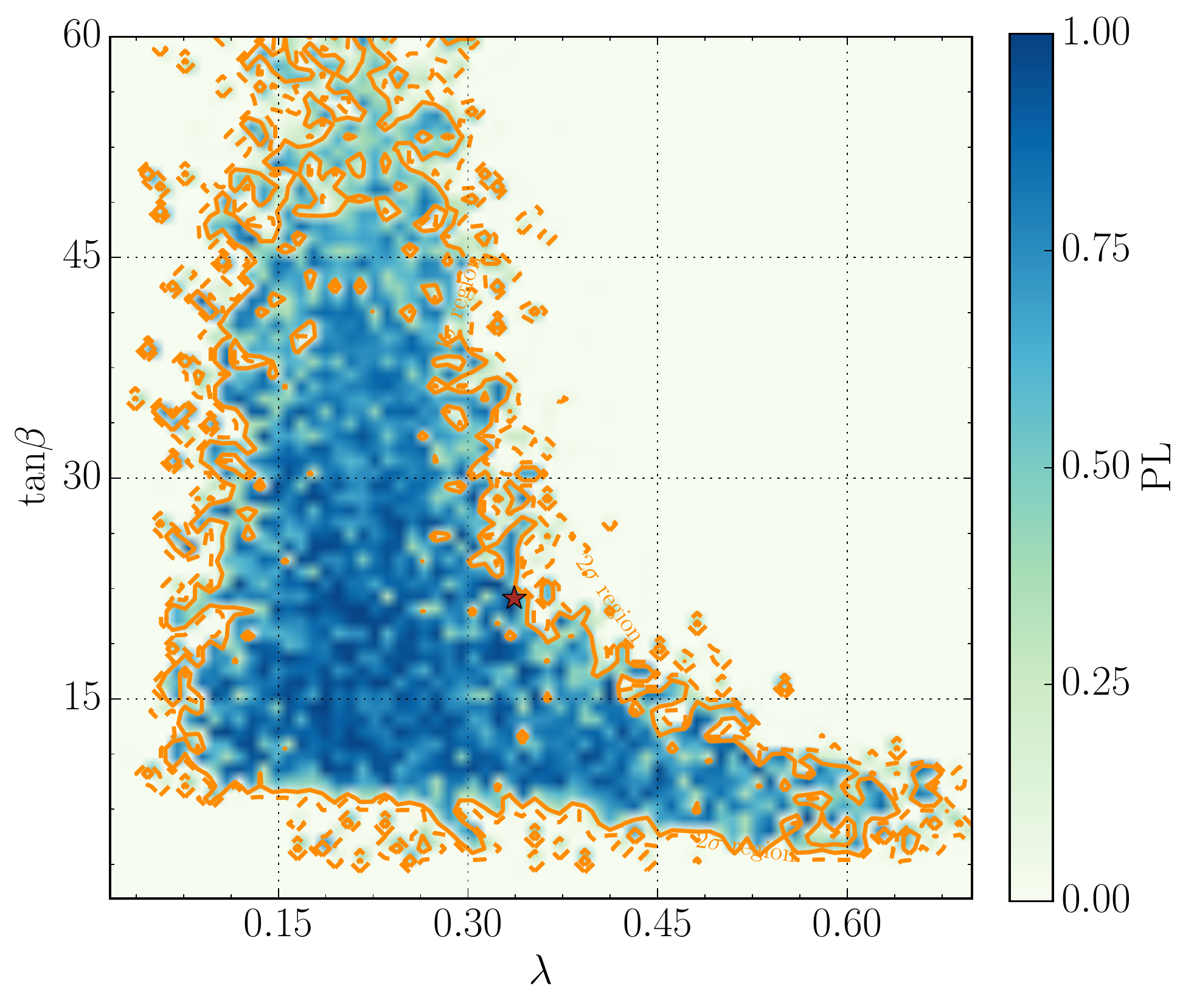}
		\includegraphics{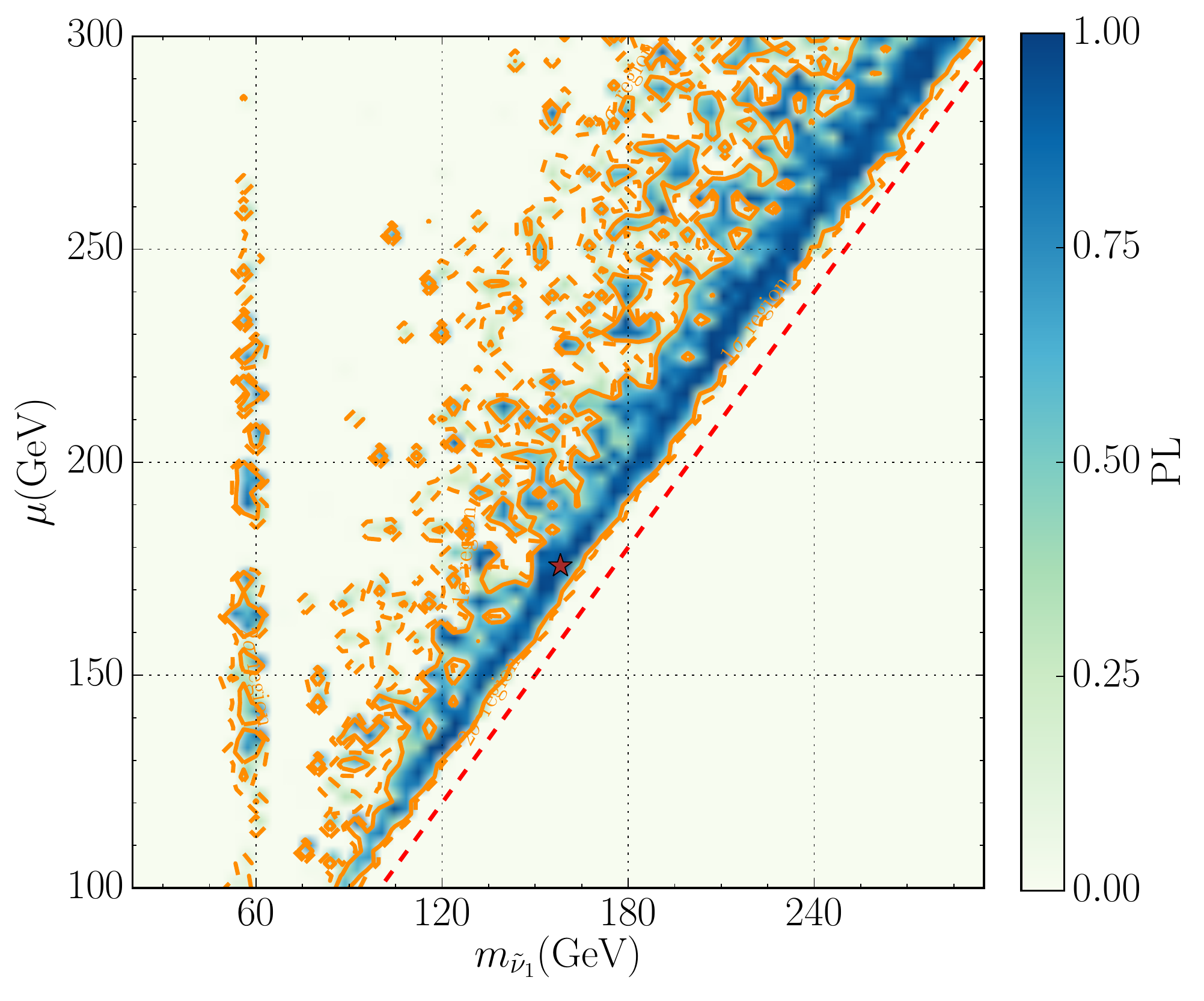}
		}
		\resizebox{0.8\textwidth}{!}{
		\includegraphics{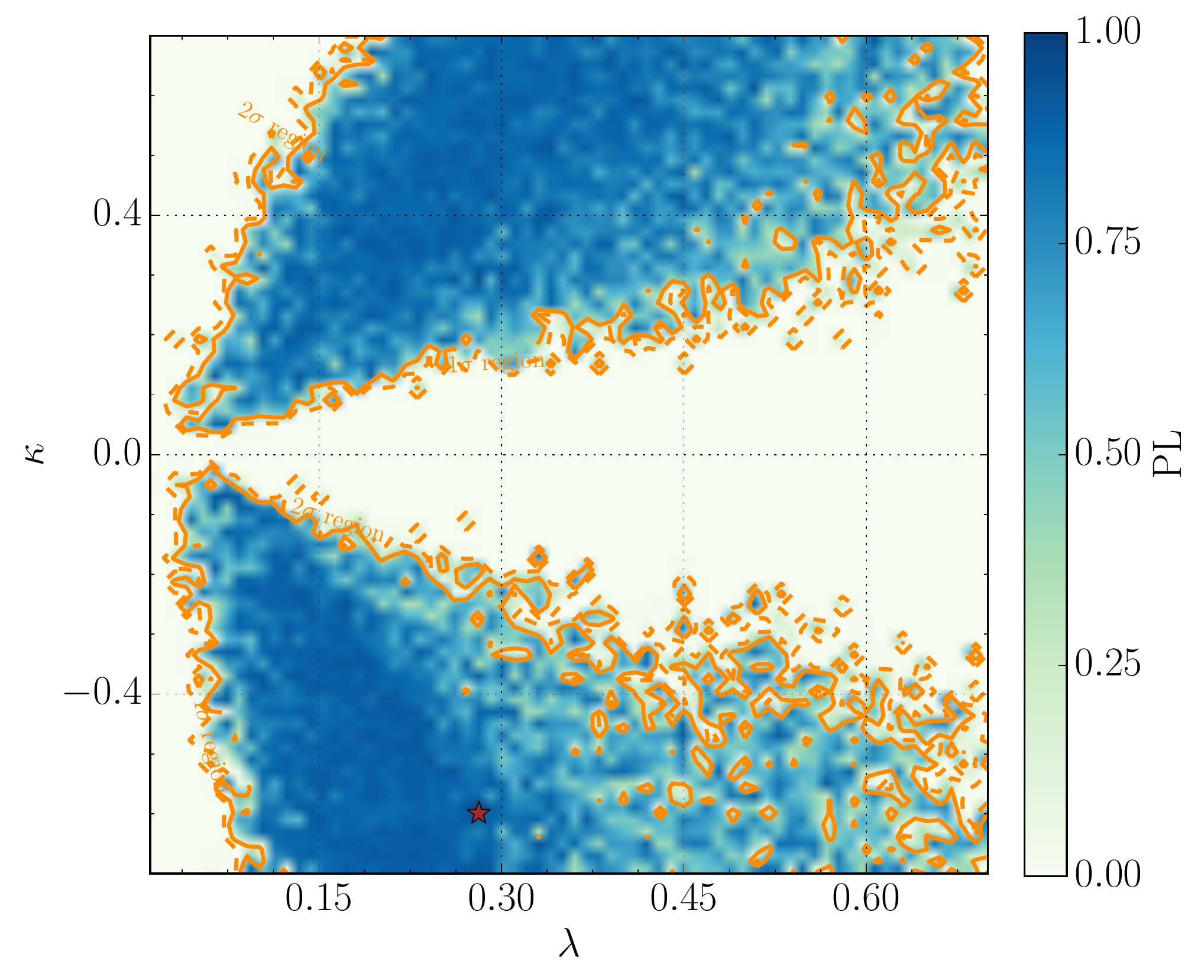}
		\includegraphics{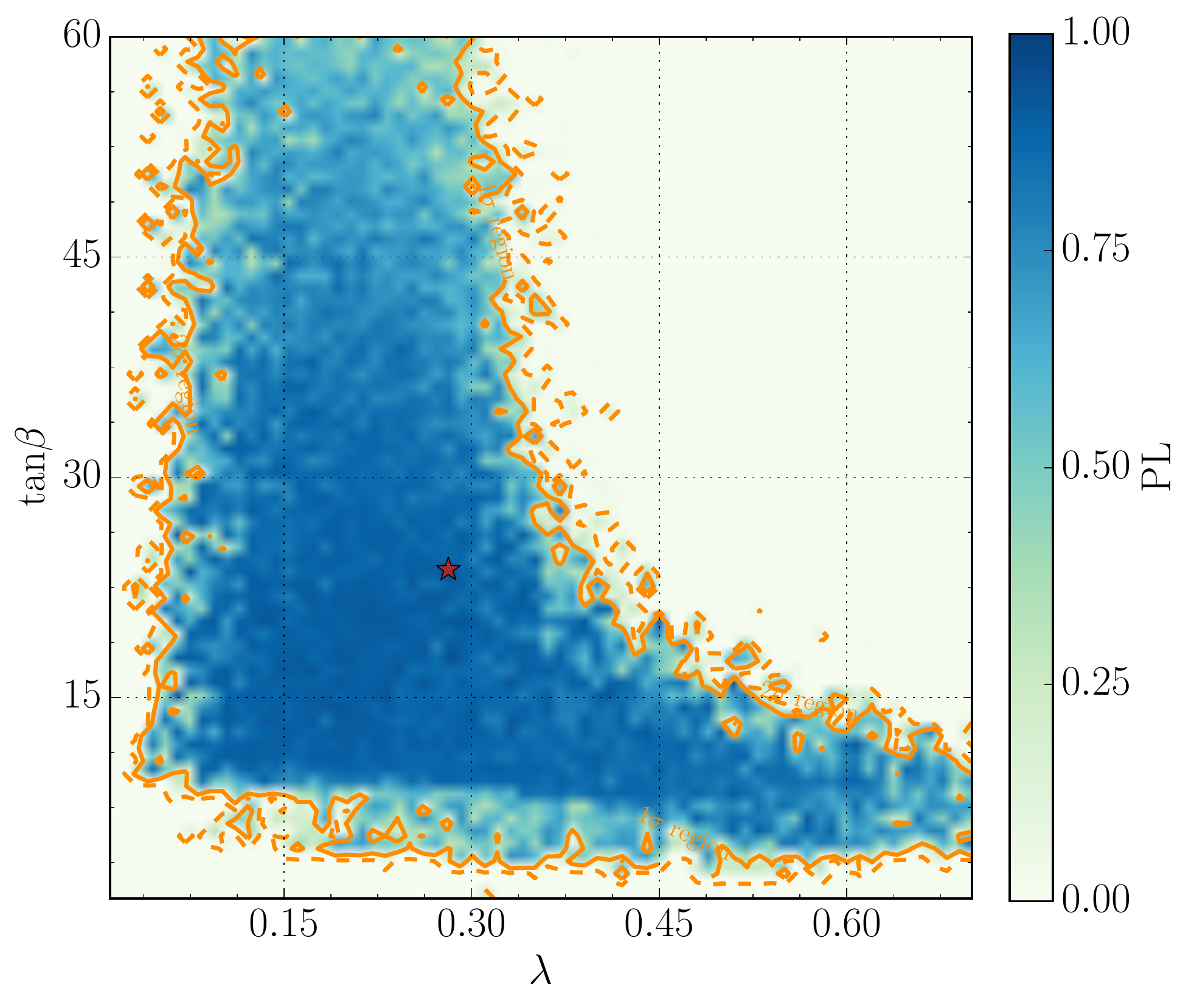}
		\includegraphics{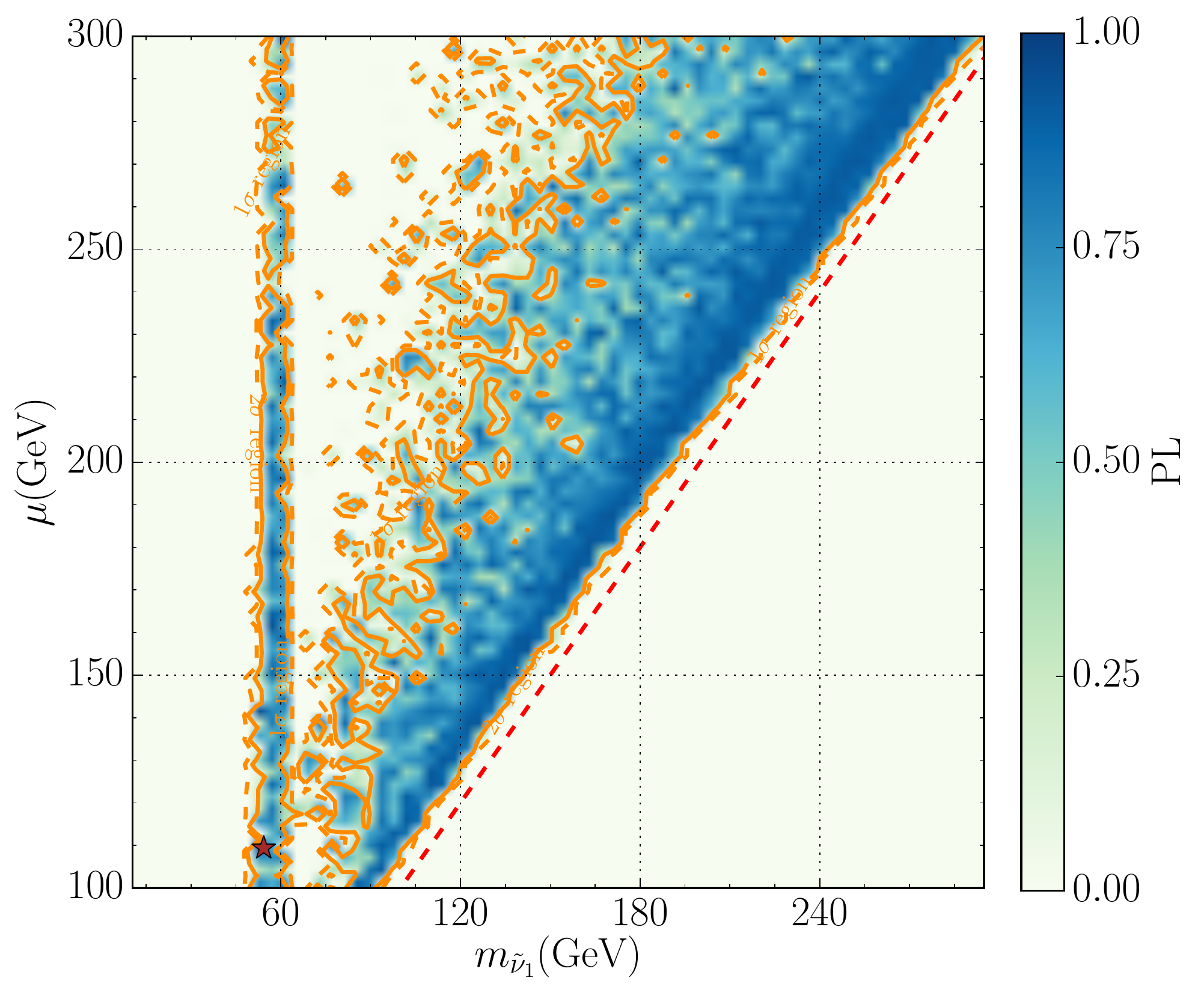}
		}
	\caption{Distinction of the 2D PLs on $\kappa-\lambda$ plane (left panels), $\tan \beta-\lambda$ plane (middle panels) and $\mu-m_{\tilde{\nu}_1}$ plane (right panels), which are
          caused by different settings of {\it nlive} for the scan in Section III.A. The top panels correspond to the setting {\it nlive} = 2000, while the bottom panels are the combined result from a scan with {\it nlive}=6000 and five independent scans with {\it nlive}=2000 for each scan.
          \label{fig25}}
	\end{figure*}

In order to determine which marginal PDF reflects the right preference of the underlying physics under current experimental data, we rewrite Eq.(\ref{stat:Bayes}) in Section III as follows
\begin{eqnarray}
P(\Theta|D) \equiv Z(\Theta)/Z \times \pi(\Theta),  \label{Alternative}
\end{eqnarray}
where $\pi(\Theta)$ is the prior PDF of the parameter $\Theta$ if one assumes that all input parameters are independent in their prior PDFs, $Z(\Theta)$ is the Bayesian evidence for a fixed $\Theta$ and obtained by integrating over the rest parameter space, and by contrast $Z$ is the evidence integrated over the whole parameter space. This formula provides an alternative way to calculate the marginal PDF $P(\Theta|D)$ by following procedures
\begin{itemize}
\item First calculate $Z(\Theta)$ for a series of $\Theta_i$;
\item Find the largest value of $Z(\Theta) \pi(\Theta)$ among the $\Theta_i$;
\item Normalize all $Z(\Theta_i) \pi(\Theta_i)$ to this largest value, and plot the normalized $Z(\Theta) \pi(\Theta)$ as a function of $\Theta$;
\item As a byproduct, one may also get the normalized $Z(\Theta)$ distribution in a similar way.
\end{itemize}
Since the evidence $Z(\Theta)$ can be calculated rather precisely and meanwhile its uncertainty can be estimated, the result of this method can serve as a criteria to judge the quality of the $P(\Theta|D)$ obtained in other ways. Moreover, by comparing the shape of the $P(\Theta|D)$ obtained directly by the code MultiNest with that of the $Z(\Theta)$ distribution, one can judge which factor plays the dominant role in deciding the $P(\Theta|D)$, the prior PDF of $\Theta$, $Z(\Theta)$, or the both.

In FIG.\ref{fig19}, we show the normalized distribution of $Z(A_\kappa)$ (left panel) and $Z(\mu)$ (right panel). These evidences are obtained by the scan in Eq.(\ref{scan-ranges1}) of Appendix A with $A_\kappa$ and $\mu$ fixed respectively. They are calculated by the code MultiNest with a relatively large {\it nlive}, {\it nlive} = 6000, and consequently the uncertainty of the evidences is below $0.1\%$. Comparing the distributions in FIG.\ref{fig19} with corresponding results in FIG.\ref{fig17} and FIG.\ref{fig18}, one can conclude that the flat prior PDF predicts rather precisely the distributions of $A_\kappa$ and $\mu$, even though they are moderately unstable with the increase of {\it nlive},  while the log prior PDF overemphasizes low $|A_\kappa|$ region so that the two distributions are affected greatly by the prior PDF.

\subsection{Differences induced by the prior PDFs \\ in Type-I extended NMSSM}

As a supplement to the results in the text, we present in this subsection the 1D marginal PDFs and PLs of $A_\kappa$ and $\mu$ obtained from the scan in Eq.(\ref{scan-ranges}) of Section III with the log prior PDF for the input parameters and the total likelihood function in Eq.(\ref{Likelihood}). In order to show the effect of the setting {\it nlive}, we carry out two independent scans with {\it nlive} =2000 and 6000 respectively. We also compare the results with those obtained in a similar way but by the flat prior PDF. The results are shown in FIG.\ref{fig20} for $A_\kappa$ and FIG.\ref{fig21} for $\mu$. From these figures, one can learn that

\begin{itemize}
\item For the flat prior PDF, the marginal posterior PDF of $A_\kappa$ extends over a broad range without any strong preference of certain regions, and its relative size in low $|A_\kappa|$ region gradually increases with the increase of {\it nlive}.  The peak of the $\mu$ posterior distribution locates around $280 {\rm GeV}$, and the location moves slowly towards to a lower value of $\mu$ with the increase of {\it nlive}.
\item The marginal posterior PDFs of $A_\kappa$ and $\mu$ from the log prior PDF are rather stable with the increase of the setting {\it nlive}, and they differ greatly from corresponding prediction of the flat prior PDF: the $A_\kappa$ distribution is sharply peaked around $A_\kappa = 0 $ region, and the $\mu$ distribution is maximized at about $160 {\rm GeV}$. As we pointed out before, these features are induced mainly by the log prior distributions of $A_\kappa$ and $\mu$.
\end{itemize}

We also compare the difference of the 2D marginal posterior PDFs on $\mu-m_{\tilde{\nu}_1}$ plane, which is the main conclusion of this work, induced by the two prior PDFs with the setting {\it nlive} = 6000. The results are presented in FIG.\ref{fig22}. From this figure, one can learn that the DM prefers to co-annihilate with the Higgsinos to get its measured relic density regardless the choice of the prior PDFs, and the difference mainly comes from wether a relatively small $\mu$ is favored or not. The fundamental reason of the difference, as we discussed above, is that the log prior PDF overemphasizes low $|A_\kappa|$ region. One can also infer that the log prior PDF is more likely to predict a light $A_1$ than the flat PDF so that there are more samples which allow the channel $\tilde{\nu}_1 \tilde{\nu}_1 \to A_1 A_1$ to happen in DM annihilation.

\section{Other related issues}

\subsection{Subtleness about the marginal PDF of $\mu$}

From the formula of marginal PDF in Eq.(\ref{stat:Bayes}) of Section III and also from the discussion in Appendix A, one can learn that the posterior PDF
depends not only on the parameter space considered in the scan, but also on the choice of $A_\lambda$. Considering that the posterior PDF of $\mu$ is relatively
sensitive to these settings among the input parameters of the theory, we further study its features in the following by changing the boundary of the space and the value of $A_\lambda$.

We first repeat the scan in Eq.(\ref{scan-ranges1}) of Appendix A with $\mathcal{L} = \mathcal{L}_{Higgs} $ and $A_\lambda = 2 {\rm TeV}$, but this time we restrict $A_\kappa$ within a narrow range
$ -200 {\rm GeV} \leq A_\kappa \leq 200 {\rm GeV}$. This case is more likely to predict a light $A_1$, and thus enhance the probability of occurrence for the annihilation $\tilde{\nu}_1 \tilde{\nu}_1 \to A_1 A_1$, which is quite similar to the prediction of the log prior PDF (see the right panel of FIG.\ref{fig22}). As a result, one may expect that a light $\mu$ is preferred. This is verified by the left panel of FIG.\ref{fig23}, where we show the marginal PDF and PL of $\mu$ for this case.

Next we reset $A_\lambda$ to be $10 {\rm TeV}$ and perform the same scan as that in Eq.(\ref{scan-ranges1}) of Appendix A with $\mathcal{L} = \mathcal{L}_{Higgs} $. We find the Bayesian evidence for the new $A_\lambda$ is reduced by a factor of $e^{1.37}$ in comparison with that for $A_\lambda = 2 {\rm TeV}$ case. This reflects the fact that $A_\lambda$ can change the distribution of the likelihood function $\cal{L}$ over the parameter space, especially the theory becomes more tuned as one increases $A_\lambda$ to coincide with the Higgs data. Correspondingly, the posterior distribution of $\mu$ may be altered significantly. The marginal PDF of $\mu$ obtained by this new $A_\lambda$ is presented in the right panel of FIG.\ref{fig23}, where one learns that moderately low values of $\mu$ are also favored.

Finally we investigate how large $\mu$ is preferred to remain consistent with all the experimental data in $A_\lambda = 2 {\rm TeV}$ case. For this end, we extend the upper bound of $\mu$ in Eq.({\ref{scan-ranges}) of Section III to $1 {\rm TeV}$, and then perform same scan as that in Section III.A. In FIG.\ref{fig24}, we show the 1D marginal PDF of $\mu$ (left panel) and 2D marginal PDF on $\mu-\tilde{\nu}_1$ plane (right panel). The left panel indicates that the marginal PDF of $\mu$ reaches its peak around $\mu \simeq 300 {\rm GeV}$, and then falls rapidly with the increase of $\mu$. Two reasons may account for this behavior. One is that although the Higgs data have weak constraints on the parameter $\mu$, the electroweak symmetry breaking disfavors a very large $\mu$, which was discussed in Appendix A. In fact, we once plotted again the right panel of FIG.\ref{fig19} by allowing  $100 {\rm GeV} \leq \mu \leq 1000 {\rm GeV}$ and found that $Z(\mu)$ decreases monotonously when $\mu$ exceeds about $350 {\rm GeV}$. The other reason is that one of the most important roles of Higgsinos in the theory is to coannihilate with sneutrino DM to get its right relic density. Since we have restricted the upper bound of the DM mass around $300 {\rm GeV}$ in the scan, there are no strong motivation for the theory to prefer a large $\mu$. This fact is illustrated in the right panel of FIG.\ref{fig24}.

\subsection{Impact of different {\it nlive}s on 2D PLs}

As a supplementary material, we show the impact of different settings of {\it nlive} on 2D PLs in this section. The phenomenon caused by the setting
and its basic reasons have been analysed in Section III.B.  In FIG.\ref{fig25}, we just show the PLs on $\kappa-\lambda$ plane (left panels),
$\tan \beta-\lambda$  plane (middle panels) and $\mu-m_{\tilde{\nu}_1}$ plane (right panels) in a comparative way, where the bottom panels are obtained with a much larger {\it nlive} than the top panels.

\end{document}